\documentclass{ws-procs975x65}
\usepackage{epsfig}
\usepackage{psfrag}
\def\be{\begin{equation}}
\def\ee{\end{equation}}
\def\bea{\begin{eqnarray}}
\def\eea{\end{eqnarray}}

\def\eps{\varepsilon}
\def\ep{\varepsilon^\prime}
\def\r#1{(\ref{#1})}
\def\nn{\nonumber\\}
\def\up{\uparrow}

\def\cal{\mathcal}
\def\half{$\frac{1}{2}$\ }

\def\bt{\tilde{\beta}}
\def\ts{\tilde{S}}
\newcommand{\ba}{BaCu$_2$Si$_2$O$_7$}
\renewcommand{\theequation}{\thesection.\arabic{equation}}

\begin{document}

\title{Applications of Massive Integrable Quantum Field\\[2mm] Theories to
  Problems in Condensed Matter Physics}
\author{Fabian H.L. Essler}
\address{The Rudolf Peierls Centre for Theoretical Physics,\\
University of Oxford, 1 Keble Road,\\ 
Oxford OX1 3NP, United Kingdom}
\author{Robert M. Konik}
\address{Department of Physics, Brookhaven National Laboratory,\\
 Upton, NY 11973-5000, USA}

\maketitle

\abstracts{
We review applications of the sine-Gordon model, the O(3) non-linear sigma
model, the U(1) Thirring model, and the O(N) Gross--Neveu model to quasi
one-dimensional quantum magnets, Mott insulators, and carbon nanotubes. 
We focus upon the determination of dynamical response functions for
these problems.  These quantities are computed by means of form factor expansions of
quantum correlation functions in integrable quantum field theories.  This approach is
reviewed here in some detail.
}

\newpage
\tableofcontents
\newpage
\noindent
{\sl In memory of Ian Kogan.}

\section{Introduction}
\setcounter{equation}{0}

The study of strongly correlated electrons in
low dimensional systems lies at the heart of much of modern
condensed matter physics.  Interest in these systems arises as
the behavior of strongly correlated 
systems is not in general adequately captured
in approximations based upon `free-particle' non-interacting
models.  In the presence of generically strong interactions, the
physics is typically much richer, exhibiting qualitatively new features.

Accessing this physics presents an imposing challenge.
In one route to comprehension, quantum field
theories can be employed.
Field theories are typically able to describe the low energy behavior of strongly
correlated systems and so are able to extract universal characteristics.  
It is these characteristics that are of greatest interest precisely because
they provide the most robust experimental signatures and 
do not, in general, depend upon particular experimental details.

In low dimensions, the study of numerous quantum field theories is aided by their integrability.
Integrable theories are characterized by an infinite number of non-trivial
conserved charges.
The existence of these charges allows for an exact characterization
of many features of these models.  Both the spectrum and the scattering matrices
of an integrable quantum field theory can be explicitly written down.
On the other hand, correlation functions in an integrable quantum field theory
cannot, in general, be exactly computed.  However there exist approaches,
based on exploiting the integrability of the model, that can be used to obtain
some information on correlation functions.  And while incomplete, this information
does reflect the non-perturbative structure of the theory.

Of these approaches, we describe in this review one based upon form-factors.  Form factors
are matrix elements of quantum fields with exact eigenstates.  Under a spectral
decomposition, all correlation functions can be written in terms of form factors.
While this decomposition is exact, its practical manipulation requires truncation
of the spectral sum.
In this review we will explore how this truncation may be done and under what
conditions exact information on the theory remains available.
As the reader will see, the behavior of correlation functions at low energy scales
fortuitously remains exactly computable.

The review is organized as follows.  In the first substantive section, Section 2, we give
in detail the form factor programme to computing correlation functions in integrable quantum 
field theories.  Our purpose here is to be pedagogical.  We begin by giving an overview of
how form factors are used to calculate zero temperature correlators.  
In Section 2.2 we then turn to describing the 
key features of an integrable model for the programme, the spectrum and its associated scattering.
Form factors are computable in an integrable model because they must satisfy a series of constraints
arising from both consistency with the underlying scattering and various assumed analyticities
in the energy-momentum.  We list these constraints in Section 2.3.  In Section 2.4
we give examples of the application of these constraints to an archetypal integrable theory,
the sine-Gordon model.  While form factors do not generically give the
behavior of correlation functions 
at all energy scales, they do provide {\it exact} information at low energies.  And at 
higher energy scales, corrective terms, as a rule, are extremely small.  We illustrate this
principle in Section 2.6 with the specific example of the Ising model.  In the final
two subsections we turn to form-factors in more involved settings: form-factors in models
with bound states and the use of form-factors to compute correlation functions at finite
temperature.

Having given an overview of the form-factor programme, we turn to specific applications.
As our first example, we consider applications of the sine-Gordon model to half-integer
spin chains.  An anisotropic Heisenberg spin-1/2 chain (in a magnetic
field) has gapless excitations. In this section we first detail the
bosonization of the spin chain and so exhibit the chain's Luttinger
liquid phase. We then consider a number of physical perturbations
under which the chain becomes gapped and is described by the massive
sine-Gordon model. These perturbations include a staggered
magnetization, a transverse magnetic field and dimerization. The
dynamical structure factor, which is measured in inelastic neutron scattering
experiments, can be determined using sine-Gordon form factors.

In Section 4, we study integer spin chains using the same techniques.  Unlike their
half-integer counter parts, an integer spin Heisenberg chain is completely gapped
and its low energy behavior is believed to be described by the O(3) non-linear sigma model without topological
term.
For pedagogical reasons we begin the section by giving the map between the two, pointing
out what approximations are used.  We then turn to a basic description of
the O(3) non-linear sigma model including its spectrum, scattering matrices, and form
factors of various physical fields.  With these form factors in hand, we next consider the computation
of various zero temperature spin-spin correlation functions that would
be measured in neutron scattering experiments.  Having considered the
correlation functions at $T=0$, we turn to their finite 
temperature counterparts.  This will be our primary example demonstrating the possibility of
a form-factor computation at $T\neq 0$ and it is done in some detail.  Using these techniques
we will address the issue of whether transport is ballistic or diffusive at finite temperatures
in integer spin chains.

In the next section, Section 5, we turn to applications of the U(1) Thirring model
to quasi-one-dimensional Mott insulating chains. Such insulators are modeled
by various extended Hubbard models which in turn are related to the
U(1) Thirring model. The latter can be bosonized in terms of a free
boson and a sine-Gordon model. Using the form factors of various
operators in the sine-Gordon model we compute both the optical
response as well as the single-particle spectral function of
half-filled and quarter filled Mott insulating chains.  While this
treatment is appropriate to one-dimensional materials, we also show how
the case of weakly coupled chains can be treated.

In the final section, Section 6, we consider our last application of form factors, that to Hubbard
ladders and armchair carbon nanotubes.
In the weak interaction limit, both materials can be described in an
RG sense by the SO(8) Gross--Neveu model, an integrable theory of four
interacting Dirac fermions.
We outline this mapping in some detail and point
out its limitations.
We then use the form factors of SO(8) Gross--Neveu to compute
a number of physical quantities including the optical response and
single particle spectral function.  We end by pointing out how 
the results are changed if small, integrable-breaking perturbations
are introduced.

\section{Correlation Functions in Integrable, Massive Quantum Field Theories}
\setcounter{equation}{0}

\def\del{\partial}
\def\rtc{\tilde{\rho}_c}
\def\tth{\tilde{\theta}}
\def\gcc{\Gamma_{cc}}
\def\la{\lambda}
\def\CO{{\cal O}}
\def\om{\omega}
\def\th{\theta}
\def\ut{{\th_{32}}}
\def\vt{{\th_{31}}}
\def\wt{{\th_{43}}}
\def\lb{{\langle}}
\def\rb{{\rangle}}
\def\dt{\frac{d\th}{2\pi}}
\def\dto{\frac{d\th_1}{2\pi}}
\def\dtt{\frac{d\th_2}{2\pi}}
\def\dttr{\frac{d\th_3}{2\pi}}
\def\dtf{\frac{d\th_4}{2\pi}}
\def\bd{{\beta\Delta}}
\def\tn{{1/T_1}}
\def\ot{{O(3) NLSM }}
\def\ots{{O(3) NLSM}}
\def\nb{{{\bf n}}}

In this section we outline the form-factor programme by which correlation functions in massive
integrable field theories can be computed.

\subsection{Computing Correlation Functions with Form Factors: Exact Results at Zero Temperature}

In general computing correlation functions in integrable field
theories is an open problem. There exists no general technique that is
able to access a generic correlation function at all energy scales.
Now it is certainly true that progress can be made in specific instances. For some integrable
models, such as Ising model variants, the sine-Gordon model at is free
fermion point, and the Bose gas with delta-function
interactions, some correlations can be determined with the aid of
Fredholm determinants, see e.g.
\cite{Jimbo80,vladb,korepin87,its90,efik95a,efik95b,efik96,ising,leclair,
goehmann98a,goehmann98b,goehmann99,kit1,kit2,kit3,cheianov04a,cheianov04b}
and references therein.
This method, however, is technically involved, and at present only
works in a handful of cases. More promising is the form-factor programme 
for the calculation of correlation functions \cite{smirnov,karowski78}.
Form factors do not allow in practice the complete determination of a
correlation function. However in a gapped (massive) field theory, they
do permit the exact determination of the low energy properties of the
corresponding spectral function.

This arises as the form-factor representation of any zero temperature correlation function
is obtained by inserting a resolution of the identity
corresponding to the basis of eigenstates.  (We will consider finite temperature correlation functions
later in this section.)
Thus, for an operator,
${\cal O}(x, \tau)$, we write the spectral decomposition schematically
($\tau$ denotes imaginary time, and $T$ time ordering)\,\footnote{\,The 
sum over $s_n$ is meant to include integrals
over the momenta of all particles and sums over particle/excitation types.}
\vbox{
\begin{eqnarray}\label{eIIi}
G^{\cal O}_T(x,\tau) &=& -\langle 0|T\bigl (
{\cal O}(x,\tau){\cal O}^{\dagger}(0,0)
\bigr)|0\rangle \nonumber \\[-3mm]
&&\\[-3mm]
&=& -\sum_{n=0}^{\infty}
\sum_{s_n}e^{- \tau E_{s_n}}
\langle 0|{\cal O}(x,0)|n;s_n\rangle 
 \langle n;s_n|{\cal O}^{\dagger}(0,0)|0\rangle,
\quad(\tau>0),~~\nonumber
\end{eqnarray}}
where $E_{s_n}$ is the energy
of an eigenstate, $|n;s_{n}\rangle$, with $n$ particles described by
quantum numbers, $\{s_n\}$.
By inserting a resolution of the identity, we have reduced the problem to one
of computing individual matrix elements.
In an integrable model the matrix elements of a physical operator between
the vacuum and the exact eigenstates can in principle be computed
exactly from the two-body S-matrix.
However the calculation of these matrix elements,
as well as the evaluation
of the sums/integrals,
$\sum_{s_n}$, becomes increasingly cumbersome as the particle
number $n$ becomes large,
so that the full expression for the correlation function
cannot be evaluated in closed form.
Often, however, a truncation of the sum at the level of
two or three particle states already provides a good approximation
to the full correlation
function \cite{les,lec,delone,deltwo,oldff1}. This may be understood
in terms of phase space arguments \cite{oldff1,mussardo.school}.  
On the other hand, this truncation is no longer necessary in a massive theory, 
if one considers the corresponding spectral function.
Only eigenstates with a fixed
energy, $\omega$, contribute to the spectral function:
\begin{eqnarray}\label{eIIii}
&&-\frac{1}{\pi}\, {\rm Im} G^{\cal O}_{T}(x, -i\omega + \delta )  \nonumber\\
&& \hskip .5in =\sum_{n=0}^{\infty}\sum_{s_n} \bigg\{ 
\langle 0|{\cal O}(x,0)|n;s_n\rangle
\langle n;s_n|{\cal O}^{\dagger}(0,0)|0\rangle  \delta(\omega- E_{s_n})
\nonumber\\
&& \hskip .5in - \epsilon \langle 0|{\cal O}^\dagger(0,0)|n;s_n\rangle
\langle n;s_n|{\cal O}(x,0)|0\rangle  \delta(\omega + E_{s_n})\bigg\}\,,
\end{eqnarray}
\noindent where $\epsilon = \pm$ 
for fields, $\cal{O}$, that are bosonic/fermionic.

In a massive theory the creation of an extra particle
in the intermediate exact eigenstate costs a finite
amount of energy, and so the sum in Eq.\,(\ref{eIIii}) is finite.
For example, when $\omega$ is smaller than the energy
of all three-particle states (i.e. when $\omega$ is below
the three-particle threshold), then only the form
factors with one and two particles $(n=1,2)$ have to be determined in order to
obtain an exact result.
All of the spectral functions we have computed in this review
are exact at sufficiently low energies for this reason.

As knowledge of the exact eigenfunctions in an integrable model is central to
computing correlations functions, we consider this in more detail in the next section.

\subsection{Spectrum and Scattering in an Integrable Model}

The key feature of an integrable system is the exact knowledge of a basis
of eigenstates of the fully interacting Hamiltonian.
At the root of integrability is
a well defined notion of ``particles'', or ``elementary
excitations'' in the fully interacting system.  These particles
scatter off each other according to two-body $S$-matrices,
that is, all particle production processes are absent
and particle number is conserved.  This is due to special
conservation laws which exist in an integrable model, preventing
the decay of these particles.  In this sense, an integrable system
is similar to a Fermi liquid.  
An additional feature is that new particles
can arise as bound states of already existing ones.
However the total number of different types of particles is
finite which makes the system analytically tractable.

Formally the elementary excitations are created and destroyed through
the Faddeev--Zamolodchikov operators, denoted by $A_{a} (\th )$.
$\th$ is termed the rapidity and it encodes the energy-momentum
carried by the excitation via 
\begin{equation}\label{eIIiii}
P = \Delta \sinh (\th ); ~~~~ E = \Delta \cosh (\th ).
\end{equation}
Two excitations will in general scatter according to a non-trivial two-body S-matrix, $S$.  
$S$ gives the amplitude of the process by which two particles
$\{a, b\}$ scatter into two potentially different particles, $\{a', b'\}$.  In terms of 
the Faddeev--Zamolodchikov operators, this scattering determines the
commutation relationship between operators
\begin{eqnarray}\label{eIIiv}
A_{a} (\th_1 ) A_{b} (\th_2 ) &=& 
S^{a'b'}_{ab} (\th_1-\th_2) A_{b'} (\th_2) A_{a'} (\th_1);\cr\cr
A^\dagger_{a} (\th_1 ) A^\dagger_{b} (\th_2 ) &=& 
S^{a'b'}_{ab} (\th_1-\th_2) 
A^\dagger_{b'} (\th_2) A^\dagger_{a'} (\th_1);\cr\cr
A_{a} (\th_1 ) A^\dagger_{b} (\th_2 ) &=& 
2\pi\delta_{ab}\delta(\th_1-\th_2)+
S^{b'a}_{ba'} (\th_1-\th_2) A^\dagger_{b'} (\th_1) A_{a'} (\th_1).
\end{eqnarray}
It is solely a function of $\th_1-\th_2 \equiv \th_{12}$ by Lorentz
invariance. 
\begin{figure}[ht]
\begin{center}
\epsfxsize=0.8\textwidth
\epsfbox{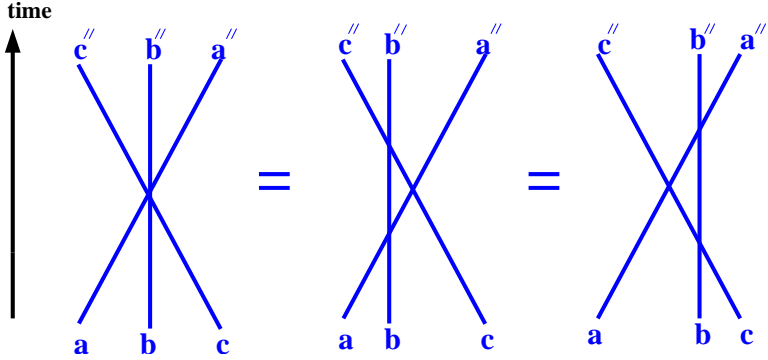}\quad
\end{center}
\caption{A graphical representation of the Yang--Baxter equation.  The left-hand figure
represents a three body process by which three particles $a,b$,
and $c$ scatter into $a'', b'',$ and $c''$.  Via integrability, this three-body process
can be factorized into two different sets of three two-body processes pictured in the central and leftmost figures.
Under the Yang-Baxter relation, these different sets of two-body scattering processes are equivalent.}
\label{figIIi}
\end{figure}

These two-body S-matrices encode all scattering information 
in the theory.  Scattering processes involving higher number of particles can
always be expressed in terms of two-body scattering matrices.  Suppose we consider
a three particle scattering process, scattering particles $\{a,b,c\}$ into
$\{a',b',c'\}$ given by the S-matrix $S^{a'b'c'}_{abc}$.  This three-body S-matrix
factorizes into a set of two-body S-matrices 
\begin{eqnarray}\label{eIIv}
S^{a'b'c'}_{abc}(\th_1,\th_2,\th_3) 
&=& S^{a'b'}_{ab}(\th_1,\th_2)S^{a''c'}_{a'c}(\th_1,\th_3)S^{b''c''}_{b'c'}(\th_2,\th_3) \cr\cr
&=& S^{b'c'}_{bc}(\th_2,\th_3)S^{a'c''}_{ac'}(\th_1,\th_3)S^{a''b''}_{a'b'}(\th_1,\th_2).
\end{eqnarray}
From the above equation we see that we can factorize the three-body S-matrix in
two different ways.  This is illustrated graphically in Fig.\,\ref{figIIi}.  
Because the theory is integrable, the different ways of
factorizing are equivalent.  This equivalence is known as the Yang-Baxter equation.
The ability to factorize the higher-body S-matrices results from the existence in the
integrable theory of conserved charges with are non-trivial powers of energy and 
momentum\cite{kulish,zam77a,Zam77,zam78,Thun77}.

As two-body scattering provides complete information on an integrable
theory, the form factors (even those involving large number of
particles) are determined by this S-matrix alone.  We will see this in
operation in the next section.

Using the Faddeev-Zamolodchikov operators, a Fock space of states can
be constructed as follows. The vacuum is defined by
\begin{equation}
A_{a}(\theta) |0\rangle=0 \ .
\end{equation}
Multiparticle states are then obtained by acting with strings of
creation operators $A_b^\dagger(\theta)$ on the vacuum
\begin{equation}
|\theta_n\ldots\theta_1\rangle_{a_n\ldots a_1} = 
A^\dagger_{a_n}(\theta_n)\ldots
A^\dagger_{a_1}(\theta_1)|0\rangle\ .
\end{equation} 
In terms of this basis the resolution of the identity is given by
\be
1\!\! 1 = |0\rangle\langle 0| \label{identity}
+ \sum_{n=1}^\infty\sum_{\{a_i\}}\int_{-\infty}^{\infty}
\frac{d\theta_1\ldots d\theta_n}{(2\pi)^nn!}
|\theta_n\ldots\theta_1\rangle_{a_n\ldots a_1}
{}^{a_1\ldots a_n}\langle\theta_1\ldots\theta_n|\ .
\ee

\subsection{Computation of Form Factors: Form Factor Axioms}

The form factors of a field $\CO$ are defined as the matrix elements
of the field with some number of particles, $A_{a} (\th )$:
\begin{equation}\label{eIIvi}
f^{\CO}_{a_1\cdots a_n} (\th_1,\cdots ,\th_n ) = 
\langle \CO (0,0) A_{a_n} (\th_n ) \cdots A_{a_1} (\th_1)\rangle.
\end{equation}
These matrix elements are constrained by a variety of requirements arising from the scattering
relations just discussed, Lorentz invariance, hermiticity, analyticity, and the locality
of the fields.  We consider each in turn.

\subsubsection{Scattering Axiom}

For the form factor 
to be consistent with two body scattering we must have
\begin{eqnarray}\label{eIIvii}
f^{\CO}_{a_1,\cdots ,a_{i+1},a_i,\cdots , a_n} 
(\th_1,\cdots,\th_{i+1},\th_i,\cdots,\th_n) &=& \cr\cr
&& \hskip -2.9in S^{{a'}_i,{a'}_{i+1}}_{a_ia_{i+1}} (\th_i -\th_{i+1})
f^{\CO}_{a_1,\cdots ,{a'}_i,{a'}_{i+1},\cdots , a_n} 
(\th_1,\cdots,\th_i,\th_{i+1},\cdots,\th_n).
\end{eqnarray}
This relation is arrived 
at by commuting the $i$-th and $i+1$-th particle and using the Faddeev--Zamolodchikov algebra
in Eq.\,(\ref{eIIiv}).

\subsubsection{Periodicity Axiom}

A second constraint 
upon the form factor can be thought of as a periodicity
axiom.  In continuing 
the rapidity, $\th$, of a particle to $\th - 2\pi i$,
the particle's energy-momentum is unchanged.  However the form-factor
is not so invariant.  We instead have 
\begin{eqnarray}\label{eIIviii}
f^{\CO}_{a_1,\cdots , a_n} 
(\th_1,\cdots ,\th_n) &=& 
f^{\CO}_{a_n,a_1,\cdots ,a_{n-1}} 
(\th_n-2\pi i,\th_1,\cdots,\th_{n-1}).
\end{eqnarray}
This constraint is derived from crossing symmetry \cite{leclair}.
It implicitly assumes that 
the field $\CO$ is local: if $\CO$ is non-local
additional braiding 
phases appear in the above relation \cite{double,smir,so8}.

These braiding phases arise when two fields are interchanged:
\begin{equation}\label{eIIix}
\psi (x,t)\CO (y,t) = R_{\psi\CO}\CO (y,t)\psi (x,t); 
~~~~~~~x < y .
\end{equation}
(On occasion, we must deal with braiding matrices, not merely phases.  But
we will not consider such a situation in this work.)
If $R_{\psi\CO}$ is non-trivial
we must alter the above periodicity axiom to read
\begin{eqnarray}\label{eIIx}
f^{\CO}_{a_1,\cdots , a_n} 
(\th_1,\cdots ,\th_n) &=& 
R_{\psi_n\CO} f^{\CO}_{a_n,a_1,\cdots ,a_{n-1}} 
(\th_n-2\pi i,\th_1,\cdots,\th_{n-1}).~~~
\end{eqnarray}
Here $\psi_n$ can be thought of as the field which creates the excitation $A_n$.

In order to employ the periodicity axiom (Eq.\,(\ref{eIIx})), 
we then need to know how to specify
the braiding of the fields.  In one approach, 
we identify both fields with their corresponding
excitations, $A_{\CO}$ and $A_n$.  
If both fields are right-moving, 
the braiding of the fields is then
encoded in the asymptotic limits of the corresponding S-matrix
(see \cite{smir}):
\begin{equation}\label{eIIxi}
R_{\CO n} = (S^{\CO n}_{\CO n} (+\infty )).
\end{equation}
If on the other hand the fields are left-moving, we find instead
$R_{\CO n} = (S^{\CO n}_{\CO n} (-\infty ))$.

\subsubsection{Annihilation Pole Axiom}

Another condition related to analyticity that a form factor must
satisfy is the annihilation pole axiom.  This condition arises in 
form factors involving a particle and its anti-particle.
Under the appropriate analytical continuation, such a combination
of particles are able to annihilate one another.  As such this
condition relates form factors with $n$ particles to those
with $n-2$ particles,  
\begin{eqnarray}\label{eIIxii}
i ~{\rm res}_{\,\th_n = \th_{n-1} +\pi i}
f(\th_1, \cdots, \th_n)_{a_1,\cdots ,a_n} 
&=& f(\th_1, \cdots, \th_{n-2})_{a'_1,\cdots ,a'_{n-2}} 
C_{a_na'_{n-1}}\nonumber\\[1.5mm]
&& \hskip -2.5in \times \bigg( \delta^{a'_1}_{a_1}
\delta^{a'_2}_{a_2}\cdots \delta^{a'_{n-2}}_{a_{n-2}}\delta^{a'_{n-1}}_{a_{n-1}}
- R_{\CO n}S^{a'_{n-1}a'_1}_{\tau_1 a_1}(\th_{n-11})
S^{\tau_1a'_2}_{\tau_2 a_2}(\th_{n-12})
\cdots \nonumber\\[1mm]
&& \hskip -1in \times S^{\tau_{n-4}a'_{n-3}}_{\tau_{n-3} a_{n-3}}(\th_{n-1n-3})
S^{\tau_{n-3}a'_{n-2}}_{a_{n-1} a_{n-2}}(\th_{n-1n-2})\bigg),
\end{eqnarray}
where $C$ is the charge conjugation matrix.
This relation as written assumes that we are normalizing our
particle states as $\lb \th | \th' \rb = 2\pi \delta(\th-\th ')$.
Note that the braiding phase plays a role in this axiom (see r.h.s. of the above equation).

\subsubsection{Lorentz Invariance}

The form factor must also 
satisfy constraints coming from Lorentz covariance.
In general, the form factor 
of a field, $\CO$, carrying Lorentz spin, s, must 
transform under a Lorentz boost, $\th_i \rightarrow \th_i + \alpha$, via
\begin{eqnarray}\label{eIIxiii}
f^{\CO}_{a_1\cdots a_n} (\th_1+\alpha,\cdots ,\th_n +\alpha) = 
e^{s\alpha} f^{\CO}_{a_1\cdots a_n} (\th_1,\cdots ,\th_n ).
\end{eqnarray}
Often in this work we will consider the correlation functions of (topological) current operators.
Let us thus consider $j_0(x,t)$, a charge density, and $j_1(x,t)$, its corresponding
conserved current.
Together
they form a Lorentz two-current.  (Here $0,1$ are
Lorentz indices.)
The form factors for $j_0$ and $j_1$ appear as
\begin{equation}\label{eIIxiv}
f^{j_\mu}_{a_1\cdots a_n} (\th_1,\cdots ,\th_n) =
\epsilon_{\mu\nu} P^\nu (\th_i ) 
f_{a_1\cdots a_n} (\th_1,\cdots ,\th_n),
\end{equation}
where
$P^0 = \sum_i \Delta \cosh (\th_i )$ and 
$P^1 = \sum_i \Delta \sinh (\th_i )$.
The function, $f_{a_1\cdots a_n}(\th_1,\cdots ,\th_n)$, on the r.h.s. of Eq.\,(\ref{eIIxiv}) 
is solely a function of $\th_i-\th_j$.

\subsubsection{Form Factor Normalization}

It is sometimes possible to determine the absolute normalization
of a set of form-factors.  (Of course, normalization of form-factors with different
particle numbers but of the same field can be fixed, among other ways, by the use
of the annihilation pole axiom.)
In the case of current operator we can rely
upon
the action of the conserved charge
$$
Q = \int dx j_0 (x,0),
$$
upon the single particle states with charge q.  We expect (with the normalization of Section 2.3.3)
\begin{equation}\label{eIIxv}
\lb \th, q | Q | \th', q \rb = 2\pi q \delta (\th -\th').
\end{equation}
Using crossing, we can relate this matrix element to a two-particle form factor:
\begin{equation}\label{eIIxvi}
\lb \th, q | Q | \th', q \rb = \lb Q |\th',q; \th-i\pi,\bar{q}\rb ,
\end{equation}
where $\bar{q}$ denotes the charge conjugate of $q$.
Thus we are able to 
fix the normalization of the two particle form factor in a natural fashion
(and so all other higher particle form factors through the annihilation pole
axiom).

With operators other than currents, we can still fix
the phase of the normalization using hermiticity.
For this purpose it is sufficient to consider 2-particle form factors.
Hermiticity then gives us
\begin{eqnarray}\label{eIIxvii}
\langle \CO (0,0) A_{a_2} (\th_2) A_{a_1} (\th_1)\rangle^*
&=& \langle A^\dagger_{a_1} (\th_1) A^\dagger_{a_2} (\th_2)
\CO^\dagger (0,0)\rangle\nonumber\\[1mm]
&=& \lb \CO^\dagger (0,0) A_{\bar{a_1}} (\th_1-i\pi) A_{\bar{a_2}} 
(\th_2-i\pi )\rangle\,,~~~~~
\end{eqnarray}
where the last line follows from crossing and so
\begin{equation}\label{eIIxviii}
f^\CO _{a_1a_2} (\th_1,\th_2 )^* = f^{\CO^\dagger}_{\bar{a_2}\bar{a_1}}
(\th_2-i\pi ,\th_1-i\pi ).
\end{equation}

\subsubsection{Minimality Principle}

These conditions do not uniquely specify the form factors.
It is easily seen that if $f(\th_1, \cdots, \th_n)_{a_1,\cdots ,a_n}$
satisfies these axioms then so does
\begin{equation}\label{eIIxix}
f(\th_1, \cdots, \th_n)_{a_1,\cdots ,a_n}\, 
\frac{P_n(\cosh(\th_{ij}))}{Q_n(\cosh(\th_{ij}))}\ ,
\end{equation}
where $P_n$ and $Q_n$ are symmetric polynomials in 
$\cosh (\th_{ij}), 1\leq i,j \leq n$, and are such that
\begin{equation}\label{eIIxx}
P_n|_{\th_n = \th_{n-1} + \pi i} = P_{n-2}\,;\quad Q_n|_{\th_n = \th_{n-1} + \pi i} = Q_{n-2}\,.
\end{equation}
To deal with this ambiguity, we employ a minimalist axiom.
We choose $P_n$ and $Q_n$ such that $P_n/Q_n$ has the minimal
number of poles and zeros in the physical strip,
$\rm {Re} (\th ) = 0$, $0< {\rm Im} \th < 2\pi$.  Additional
poles are only added in accordance with the theory's bound
state structure.
Using this minimalist
ansatz, one can determine $P_n/Q_n$ up to a constant.

\subsection{Simple Example: Form Factors for the Sine-Gordon Model in
the Repulsive Regime} 

In this section we consider form factors of fields in the sine-Gordon
model in its repulsive regime.  The sine-Gordon model is described by
the following action
\begin{equation}\label{eIIxxi}
S = \frac{1}{16\pi} \int dx d\tau \left[ \partial_\mu\Phi\partial^\mu\Phi
- \mu\cos(\beta\Phi)\right].
\end{equation}
For $\beta > 1/\sqrt{2}$ this theory describes repulsively interacting
solitons alone, i.e. there are no bound states.  
Classically these solitons arise as interpolations of the field
$\Phi(x,t)$ between minima of the cosine potential.  The two solitons are
characterized by a topological $U(1)$ charge, +/-.

As it is integrable, the model is characterized solely by a two particle
S-matrix.  Its nonzero, $U(1)$-conserving elements are given by
\begin{eqnarray}\label{eIIxxii}
S^{++}_{++}(\th ) &=& S^{--}_{--}(\th) \equiv S_0(\th) = 
-\exp\bigg[\int^\infty_0\frac{dx}{x}\,\sinh\Big(\frac{x\th}{i\pi}\Big)
\frac{\sinh((\frac{1}{2}-\frac{\xi}{2})x)}{\cosh(\frac{x}{2})\sinh(\frac{x\xi}{2})}
\bigg];\cr\cr
S^{+-}_{+-}(\th ) &=& S^{-+}_{-+}(\th) = -\frac{\sinh(\frac{\th}{\xi})}{\sinh(\frac{\th-\pi i}{\xi})}\,S_0(\th);\nonumber\\[1mm]
S^{+-}_{-+}(\th ) &=& S^{-+}_{+-}(\th) = -\frac{i\sin(\frac{\pi}{\xi})}{\sinh(\frac{\th-\pi i}{\xi})}\,S_0(\th),
\end{eqnarray}
where
\be
\xi=\frac{\beta^2}{1-\beta^2}\ .
\ee
With this brief description in hand we go on to compute 
two particle form-factors in this theory.

\subsubsection{Soliton-Antisoliton Form Factor for the Current Operator}

\def\fsg{f^\mu_{\epsilon_1\epsilon_2}(\th_1,\th_2)}
\def\fsgo{f^\mu_{\epsilon_2\epsilon_1}(\th_1,\th_2)}
\def\fsgtw{f^\mu_{\epsilon_2\epsilon_1}(\th_2,\th_1)}
\def\fsgth{f^\mu_{\epsilon_1\epsilon_2}(\th_1-2\pi i,\th_1)}

We will first consider the form factor of the current correlator with a two soliton
state.  The (topological) current operator in sine-Gordon
is given by
\begin{equation}\label{eIIxxiii}
j^{\mu}= \epsilon^{\mu\nu}\partial_{\nu}\Phi\, ,
\end{equation}
(where $\epsilon^{01}=1$).
As the operator itself carries no $U(1)$ charge, it couples to a soliton-anti-soliton
pair.  The matrix element has the general form
\begin{equation}\label{eIIxxiv}
f^\mu_{\epsilon_1\epsilon_2} (\th_1,\th_2)
= \langle 0|j^{\mu}(0,0)|A^\dagger_{\epsilon_2}(\th_2)A^\dagger_{\epsilon_1}(\th_1) \rangle\,.
\end{equation}
To determine $\fsg$, we note that
the current couples anti-symmetrically to the soliton-anti-soliton pair (as is clear if
one examines the limit $\beta^2 = 1/2$ where the model reduces to free massive fermions).
Thus $\fsg$ takes the form,
\begin{equation}\label{eIIxxv}
\fsg = \epsilon_{\epsilon_1\epsilon_2} f^\mu (\th_1,\th_2)\, ,
\end{equation}
where $\epsilon$ is the anti-symmetric tensor.
We can scalarize the above by explicitly exhibiting the piece of the form factor
satisfying Lorentz covariance
\begin{equation}\label{eIIxxvi}
f_\mu (\th_1,\th_2) =
\bigg(
e^{(\th_1+\th_2)/2} - (-1)^\mu e^{-(\th_1+\th_2)/2}\bigg)
f(\th_{12})\, .
\end{equation}
Having so constrained the form of $\fsg$, we now apply the scattering axiom.
Using the anti-symmetry of $\fsg$ in the indices $\epsilon_1,\epsilon_2$, we find
\begin{equation}\label{eIIxxvii}
\fsgtw = S(\th_{12} )\fsgo\,,
\end{equation}
where
\begin{eqnarray}\label{eIIxxviii}
\hspace{-10mm}
S(\th ) &=& 
\frac{\sinh(\frac{\th}{\xi})-i\sin(\frac{\pi}{\xi})}{\sinh(\frac{\th-\pi
i}{\xi})}\,S_0(\th)
=-\exp\bigg(\!\int^\infty_0\!\frac{dx}{x}\sinh\Big(\frac{x\th}{i\pi}\Big)\,G_c(x)\bigg)\,,~~
\end{eqnarray}
\begin{eqnarray}\label{eIIxxix}
&&\hspace{-8mm}G_c(x) \equiv G_{c1}(x) + G_{c2}(x);\cr\cr
&&\hspace{-8mm}G_{c1}(x)\! =\! \frac{\sinh((\frac{\xi}{2}\!-\!1)x)}{\sinh(\frac{x\xi}{2})}
\!-\!\frac{\cosh((\frac{\xi}{2}\!-\!1)x)}{\cosh(\frac{\xi x}{2})}\,;
~~G_{c2}(x)\!=\!
\frac{\sinh((\frac{1}{2}\!-\!\frac{\xi}{2})x)}{\cosh(\frac{x}{2})\sinh(\frac{x\xi}{2})}\,.~~~
\end{eqnarray}
We note that $S(\th)$ tends to $-1$ as $\th$ goes to zero.  This implies the form factor
will vanish in the same low-energy limit.
As the current operator is local and bosonic, the braiding here is trivial.
The periodicity axiom reading,
\begin{equation}\label{eIIxxx}
\fsgtw = \fsgth
\end{equation}
thus reduces to
\begin{equation}\label{eIIxxxi}
f(-\th) = f(\th - 2\pi i).
\end{equation}
A minimal solution satisfying these constraints on the form factor
is
\begin{eqnarray}\label{eIIxxxii}
\fsg &=& iA \epsilon_{\epsilon_1,\epsilon_2}
(e^{(\th_1+\th_2)/2} - (-1)^\mu e^{-(\th_1+\th_2)/2})
s(\th_{12}/2) \cr\cr
&& \hskip .4in \times\exp\bigg[\int^\infty_0 \frac{dx}{x} \frac{G_c(x)}{\sinh (x)}
\sin^2(\frac{x}{2\pi}(i\pi + \th_{12}))\bigg],~~~
\end{eqnarray}
where $i A$ is some normalization with mass dimension
$[m]$.
The phase of $A$ is determined through the hermiticity condition
\begin{equation}\label{eIIxxxiii}
\fsg^* = f^{\mu}_{\epsilon_2\epsilon_1}(\th_2 - i\pi,\th_1-i\pi) .
\end{equation}
This implies that $A$ is real.

We note that 
\begin{equation}\label{eIIxxxiv}
\exp\bigg[\int^\infty_0 \frac{dx}{x} \frac{G_{c1}(x)}{\sinh (x)}
\sin^2(\frac{x}{2\pi}(i\pi + \th_{12}))\bigg] = \frac{1}{\cosh(\frac{\th_{12}+i\pi}{2\xi})}\,.
\end{equation}
The form factor can then be written in the same form appearing in Ref. \cite{smirnov}:
\begin{eqnarray}\label{eIIxxxv}
\fsg &=& iA \epsilon_{\epsilon_1,\epsilon_2}
(e^{(\th_1+\th_2)/2} - (-1)^\mu e^{-(\th_1+\th_2)/2})
\frac{s(\th_{12}/2)}{\cosh(\frac{\th_{12}+i\pi}{2\xi})} \nonumber\\[1mm]
&& \hskip .4in \times\exp\bigg[\int^\infty_0 \frac{dx}{x} \frac{G_{c2}(x)}{\sinh (x)}
\sin^2(\frac{x}{2\pi}(i\pi + \th_{12}))\bigg],~~~
\end{eqnarray}

\subsubsection{Soliton-Antisoliton Form Factor for Non-Local 
Operators}

We now consider the two soliton form factor of the non-local operator,
$e^{\pm i\frac{\beta}{2}\Phi}$.  This operator's non-locality can be seen if
we relate the field $\Phi$ to the current operator 
\begin{eqnarray}\label{eIIxxxvi}
e^{\pm i\frac{\beta}{2}\Phi} = e^{\pm i \frac{\beta}{2} \int^x_{-\infty}dx'j^0(x,t)}.
\end{eqnarray}
This non-locality will lead to a non-trivial braiding relation for the field.

Again these operators carry no $U(1)$ charge and so will couple to a soliton-anti-soliton
pair.  It will prove to be convenient to consider form-factors involving
symmetric and anti-symmetric combinations
of the solitons:
\begin{eqnarray}\label{eIIxxxvii}
f^\pm_S (\th_1,\th_2)
&=& \langle 0|e^{\pm i\frac{\beta}{2}\Phi}|A^\dagger_{-}(\th_2)A^\dagger_{+}(\th_1)\rangle +
\langle 0|e^{\pm i\frac{\beta}{2}\Phi}|A^\dagger_{+}(\th_2)A^\dagger_{-}(\th_1)\rangle;\cr\cr
f^\pm_A (\th_1,\th_2)
&=& \langle 0|e^{\pm i\frac{\beta}{2}\Phi}|A^\dagger_{-}(\th_2)A^\dagger_{+}(\th_1)\rangle -
\langle 0|e^{\pm i\frac{\beta}{2}\Phi}|A^\dagger_{+}(\th_2)A^\dagger_{-}(\th_1)\rangle\,.~~~~
\end{eqnarray}
As the operators $e^{\pm i\frac{\beta}{2}\Phi}$ are Lorentz scalars, these form factors
are solely a function of $\th_{12}=\th_1-\th_2$.  Applying the scattering axiom to the
form factors we find
\begin{eqnarray}\label{eIIxxxviii}
f^\pm_S(-\th ) &=& S_S(\th )f^\pm_S(\th )\,;\nonumber\\[1mm]
f^\pm_A(-\th ) &=& S_A(\th )f^\pm_A(\th )\,;
\end{eqnarray}
where
\begin{eqnarray}\label{eIIxxxix}
S_S(\th) &=& \frac{\sinh(\frac{\th+\pi i}{2\xi})}{\sinh(\frac{\th-\pi i}{2\xi})}
\exp\bigg(\int^\infty_0\frac{dx}{x}\sinh(\frac{x\th}{i\pi})G_{c2}(x)\bigg);
\nonumber\\[1mm]
S_A(\th) &=& -\frac{\cosh(\frac{\th+\pi i}{2\xi})}{\cosh(\frac{\th-\pi i}{2\xi})}
\exp\bigg(\int^\infty_0\frac{dx}{x}\sinh(\frac{x\th}{i\pi})G_{c2}(x)\bigg),
\end{eqnarray}
where $G_{c2}(x)$ is as for the current form-factor.
The non-locality of the fields, $e^{\pm i\frac{\beta}{2}\Phi}$, relative to the solitons
implies that $R$ should be taken to be $-1$ in the periodicity axiom (see Eq.\,(\ref{eIIx})).  
Thus we have
\begin{eqnarray}\label{eIIxl}
f^\pm_{+-}(-\th ) &=& -f^\pm_{-+}(\th - 2\pi i );\nonumber\\[1mm]
f^\pm_{-+}(-\th ) &=& -f^\pm_{+-}(\th - 2\pi i ).
\end{eqnarray}
In terms of the symmetric and anti-symmetric combinations we obtain
\begin{eqnarray}\label{eIIxli}
f^\pm_{S}(-\th ) &=& -f^\pm_{S}(\th - 2\pi i );\nonumber\\[1mm]
f^\pm_{A}(-\th ) &=& f^\pm_{A}(\th - 2\pi i ).
\end{eqnarray}
Having expressed the constraints on the form factors in this way, we can readily
write down a minimal solution
\begin{eqnarray}\label{eIIxlii}
f^\pm_{S}(\th_1,\th_2) &=& A^\pm_S 
\frac{\sinh (\th_{12}/2)}{\sinh(\frac{\th_{12}+\pi i}{2\xi})} 
\exp\Big[\int^\infty_0 \frac{dx}{x} \frac{G_{c2}(x)}{\sinh (x)}
\sin^2(\frac{x}{2\pi}(i\pi + \th_{12}))\Big];~~~\nonumber\\[-2mm]
&&\\[-2mm]
f^\pm_{A}(\th_1,\th_2) &=& A^\pm_A
\frac{\sinh(\th_{12}/2)}{\cosh(\frac{\th_{12}+\pi i}{2\xi})} 
\exp\Big[\int^\infty_0 \frac{dx}{x} \frac{G_{c2}(x)}{\sinh s(x)}
\sin^2(\frac{x}{2\pi}(i\pi + \th_{12}))\Big],~~~\nonumber
\end{eqnarray}
where $A^\pm_{S/A}$ are normalization constants.
We now turn to their determination.

From the hermiticity condition,
\begin{equation}\label{eIIxliii}
f^{\pm}_{S/A}(\th_1,\th_2)^* = f^{\mp}_{S/A}(\th_2 - i\pi,\th_1-i\pi),
\end{equation}
we see that $A_S^{\pm*}=A_S^{\mp}$ and $A_A^{\pm*}=-A_A^{\mp}$.
The operators
$e^{\pm i\frac{\beta}{2}\Phi}$ have a nonzero vacuum expectation value.
Using the annihilation pole axiom it is possible to relate the value
of $A^{\pm}_S$ to this expectation value.  In this case it reads
\begin{equation}\label{eIIxliv}
i \text{res}_{\th_1=\th_2-\pi i}f^{\pm}_S(\th_1,\th_2) = 
4\langle e^{\pm i\frac{\beta}{2}\Phi}\rangle.
\end{equation}
We thus obtain $A^\pm_S = \frac{2}{\xi} \langle e^{\pm
i\frac{\beta}{2}\Phi}\rangle$. Using the transformation properties
under charge conjugation we conclude that $A_S^+=A_S^-$. To relate
$A^\pm_S$ and $A^\pm_A$ we appeal to the explicit calculation in Ref.\cite{smirnov}
which shows that $A^+_S=A^+_A$.

\subsection{Relevance of Higher Particle Form Factors: Quantum Ising Model as an Example}

We have argued in Section 2.2 that to obtain exact information on spectral functions
at low energies it is enough to include form factors involving only a few particles
or excitations.  However as a rule of thumb, form factor calculations fare even better.
In practice, form factor sums have been found to be strongly convergent
for operators in massive theories \cite{oldff1,delone,deltwo}.
To obtain a good approximation to correlators involving such fields {\it at all energies},
only the first few terms need to be kept.
Even in massless theories where there are no explicit
thresholds, convergence is good provided the 
engineering dimension of the operator matches its anomalous
dimension \cite{les,lec}.
At high energies where higher particle form factors do begin to contribute, their
contributions are progressively (often exponentially) smaller.  
We will illustrate this with a simple example,
the spectral function of the spin-spin correlator in the Ising model.

The quantum Ising model is a model of one dimensional spins governed by the Hamiltonian
$$
{\cal H} = -J\sum_i (\sigma^z_i\sigma^z_{i+1} + g\sigma^x_i),
$$
where $\sigma^z$ and $\sigma^x$ are Pauli matrices and $J$ is assumed to be positive.
At zero temperature, this model undergoes a $Z_2$ phase transition as a function of $g$ between
an ordered and a paramagnetic state.  
This transition is in the same universality class as a classical two dimensional Ising model.
In the continuum limit, it is thus described by a massive
Majorana (real) fermion:
\begin{equation}\label{eIIxlv} 
S = \frac{1}{8\pi} \int dx d\tau 
\bigg(\psi \partial_{\bar z} \psi + 
\bar{\psi} \partial_{z} \bar{\psi} + 
2im \psi\bar{\psi}\bigg) ,
\end{equation}
where $z/\bar{z} = (\tau \pm ix)/2$.
$\psi$ and $\bar\psi$ are the left and right moving self-conjugate components of the Majorana
spinor.
The mass of the fermion corresponds to the distance from criticality $m \sim (g_c-g)$
which we will assume to be positive (the model is in its ordered phase).
The fundamental excitation, $A_a(\th )$, of the model arises from the mode expansions of
the Fermi fields:
\begin{eqnarray}\label{eIIxlvi}
\hspace{-5mm}\psi &=& \sqrt{m} \int^\infty_{-\infty} \frac{d\theta}{\sqrt{2 \pi} i}
e^{-\th /2} \bigg( A(\th ) e^{-m(ze^{-\th } +\bar z e^{\th })} -
A^\dagger (\th ) e^{m(ze^{-\th } +\bar z e^{\th })} \bigg) \nonumber\\[1mm]
\hspace{-5mm}\bar{\psi} &=& -i \sqrt{m} \int^\infty_{-\infty} \frac{d\theta}{\sqrt{2 \pi} i}
e^{\th /2} \bigg( A(\th ) e^{-m(ze^{-\th } +\bar z e^{\th })} +
A^\dagger (\th ) e^{m(ze^{-\th } +\bar z e^{\th })} \bigg) .~~~
\end{eqnarray}
The defining commutation relations of $A$ and $A^\dagger$ are
$\{ A(\th ),A^\dagger (\th')\} = 2\pi \delta (\th-\th')$.

While the form factors of the Fermi fields are trivial (only the one particle
form factors are non-zero), the form factors of the order parameter field, $\sigma_z (x,t)$,
are more complicated (ultimately a result of the spin and Fermi fields being 
mutually non-local).  From Refs. \cite{berg,ising,yurov,cardmuss}, we have
\begin{eqnarray}\label{eIIxlvii}
\langle 0|\sigma_z (0,0)|A(\th_{2n})\cdots A(\th_1)\rangle &\equiv& 
f(\th_1,\cdots,\th_{2n})\nonumber\\[1mm]
&=& 
i^{n} \prod_{i<j}^{2n} \tanh \Big(\frac{\th_i-\th_j}{2}\Big)\,.
\end{eqnarray}
Only even numbers of fermions couple to the spin field (in contrast, odd numbers
of fermions couple to the dual disorder field).
The fermion S-matrix of this model is $S=-1$.  The mutual non-locality of the
spin, $\sigma_z$, and Fermi, $\psi$, fields is encoded in the braiding phase,
$R_{\sigma_z\psi}=-1$.  Knowing $S=R_{\sigma_z\psi}=-1$, it is easy to check that the
form factors in Eq.\,(\ref{eIIxlvii}) satisfy the axioms of Section 2.3.

\begin{figure}[ht]
\vskip -.2cm
\begin{center}
\epsfxsize=0.5\textwidth
\rotatebox{270}{\epsfbox{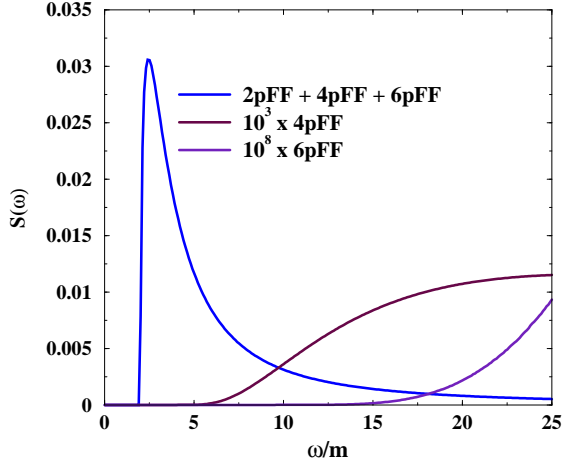}}
\end{center}
\caption{Plots of the response function derived from
the spin-spin correlator in the Ising model.
In the first plot, the contributions from the two through six particle form factors
are included.  In addition, we plot the individual contributions of the four and 
six particle form factors.  We see that the four and six particle form factors,
to be seen, require a magnified scale.}
\label{figIIii}
\end{figure}

The spectral function of the spin-spin correlator is defined as 
\begin{equation}\label{eIIxlviii}
S(\omega ,k )
= -\text{Im}\bigg[\frac{1}{\pi} \int^\infty_{-\infty} dx \int^\infty_{-\infty}d\tau 
e^{i\omega \tau-i k x}
\big(-\langle T (\sigma_z(x,\tau)\sigma_z (0,0))
\rangle\big)|_{\omega \rightarrow -i\omega + \epsilon}\bigg].
\end{equation}
Here $\langle T (\sigma_z(x,\tau)\sigma_z (0,0))\rangle$ is the time ordered correlator
in imaginary time.
If we insert a resolution of the identity in between the two fields, do the
Fourier transforms, and take the appropriate analytic continuation,
we find at positive frequencies:
\begin{eqnarray}\label{eIIxlix}
S (\omega > 0 ,k) &=&
2\pi\sum^\infty_{n=1} \frac{1}{2n!}\prod^{2n}_{i=1}\int^\infty_{-\infty}
d\hat\th_i ~
\delta(k-m\sum^{2n}_{i=1}\sinh(\th_i))\cr\cr
&& \hskip .4in \times \delta(\omega-m\sum^{2n}_{i=1}\cosh(\th_i))
~|f(\th_1,\cdots,\th_{2n}|^2,
\end{eqnarray}
where $d\hat\th \equiv d\th/(2\pi)$.
Integrating over $\th_1$ and $\th_2$ we obtain
\begin{eqnarray}\label{eIIl}
S (\omega > 0 ,k) &=&
\frac{2}{\pi}\sum^\infty_{n=1} \frac{1}{2n!}\prod^{2n}_{i=3}\int^\infty_{-\infty}
d\hat\th_i \frac{1}{\sqrt{\omega'^2-k'^2-4m^2}}\frac{1}{\sqrt{\omega'^2-k'^2}}
\nonumber\\[1mm]
&& \times \Theta(\omega'-\sqrt{k'^2+4m^2}) |f(\th_1,\cdots,\th_{2n}|^2,
\end{eqnarray}
where $\th_1$ and $\th_2$ are given by
\begin{eqnarray}\label{eIIli}
\omega' &=& \omega - m\sum^{2n}_{i=3} \cosh(\th_i);\cr\cr
k' &=& k - m\sum^{2n}_{i=3} \sinh(\th_i);\cr\cr
\th_1 &=&
\begin{cases}
\cosh^{-1}\bigg(\frac{1}{2m}(\omega'+|k'|\sqrt{1-\frac{4m^2}{\omega'^2-k'^2}})\bigg)
~~ \text{if $k' > 0$};\cr
-\cosh^{-1}\bigg(\frac{1}{2m}(\omega'+|k'|\sqrt{1-\frac{4m^2}{\omega'^2-k'^2}})\bigg)
~~ \text{if $k' < 0$};
\end{cases}\cr\cr
\th_2 &=& \sinh^{-1}\bigg(\frac{k'}{m}-\sinh(\th_1)\bigg).
\end{eqnarray}
We plot as a function of energy, $\omega$,
the first three non-zero contributions (the 2-particle, the 4-particle,
and the 6-particle) to the spectral function in Fig.\,\ref{figIIii}.  
We first note that the 2n-particle form-factor only contributes for
energies $\omega > 2nm$.  Thus the plot of these contributions is exact for $\omega < 16m$.  We
also see that for energies beyond $4m$ the higher particle contributions 
barely register.  The 4-particle contribution, at its peak, is still $1/3000$
of the 2-particle contribution.  And the 6-particle contribution is considerably
smaller yet.  To be seen, it must be plotted on a scaled magnified by $10^8$.
We thus learn that to compute the spectral function to energies far above
the mass gap, it would seem to be sufficient to consider only the two-particle contribution.
While we have only shown this to be the case in the off-critical Ising model, it is
a generic feature of all massive integrable models.  We will see other examples in
Sections 4 and 5 with the spin spectral function of the $O(3)$
non-linear sigma model and the optical conductivity in the U(1)
Thirring model.

\subsection{Form Factors for Integrable Models with Bound States}

When there are bound states in an integrable model, additional analytical requirements
are imposed upon the form factors.  Suppose we have an excitation, $A_c$, which is
a bound state of excitations, $A_a$ and $A_b$.  The presence of this bound
state is indicated by the analytic structure of the two-body S-matrix between
particles $A_a$ and $A_b$.  The S-matrix must have a pole at $iu^c_{ab}$ of the form\\[-6mm]~
\begin{equation}\label{eIIlii}
S^{ab}_{ab}(\th ) \sim i \,\frac{g^c_{ab}g_c^{ab}}{\th - iu^c_{ab}}\,.
\end{equation}
The coefficients forming the residue of the pole, $g^c_{ab}$, effectively measure
the strength with which the two particles bind.

We can formally indicate that $A_a$ and $A_b$ have a bound state $A_c$ via
\begin{equation}\label{eIIliii}
i g^c_{ab} A_c (\th ) ={\rm res}_{\delta = 0}~
A_{a}(\th + \delta + i\bar{u}^{\bar{b}}_{a\bar{c}})
A_{b}(\th - i\bar{u}^{\bar{a}}_{b\bar{c}})\,.
\end{equation}
Here $\bar{u}^{\bar{a}}_{b\bar{c}} = \pi - u^{\bar{a}}_{b\bar{c}}$
and $u^{\bar{a}}_{b\bar{c}}$ is the location of the pole in the two-body 
S-matrix between $A_b$ and $A_{\bar{c}}$ indicative of the formation of the
bound state, $A_{\bar{a}}$.  The locations, $u$, of the various poles satisfy a
simple relation
\begin{equation}\label{eIIliv}
2\pi = u^{c}_{ab} + u^{\bar{a}}_{b\bar{c}} + u^{\bar{b}}_{a\bar{c}}\,.
\end{equation}
which can be proven on the basis of kinematical considerations.  If $m_a$ and
$m_b$ are the masses of the particles $A_a$ and $A_b$, then the mass of the
bound state, $m_c$, satisfies $m_c^2 = m_a^2+m_b^2+2m_am_b\cos (u^c_{ab})$.

The form of Eq.\,(\ref{eIIliii}) indicates that we are able to relate $n-1$ particle
to $n$-particle form factors in the presence of bound states.  In particular we have
\begin{eqnarray}\label{eIIlv}
ig^c_{ab}f(\th_1,\cdots,\th_{n-2},\th_{n-1})_{a_1,\cdots ,a_{n-2},c} &=& \nonumber\\
&& \hskip -2.25in {\rm res}_{\delta = 0} 
f(\th_1,\cdots,\th_{n-2},\th_{n-1} - i\bar{u}^{\bar{a}}_{b\bar{c}},\th_{n-1} + \delta + 
i\bar{u}^{\bar{b}}_{a\bar{c}})_{a_1,\cdots ,a_{n-2},a,b}\,.~~~~~
\end{eqnarray}
Thus with theories with bound states, consistency demands that form-factors come equipped
with poles indicative of the bound states together with constraints upon the corresponding residues.

\subsection{Form Factors in Finite Temperature Correlators}

Up to this point we have only considered correlation functions at zero temperature.  
However the form factor approach can also be fruitful at finite temperatures.
A finite temperature expansion of correlators is given in
terms of a trace over the Boltzmann density matrix,
\begin{eqnarray}
\label{eIIlvi}
G^{\cal O} (x,t) &=& \frac{1}{\cal Z} \,
{\rm Tr}(e^{-\beta H} {\cal O}(x,t) {\cal O}(0,0))\cr\cr
&=& 
\frac{\sum_{n s_n} e^{-\beta E_{s_n}}
\langle n,s_n|{\cal O}(x,t){\cal O}(0,0)|n,s_n\rangle}{\sum_{n s_n} e^{-\beta E_{s_n}}
\langle n,s_n|n,s_n\rangle}\, .
\end{eqnarray}
Here the state, $|n,s_n\rb$, denotes a set of n-particles carrying
quantum numbers, $\{s_n\}$.
Inserting a resolution of the identity between the two fields as we did in the
zero temperature case then
leads us to a double sum,
\hskip .1in
\begin{equation}\label{eIIlvii}
G^{\cal O} (x,t) =  \frac{\sum_{\genfrac{}{}{0pt}{}{n s_n}{m s_m}} e^{-\beta E_{s_n}}
\langle n,s_n|{\cal O}(x,t)|m,s_m\rangle
\langle m, s_m |{\cal O}(0,0)|n,s_n\rangle}
{\sum_{n s_n} e^{-\beta E_{s_n}}
\langle n,s_n|n,s_n\rangle}\,.
\end{equation}
We thus again have reduced the evaluation of the finite temperature correlator to the evaluation of 
form factors.

To evaluate this expression for the correlator, 
we again focus on the associated spectral function, $G^\CO (k,\om )$.
In computing $G^\CO (k,\om)$, only terms
in the form factor sum with a given energy, $\om $, and momentum, $k$,
contribute to the sum, i.e.
\begin{eqnarray}\label{eIIlviii}
G^\CO (k,\om ) &=& \frac{1}{\cal Z}
\sum_{\genfrac{}{}{0pt}{}{n s_n}{m s_m}}\delta (\om - E_{s_n}+E_{s_m})
\delta (k - p_{s_n}+p_{s_m})\cr
&&  \hskip .2in \times e^{-\beta E_{s_n}}
{\langle n,s_n|{\cal O}(0,0)|m,s_m\rangle
\langle m, s_m |{\cal O}(0,0)|n,s_n\rangle},~~~~
\end{eqnarray}
as enforced by the presence of the two delta functions.
For any $\om , k$, this dramatically reduces the number of matrix elements
one must compute.\footnote{\,Here $G^\CO$ is simply the Fourier transform
of $G^\CO (x,t)$, but similar considerations also apply to the corresponding
retarded correlator.\\[-3mm]~} 
This reduction nonetheless usually leaves a difficult
computation.  However we can exploit the gapped nature of the spin chain
to make the problem more tractable.  Because the theory is gapped or massive
(with gap, $\Delta$), the correlator admits a low temperature expansion
of the form,
\begin{equation}\label{eIIlix}
G^\CO (k,\om ) = {\sum_n} \alpha_n (k,\om ) e^{-n\beta \Delta} .
\end{equation}
For the particular correlators of concern in this review and for
the range of $\om $ and $k$ in which we are interested, each $\alpha_n$
is determined by a single matrix element.  Because we can compute these
matrix elements, we obtain an {\it exact} low temperature expansion.

The ability to compute such an expansion should be compared with
the approach taken by LeClair and Mussardo \cite{leclaira} (following Ref. \cite{leclaira1}).  
These authors argued that
it was possible to use the same form-factors we employ here to compute
finite temperature correlators.  However rather than directly evaluate
individual terms in the sum appearing in Eq.~(\ref{eIIlvii}), they first conjectured an ansatz
involving a resummation of terms in the sum.  
This procedure was criticized in Ref. \cite{salrep} (see also Ref.\cite{castro}).  There it
was argued that while this worked for the computation of one-point functions,
it was problematic for two-point functions.\footnote{\,That such a procedure is valid for one
point functions is also supported by the study in Ref. \cite{schoutens} where form-factors
are used to compute CFT correlators.}
Rather Ref \cite{salrep} put forth the view that
such problems should be attacked through the use of form factors
computed against a thermalized vacuum \cite{fred,fred1,vladb}.
However the counterexample cited in \cite{salrep}, 
a computation involving interacting
quantum Hall edge states, involved a gapless theory, and so is
in a different class than the models considered in this review.  (Without
a gap, the low temperature expansion we consider above ceases to
make sense.)  The results of Section 4 show that it is indeed possible, at least
in certain cases, to make sense of the form-factor expansion of two
point functions at finite temperature.  
But while one can make sense of this expansion, it will not be possible
to directly compare computations in this review to the ansatz posited in \cite{leclaira}.
Their ansatz as is applies only to diagonal theories where scattering
does not permute internal quantum numbers, contrary, for example, 
to the case of the O(3) non-linear
sigma model considered in Section 4.

\subsubsection{Regularization of Form Factors}

Unlike in the zero temperature case, at finite temperature we must compute
form factors with particles both to the
right and to the left of the field, $\CO$,
$$
\langle A_{b_m} (\tth_m ) \cdots A_{b_1} (\tth_1 )
\CO (0,0) A_{a_n} (\th_n ) \cdots A_{a_1}(\th_1 )\rangle .
$$
To understand such an object we must contend with the possibility
that $\tth_i = \th_j$, $a_i=b_j$ for some $i,j$.  From the
algebra of the Faddeev--Zamolodchikov operators (Eq.\,(\ref{eIIiv})),
we know the commutation relations involve $\delta$-functions,
i.e.
\begin{equation}\label{eIIlx}
A_{a_i}(\tilde{\th}_i) A^\dagger_{b_j}(\th_j) = 
2\pi \delta (\tilde{\th}_i-\th_j ) \delta_{a_ib_j}
+ \cdots .
\end{equation}
It is crucial to include the contributions of the 
$\delta$-functions to the correlators. 
To do so,
we must understand the above form factor to equal
\begin{eqnarray}\label{eIIlxi}
\langle A_{b_m} && (\tth_m ) \cdots A_{b_1} (\tth_1 )
\CO (0,0) A_{a_n} (\th_n ) \cdots A_{a_1}(\th_1 )\rangle  \nonumber\\[1mm]
&& =\sum_{\genfrac{}{}{0pt}{}{{\{a_i\} = A_1 \cup A_2}}{{\{b_i\} = B_1 \cup B_2}}}
S_{A,A_1} S_{B,B_1} \langle B_1 | A_1\rangle
\langle B_2 |\CO (0,0) |A_2\rangle_{\rm connected}.
\end{eqnarray}
The sum in the above is over all possible subsets of 
$\{ a_i \}$ and $\{ b_i \}$.  
The S-matrix $S_{A,A_1}$ arises from the commutations necessary
to rewrite $A_{a_n}(\th_n ) \cdots A_{a_1} (\th_1 ) |0\rangle$
as $A_2 A_1 |0\rangle$ and similarly for $S_{B,B_1}$.
The matrix element,
$\langle B_1 | A_1 \rangle$, is evaluated using the Faddeev--Zamolodchikov
algebra.
In this way (ill-defined) terms proportional
to $\delta (0)$ are produced but which cancel 
similarly ill-defined terms arising
from the evaluation of the partition function.

The `connected' form factor appearing in the above expression is
to be understood as follows.  Using crossing symmetry, the form
factor can be rewritten as
\begin{eqnarray}\label{eIIlxii}
\langle B_2 |\CO (0,0) |A_2\rangle_{\rm connected} &&\nonumber\\[1mm]
&&\hskip -1.5in = \langle A_{b'_{i_k}} (\tth_{i_k} ) \cdots A_{b'_{i_1}} (\tth_{i_1} )
\CO (0,0) A_{a'_{j_q}} (\th_{j_q} )
\cdots A_{a'_{j_1}}(\th_{j_1} )\rangle_{\rm connected}\nonumber\\[1mm]
&&\hskip -1.5in = \langle \CO (0,0) A_{a'_{j_q}} (\th_{j_q} )
\cdots A_{a'_{j_1}}(\th_{j_1} )
A_{\bar{b}'_{i_k}} (\tth_{i_k} -i\pi ) 
\cdots A_{\bar{b}'_{i_1}} (\tth_{i_1}-i\pi )
\rangle_{\rm connected} \nonumber\\[1mm]
&&\hskip -1.5in = f^\CO _{\bar{b}'_{i_1}\cdots \bar{b}'_{i_k}a'_{j_1}\cdots a'_{j_q}}
(\tth_{i_1}-i\pi,\cdots ,\tth_{i_k}-i\pi,\th_{j_1},\cdots 
,\th_{j_q})_{\rm connected},
\end{eqnarray}
where the last relation holds provided we do not have $\th_i = \tth_j$,
$a_i = b_j$ for any $i,j$.  If this does occur we see from the
annihilation pole axiom that the form factor is not well defined,
having a pole at $\th_i = \tth_j$.  In such cases the form factor
requires regulation. 

To regulate the form factor, we employ a scheme suggested by Balog
\cite{balog1}
and used by LeClair and Mussardo\cite{leclaira}.  We define
\begin{eqnarray}\label{eIIlxiii}
&& f^\CO_{\bar{b}'_{i_1}\cdots 
\bar{b}'_{i_k}a'_{j_1}\cdots a'_{j_q}}
(\tth_{i_1}-i\pi,\cdots ,
\tth_{i_k}-i\pi,\th_{j_1},\cdots ,\th_{j_q})_{\rm connected}\nonumber\\[-2mm]
&&\\[-2mm]
&& =\begin{array}{c} 
\mbox{\scriptsize\rm finite}\\[-2mm] \mbox{\scriptsize\rm piece~of}\end{array} \lim_{\eta_i \rightarrow 0}
f^\CO _{\bar{b}'_{i_1}\cdots \bar{b}'_{i_k}a'_{j_1}\cdots a'_{j_q}}
\!(\tth_{i_1}\!-\!i\pi\!+\!i\eta_1,\cdots ,
\tth_{i_k}\!-\!i\pi\!+\!i\eta_k,\th_{j_1},\cdots ,\th_{j_q})\,.\nonumber
\end{eqnarray}
In taking the finite piece of $f^\CO$, we discard terms proportional
to $\eta_i^{-p}$ as well as terms proportional to $\eta_i/\eta_j$.
In this way the connected piece is independent of the way
the various limits $\eta_i \rightarrow 0$ are taken.
Balog \cite{balog1} has already used this prescription to compute one point
functions and successfully compare them to TBA calculations.
In Ref. \cite{balog1} it was argued that the delta functions leading to 
such terms arise from the
use of infinite volume wavefunctions.  If such wavefunctions
are replaced instead with finite volume counterparts, the delta
functions are regulated.  For example, a pole in $\eta$ is changed
as follows
\begin{equation}\label{eIIlxiv}
\frac{1}{i\eta} = \int d\th \,\frac{\delta (\th )}{\th + i\eta}
\rightarrow \int d\th \,\frac{f(\th )}{\th + i\eta}\, ,
\end{equation}
where $f(\th )$ is some sharply peaked function about $\th = 0$
which in the infinite volume limit evolves into a $\delta$-function.
However the principal value of this regularized integral
is zero.  
Thus discarding the pole terms is justified in this sense.
For terms that are ratios of infinitesimals, Balog also demonstrates
that such terms, once regularized, disappear in the infinite volume
limit.

\section{Sine-Gordon Model and Spin-\boldmath{$\frac{1}{2}$} Quantum Magnets}
\setcounter{equation}{0}
The first set of applications of massive integral QFTs we will discuss 
involves the quantum sine-Gordon model (SGM). It describes the low
energy limit of a variety of gapped ``deformations'' of the spin-\half
Heisenberg chain. In order to fix notations and set the stage we first
review the low-energy limit of the anisotropic spin-\half Heisenberg
model.

\subsection{The Spin-$\frac{1}{2}$ Heisenberg Chain in a Uniform
  Magnetic Field}
\label{XXZinfield}
The anisotropic spin-\half Heisenberg chain in a uniform 
``longitudinal'' magnetic
field $H$ is defined by the Hamiltonian 
\bea
{\mathcal H}_{\rm XXZ}=J\sum_{j}{S}^x_{j}{S}^x_{j+1}+{S}^y_{j}{S}^y_{j+1}+
\delta {S}^z_{j}{S}^z_{j+1}-H\sum_j S^z_j,
\label{HXXZ}
\eea
and is exactly solvable by Bethe's ansatz \cite{Bethe31}. In what
follows we will consider only the region 
\be
-1<\delta\leq 1,
\label{Delta}
\ee
which corresponds to an ``XY''-like exchange anisotropy. Ground state
properties, the excitation spectrum in the thermodynamic limit,
thermodynamic properties and the large-distance asymptotics of
correlation functions have been determined from the Bethe ansatz
solution \cite{hulthen,Yang66a,Yang66b,Yang66c,Taka71,Taka72,Taka74,JMC,
Klumper91}. Dynamical correlation functions of the lattice model at zero
temperature and zero magnetic field have been evaluated in form of a
spectral sum using the results of \cite{MiwaJimbo} in Refs.
\cite{Bougourzi96,Abada97,Bougourzi97,karbach}. 

\subsubsection{Excitation Spectrum of the Lattice Model}
\begin{figure}[ht]
\begin{center}
\epsfxsize=0.7\textwidth
\epsfbox{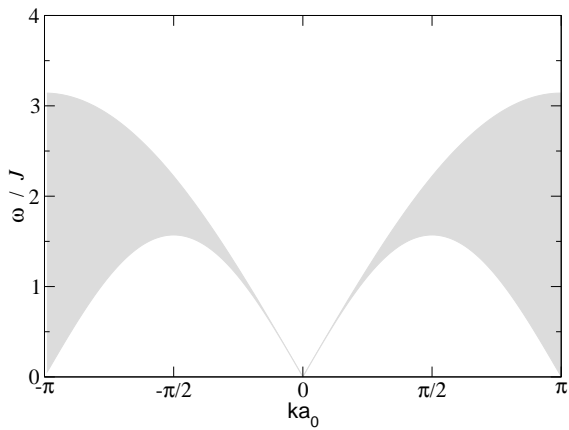}
\end{center}
\vskip -3mm
\caption{Two-spinon continuum in the isotropic spin-\half Heisenberg
  chain ($\delta=1$) in zero field. The triplet $S=1$ and singlet
  $S=0$ states are degenerate. }
\label{fig:trans0}
\end{figure}

The spectrum of low-lying excitations for the isotropic Heisenberg
chain in zero field was derived from Bethe's equations \cite{Bethe31}
in Refs. \cite{Faddeev81,Faddeev84}. It was shown that for even chain
lengths the spectrum is given in terms of scattering states of an even
number of gapless, elementary excitations carrying spin $\frac{1}{2}$,
called ``spinons''. There are no single-particle excitations: spinons
occur only in pairs. The two-spinon continuum is shown in
Fig.\,\ref{fig:trans0}. 
\vskip -5mm

Application of  magnetic field leads to a change in the ground state
properties. For any $H>0$ there is a finite magnetization per site
(see below). At a critical field 
\be
H_c=J(1+\delta)\ ,
\ee
the ground state becomes fully polarized. At $H_c$ a phase transition
between a gapped, commensurate ($H>H_c$) phase and a gapless,
incommensurate ($H<H_c$) phase occurs, which is in the universality
class of the commensurate-incommensurate transition
\cite{DN78,Pokrovski79,Schulz80,Haldane82b}. In the following we will
consider only the gapless regime $H<H_c$.

\begin{figure}[ht]
\begin{center}
\epsfxsize=0.6\textwidth
\epsfbox{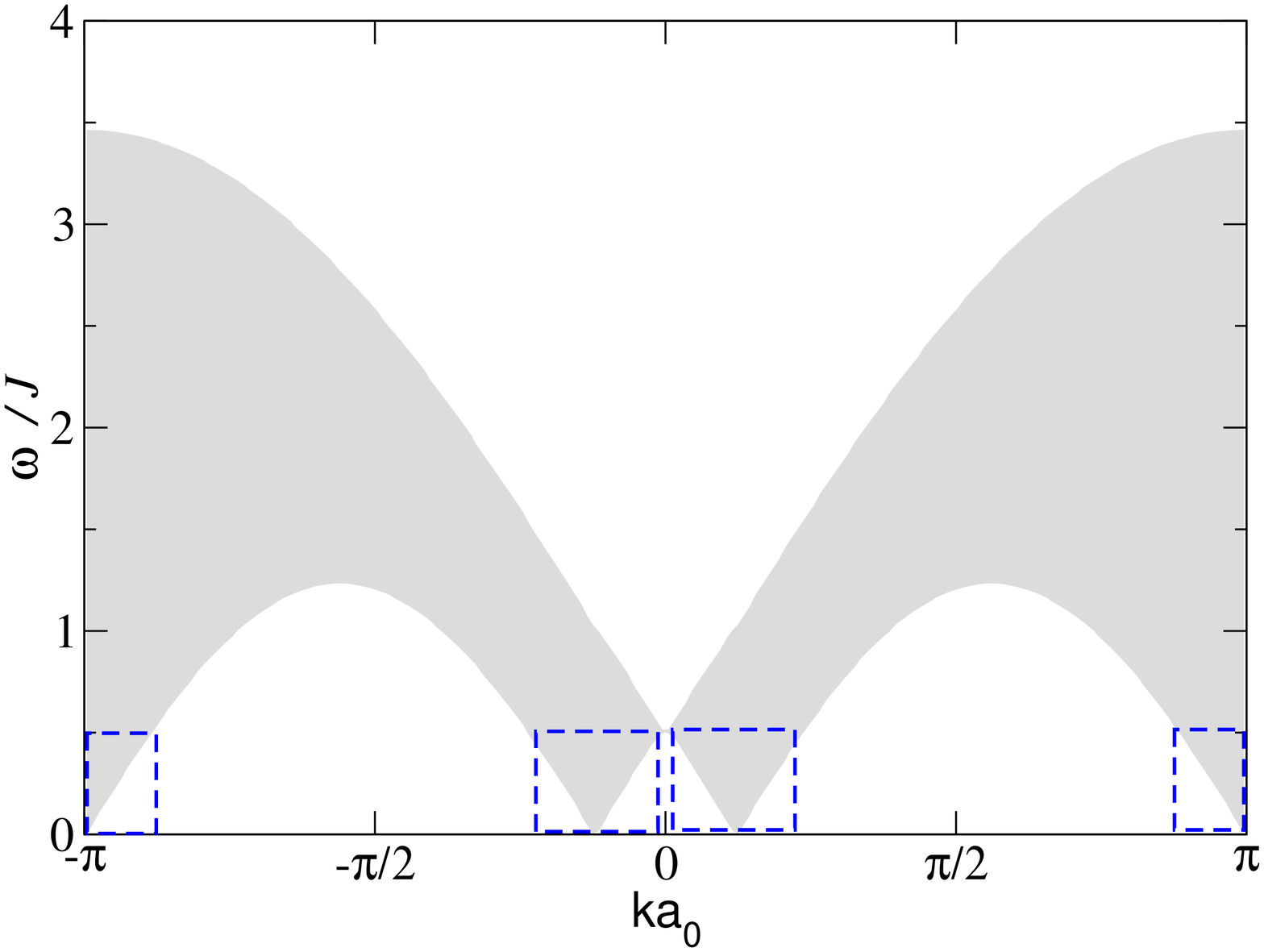}\vskip -0.2cm
\epsfxsize=0.6\textwidth
\epsfbox{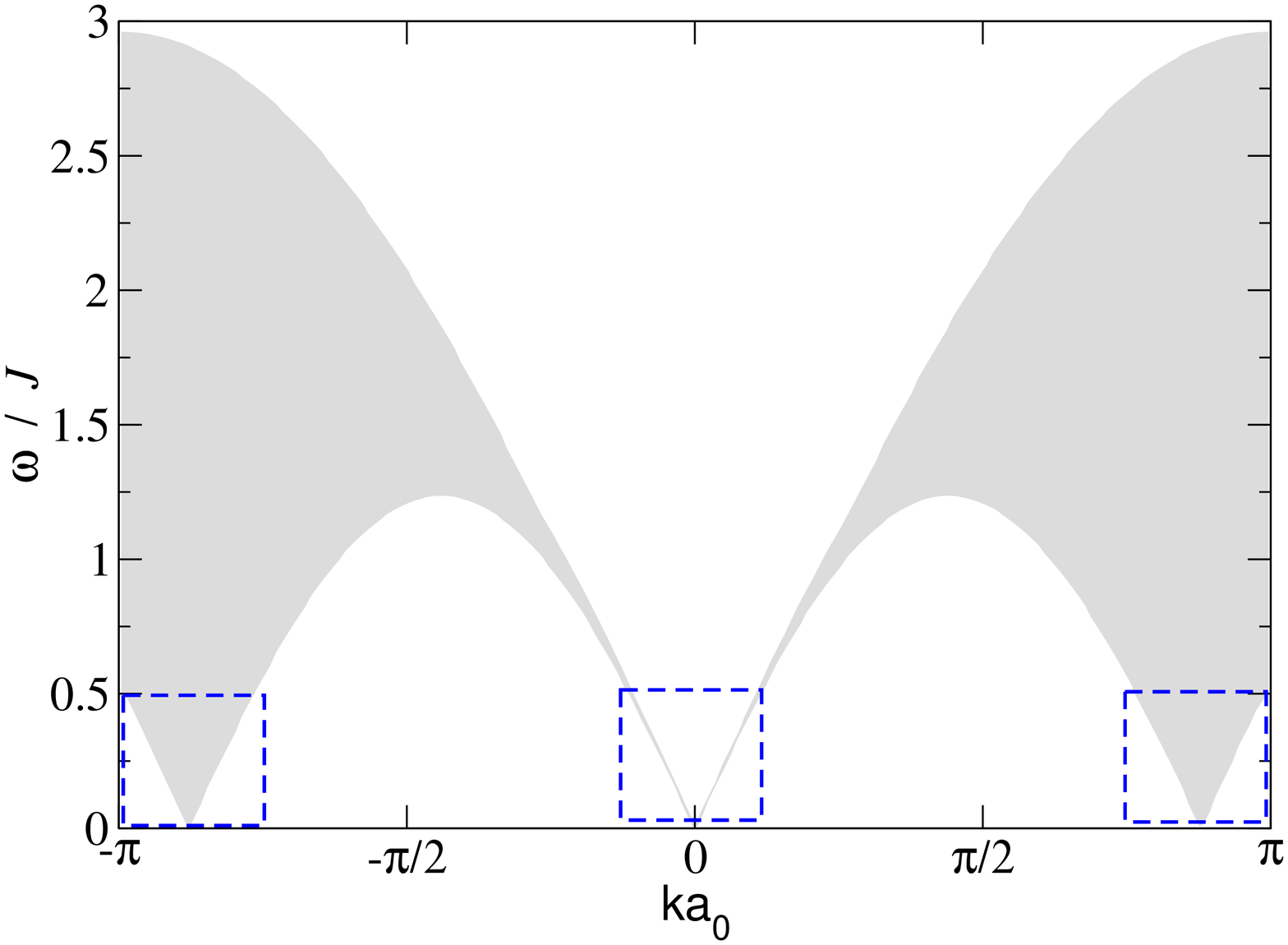}\
\end{center}
\vskip -0.2cm
\caption{Two-particle continua in the isotropic spin-\half Heisenberg
  chain ($\delta=1$) in an applied field $H=0.5J$. (top figure) States
  with $\delta S^z=1$, which are probed by the transverse correlation
  functions $\langle S^+_n(t) S^-_1(0)\rangle$. (bottom figure) States with $\delta
  S^z=0$ relevant to longitudinal correlation functions $\langle
  S^z_n(t)S^z_1(0)\rangle$. The boxes indicate the low-energy regime,
  which is described by the field theory approximation.} 
\label{fig:transH}
\end{figure}

In the presence of a magnetic field the spectrum is
more complicated \cite{Mueller81,TalstraHaldane96,Karbach02}. Now
single-particle excitations exist,\footnote{\,They correspond to
  ``string solutions'' of the Bethe   ansatz equations.} but they
have spectral gaps. The most relevant 
excitations at low energies are two-fold parametric and are similar in
nature to the two-spinon excitations in zero field. We therefore
call the elementary excitations giving rise to the gapless
two-particle continua for $H>0$ spinons.\footnote{\,In the literature
various other terminologies have been used to denote these
excitations.}
The two-spinon excitations with $\delta S^z=0$ and $\delta S^z=1$ are
shown in Fig.\,\ref{fig:transH}. The $\delta S^z=1$ excitation remains
gapless at $P=\pi/a_0$, whereas the $\delta S^z=0$ excitation becomes
gapless at an incommensurate wave number for $H>0$.
Several variants of \r{HXXZ} are of considerable interest as they
describe situations of direct experimental relevance. However, unlike
the Heisenberg model itself these lattice models are not solvable by
Bethe's ansatz. Interestingly, the low energy limits of these models
are described by integrable massive quantum field theories, which
allows for the calculation of the low-energy behavior of dynamical
correlation functions. The starting point for all these considerations
is the low-energy limit of the Heisenberg model \r{HXXZ}, which we
derive next following a well established procedure
\cite{Luther75,Haldane81a,Haldane81b,Affleck89b}.

\subsubsection{Continuum Limit}
\label{contlimit}
The Heisenberg chain \r{HXXZ} is equivalent to a model of spinless
fermions as can be shown by means of the Jordan--Wigner transformation
\be
S^z_j=c^\dagger_jc_j-\frac{1}{2}\ ,\quad
S^+_j=S^x_j+iS^y_j=c^\dagger_j\ e^{-i\pi\sum_{k<j}c^\dagger_kc_k}\ ,
\label{JW}
\ee
which maps \r{HXXZ} onto
\be
{\mathcal H}_{\rm  XXZ}
=
J\sum_{j}\frac{1}{2}\left[
c^\dagger_{j}c_{j\!+\!1}\!+\!c^\dagger_{j\!+\!1}c_{j}\right]
\!+\!\delta \left[c^\dagger_jc_j\!-\!\frac{1}{2}\right]
\left[c^\dagger_{j\!+\!1}c_{j\!+\!1}\!-\!\frac{1}{2}\right]
\!-\!H\sum_j c^\dagger_j c_j\!-\!\frac{1}{2}.
\label{HXXZfermion}
\ee
It is instructive to analyze \r{HXXZfermion} for small values of
$\delta$. Denoting normal ordering by
\be
:\!c^\dagger_jc_j\!:\,=c^\dagger_jc_j-\langle c^\dagger_jc_j\rangle,
\ee
we rewrite \r{HXXZfermion} as
\bea
{\mathcal H}_{\rm  XXZ}&=&J\sum_{j}\frac{1}{2}\left[
c^\dagger_{j}c_{j+1}+c^\dagger_{j+1}c_{j}\right]
+\delta :\!c^\dagger_jc_j\!: :\!c^\dagger_{j+1}c_{j+1}\!:\nn
&&-\left[H+J\delta (1-2\langle c^\dagger_jc_j
\rangle)\right]\sum_j c^\dagger_j c_j +{\rm const}.
\label{HXXZfermion2}
\eea
Neglecting the interaction term, we may diagonalize the quadratic part
of \r{HXXZfermion2} by Fourier transform and determine the fermion
number self-consistently. For small $\delta$ the low-energy degrees of
freedom are then found to occur in the vicinity of $\pm k_F^0$, where
\be
k_F^0a_0={\rm  arccos}\Bigl(\frac{H}{J}\Bigr)
\left[1-\frac{2\delta}{\pi\sqrt{1-H^2/J^2}}\right]+
\frac{\delta}{\sqrt{1-H^2/J^2}}+{\mathcal O}(\delta^2).
\ee
Hence the continuum limit of
\r{HXXZfermion2} is obtained by keeping only Fourier modes with
$k\approx \pm k_F^0$, i.e. 
\be
c_j\longrightarrow\sqrt{a_0}\left[e^{-ik_F^0x}R(x)+e^{ik_F^0x}L(x)\right].
\label{ctoRL}
\ee
Substituting \r{ctoRL} into \r{HXXZfermion2} we arrive at the
low-energy fermion Hamiltonian
\bea
{\mathcal H}&=&v_F\int dx \left[-iR^\dagger\partial_x R
+iL^\dagger\partial_xL\right]\nn
&&+Ja_0\delta\int dx \ (R^\dagger R+L^\dagger L)
(R^\dagger R+L^\dagger L)\nn
&&+Ja_0\delta\int dx \ (R^\dagger L e^{2ik_F^0x}+{\rm h.c.}
)(R^\dagger Le^{2ik_F^0(x+a_0)}+{\rm h.c.}).
\label{spinlessfermion}
\eea
Here $v_F=Ja_0\sin(k_F^0a_0)$ and point splitting and normal ordering are
implicit. Finally we may bosonize \r{spinlessfermion} using
\be
R^\dagger(x)=\frac{1}{\sqrt{2\pi}}
\exp\left(-\frac{i}{\sqrt{2}}\varphi'(x)\right)\ ,\quad
L^\dagger(x)=\frac{1}{\sqrt{2\pi}}
\exp\left(\frac{i}{\sqrt{2}}\bar{\varphi}'(x)\right),
\label{RLFF}
\ee
where we choose the chiral bosons $\varphi$ and $\bar\varphi$ to
be normalized such that
\bea
\left\langle
\exp\left(i\alpha\varphi'(\tau,x)\right)\ 
\exp\left(-i\alpha\varphi'(0)\right)\right\rangle
&=&\frac{1}{(v\tau-ix)^{2\alpha^2}}\,,\nn
\left\langle
\exp\left(i\alpha\bar\varphi'(\tau,x)\right)\ 
\exp\left(-i\alpha\bar\varphi'(0)\right)\right\rangle
&=&\frac{1}{(v\tau+ix)^{2\alpha^2}}\ .
\label{normalizationcond}
\eea
The bosonized form of \r{spinlessfermion} is 
\bea
{\mathcal H}&=&\frac{v_F}{16\pi}\int dx \left[
(1+\frac{4\delta\sin(k_F^0a_0)}{\pi})(\partial_x\Phi')^2
+(\partial_x\Theta')^2\right]\nn
&+&\frac{J\delta\sin(2k_F^0a_0)}{\pi^2\sqrt{8}}
\int dx\ \partial_x\Phi'
-g\int dx\ \cos\left(\sqrt{2}\Phi'-4k_F^0[x-\frac{a_0}{2}]\right),\nn
\label{Hbosonprime}
\eea
where $g=J\delta a_0^3/2\pi^2$ 
and the scalar field $\Phi'$ and its dual $\Theta'$ are
defined as 
\be
\Phi'(x)=\varphi'(x)+\bar\varphi'(x)\ ,\quad
\Theta'(x)=\varphi'(x)-\bar\varphi'(x).
\ee
We emphasize that the normalization condition \r{normalizationcond}
implies that the operator $\cos(\sqrt{2}\Phi')$ has dimension 
${\rm length}^{-4}$ and hence the units in Eq.\,\r{Hbosonprime} are
correct. The third term in \r{Hbosonprime} is important only in the
vicinity of the isotropic limit $\delta\to 1$ in zero magnetic
field. In this limit it is marginally irrelevant and gives rise to
logarithmic corrections. These have been calculated by renormalization
group improved perturbation theory in Refs.
\cite{Affleck98,LukyanovXXZ,oa2,BarzykinAffleck,Barzykin00a,LukTer}. The second term 
in \r{Hbosonprime} is removed by shifting the field $\Phi'$
\be
\Phi'\longrightarrow\Phi'-\frac{4\sqrt{2}\delta\cos(k_F^0a_0)}{\pi
  a_0}\,x\,. 
\ee
It follows from \r{RLFF} and \r{ctoRL} that the effect of this shift
is to change the Fermi wave number to
\be
k_F=k_F^0+2\frac{\delta H}{\pi Ja_0}\ .
\ee

\begin{figure}[ht]
\begin{center}
\epsfxsize=0.7\textwidth
\epsfbox{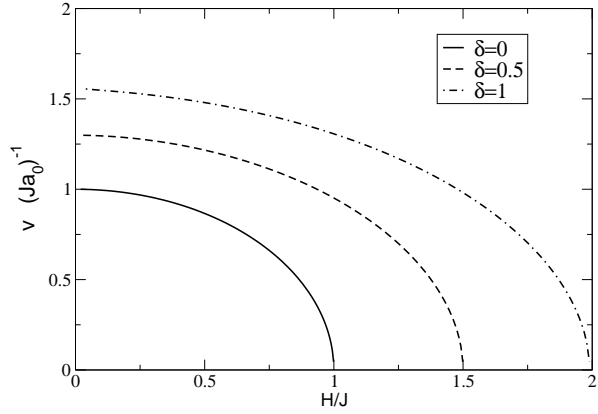}\quad
\end{center}
\vskip -0.5cm
\caption{Spin velocity as a function of magnetic field for different
values of $\delta$. Note that the spin velocity goes to zero as the
magnetic field approaches $H_c=J(1+\delta)$.}
\label{fig:vf}
\end{figure}

\begin{figure}[ht]
\begin{center}
\epsfxsize=0.7\textwidth
\epsfbox{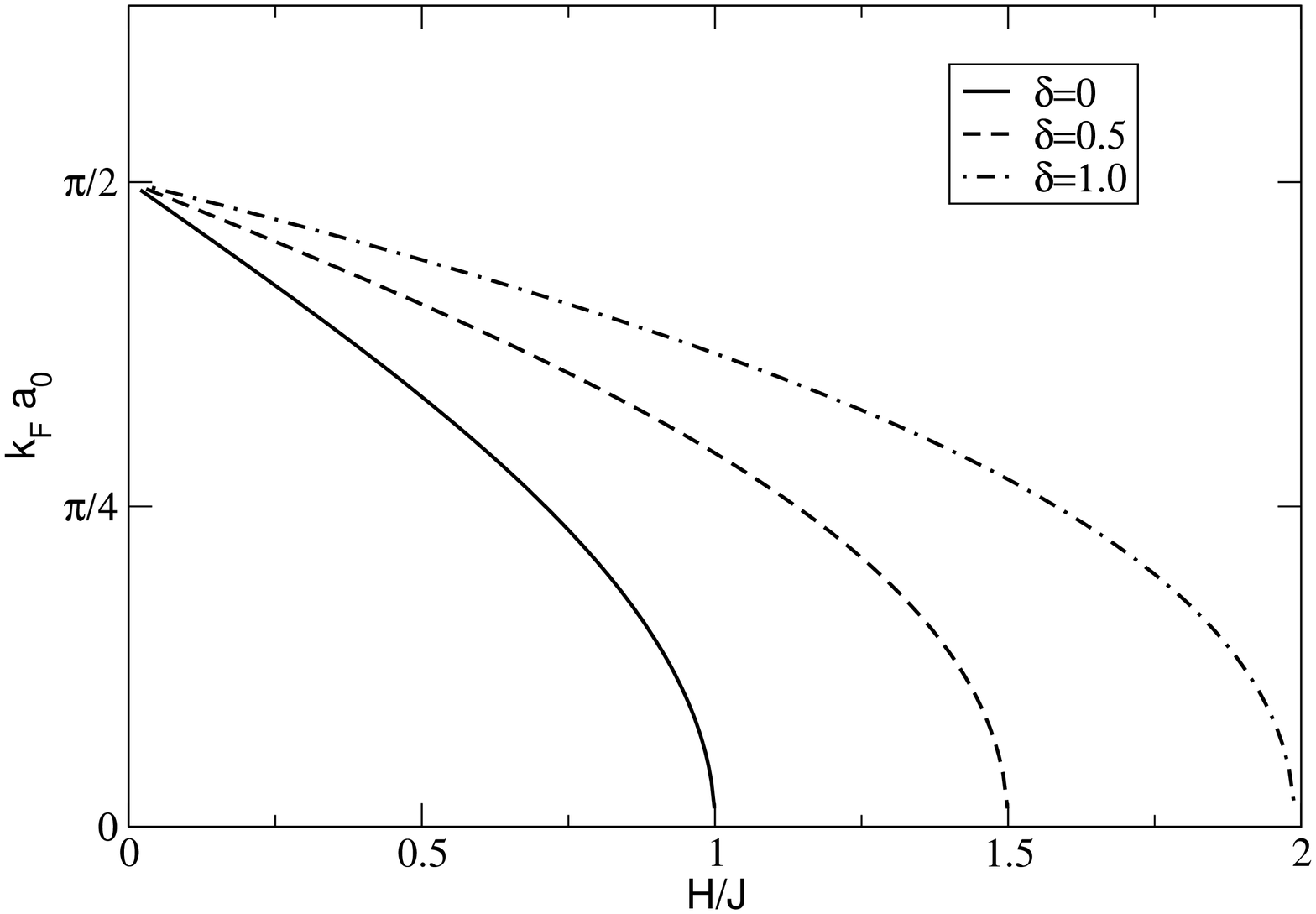}\quad
\end{center}
\vskip -0.5cm
\caption{``Fermi momentum'' $k_F$  as a function of magnetic field for
different values of $\delta$. We note that $k_F$ goes to zero for
$H\to H_c$.}
\label{fig:kf}
\vskip -0.5cm
\begin{center}
\epsfxsize=0.7\textwidth
\epsfbox{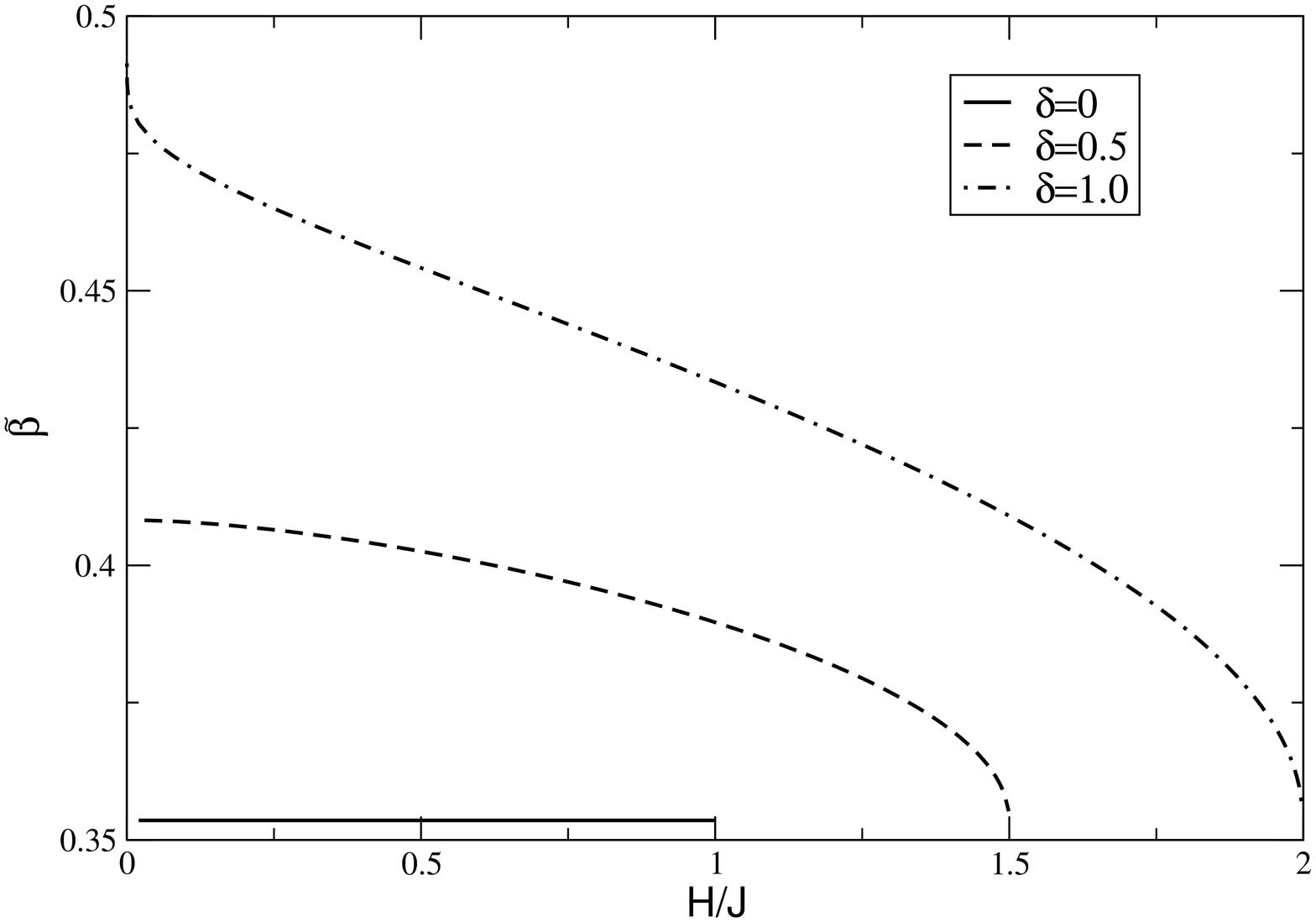}\quad
\end{center}
\vskip -0.5cm
\caption{Parameter $\bt$ as a function of magnetic field for different
  values of $\delta$. We note that
$\tilde{\beta}\to\frac{1}{\sqrt{8}}$ for $H\to H_c$.}
\label{fig:beta}
\end{figure}

The canonical transformation
\bea
\Phi'&=&\left[1+\frac{4\delta\sin(k_F^0a_0)}{\pi}\right]^{-\frac{1}{4}}\Phi
=\frac{1}{\sqrt{8}\bt}\Phi+{\mathcal O}(\delta^2)\ ,\nn
\Theta'&=&
\left[1+\frac{4\delta\sin(k_F^0a_0)}{\pi}\right]^\frac{1}{4}\Theta
=\bt\sqrt{8}\Theta+{\mathcal O}(\delta^2),
\label{rescaling}
\eea
brings the Hamiltonian to the standard Gaussian form
\bea
{\mathcal H}&=&\frac{v}{16\pi}\int dx \left[
(\partial_x\Phi)^2+(\partial_x\Theta)^2\right],
\label{GaussianModel}
\eea
where
\be
v=v_F\left(1+\frac{2\delta\sin(k_Fa_0)}{\pi}\right)
+{\mathcal O}(\delta^2).
\ee
Here we have made use of the fact that we are working in the
small$-\delta$ limit and all formulas are valid only to lowest order
in $\delta$. Importantly, the field $\Theta'$ and hence also $\Theta$
is compactified, i.e.   
\be
\Theta(x)\equiv\Theta(x)+\frac{2\pi}{\bt}\ .
\label{compactTheta}
\ee
This is a consequence of the U(1) symmetry 
$c_j\longrightarrow e^{i\alpha}c_j$ of the fermion Hamiltonian
\r{HXXZfermion}, under which $\Phi'$ is invariant
whereas $\Theta'$ transforms as $\Theta'\longrightarrow
\Theta'+2\sqrt{2}\alpha$. The identification \r{compactTheta} 
follows by setting $\alpha=2\pi$ and using that the corresponding
$U(1)$ transformation is the identity. Similarly, the field $\Phi'$
and hence $\Phi$ is compactified 
\be
\Phi\equiv \Phi+8\pi\bt.
\label{compactPhi}
\ee
It turns out that the above considerations can be generalized to the
entire range \r{Delta} of $\delta$. This may be done by comparing the
${\mathcal O}(L^{-1})$ spectrum of the Heisenberg Hamiltonian
calculated directly from the Bethe ansatz with the spectrum of the
compactified boson \r{GaussianModel}. They are found to agree if the
compactification radii fulfill \r{compactPhi}, \r{compactTheta} and
\r{vbeta} and the velocity $v$ and Fermi wave number $k_F$ are chosen
as follows \cite{vladb}. Let us introduce a dressed energy $\eps(\lambda)$,
dressed momentum $p(\lambda)$, dressed density $\rho(\lambda)$ and
dressed charge $Z(\lambda)$ through the Fredholm integral equations
\bea
\eps(\lambda)&-&\int_{-A}^A \frac{d\mu}{2\pi}\
K(\lambda-\mu)\ \eps(\mu) = H-\frac{J\sin^2\gamma}{\cosh 2\lambda
-\cos\gamma}\ ,\nn
p(\lambda)&=&\frac{2\pi}{a_0}\int_0^\lambda d\mu\ \rho(\mu)\ ,\nn
\rho(\lambda)&-&\int_{-A}^A \frac{d\mu}{2\pi}\
K(\lambda-\mu)\ \rho(\mu) = \frac{2\sin\gamma}{2\pi[\cosh 2\lambda
-\cos\gamma]}\ ,\nn 
Z(\lambda)&-&\int_{-A}^A \frac{d\mu}{2\pi}\
K(\lambda-\mu)\ Z(\mu) = 1\ ,
\label{inteqs}
\eea
where $\delta=\cos(\gamma)$ and the integral kernel is given by 
\be
K(\lambda)=-2\sin 2\gamma/(\cosh2\lambda -\cos 2\gamma)\,.
\ee
The integration boundary $A$ is fixed by the condition
\be
\eps(\pm A)=0\ .
\ee
The physical meaning of the various quantities is as follows:
$\eps(\lambda)$ and $p(\lambda)$ are the energy and momentum of an
elementary ``spinon'' excitation carrying spin $S^z=\pm\frac{1}{2}$.
We note that spinons can only be excited in pairs. The magnetization
per site in the ground state is given in terms of the ground state
root density $\rho(\lambda)$ as
\be
\langle S^z_j\rangle=\frac{1}{2}-\int_{-A}^Ad\lambda\ \rho(\lambda)
\ee
The Fermi momentum is equal to
\be
k_F=p(A)=\frac{2\pi}{a_0}\int_0^A d\lambda\ \rho(\lambda)
=\frac{\pi}{a_0}\left[\frac{1}{2}-\langle S^z_j\rangle\right],
\ee
where we have used that $\rho(-\lambda)=\rho(\lambda)$.
The spin velocity is equal to the derivative of the spinon energy
with respect to the momentum at the Fermi points
\be
v=\frac{\partial\epsilon(\lambda)}{\partial
p(\lambda)}\bigg|_{\lambda=A}=\frac{\partial\epsilon(\lambda)/\partial\lambda}
{2\pi\rho(\lambda)}\bigg|_{\lambda=A} a_0\ .
\label{velocity}
\ee
Finally, the dressed charge is related to $\bt$ by
\be
\bt= \frac{1}{\sqrt{8}Z(A)}\ .
\label{beta}
\ee
In order to determine $v$ and $\bt$ we solve \r{inteqs}
numerically, which is easily done to very high precision as the
equations are linear. The results are shown in Fig.\,\ref{fig:vf},
\ref{fig:kf} and \ref{fig:beta}.
\vskip -0.5cm
For zero magnetic field we have $k_F=\pi/2a_0$ and
\bea
\bt^2=\frac{1}{4\pi}\,{\rm arccos}(-\delta)\ ,\quad
v=\frac{Ja_0}{2}\,\frac{\sin4\pi\bt^2}{1-4\bt^2}\,.
\label{vbeta}
\eea
For later convenience we define a dimensionless spin velocity by
\be
\tilde{v}=\frac{v}{Ja_0}\,.
\label{vtilde}
\ee
In what range of energies do we expect the continuum limit to
provide a good description of the physics of the underlying lattice
model? In zero field the answer is simply that the energy at which we
probe the system must be small compared to the cutoff of the continuum
theory, which is equal to the exchange integral $J$. In the presence
of a magnetic field, the cutoff is actually provided by $H$ rather
than $J$.

\subsubsection{Operators and their Normalizations}
For small values of $\delta$ the lattice spin operators are related to
the scalar fields $\Phi$, $\Theta$ through formulas \r{JW}, \r{ctoRL},
\r{RLFF} and \r{rescaling}. A straightforward calculation gives
\bea
S^z_j&\sim&\frac{a_0}{8\pi\bt}\,\partial_x\Phi-a(H)
a_0^\frac{1}{8\bt^2} 
\sin\left(\frac{\Phi}{4\bt}-2k_Fx\right)+\ldots\ ,\nn
S^+_j&\sim&
{\cal  A}(H)a_0^\frac{16\bt^2+\bt^{-2}}{8}(-1)^j\bigg\{
e^{2ik_Fx}\ e^{-i(\bt\Theta+\frac{1}{4\bt}\Phi)}
+e^{-2ik_Fx}\ e^{-i(\bt\Theta-\frac{1}{4\bt}\Phi)}\biggr\}\nn
&&+(-1)^jc(H)a_0^{2\bt^2}\ e^{-i\bt\Theta}+\ldots \ .
\label{Sboso}
\eea
The expressions \r{Sboso} turn out to hold for $-1<\delta\leq 1$ and
the dots indicate operators with larger scaling dimensions, which
contribute to the subleading terms in the large distance asymptotics
of the spin correlation functions. In order to obtain the bosonized
expression for $S^+_j$ it is necessary to symmetrize the Jordan--Wigner
string operator as \cite{Affleck89b}
\be
e^{-i\pi\sum_{k<j}c^\dagger_jc_j}=\frac{1}{2}\left[
e^{-i\pi\sum_{k<j}:c^\dagger_jc_j:+\langle c^\dagger_jc_j\rangle}
+{\rm h.c.}\right],
\ee
and use that $\langle c^\dagger_j c_j\rangle=1-\frac{k_Fa_0}{\pi}$.

Interestingly, the coefficients $a(H=0)$, $c(H=0)$, ${\cal A}(H=0)$
are known exactly \cite{LukyanovXXZ,LukTer}:
\bea
I(a,b,c)&=&\!\!
\left[\frac{\Gamma\bigl(\frac{\eta}{2-2\eta}\bigr)}
{2\sqrt{\pi}\Gamma\bigl(\frac{1}{2-2\eta}\bigr)}\right]^a
\!\!\exp\Bigl[\int_0^\infty\!\!\frac{dt}{2t}\left[
\frac{\sinh(bt)}{\sinh(ct)\cosh([1-\eta]t)}
-\frac{b}{c}e^{-2t}\right]\Bigr],\nn
a(0)&=&\frac{2}{\pi}
I\Bigl(\frac{1}{2\eta},{2\eta-1},\eta\Bigr)\label{a0}\ ,\\
c(0)&=&\frac{1}{2(1-\eta)}
I\Bigl(\frac{\eta}{2},\eta,-1\Bigr)
\label{c0}\ ,\\
{\cal A}(0)&=&\frac{\sqrt{\pi}}{2}a(0)c(0)
\frac{\Gamma\bigl(1+\frac{\eta}{2-2\eta}\bigr)}
{\Gamma\bigl(\frac{3}{2}+\frac{\eta}{2-2\eta}\bigr)},
\label{A0}
\eea
where $\eta=4\tilde{\beta}^2$. We note that the limit
$\tilde{\beta}\to\frac{1}{2}$, which corresponds to the 
isotropic spin-$\frac{1}{2}$ chain $\delta\to 1$ is singular. This is
because in this limit a marginally irrelevant interaction of spin
currents is present in the Hamiltonian, which gives rise to
logarithmic corrections in spin correlation functions. The isotropic 
limit $\delta\to 1$ is discussed in detail in Refs.
\cite{Affleck98,LukyanovXXZ}.  

For finite magnetic fields the amplitudes have been determined
numerically in \cite{HF01,EFH03}. In Table \ref{tab:Amp} we list the
results for the case of the isotropic spin-$\frac{1}{2}$ chain
($\delta=1$ in \r{HXXZ}).
\begin{table}
\label{tab:Amp}
\tbl{
Amplitudes $a$ and $c$, dimensionless spin velocity $\tilde{v}$,
coupling $\bt$ and magnetic field $H$ as functions of the
magnetization $m$ for the isotropic Heisenberg chain ($\delta=1$). The
amplitudes are determined numerically  except for $m = 0.5$ where
exact values are shown. The figures in parentheses for $a$ and $c$
indicate the error  on the last quoted digits.
}{\normalsize
\begin{tabular}{c|c|c|c|c|c}
$m$ & $a$ & $c$ & $\tilde{v}$ & $\bt$ & $H/J$ \\
\hline
  0.02 &  0.591(3)  &   0.4937(3) &  1.54271 &  0.46879 &  0.17599 \\
  0.04 &  0.550(5)  &   0.4883(2) &  1.51707 &  0.46095 &  0.34214 \\
  0.06 &  0.520(4)  &   0.4863(2) &  1.48415 &  0.45427 &  0.50013 \\
  0.08 &  0.4947(6) &   0.4853(2) &  1.44425 &  0.44812 &  0.65001 \\
  0.10 &  0.475(1)  &   0.4847(2) &  1.39796 &  0.44229 &  0.79164 \\
  0.12 &  0.454(1)  &   0.4842(2) &  1.34593 &  0.43670 &  0.92489 \\
  0.14 &  0.437(2)  &   0.4835(2) &  1.28879 &  0.43129 &  1.04965 \\
  0.16 &  0.422(2)  &   0.4825(2) &  1.22720 &  0.42604 &  1.16589 \\
  0.18 &  0.4070(7) &   0.4810(2) &  1.16178 &  0.42094 &  1.27360 \\
  0.20 &  0.3938(8) &   0.4790(2) &  1.09314 &  0.41597 &  1.37287 \\
  0.22 &  0.3813(6) &   0.4764(2) &  1.02184 &  0.41112 &  1.46380 \\
  0.24 &  0.3700(8) &   0.4731(2) &  0.94844 &  0.40639 &  1.54656 \\
  0.26 &  0.3596(7) &   0.4690(2) &  0.87347 &  0.40177 &  1.62134 \\
  0.28 &  0.3499(4) &   0.4639(2) &  0.79741 &  0.39725 &  1.68839 \\
  0.30 &  0.3406(4) &   0.4578(2) &  0.72074 &  0.39284 &  1.74794 \\
  0.32 &  0.3330(2) &   0.4504(2) &  0.64387 &  0.38852 &  1.80030 \\
  0.34 &  0.3262(2) &   0.4416(2) &  0.56722 &  0.38430 &  1.84575 \\
  0.36 &  0.3200(3) &   0.4310(2) &  0.49116 &  0.38017 &  1.88462 \\
  0.38 &  0.3145(4) &   0.4183(2) &  0.41602 &  0.37612 &  1.91723 \\
  0.40 &  0.3094(2) &   0.4029(1) &  0.34212 &  0.37216 &  1.94390 \\
  0.42 &  0.3070(8) &   0.3841(1) &  0.26973 &  0.36828 &  1.96497 \\
  0.44 &  0.3058(2) &   0.3601(1) &  0.19912 &  0.36449 &  1.98079 \\
  0.46 &  0.3062(6) &   0.3284(1) &  0.13049 &  0.36077 &  1.99168 \\
  0.48 &  0.309(1)  &   0.2802(1) &  0.06407 &  0.35712 &  1.99797 \\
  0.50 &  0.3183    &   0         &  0       &  0.35355 &  2       \\
\end{tabular}}
\end{table}

\subsubsection{Dynamical Spin Correlation Functions}
It is now a straightforward exercise\,\footnote{\,We exclude the limit
  $H\to 0$, $\delta\to 1$ from our discussion, see Refs.
  \cite{Affleck98,LukyanovXXZ} for a discussion of the latter case.} 
to determine the dynamical spin
correlation functions at low energies, i.e. in the regions of the
$(\omega,k)$-plane marked in Fig.\,\ref{fig:transH}. Using the
bosonization dictionary \r{Sboso} we find in imaginary time
($\tau=it$, $x=ja_0$) 
\bea
&&\langle T_\tau\ S^+_{j+1}(\tau)\ S^-_1(0)\rangle\sim\nn
&&\quad{\cal A}^2(H)a_0^{4\bt^2+\frac{1}{4\bt^2}}(-1)^j\biggl\{
e^{2ik_Fx}\bigl\langle T_\tau\ 
e^{-i\bigl(\bt\Theta+\frac{1}{4\bt}\Phi\bigr)(\tau,x)}
e^{i\bigl(\bt\Theta+\frac{1}{4\bt}\Phi\bigr)(0,0)}
\bigr\rangle\nn
&&\hskip3.5cm +e^{-2ik_Fx}\bigl\langle T_\tau\ 
e^{-i\bigl(\bt\Theta-\frac{1}{4\bt}\Phi\bigr)(\tau,x)}
e^{i\bigl(\bt\Theta-\frac{1}{4\bt}\Phi\bigr)(0,0)}
\bigr\rangle\biggr\}\nn
&&\quad+c^2(H)(-1)^ja_0^{4\bt^2}\bigl\langle T_\tau\ 
e^{-i\bt\Theta(\tau,x)}\ e^{i\bt\Theta(0,0)}\bigr\rangle\nn
&&={\cal A}^2(H)(-1)^j
\left[\frac{a_0^2}{x^2+v^2\tau^2}\right]^{2\bt^2+1/8\bt^2} 
\left\{e^{2ik_Fx}\frac{v\tau+ix}{v\tau-ix}+{\rm h.c.}
\right\}\nn
&&\quad+c^2(H)(-1)^j\left[\frac{a_0^2}{x^2+v^2\tau^2}\right]^{2\bt^2}\ ,
\eea
\bea
\langle T_\tau\ S^z_{j+1}(\tau)\ S^z_1(0)\rangle&\sim&
\left[\frac{a_0}{8\pi\bt}\right]^2\langle T_\tau\ 
\partial_x\Phi(\tau,x)\ \partial_x\Phi(0,0)\rangle\nn
&+&a^2(H)a_0^\frac{1}{4\bt^2}
\bigl\langle T_\tau\ 
\sin\bigl(\frac{\Phi(\tau,x)}{4\bt}-2k_Fx\bigr)
\sin\bigl(\frac{\Phi(0,0)}{4\bt}\bigr)\bigr\rangle\nn
&=&\frac{1}{(8\pi\bt)^2}\frac{2a_0^2}{x^2+v^2\tau^2}
\left[\frac{v\tau+ix}{v\tau-ix}+{\rm h.c.}
\right]\nn
&&+\frac{a^2(H)\cos(2k_Fx)}{2}
\left[\frac{a_0^2}{x^2+v^2\tau^2}\right]^\frac{1}{8\bt^2}.
\eea
We see that the slowest decay of correlations occurs with an
oscillating factor $(-1)^j$ in the transverse correlations and with
factors $\exp(\pm 2ik_Fx)$ in the longitudinal ones. Concomitantly 
their Fourier transforms will be most singular at these wave numbers,
which in turn makes $k\approx \pm 2k_F$ and $k\approx\frac{\pi}{a_0}$
the most interesting regions in the Brillouin zone from an
experimental point of view.  
The inelastic neutron scattering intensity (at zero temperature) is
proportional to 
\be
I(\omega,{\bm k})\propto \sum_{\alpha,\gamma}
\left(\delta_{\alpha\gamma}-\frac{k_\alpha k_\gamma}{{\bm k}^2}
\right)\ S^{\alpha\gamma}(\omega,{\bm k})\ ,
\label{intensity}
\ee
where $\alpha, \gamma = x,y,z$ and the dynamical structure factor
(DSF) $S^{\alpha\gamma}$ is defined by
\begin{eqnarray}
S^{\alpha \gamma} (\omega, k)&=& -\frac{1}{\pi}\ {\rm Im}\
\chi^{\alpha\gamma}(\omega,k)\ ,\\
\chi^{\alpha\gamma}(\omega,k)&=&-i
\sum_{l}\int_{0}^{\infty} dt\ e^{-i kla_0 + i \omega t}
\langle [S^{\alpha}_{l+1} (t), S^{\gamma}_{1} (0)]\rangle\,.
\label{Dsus}
\end{eqnarray}
Here $k$ denotes the component of $\bm{k}$ along the chain direction
and $\chi^{\alpha\gamma}(\omega,k)$ are the components of the dynamical
susceptibility. We note that in \r{intensity} only the symmetric
combinations 
$
S^{(\alpha \gamma)} (\omega, k)=
S^{\alpha \gamma} (\omega, k)+
S^{\gamma\alpha} (\omega, k)
$
enter. Using a spectral representation one may show that
for positive frequencies $\omega>0$
\be
S^{(\alpha \gamma)} (\omega, k)=
\sum_{l}\int_{-\infty}^{\infty} \frac{dt}{2\pi}\ 
e^{-i kla_0 + i \omega t}
\left[\langle S^{\alpha}_{l+1} (t) S^{\gamma}_{1} (0)
\rangle+
\langle S^{\gamma}_{l+1} (t) S^{\alpha}_{1} (0)\rangle\right].
\ee
We evaluate the retarded correlator in \r{Dsus} by Fourier
transforming the corresponding time-ordered correlation function in
imaginary time and then analytically continuing to real frequencies.
We arrive at the following results for the DSF
in the vicinity of $2k_F$ and $\pi/a_0$ respectively
\bea
S^{\rm xx}(\omega,\frac{\pi}{a_0}+q)&=&
S^{\rm yy}(\omega,\frac{\pi}{a_0}+q)=\frac{c^2(H)}{2}\
F\bigl(\omega^2-v^2q^2,2\bt^2\bigr)\ ,\nn
S^{\rm zz}(\omega,2k_F+q)&=&\frac{a^2(H)}{4}\
F\Bigl(\omega^2-v^2q^2,\frac{1}{8\bt^2}\Bigr)\ ,\nn
F(s^2,\alpha)&=&\frac{\Gamma(1-\alpha)}{\Gamma(\alpha)}\,
\frac{\sin(\pi\alpha)}{2\tilde{v}J}
\left[\frac{(2\tilde{v}J)^2}{s^2}\right]^{1-\alpha}.
\label{DSFXXZ}
\eea
We see that the components of the DSF exhibit power-law
singularities. The exponents vary with the applied field and also
depend on the exchange anisotropy. However, the transverse components
are always more singular than the longitudinal one.

\subsection{Massive Perturbations and Window of Applicability of
  MIQFT}
As discussed above, the spin-$\frac{1}{2}$ Heisenberg chain is {\sl
quantum critical} in the sense that the elementary spinon excitations
are gapless and spin correlation functions decay as power laws.
In what follows we consider several perturbations of the Heisenberg
chain, in which excitations have a gap and spin correlation functions
decay exponentially with distance. In particular, we will determine
how the DSF is changed compared to the critical form \r{DSFXXZ}. One
issue to keep in mind that the field theory approach we employ has a
limited window of applicability. Field theory becomes exact in
particular scaling limits of the underlying lattice models. However,
in experimentally relevant situations one usually is at some distance
in parameter space from the scaling limit and a practical criterion is
needed to judge, whether field theory will provide good approximations
to the results for the underlying lattice model. A simple such
criterion is that the spectral gap should be small compared to the
cutoff set by the lattice model. The latter is simply ${\rm min(H,J)}$
for the Heisenberg chain in a magnetic field.  Another possible criterion is to demand
that the low energy excitations are relativistic up to some energy scale
beyond that of concern (see, for example, Ref. \cite{anis}).

\subsection{Field-Induced Gap Problem}
\label{FIGP}
In many materials with low-symmetry crystal structure the coupling of
a uniform magnetic field to spin degrees of freedom is of a tensorial
nature. In the case where there are two spins per unit cell the
$g$-tensor has generally both uniform and staggered components (see
e.g. \cite{oshima76,oshima78})
\be
{\cal H}_{\rm mag}=-\mu_B H_a\ \sum_j [g^u_{ab}+(-1)^j g^s_{ab}]S^b_j\,.
\ee
A low crystal symmetry also frequently results in the presence of a 
Dzyaloshinskii--Moriya (DM) \cite{Dzyaloshinskii58,Moriya60}
interaction  
\be
H_{\rm DM}=\sum_j \ {\bf D}_j\cdot\left({\bf S}_{j-1}\times {\bf
  S}_j\right). 
\ee
The direction of the DM vector ${\bf D}_j$ is constrained by the
crystal symmetries. Of particular interest is the case when ${\bf
  D}_j$ has a staggered component, i.e. ${\bf   D}_j=(-1)^j{\bf D}$.
It was shown by Oshikawa and Affleck in \cite{oa} that in this case
the application of a uniform magnetic field ${\bf H}$ leads to the
generation of a staggered field perpendicular to the direction of
${\bf H}$. The effects of having a staggered component of the
$g$-tensor and a staggered DM-interaction are captured by the
following model \cite{oa}
\bea
{\mathcal H}=\sum_jJ{\bf S}_{j}\cdot{\bf S}_{j+1}-HS^z_j-h(-1)^jS^x_j\ .
\label{StaggField}
\eea
Here the staggered field is proportional to the uniform field
\be
h=\gamma H\ , 
\label{hH}
\ee
where $\gamma\ll 1$ depends on the details of the
material under investigation. The model \r{StaggField} has been applied
successfully to experiments on several quasi-1D spin-$\frac{1}{2}$
Heisenberg magnets: 
Copper Benzoate [${\rm Cu(C_6D_5COO)_2\cdot3D_2O}$]
\cite{magn,dender,Asano00,Ajiro00,Asano02},
CDC [${\rm CuCl_2\cdot 2((CD_3)_2SO)}$]\cite{Kenzelmann},
Copper-Pyrimidine [$({\rm PM\cdot Cu(NO_3)_2\cdot(H_2O)_2})_n$
(PM=pyrimidine)] \cite{feyer,Wolter03a,Wolter03b}
and ${\rm Yb_4As_3}$ \cite{Oshikawa99,kohgi}.
Theoretical calculations have been carried out for the excitation
spectrum \cite{oa,ET97,Lou02}, the dynamical structure factor
\cite{ET97,EFH03}, the specific heat \cite{Essler99}, the magnetic
susceptibility \cite{oa2,Wolter03b} and the electron-spin resonance
lineshape \cite{OshikawaAffleck02}. 

Bosonizing the Hamiltonian \r{StaggField} by means of the identities
\r{Sboso} yields a sine-Gordon model \cite{oa}
\bea
{\mathcal H}=\frac{v}{16\pi}\int dx\left[\left(\partial_x\Phi\right)^2
+\left(\partial_x\Theta\right)^2\right]-\mu(h)\int dx \cos(\beta\Theta)\,.
\label{SGM}
\eea
where in our normalizations $\mu(h)$ is a dimensionful quantity
\be
\mu(h)\simeq h\ c(H)\ a_0^{2\beta^2-1}.
\ee
In our approach the velocity $v$ and the parameter $\beta$ in \r{SGM}
are determined for the lattice Hamiltonian \r{StaggField} with
$h=0$. This is a good approximation as long as $h$ is small compared
to $H$ and $J$. For given $J$ and $H$, $\beta$ is then calculated from
the Bethe ansatz solution of the Heisenberg model in a uniform field,
see Eq.\,\r{beta} and Fig.\,\ref{fig:beta}, and
\be
\beta\equiv\bt\ .
\ee
Similarly, the spin velocity is given by \r{velocity} and
Fig.\,\ref{fig:vf}. The parameters entering the effective SGM are
summarized in Table \ref{tab:Amp}.

\subsubsection{Spectrum of the SGM and Quantum Numbers}
It is useful to define a parameter
\be
\xi=\frac{\beta^2}{1-\beta^2}\,.
\ee
The spectrum of the sine-Gordon model \r{SGM} in the relevant range of
$\beta$ consists of a soliton-antisoliton doublet and several
soliton-antisoliton bound states called ``breathers.'' There are
altogether $[1/\xi]$ breathers, where $[x]$ denotes the integer part
of $x$.
The breather gaps \cite{Dashen75b,FaddeevKorepin78} can be determined
for example from the Bethe ansatz solution of the massive Thirring
model \cite{Korepin79,Bergknoff} and are given by 
\be
\Delta_n=2\Delta\sin\left(\frac{\pi\xi n}{2}\right),\quad
 n=1,\ldots,
\left[\frac{1}{\xi}\right]. 
\label{breathermass}
\ee
Here $\Delta$ is the gap for soliton and antisoliton. The parameter
$\xi$ depends on the applied uniform field $H$ through $\beta$.
We introduce labels $B_n$ for the n$^{\rm th}$ breather, $s$ for the
soliton and $\bar{s}$ for the antisoliton. A convenient basis for the
SGM \r{SGM} can be constructed in terms of scattering states of
(anti)solitons and breathers by means of the Faddeev--Zamolodchikov
algebra (see Section 2.2). The action of the charge conjugation
operator $C$ on these basis states states follows from 
\bea
C|0\rangle&=&|0\rangle\, ,\nonumber\\[1mm]
CA_s^\dagger(\theta)C^{-1}&=&A^\dagger_{\bar{s}}(\theta)\,,\nonumber\\[1mm]
CA^\dagger_{B_n}(\theta)C^{-1}&=&(-1)^nA^\dagger_{B_n}(\theta)\,.
\label{chargeconj}
\eea
The topological charge
\be
Q=\frac{\beta}{2\pi}\int_{-\infty}^\infty dx\ \partial_x\Theta\, ,
\label{Q}
\ee
is a conserved quantity in the sine-Gordon model. We will use
conventions in which soliton and antisoliton have topological charges
$-1$ and $1$ respectively. Breathers have topological charge zero and
so are neutral.

The soliton gap as a function of the parameters $H$ and $h$ of the
underlying spin chain was determined in Ref. \cite{oa2} in the regime
$\Delta\ll H$, where 
\bea
\frac{\Delta}{J}\simeq\left(\frac{h}{J}\right)^{(1+\xi)/2}
\left[B \left(\frac{J}{H}\right)^{\frac{1}{2}-2\beta^2}
\left(2-8\beta^2\right)^{1/4}\right]^{-(1+\xi)/2},
\label{deltaRG}
\eea
with $B=0.422169$. Equation \r{deltaRG} is applicable as long as $H$
is sufficiently smaller than $J$ or more precisely as long as the
magnetization is small. In the derivation of \r{deltaRG} both the
magnetic field dependences of spin velocity and the normalization
constants $a$ and $c$ of the spin operators in \r{Sboso} have been
neglected, which is justified in weak fields. An additional
complication in weak fields is the presence of a only slightly
irrelevant perturbation to the free bosonic effective Hamiltonian
describing the low energy physics of the Heisenberg chain (see 
Eq.\,\r{Hbosonprime}). By taking this perturbation into account,
renormalization group improved perturbation theory essentially leads
to the result \r{deltaRG}.

\vskip -.5cm
\begin{figure}
\begin{center}
\noindent
\epsfxsize=0.7\textwidth
\epsfbox{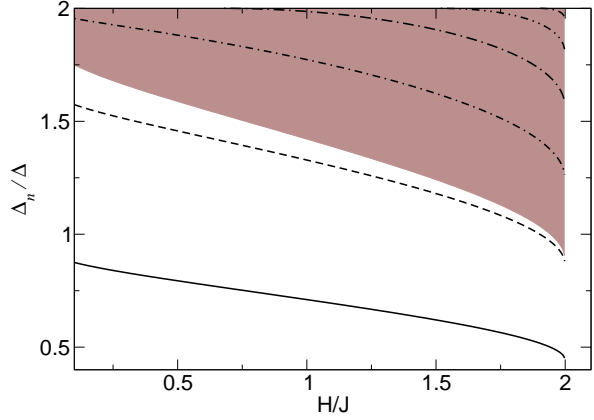}
\end{center}
\vskip -4mm
\caption{\label{fig:breatherM}%
Breather gaps $\Delta_n$ is units of the soliton gap $\Delta$ as
functions of the magnetic field. We note that $\Delta$ itself depends
on $H$ as well. The shaded area is the $B_1$-$B_1$ two-breather
continuum. As the field is increased more breathers split off from
the soliton-antisoliton continuum with threshold $2\Delta$.
}
\end{figure}

For magnetic fields comparable to $J$ it is necessary to take into
account the magnetic field dependences of the spin velocity and the
normalization $c(H)$ of the spin operator, whereas the effects of the
irrelevant perturbations to the free boson Hamiltonian may be
neglected. Using the results of Ref. \cite{zam} we obtain the
following expression for the gap in the regime of $H$ comparable to
$J$ (but still $h\ll J$)
\bea
\frac{\Delta}{J}&\simeq&\frac{2\tilde{v}(H)}{\sqrt{\pi}}\,
\frac{\Gamma(\frac{\xi}{2})}
{\Gamma(\frac{1+\xi}{2})}
\left[\frac{c(H)\pi}{2\tilde{v}(H)}\,
\frac{\Gamma(\frac{1}{1+\xi})}{\Gamma(\frac{\xi}{1+\xi})}\,
\frac{h}{J}\right]^{(1+\xi)/2} .
\eea
Here  $\tilde{v}=v/(Ja_0)$ is the ``dimensionless spin velocity''.

\subsubsection{Thermodynamics}

\begin{figure}[ht]
\begin{center}
\noindent
\epsfxsize=0.75\textwidth
\epsfbox{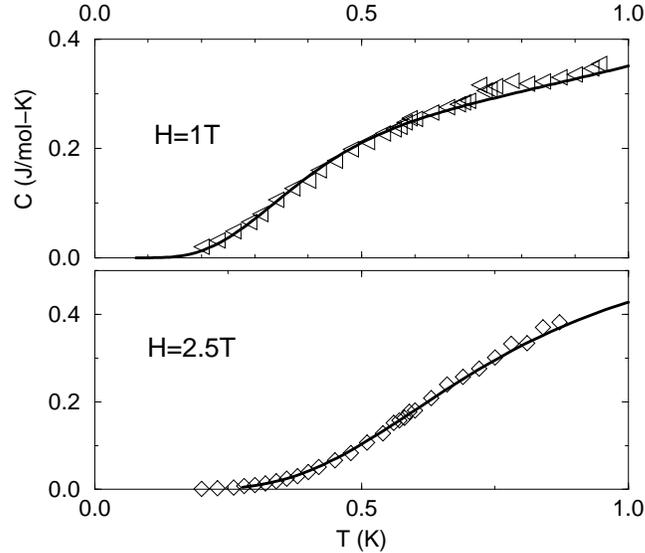}
\end{center}
\vskip -7mm 
\caption{\label{fig:caxis1}%
Specific heat as a function of temperature for fields of
H=1T and H=2.5T applied along the $c^{\prime\prime}$ axis.}
\end{figure}

The thermodynamics of the SGM is most efficiently studied \cite{ddv95}
{\sl via} the recently developed Thermal Bethe Ansatz approach
\cite{Suzuki85,Koma87,Taka90,Klumper91,ddv92}, which circumvents
problems associated with solving the infinite number of coupled
nonlinear integral equations that emerge in the standard approach
based on the string hypothesis \cite{Fowler81,Fowler82} (note that the
coupling constant $\beta$ in our problem is a continuously varying
quantity and no truncation to a finite number of coupled equations is
possible). It was shown in \cite{ddv95} that the free energy of the
SGM can be expressed in terms of the solution of a single nonlinear
integral equation for the complex quantity $\eps(\theta)$ (we set
$k_B=1$) 
\bea 
\eps(\theta) &=& -i \frac{\Delta}{T}\sinh(\theta+i\eta^\prime)
- \int_{-\infty}^\infty d\theta^\prime
G_0(\theta-\theta^\prime) 
\ln\left(1+\exp\left[-\eps(\theta^\prime)\right]\right)\nn
&&+\int_{-\infty}^\infty d\theta^\prime
G_0(\theta-\theta^\prime+2i\eta^\prime) 
\ln\left(1+\exp\left[-\bar{\eps}(\theta^\prime)\right]\right),
\label{inteq}
\eea
where $\Delta$ is the soliton mass and
\bea
G_0(\theta) &=&\int_0^\infty \frac{d\omega }{\pi^2}
\frac{\cos(2\omega\theta/\pi)
\sinh(\omega(\xi-1))}{\sinh(\omega\xi)\cosh(\omega)}\ .
\eea
The free energy density is given by
\bea
f(T)&=&-\frac{T\Delta}{\pi v}\ {\rm Im}
\int_{-\infty}^\infty d\theta\ 
\sinh(\theta+i\eta^\prime) \ln\left[1+e^{-\eps(\theta)}\right].
\eea
As we are interested in the attractive regime of the SGM we have
\be
0<\eta^\prime<\pi\xi/2\ .
\label{etap}
\ee
Note that the free energy does not depend on the value of
$\eta^\prime$ as long as it is chosen in the interval \r{etap}.
The set \r{inteq} of two coupled nonlinear integral equations is
solved by iteration. For $T\to 0$ the first iterations can be
calculated analytically and the corresponding contributions to the
free energy density are seen to be of the form
\bea
f(T)&\sim& -\frac{2T\Delta}{\pi v}\sum_{n=1}^\infty 
\frac{(-1)^{n+1}}{n}K_1\left(\frac{n\Delta}{T}\right)
-\frac{T\Delta_1}{\pi v}K_1\left(\frac{\Delta_1}{T}\right)+\ldots\ ,~~~
\eea
where $\Delta_1=2\Delta\sin\frac{\pi\xi}{2}$ is the mass of the first
breather and $K_1$ is a modified Bessel function. The first term is the
contribution of soliton-antisoliton scattering states to the free
energy, whereas the second term is the contribution of the first
breather. Both terms have the form characteristic of massive
relativistic bosons. The contributions of the heavier breathers are
found in higher orders of the iterative procedure employed in solving
\r{inteq}. The specific heat is obtained from the free energy
\be
C=T\,\frac{\partial^2f(T)}{\partial T^2}\ .
\label{specH}
\ee
At low temperatures it is found to be of the form
\bea
\hspace{-8mm}
C&\approx& \sum_{\alpha=0}^{[1/\xi]}
\frac{(1+\delta_{\alpha 0})\Delta_\alpha}
{\sqrt{2\pi}v}\left[1+\frac{T}{\Delta_\alpha}+
\frac{3}{4}\left(\frac{T}{\Delta_\alpha}\right)^2\right]
\left(\frac{\Delta_\alpha}{T}\right)^{\frac{3}{2}}
\exp\left( -\frac{\Delta_\alpha}{T}\right),~~~
\eea
where $\Delta_0\equiv\Delta$. In order to compare
theoretical predictions based on the SGM with the specific heat data
of \cite{dender} we need the free energy at ``intermediate''
temperatures and thus have to resort to a numerical solution of
\r{inteq} by iteration. A detailed comparison of the specific heat
calculated in the framework of the low-energy effective sine-Gordon
theory to experimental data on Copper Benzoate was carried out in
Ref. \cite{Essler99}. In Fig.\,\ref{fig:caxis1} we show a fit of the
specific heat calculated from \r{specH} \cite{Essler99} to data taken
for a particular orientation of the magnetic field on Copper Benzoate
\cite{dender}.  
\vskip -4mm
\vskip -7mm

\subsubsection{Normalization of Form Factors in the Sine-Gordon Model}
The short-distance behavior of correlation function in the
sine-Gordon model is governed by the operator product expansion
(in imaginary time)
\be
\lim_{x^2+v^2\tau^2\to 0}\ 
e^{i\alpha\Phi(\tau,x)}\ e^{i\alpha'\Phi(0,0)}
= |x^2+v^2\tau^2|^{2\alpha\alpha'}\ e^{i(\alpha+\alpha')\Phi(0,0)}. 
\label{opesg}
\ee
Given the normalization implied by \r{opesg}
the large-distance asymptotics is\,\footnote{\,Here we do not consider
  the connected correlation function.}
\be
\lim_{x^2+v^2\tau^2\to \infty}\ 
\bigl\langle e^{i\alpha\Phi(\tau,x)}\ e^{i\alpha'\Phi(0,0)}\bigr\rangle
= \bigl\langle e^{i\alpha\Phi}\bigr\rangle\ \bigl\langle
e^{i\alpha'\Phi}\bigr\rangle\, .
\label{LDAsg}
\ee
where the one-point functions have been determined in \cite{LZ}
\bea
{\cal G}_\alpha&=&a_0^{2\alpha^2}\langle e^{i\alpha\Theta(0,0)}\rangle=
\left(\frac{\Delta}{J\tilde{v}}\frac{\sqrt{\pi}}{2}
\frac{\Gamma\left((1+\xi)/2\right)}{\Gamma(\xi/2)}
\right)^{2\alpha^2}\nonumber\\[1mm]
&\times&
\exp\left[\int_0^\infty\frac{dt}{t}\left(\frac{\sinh^2(\alpha\beta t)}
{2\sinh(\frac{\beta^2}{2} t)\sinh(\frac{t}{2})\cosh[\frac{1-\beta^2}{2}t]}
 -2\alpha^2e^{-t}\right)\right].\quad
\label{Galpha}
\eea
Equation \r{LDAsg} can be used to fix the overall constant factor in the form
factor expansion of correlation functions.
\subsubsection{Dynamical Structure Factor}
\label{DSF}
\begin{figure}
\begin{center}
\noindent
\epsfxsize=0.6\textwidth
\epsfbox{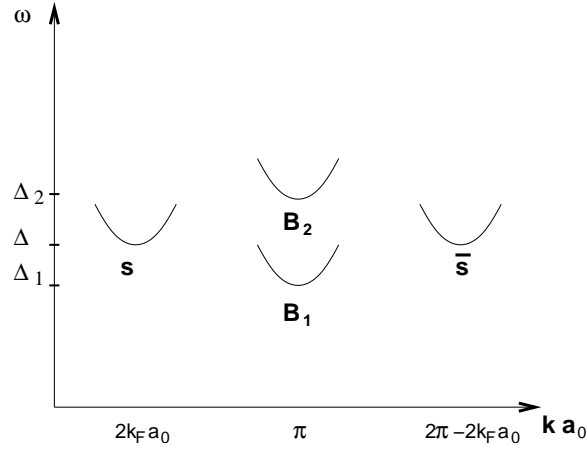}
\end{center}
\caption{\label{fig:modes}%
Schematic structure of the lowest-energy excited states relevant to
neutron scattering experiments. Soliton and antisoliton occur in the
vicinity of the incommensurate wave numbers $2k_F$,
$2(\frac{\pi}{a_0}-k_F)$ and are seen in $S^{\rm zz}$,  whereas the
first breather $B_1$ occurs in the vicinity of $\pi/a_0$ and
contributes to $S^{\rm yy}$. At higher energies further breather bound
states as well as multi-particle scattering continua are present.}  
\end{figure}

Let us now turn to the issue how the staggered field modifies the
dynamical structure factor. Without a staggered field one simply has
the result \r{DSFXXZ}. At low energies the field theory description
can be used to determine the dynamical spin-spin correlation
functions. We substitute the decomposition \r{Sboso} into the
expression \r{Dsus} for the dynamical susceptibility. Neglecting the
contributions of rapidly oscillating terms to the integrals, we find
that 
\bea
\chi^{\rm xx}(\omega,\frac{\pi}{a_0}+q)&\simeq&
-ic^2(H)a_0^{4\beta^2-1}\int_{-\infty}^\infty dx\int_0^\infty dt\
e^{i\omega t-iqx}\nn
&&\qquad\qquad\times\
\langle[\cos(\beta\Theta(t,x)),\cos(\beta\Theta(0)]\rangle ,\\[1mm]
\chi^{\rm yy}(\omega,\frac{\pi}{a_0}+q)&\simeq&
-ic^2(H)a_0^{4\beta^2-1}\int_{-\infty}^\infty dx\int_0^\infty dt\
e^{i\omega t-iqx}\nn
&&\qquad\qquad\times\
\langle[\sin(\beta\Theta(t,x)),\sin(\beta\Theta(0)]\rangle ,\\[1mm]
\chi^{\rm zz}(\omega,\pm 2k_F+q)&\simeq&\pm 
\frac{a^2(H)a_0^{\frac{1}{4\beta^2}-1}}{2}
\int_{-\infty}^\infty dx\int_0^\infty dt\
e^{i\omega t-iqx}\nn
&&\quad\quad\times\
\biggl\langle\Big[\exp\left(\mp i\frac{\Phi(t,x)}{4\beta}\right),
\sin\left(\frac{\Phi(0)}{4\beta}\right)\Big]\biggr\rangle .~~~
\label{susceptibilities}
\eea
Using a spectral representation in terms of scattering states of
solitons, antisolitons and breathers \r{eIIii} we have e.g.
\bea
&&\chi^{\rm xx}(\omega,\frac{2\pi}{a_0}+q)= 2\pi c^2(H)a_0^{4\beta^2-1}
\sum_{n=1}^\infty\sum_{\eps_i}\!\int\!
\frac{d\theta_1\ldots d\theta_n}{(2\pi)^nn!}\nn
&&\hskip1cm\times\left|\langle 0|\cos\beta\Theta(0)
|\theta_n,\ldots,\theta_1\rangle_{\eps_n,\ldots,\eps_1}\right|^2\nn
&&\hskip1cm\times\bigg\lbrace
\frac{\delta(q-\sum_j\frac{\Delta_{\epsilon_j}}{v}\sinh\theta_j)}{\omega
  - \sum_j \Delta_{\epsilon_j}\cosh\theta_j +i\epsilon}
-\frac{\delta(q+\sum_j\frac{\Delta_{\epsilon_j}}{v}\sinh\theta_j)}{\omega +
  \sum_j \Delta_{\epsilon_j}\cosh\theta_j +i\epsilon}\bigg\rbrace .~~~
\label{dcos}
\eea
Here the indices $\eps_j\in\{s,\bar{s},B_1,B_2,\ldots,B_{[1/\xi]}\}$
and $\Delta_s=\Delta_{\bar{s}}\equiv\Delta$,
$\Delta_{B_n}\equiv\Delta_n$. Many of the form factors
actually vanish for the operators we are interested in here. Using the
transformation property of the scalar field $\Theta$ under charge
conjugation
\be
C\Theta C^{-1}=-\Theta ,
\ee
we observe see that by virtue of \r{chargeconj}
\bea
\langle 0|\cos(\beta\Theta(0))|\theta\rangle_{B_{2n-1}} &=&0\ ,\nn[1mm]
\langle 0|\sin(\beta\Theta(0))|\theta\rangle_{B_{2n}} &=&0\ ,\nn[1mm]
\langle 0|\sin(\beta\Theta(0))|\theta_2,\theta_1\rangle_{B_n,B_n}
&=&0\ ,
\label{statesxx}
\eea
and so on. On the other hand, the $\pm 2k_F$ components of $S^z$ carry
topological charge $\pm 1$. This follows from their commutators with
the operator \r{Q}
\be
\left[Q,\exp\left(\pm i\frac{n}{4\beta}\Phi\right)\right]=\mp n
\exp\left(\pm i\frac{n}{4\beta}\Phi\right).
\label{QVO}
\ee
As a result the only non-vanishing form factors are of the form
\bea
\hspace{-6mm}
\langle 0|e^{-\frac{i}{4\beta}\Phi(0)}|\theta\rangle_{s} \, ,\
\langle 0|
e^{-\frac{i}{4\beta}\Phi(0)}|\theta_1,\theta_2\rangle_{s,B_m} \, ,\
\langle 0|
e^{-\frac{i}{4\beta}\Phi(0)}|\theta_1,\theta_2,\theta_3\rangle_{s,s,\bar{s}}
\,,
\label{stateszz}
\eea
and so on.
Combining \r{statesxx}, \r{stateszz} and \r{susceptibilities} we can
identify which excited states inelastic neutron scattering
experiments probe at different momentum transfers. This is summarized
in Fig.\,\ref{fig:modes}.

The form factors required to evaluate the first few terms in the
spectral representation \r{dcos} have been determined in
\cite{smirnov86b,smirnov,lukyanov97,babujian99,LukZam01,
babujian02a,babujian02b}. Determining the breather form factors
via the bootstrap axiom \r{eIIlv} one obtains \cite{ET97}\\[-5mm]~
\bea
&&S^{\rm xx}(\omega,\frac{\pi}{a_0}+q)={\cal C}_x(H)\
{\rm Re}\Bigg\{
2\pi\sum_{n=1}Z_{2n}\ \delta(s^2 - \Delta_{2n}^2) \nonumber\\[1mm]
&&\qquad
+\frac{|F^{\cos(\beta\Theta)}_{11}
[\theta(\Delta_1,\Delta_1,s)]|^2}{s\sqrt{s^2 - 4\Delta_1^2}}
+ 2\frac{|F^{\cos(\beta\Theta)}_{+-}
[\theta(\Delta,\Delta,s)]|^2}{s\sqrt{s^2-4\Delta^2}}+
\ldots\Biggr\}, \qquad
\label{cos}
\eea
where 
\be
\theta(m_1,m_2,s) = {\rm
arccosh}\left(\frac{s^2-m_1^2-m_2^2}{2m_1m_2}\right)\ ,
\label{thetamms}
\ee
and the overall normalization is given by
\bea
&&{\cal C}_x(H)=\frac{1}{\pi}\,c^2(H)\tilde{v}(H)J\
{\cal G}_\beta^2\,.
\label{cofH}
\eea
Here ${\cal G}_\beta$ is given by \r{Galpha}. As we are dealing with a
two point function of a Lorentz scalar, the susceptibility depends
only on the Mandelstam variable
\be
s=\sqrt{\omega^2-v^2q^2}.
\label{MandelstamS}
\ee
The first terms in \r{cos}
correspond to single-particle breather states and are given by
\cite{ET97}
\bea
Z_2&=& \frac{2(\sin 2\pi\xi)^2}{\cot\pi\xi}\ 
\exp\left[-2\int_0^\infty\frac{dx}{x}\ 
\frac{(\sinh{2\xi x})^2 \sinh x(1-\xi)}{\cosh x\ \sinh 2x\ \sinh\xi
x}\right] ,
\label{Z2}\\[1mm]
Z_4&=& \frac{2(\sin 4\pi\xi)^2}{(\cot\pi\xi)^2(\cot 3\pi\xi/2)^2\cot(2\pi\xi)}\ \cr\cr
&& \hskip .5in \times\exp\left[-2\int_0^\infty\frac{dx}{x}\ 
\frac{(\sinh{4\xi x})^2 \sinh x(1-\xi)}{\cosh x\ \sinh 2x\ \sinh\xi
x}\right].
\eea
The soliton-antisoliton contribution is
\bea
|F^{\cos(\beta\Theta)}_{+-}(\theta)|^2&=&\frac{(2\cot\pi\xi/2\
\sinh\theta)^2}{\xi^2} \frac{\cosh\theta/\xi+\cos\pi/\xi}
{\cosh 2\theta/\xi-\cos 2\pi/\xi}E(\theta),
\label{fcospm}\\[1mm]
E(\theta)&=&
\exp\left[\int_0^\infty\frac{dx}{x}
\frac{[\cosh 2x \cos2x\theta/\pi -1] \sinh x(\xi-1)}{\cosh x\ \sinh 2x\
\sinh\xi x}\right],~~~~~\quad
\label{Eoftheta}
\eea
and the $B_1B_1$ breather-breather contribution is
\bea
&&\hspace{-1.1cm}|F^{\cos(\beta\Theta)}_{11}(\theta)|^2=
\left[2\cos\frac{\pi\xi}{2}\sqrt{2\sin\frac{\pi\xi}{2}}\right]^4
\exp\left[-4\int_0^{\pi\xi}\frac{dx}{2\pi}\frac{x}{\sin x}\right]\nonumber\\[1mm]
&&\hspace{-1.1cm}\times\frac{(\sinh\theta)^2}{(\sinh\theta)^2+ (\sin\pi\xi)^2}
\exp\left[-4\int_0^\infty\!\frac{dx}{x}
\frac{\cos\frac{2x\theta}{\pi} \sinh\xi x \sinh x(1+\xi)}{
\sinh 2x\cosh x}\right].
\label{fcos11}
\eea
The next most important contribution ({\sl i.e.} the term with the
lowest threshold) in the expansion \r{cos} comes from $B_1B_3$
breather-breather states. It will contribute at energies larger than
$\Delta_1+\Delta_3$, where $\Delta_{1,3}$ are given by \r{breathermass}. 
\begin{figure}[ht]
\vskip -5mm
\begin{center}
(a)
\epsfxsize=0.75\textwidth
\epsfbox{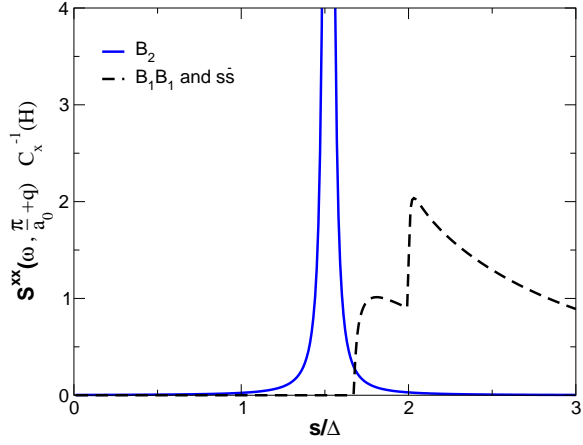}\\
(b)
\epsfxsize=0.75\textwidth
\epsfbox{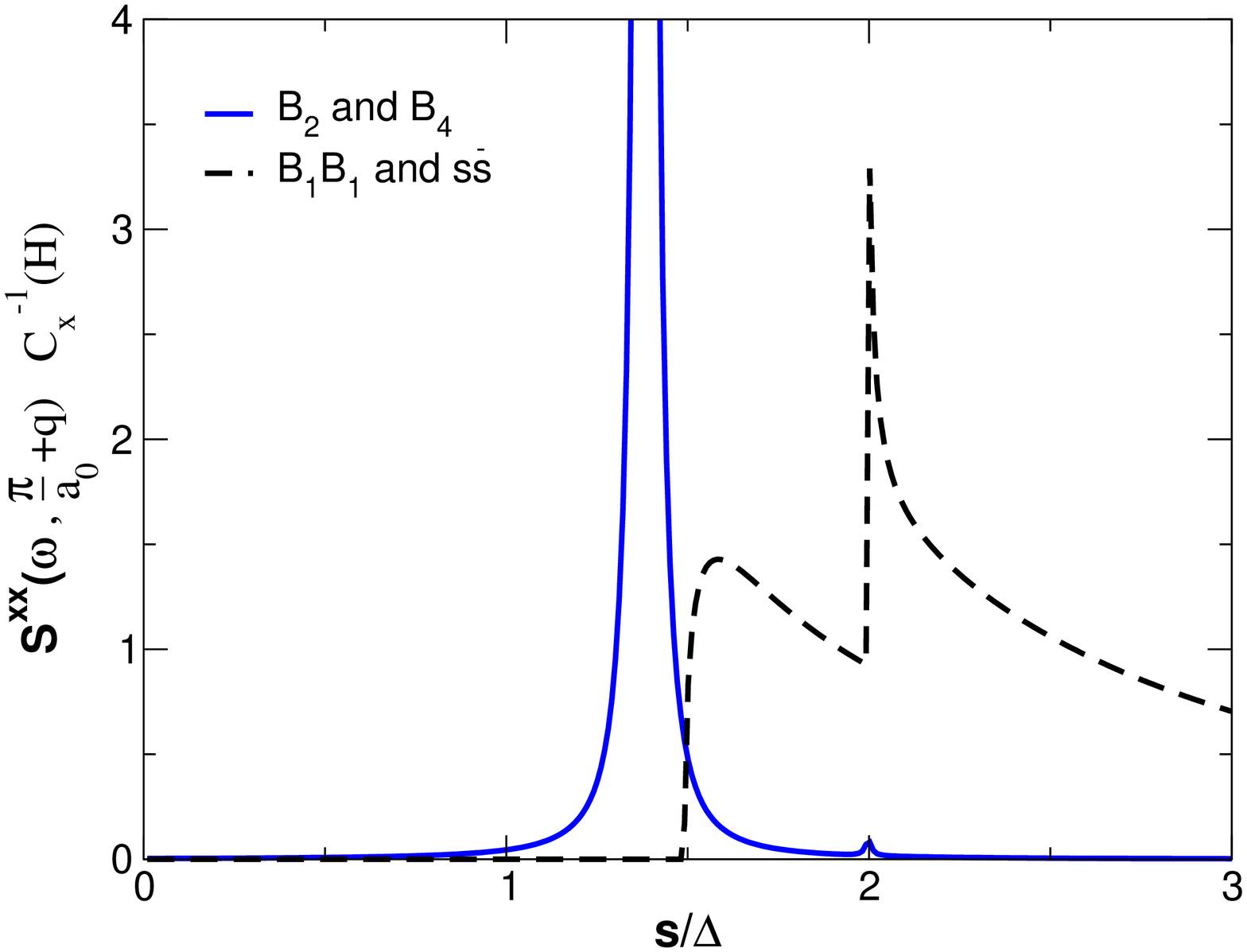}
\end{center}
\caption{{\rm xx} component of the dynamical structure factor for
magnetizations (a) $m=0.03$ and (b) $m=0.1$. The delta-functions
corresponding to the second breather have been broadened by a
Lorentzian to make them visible.}
\label{fig:sxx003}
\end{figure}
In Fig.\,\ref{fig:sxx003} we plot $S^{\rm
xx}(\omega,\frac{\pi}{a_0}+q)$ as a function of
$s=\sqrt{\omega^2+v^2q^2}$ for two values of the magnetization $m$. We
see that the coherent particle peak due to the second breather carries
most of the spectral weight. For $m=0.1$ the contribution due to the
soliton-antisoliton continuum is considerably sharper above the
threshold than for $m=0.03$. This is because the $s\bar{s}$ continuum
has just given birth to the fourth breather. The latter has a very
small binding energy and carries very little spectral weight. In
general the $s\bar{s}$ continuum becomes singular at the threshold,
whenever a new breather peak splits off from it.

\begin{figure}[ht]
\begin{center}
(a)
\epsfxsize=0.75\textwidth
\epsfbox{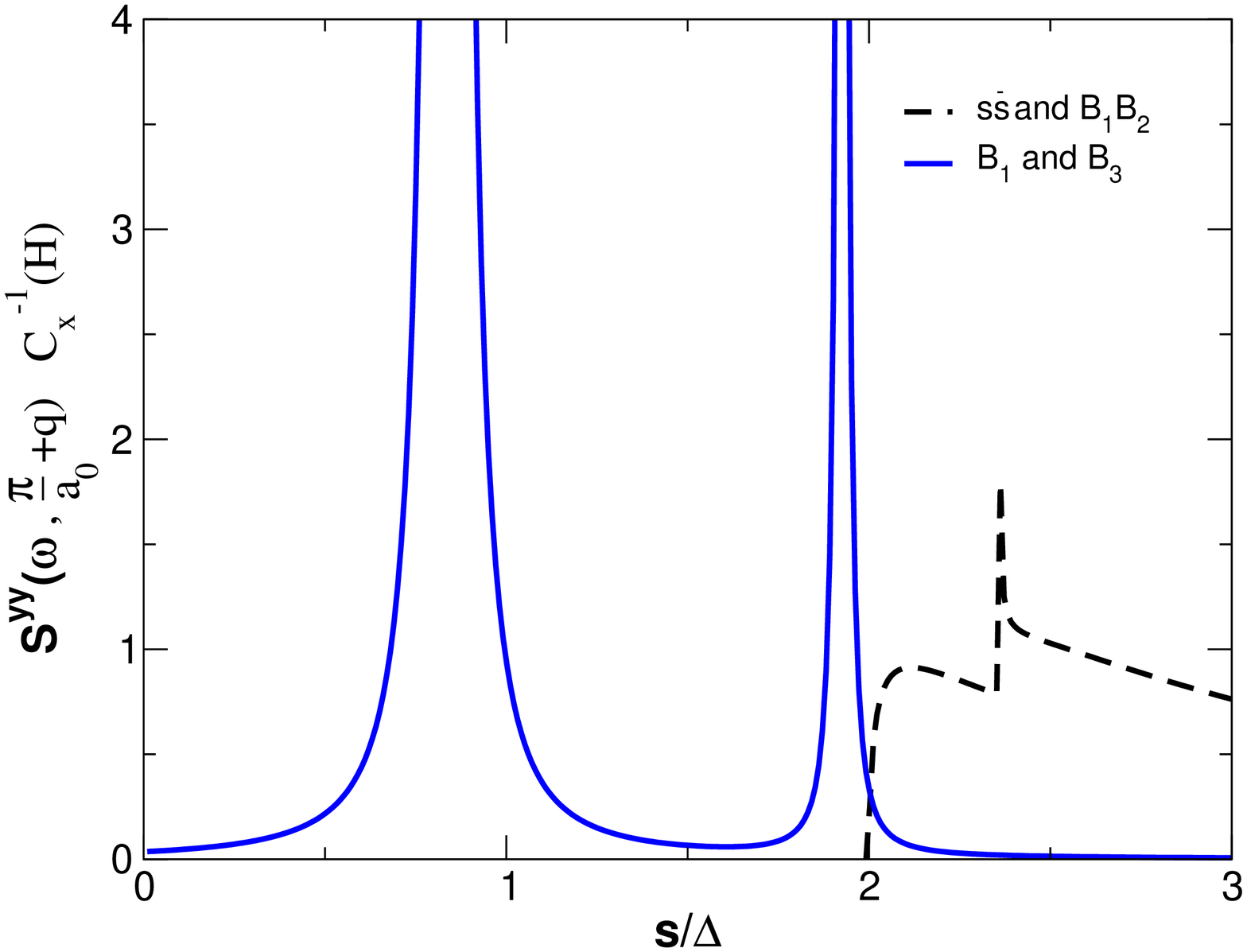}\\
(b)
\epsfxsize=0.75\textwidth
\epsfbox{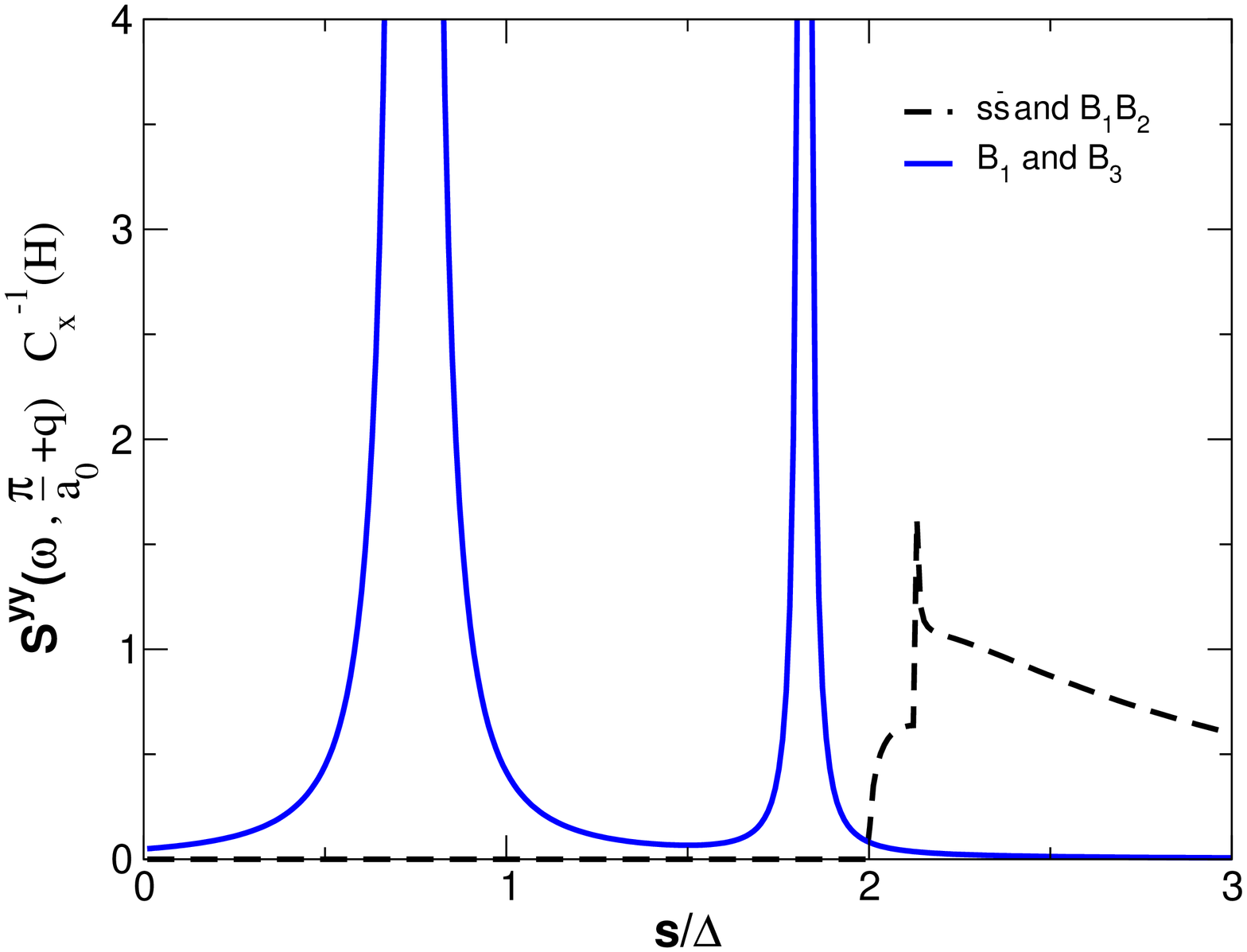}
\end{center}
\caption{{\rm yy} component of the dynamical structure factor for
magnetizations (a) $m=0.03$ and (b) $m=0.1$. The delta-functions
corresponding to the first and third breathers have been broadened by a
Lorentzian to make them visible.}
\label{fig:syy003}
\end{figure}

Let us now turn to the $\rm yy$ component of the dynamical susceptibility.
As the U(1) spin rotational symmetry around the axis of the uniform
field $H$ is broken by the staggered field $h$, $\chi^{\rm yy}$ must
be different from $\chi^{\rm xx}(\omega,q)$. The form factor expansion
yields  
\bea
\hspace{-10mm}&&
S^{\rm yy}(\omega,\frac{\pi}{a_0}+q)=
{\cal C}_x(H)\ {\rm Re}\Bigg\{
2\pi\sum_{n=1}Z_{2n-1}\ \delta(s^2 - \Delta_{2n-1}^2)\nn
\hspace{-10mm}&&~~+\frac{2|F^{\sin(\beta\Theta)}_{+-}
[\theta(\Delta,\Delta,s)]|^2}{s\sqrt{s^2 - 4\Delta^2}}
+\frac{2|F^{\sin(\beta\Theta)}_{12}
[\theta(\Delta_1,\Delta_2,s)]|^2}
{\sqrt{(s^2\!-\!\Delta_1^2\!-\!\Delta_2^2)^2\!-\!4\Delta_1^2\Delta_2^2}}+ 
\ldots\Biggr\}.
\label{sin}
\eea
Here the breather contributions are
\bea
Z_1&=& \frac{8(\cos\frac{\pi\xi}{2})^4}{\cot\frac{\pi\xi}{2}}
\exp\left[-2\int_0^\infty\frac{dx}{x}\ 
\frac{\sinh\xi x\ \sinh x(1-\xi)}{\cosh x\ \sinh 2x}\right] ,
\label{Z1}\\[1mm]
Z_3&=& \frac{4\sin(3\pi\xi) (\sin\frac{3\pi\xi}{2})^2}{(\cot\pi\xi)^2}\cr
&& \hskip .5in \times\exp\left[-2\int_0^\infty\frac{dx}{x}\ 
\frac{(\sinh{3\xi x})^2 \sinh x(1-\xi)}{\cosh x\ \sinh 2x\ \sinh\xi
x}\right].\quad
\eea
The soliton-antisoliton contribution is
\bea
&&|F^{\sin(\beta\Theta)}_{+-}(\theta)|^2=\frac{(2\cot\pi\xi/2\,
\sinh\theta)^2}{\xi^2}\, \frac{\cosh\theta/\xi-\cos\pi/\xi}
{\cosh 2\theta/\xi-\cos 2\pi/\xi}\,E(\theta)\,,~~~
\label{fsinpm}
\eea
where $E(\theta)$ is given by \r{Eoftheta}.
Finally we take into account the $B_1B_2$ breather-breather state
which contributes
\bea
&&|F^{\sin(\beta\Theta)}_{12}(\theta)|^2=\frac{\tan\pi\xi}{2}
\left|\frac{\sinh^2(\theta-i\frac{\pi\xi}{2})}
{\sinh^2(\theta-i\frac{\pi\xi}{2})+\sin^2\pi\xi}\right|^2
\left[2\cos\frac{\pi\xi}{2}\sqrt{2\sin\frac{\pi\xi}{2}}\right]^6\nn
&&\times
\exp\left[-6\int_0^{\pi\xi}\frac{dx}{2\pi}\frac{x}{\sin x}\right]
\left[1+\frac{1}{4\cos
\frac{\pi\xi}{2}(\cosh\theta+\cos\frac{\pi\xi}{2})}\right]^2\nn
&&\times\exp\left[-4\int_0^\infty\frac{dx}{x}
\frac{[\cos\frac{x\theta}{\pi} \sinh \xi x
+\cosh x\xi \sinh\bigl(\frac{\xi x}{2}\bigr)]
\sinh\bigl(x\frac{1+\xi}{2}
\bigr)}{\sinh x \cosh\bigl(\frac{x}{2}\bigr)}\right].\
\label{fsin12}
\eea
The next most important contribution to \r{sin} is due to $B_1B_1B_1$ 
three breather states with a threshold at $3\Delta_1$. In
Fig.\,\ref{fig:syy003} we plot $S^{\rm yy}(\omega,\frac{\pi}{a_0}+q)$
as a function of $s=\sqrt{\omega^2+v^2q^2}$ for two values of the
magnetization $m$. The two coherent single-particle peaks due to the first
and third breathers carry most of the spectral weight. The incoherent
scattering continua are always weak and become less pronounced as the
magnetization increases.

The leading contributions to the longitudinal structure factor at the
incommensurate wave numbers $2k_F$, $\frac{2\pi}{a_0}-2k_F$ are due to
soliton and antisoliton single-particle states. The form factor 
$\langle 0|e^{-\frac{i}{4\beta}\Phi(0)}|\theta\rangle_{s}$ is a
rapidity-independent constant (as the operator is a Lorentz scalar),
which has been determined by Lukyanov and Zamolodchikov
Ref.\cite{LukZam01}. A simple calculation then gives \cite{EFH03}
\bea
S^{\rm zz}\left(\omega,2k_F+q\right)&=&
{\cal C}_z(H)\ 2\pi\ {\cal Z}_1\ \delta\left(\omega^2-(vq)^2-\Delta^2\right),
\eea
where $
{\cal C}_z(H)=\frac{\tilde{v}J}{4\pi}\ a^2(H)\ ,$
\bea
{\cal Z}_n&=&\left(\frac{{\cal C}_2}{2{\cal C}_1^2}\right)^\frac{n}{2}
\left(\frac{16}{\xi{\cal C}_2}\right)^\frac{n^2}{4}
\left(\frac{\sqrt{\pi}\ \Delta\Gamma\left(\frac{3}{2}+\frac{\xi}{2}\right)}
{J\tilde{v}\Gamma\left(\frac{\xi}{2}\right)}\right)^\frac{n^2}{4\beta^2}\nn
&\times&\exp\biggl[
\int_0^\infty\frac{dt}{t}\biggl(\frac{\exp[-(1+\xi)nt]-1}
{2\sinh(\xi t)\sinh[(1+\xi)t]\cosh(t)}
 +\frac{n}{2\sinh(t\xi)}-\frac{n^2e^{-2t}}{4\beta^2}
\biggr)\biggr].\nn
\label{Zn}
\eea
Here the constants ${\cal C}_{1,2}$ are given by \cite{LukZam01}
\bea
{\cal C}_1&=&\exp\left(-\int_0^\infty \frac{dt}{t}
\frac{\sinh^2(t/2)\ \sinh[t(\xi-1)]}{\sinh(2t)\ \sinh(\xi t)
\ \cosh(t)}\right),\label{C1}\\[1mm]
{\cal C}_2&=&\exp\left(4\int_0^\infty \frac{dt}{t}
\frac{\sinh^2(t/2)\ \sinh[t(\xi-1)]}{\sinh(2t)\ \sinh(\xi t)}\right).
\eea
As was pointed out in Ref.\cite{oa2}, at $H=0$ the low-energy
effective theory of \r{StaggField} is SU(2) symmetric. In our notations
this implies that
\bea
\lim_{H\to 0}\frac{2\pi{\cal C}_xZ_1}{c^2(H)} 
= 2\lim_{H\to 0}\frac{2\pi{\cal C}_z{\cal Z}_1}{a^2(H)}\ . 
\label{su2}
\eea
Equation \r{su2} is easily verified numerically. 

\subsection{Transverse-Field Model}
Some quasi-1D spin-$\frac{1}{2}$ antiferromagnets are described by
Heisenberg models with exchange anisotropy. One example is
${\rm Cs_2CoCl_4}$ \cite{Kenzelmann02}, where the Co spin-$\frac{3}{2}$
multiplets are split by a strong single-ion anisotropy. Projecting to
the low-lying spin-$\frac{1}{2}$ doublet leads to a Hamiltonian of the
form \r{transF} with $\delta=\frac{1}{4}$ and $H=0$. 
Anisotropic Heisenberg models can also be used to describe
spin chains with Dzyaloshinskii--Moriya
\cite{Dzyaloshinskii58,Moriya60} interaction. The bulk DM interaction
can be removed by a gauge transformation (local rotation of the
coordinate system in spin space); the resulting Hamiltonian has an
exchange anisotropy and twisted boundary conditions (in a ring
geometry)\cite{Barry68,Alcaraz89}.

When applying a magnetic field to an anisotropic Heisenberg model we
have to distinguish between two cases: 
\begin{itemize}
\item[(i)] the magnetic field is along the direction singled out by
  the anisotropy and leaves the spin rotational $U(1)$ symmetry of the
  Hamiltonian unchanged;
\item[(ii)] the magnetic field is at an angle to the direction singled
  out by the anisotropy and breaks the spin rotational $U(1)$ symmetry of the
  Hamiltonian.
\end{itemize}
In case (i) the model remains exactly solvable and has been discussed
in section \ref{XXZinfield}. The simplest realization of case (ii) is
when the field is perpendicular to the anisotropy
\bea
{\mathcal H}=\sum_jJ[S^x_j S^x_{j+1}+S^y_j S^y_{j+1}+
\delta S^z_j S^z_{j+1}]-HS^x_j\ ,
\label{transF}
\eea
where $\delta< 1$. The model \r{transF} has been studied in Refs.
\cite{Dmitriev02a,Dmitriev02b,CEL,capraro}. At low fields there are
both uniform and staggered ordered moments and excitations are
gapped. At a critical field $H_c(\delta)$ there is quantum phase
transition in the universality class of the 2D Ising model, at which
the staggered magnetization vanishes. In high fields only the uniform
magnetization remains and excitations are again gapped. 
The simplest case to deal with by field theory methods is the one
where the magnetic field $H$ is much stronger than the anisotropy
$J(1-\delta)$, but is smaller than $H_c(\delta)$. The field theory
limit in this case has been studied in Refs.
\cite{Essler99,OshikawaAffleck02,CEL}. It is useful to change
coordinate system and rewrite the Hamiltonian \r{transF} as  
\bea
{\cal H}_{\rm ZXX,H}=\sum_{j}J
{\bf S}_j\cdot{\bf S}_{j+1}-H S^z_j
+J(\delta-1)\sum_j S^y_jS^y_{j+1}\equiv{\cal H}_0+{\cal H}_1\, .~~~~
\label{hxxz}
\eea
Our strategy in the regime $H\gg J(1-\delta)$ is to bosonize at the
point $\delta=1$ at a finite value for the magnetization per site
and then to treat the exchange anisotropy ${\cal H}_1$ as a
perturbation. The continuum limit of ${\cal H}_0$ is constructed in
section \r{contlimit} and is given by a compactified boson
\r{GaussianModel}. The perturbing Hamiltonian ${\cal H}_1$ can then be 
bosonized using \r{Sboso}. Fusion of the y-components of the staggered
magnetizations gives a contribution
\bea
{\cal O}_j=S^y_jS^y_{j+1}\longrightarrow {\cal C}\, a_0^{8\bt^2}\cos(2\bt\Theta)
+\ldots \ .
\label{normalisation}
\eea
In addition there is a small marginal contribution that shifts the
compactification radius. For simplicity we neglect it here. 
Putting everything together we conclude that at low energies compared
to the scale set by the applied field $H<2J$, the effective
Hamiltonian is given by a SGM
\bea
{\cal
H}=\frac{v}{16\pi}[(\partial_x\Phi)^2+(\partial_x\Theta)^2]
-\mu(\delta)\ \cos(\beta\Theta)\ ,
\label{SGM2}
\eea
where 
\be
\beta=2\bt\ .
\label{bbtilde}
\ee
The cosine term in the SGM is relevant and
generates a spectral gap. As $\beta>\frac{1}{\sqrt{2}}$ (see
Fig.\,\ref{fig:beta}) the SGM is in the {\sl repulsive} regime and the
the spectrum consists of soliton and antisoliton only. No breather
bound states exist. The magnetic field dependence enters both via the
prefactor and via the $H$-dependence of $\beta$. In order to calculate
the prefactor as well as quantities like the magnetization we need to
know the normalization ${\cal C}$ in \r{normalisation} in the
Heisenberg chain in a field, i.e. the Hamiltonian \r{hxxz} with
$\delta=1$. In Ref. \cite{HF04} ${\cal C}$ was estimated numerically
from the large distance asymptotics of an appropriately chosen
four-point function and found to be very small. An independent method
for determining ${\cal C}$ from the staggered magnetization was
suggested in \cite{CEL}. The gap is given by 
\bea
\frac{\Delta}{J}&=&\frac{2\tilde{v}}{\sqrt{\pi}}
\frac{\Gamma\left(\frac{\beta^2}{2-2\beta^2}\right)}
{\Gamma\left(\frac{1}{2-2\beta^2}\right)}
\left(
\frac{(1-\delta){\cal C}\pi}{2\tilde{v}}
\frac{\Gamma\left(1-\beta^2\right)}{\Gamma\left(\beta^2\right)}
\right)^\frac{1}{2-2\beta^2}\ ,
\eea
where $\tilde{v}$ is the dimensionless spin velocity
$\tilde{v}=\frac{v}{Ja_0}$. As we have mentioned before, in the low
field phase the staggered magnetization in x-direction
is nonzero in the presence of a transverse field. One has
\bea
\langle(-1)^n S^x_n\rangle=c(H)\, \bigl\langle\cos\bigl(
\frac{\beta}{2}\Theta\bigr)\bigr\rangle =c(H){\cal G}_\frac{\beta}{2}\,,
\label{SGMSMag}
\eea
where ${\cal G}_\frac{\beta}{2}$ is given by \r{Galpha} (which in turn depends
upon $\frac{\Delta}{J}$).  Knowledge of the staggered magnetization as a function of $H$
(say from numerics) then will yield ${\cal C}$.

Next we turn to the low-energy behavior of dynamical correlation
functions. We start with the transverse correlations and concentrate on
momenta close to $\pi/a_0$. The staggered components of the spin
operators are given by \r{Sboso} and calculating their correlation
functions reduces to the calculation of appropriate correlators in the 
SGM \r{SGM2}. The leading contribution to the yy-component of the
dynamical structure factor is due to soliton-antisoliton two-particle
intermediate states. One finds \cite{CEL}
\bea
S^{\rm yy}(\omega,\frac{\pi}{a_0}+q)&=&\frac{\tilde{v}Jc^2(H)
{\cal G}_{\beta/2}^2}
{\pi\xi^2\Delta^2}\,\frac{\sqrt{s^2-4\Delta^2}}{s}\,
\frac{E(\theta(\Delta,\Delta,s))\theta_{H}\left(\frac{s}{\Delta}-2\right)
}{\cosh\bigl(\frac{\theta(\Delta,\Delta,s)}{\xi}
\bigr)+\cos\bigl(\frac{\pi}{\xi}\bigr)}\nonumber\\[1mm]
&&+\,{\rm contributions\ from\ 4,6,\ldots\ particles.} 
\eea
where $\theta_H(x)$ is the Heaviside function, $\theta(\Delta,\Delta,s)$,
${\cal G}_\alpha$ and $E(\theta)$ are given by\r{thetamms},
\r{Galpha} and \r{Eoftheta} respectively and $s=\sqrt{\omega^2-v^2q^2}$.
The analogous result for $S^{\rm xx}(\omega,\frac{\pi}{a_0}+q)$ is 
\bea
S^{\rm xx}(\omega,\frac{\pi}{a_0}+q)&=&\frac{\tilde{v}Jc^2(H)
{\cal G}_{\beta/2}^2}
{\pi\xi^2\Delta^2}\,\frac{\sqrt{s^2-4\Delta^2}}{s}\,
\frac{E(\theta(\Delta,\Delta,s))\theta_{H}\left(\frac{s}{\Delta}-2\right)
}{\cosh\bigl(\frac{\theta(\Delta,\Delta,s)}{\xi}
\bigr)-\cos\bigl(\frac{\pi}{\xi}\bigr)}\nonumber\\[1mm]
&&+\,{\rm contributions\ from\ 4,6,\ldots\ particles.} 
\eea
Finally we determine the longitudinal structure factor
$S^{\rm zz}(\omega,k)$ in the vicinity of $k=2k_F$ and
$k=\frac{2\pi}{a_0}-2k_F$. Recalling that the sine-Gordon coupling
constant is given by \r{bbtilde}, we see from the bosonized
expressions for the lattice spin operators \r{Sboso} that we need
to consider two-point functions of
$\exp\left(\pm i\frac{1}{2\beta}\Phi\right)$
in the SGM \r{SGM2}. By \r{QVO} these operators have topological
charge $\mp 2$ respectively. Hence the leading contribution to the
zz-component of the dynamical structure factor is due to intermediate
states with two solitons or two antisolitons. The corresponding form
factors are readily calculated from the axioms
\r{eIIvi}-\r{eIIxiii}. The normalization has been determined in
Ref. \cite{LukZam01}. A short calculation then gives
\bea
S^{\rm zz}(\omega,2k_F+q)&=&
\frac{\tilde{v}Ja^2(H){\cal Z}_2{\cal C}_1^2}{16\pi\Delta^2}\,
\frac{\sqrt{s^2-4\Delta^2}}{s}\,E(\theta(\Delta,\Delta,s))
\theta_{H}\left(\frac{s}{\Delta}-2\right)\nonumber\\[2mm]
&& +\ {\rm contributions\ from\ 4,6,\ldots\ particles},
\eea
where ${\cal Z}_2$, $E(\theta)$, ${\cal C}_1$ are given by \r{Zn},
\r{Eoftheta} and \r{C1} respectively.
We see that the leading contributions to the various components of the
dynamical structure factor are always due to two particles. The
structure factor is entirely incoherent and always vanishes at the
threshold (as $\beta$ is always strictly larger than
$\frac{1}{\sqrt{2}}$), which occurs at $\omega=2\Delta$. 

\subsection{Dimerized Chain}
\label{Dimer}
\begin{figure}[ht]
\vskip -5mm
\begin{center}
\epsfxsize=0.75\textwidth
\epsfbox{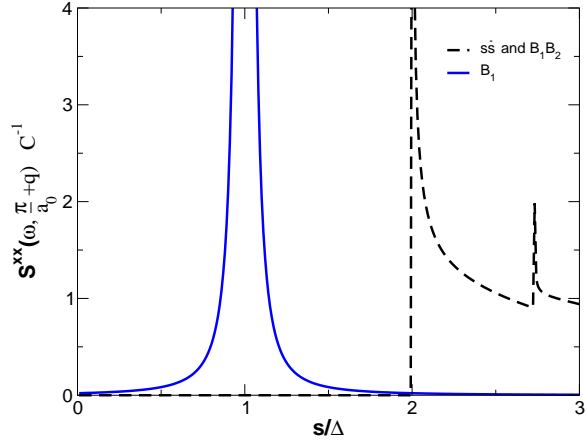}\qquad
\end{center}
\vskip -5mm
\caption{Dynamical structure factor of the dimerized chain. Most of
  the spectral weight is carried by the coherent triplet excitations,
  which are identified as soliton, antisoliton and the first breather
  in the SGM. The delta function corresponding to the coherent
  triplet excitation has been broadened by a Lorentzian to make it
  visible. }
\label{fig:SP}
\end{figure}

In actual materials the spin degrees of freedom are always coupled to
phonons. In many cases this coupling is small and can be ignored,
particularly at low temperatures. However, in quasi-1D compounds such
as ${\rm CuGeO_3}$ the magneto-elastic coupling leads to a phase
transition, known as the spin-Peierls transition. In the
low-temperature phase the lattice is then {\sl dimerized}. This is
turn leads to a modulation of the Heisenberg exchange and an
appropriate model Hamiltonian is
\bea
{\cal H}&=&J\sum_n\left[1-(-1)^n\delta\right]{\bf S}_n\cdot{\bf
  S}_{n+1}\ .
\label{Hdimer}
\eea
For $\delta\ll 1$ one may study \r{Hdimer} as a perturbation of the
isotropic Heisenberg chain in the field theory regime. The bosonized
expression for the staggered energy density $\varepsilon_n=
(-1)^n{\bf S}_n\cdot{\bf  S}_{n+1}$ can be determined from the
bosonized expressions of the spin operators \r{Sboso} for zero
magnetic field. The relevant piece arises from the fusion of the
smooth with the staggered components and is given by \cite{Fukuyama}
\be
\varepsilon(x)={\cal D}\sqrt{a_0}\
\cos\Bigl(\frac{1}{2}\,\Phi(x)\Bigr).
\label{dimerization}
\ee
The value of the constant ${\cal D}$ is presently not known analytically. In
the aforementioned fusion procedure subleading (i.e. very irrelevant)
terms in the bosonized expressions for the spin operators \r{Sboso}
contribute to the value of ${\cal D}$, so that it cannot be simply calculated
from the known expressions \r{a0}, \r{c0}, \r{A0} for $a(0)$, $c(0)$,
and ${\cal A}(0)$. However, it was found in \cite{Orignac} by
comparing to numerical results that ${\cal D}$ is well approximated by 
\be
{\cal D}\approx\frac{3}{\pi^2}\left(\frac{\pi}{2}\right)^\frac{1}{4}\ .
\label{constC}
\ee
Bosonizing the $\delta=0$ part of \r{Hdimer} and then using
\r{dimerization} we find that the field theory limit of the dimerized
chain is
\bea
{\mathcal H}
=\!\int\! dx\,\Biggl\{ \frac{v}{16\pi}\left[
(\partial_x\Phi)^2\!+(\partial_x\Theta)^2\right]
-\mu\cos\Bigl(\frac{1}{2}\,\Phi\Bigr)\Biggr\},
\label{Hdimer2}
\eea
where $\mu=a_0^{-\frac{1}{2}}{\cal D}J\delta$ and where we have dropped a
marginally irrelevant interaction of spin currents present in the field theory
limit of the isotropic Heisenberg chain \footnote{\,The bare value of
$\gamma$ has been estimated e.g. in \cite{Affleck89}} 
\be
H_{\rm marg.}=\gamma\int
dx\left[\cos(\Phi)+\frac{1}{16}\left[(\partial_x\Theta)^2-
(\partial_x\Phi)^2\right]\right].
\label{HJJ}
\ee
Hence the dimerized chain at
small $\Delta$ is described by a SGM with $\beta=\frac{1}{2}$.
It is important to note that the field theory respects the
spin rotational SU(2) symmetry of the original lattice Hamiltonian:
the SGM \r{Hdimer2} is SU(2) invariant \cite{haldane82}. This non-obvious fact is
most easily understood using non-Abelian bosonization, see section VI of
Ref. \cite{oa2}. The spectrum of the SGM \r{Hdimer2} consists of four
particles: soliton, antisoliton and a light breather forming a SU(2)
triplet with gap $\Delta$ and a heavy breather with gap
$\sqrt{3}\Delta$. The latter is a SU(2) singlet. 
In contrast to the field induced gap problem, the
nonlinear term is now the cosine of the bosonic field $\Phi$ rather
than the dual field $\Theta$. Hence the topological charge is the dual
of \r{Q} for $\beta=\frac{1}{2}$, which is related to the z-component
of the total spin by \r{Sboso}
\be
Q=\frac{1}{4\pi}\int_{-\infty}^\infty dx\ \partial_x\Phi
\equiv \sum_j S^z_j\, .
\label{Q2}
\ee
The relation of the topological charge to the z-component of the spin
agrees with our assertion that the breathers have $S^z=0$ and
soliton/antisoliton spin $\mp 1$.
A consequence of the SU(2) symmetry is that the following two-point
functions are identical in the SGM \r{Hdimer2}
\bea
\hspace{-8mm}
\bigl\langle\cos\bigl(\textstyle{\frac{1}{2}}\Theta(\tau,x)\bigr)
\cos\bigl(\textstyle{\frac{1}{2}}\Theta(0,0)\bigr)\bigr\rangle
&=&\bigl\langle\sin\bigl(\textstyle{\frac{1}{2}}\Theta(\tau,x)\bigr)
\sin\bigl(\textstyle{\frac{1}{2}}\Theta(0,0)\bigr)\bigr\rangle\nonumber\\[2mm]
&=&\bigl\langle\sin\bigl(\textstyle{\frac{1}{2}}\Phi(\tau,x)\bigr)
\sin\bigl(\textstyle{\frac{1}{2}}\Phi(0,0)\bigr)\bigr\rangle\, .
\label{SU2}
\eea
Equations \r{SU2} imply various relations between form factors, which can
be verified by direct calculation (see e.g. \cite{ETD}). In order to
determine the dynamical structure factor it is sufficient to consider
the two-point function of the operator $\sin(\frac{1}{2}\Phi)$ in the
SGM \r{Hdimer2}. Apart from the overall constant factor, the latter is
the same as the result for the field induced gap problem \r{sin} with
$H$ set to zero\,\footnote{\,At $H=0$ we have   $\beta=\frac{1}{2}$,
$\tilde{v}=\frac{\pi}{2}$ as required.} and an appropriate expression
substituted for the soliton gap. Using \r{constC} and taking the
marginally irrelevant interaction of spin currents into account in
renormalization group improved perturbation theory, the soliton gap is
found to be \cite{Orignac} 
\bea
\frac{\Delta}{J}\approx
\frac{\sqrt{\pi}\Gamma(\frac{1}{6})}{\Gamma(\frac{2}{3})}
\left[\frac{3}{\pi^2}\frac{\Gamma(\frac{3}{4})}{\Gamma(\frac{1}{4})}
\left[\frac{\pi}{2}\right]^\frac{1}{4}\delta\right]^\frac{2}{3}
\frac{1}{\sqrt{1+\frac{2}{3}|\lambda|
\ln\bigl|\frac{\lambda}{1.3612\delta}\bigr|}}\ ,
\label{gapDimer}
\eea
where $\lambda\approx -0.22$. The dynamical structure factor is then
given by
\bea
&&\hspace{-5mm}S^{\rm \alpha\alpha}(\omega,\frac{\pi}{a_0}+q)=
C\ {\rm Re}\Bigg\{
2\pi Z_1\ \delta(s^2 - \Delta^2)\nn
&&\hspace{-5mm}~ +\frac{2|F^{\sin\bigl(\frac{1}{2}\Theta\bigr)}_{+-}
[\theta(\Delta,\Delta,s)]|^2}{s\sqrt{s^2 - 4\Delta^2}}
+\frac{2|F^{\sin\bigl(\frac{1}{2}\Theta\bigr)}_{12}
[\theta(\Delta,\sqrt{3}\Delta,s)]|^2}
{\sqrt{(s^2-4\Delta^2)^2-12\Delta^4}}+ 
\ldots\Biggr\},\qquad
\label{sSP}
\eea
where $Z_1$, $\theta(a,b,c)$, $F^{\rm sin}_{+-}(\theta)$ and $F^{\rm
sin}_{12}(\theta)$ are given by \r{Z1}, \r{thetamms}, \r{fsinpm} and \r{fsin12} with
$\beta=\frac{1}{2}$ ($\xi=\frac{1}{3}$) respectively. We have
denoted the analog of the overall factor ${\cal C}(H)$ in \r{sin} by $C$. 
We conjecture that
\be
C=\frac{J}{(2\pi)^\frac{3}{2}}\,{\cal G}^2_\frac{1}{2}\,,
\ee
where ${\cal G}_\frac{1}{2}$ is given by \r{Galpha} with
$\beta=\frac{1}{2}$ and $\tilde{v}=\frac{\pi}{2}$.
We plot the DSF \r{sSP} in Fig.\,\ref{fig:SP}. 
The DSF is dominated by the coherent triplet modes with gap
$\Delta$ given by \r{gapDimer}. Starting at twice the triplet gap
there is an incoherent scattering continuum. The latter is singular at
its threshold $2\Delta$, as is the $B_1B_2$ scattering continuum which
occurs above $\Delta+\sqrt{3}\Delta$.
The heavy breather $B_2$ has total spin zero and is invisible to
neutron scattering experiments.  However, it contributes to the
two-point function of energy densities $\varepsilon(x)$ and hence
should be observable in Raman scattering experiments. An expression
for the two-point function of energy densities was derived in
\cite{ETD}. 

\subsection{Quasi-1D Spin-\half Antiferromagnets in the Ordered Phase}
A generic feature of quasi-1D spin-$\frac{1}{2}$ materials is that at
low temperatures they develop long range N\'eel order. This is
a consequence of the residual coupling between the 1D chains 
as spontaneous symmetry breaking is forbidden in one spatial
dimension. At low energies in the ordered phase the physics is
dominated by Goldstone modes: if the N\'eel order is along the z-axis
in spin space, there will be two transversely (in the xy-plane)
polarized spinwave modes (one mode each for the ``broken'' symmetry
generators $S^x$ and $S^y$). Their quantum numbers are $S^z=\pm 1$.
In addition to the transverse Goldstone modes there a priori also must
be gapped {\sl longitudinal} (polarized along the z-axis) excitations,
which correspond to fluctuations in the size of the ordered moment.

On the other hand it is clear that at temperatures or energies that
are large compared to the interchain coupling one has to recover the
1D spin chain  physics.

A long-standing question is how the system crosses over from
the 3D physics of the ordered phase to the 1D physics of the chains,
as one varies the energy scale at which it is probed. This issue was
addressed theoretically in Refs. \cite{Schulz96,AffleckHalperin,ETD,andrey} 
and experimentally in \cite{Bella} for ${\rm  KCuF_3}$ (see also \cite{satija,Tennant95a,Tennant95b}) and in
\cite{Zheludev02,andrey} for \ba .  

Our starting point is the Hamiltonian for an ensemble of weakly
coupled chains. For definiteness we discuss the form appropriate for
\ba \cite{Zheludev00,KenzelmannZheludev01,ZheludevKenzelmann01}, which reads
  \bea
  H&=&J\sum_{i,j,n}{\bf S}_{i,j,n}\cdot{\bf S}_{i,j,n+1}\ +H',\nn
  H'&=&\sum_{i,j,n} J_x\ {\bf S}_{i,j,n}\cdot{\bf S}_{i+1,j,n}+J_y\
{\bf S}_{i,j,n}\cdot{\bf S}_{i,j+1,n}\nn 
&&\qquad +J_3\   {\bf S}_{i,j,n}\cdot\left({\bf S}_{i+1,j+1,n}+{\bf
S}_{i+1,j-1,n}\right).
  \eea
where
$J_x=-0.460(7)~\mathrm{meV}$, $J_y=0.200(6)~\mathrm{meV}$,
$2J_3=0.152(7)~\mathrm{meV}$ and $J=24.1$~meV. The Fourier transform
of the inter-chain coupling is defined as 
\bea
J'({\bf p})= J_x\cos(p_x)+J_y\cos(p_y)
+J_3\left[\cos(p_x+p_y)+\cos(p_x-p_y)\right].~~~\quad
\label{JQ}
\eea
As some of the exchange constants are positive and some negative
it is convenient to introduce new spin variables $\widetilde{S}^\alpha$: 
\bea 
S^x_{i,j,n}=\widetilde{S}^{x}_{i,j,n}\ ,\
S^{\alpha}_{i,j,n}=(-1)^j\widetilde{S}^{\alpha}_{i,j,n}\ ,\quad \alpha=y,z\,.
\label{stilde}
\eea
The transformation (\ref{stilde}) leaves the Hamiltonians of the 1D
chains invariant,  but flips the signs of $J_y$ and $J_3$ in the
interaction of the $y$ and $z$ components of the spin operators in
$H'$. In the new spin variables the interchain coupling is
ferromagnetic so that we may take the staggered magnetization at $T=0$
to be
\bea 
\langle \ts^\alpha_{i,j,n}\rangle&=& \delta_{\alpha,z}\
(-1)^{n}\ m_0\,. 
\eea

\subsubsection{Mean-Field Theory}
As a first step we take the long range N\'eel order into account
in a self-consistent mean-field approximation
\cite{RPA,Schulz96}. Writing 
\bea 
\ts^\alpha_{i,j,n}=\langle\ts^\alpha_{i,j,n}\rangle +
\delta\ts^\alpha_{i,j,n}\ , 
\label{decoupling}
\eea
where $\delta\ts^\alpha_{i,j,n}$ denote (small) fluctuations around the
expectation value, and then substituting (\ref{decoupling}) in $H'$ we
obtain 
\bea 
H_{\rm MF}&=&\sum_{i,j}\sum_n J{\bf\ts}_{i,j,n}\cdot{\bf
  \ts}_{i,j,n+1} -h(-1)^n\ \ts^z_{i,j,n}\, ,\nn 
h&=&2(J_y-J_x+2J_3)\ m_0\equiv J^\prime m_0\, . \label{HMF}
\eea
  The Hamiltonian (\ref{HMF}) describes an ensemble of {\sl
uncoupled} spin-$\frac{1}{2}$ Heisenberg chains in a staggered
magnetic field and is a special case of the field-induced gap problem
we studied in section \ref{FIGP}. At low energies, the
mean-field theory Hamiltonian \r{HMF} reduces to a SGM for the Bose
field $\Phi$ (rather than the dual field $\Theta$ as was the case in
section \ref{FIGP})
\bea
{\mathcal H}_{\rm MF}
=\int dx\,\Biggl\{ \frac{v}{16\pi}\left[
(\partial_x\Phi)^2+(\partial_x\Theta)^2\right]
-\mu\cos\Bigl(\frac{1}{2}\,\Phi\Bigr)\Biggr\},
\label{MFSGM}
\eea
where $\mu=cha_0^{-\frac{1}{2}}$ with some constant $c$ and\,%
\footnote{\,$a_0$ is the period 
of the spin chains that for \ba\ is equal to half the lattice constant
in c-direction.}  
\bea
v=\frac{\pi Ja_0}{2}\,.
\eea
The form \r{MFSGM} is obtained as follows: bosonizing the Hamiltonian
\r{HMF} in the absence of the staggered field gives a Gaussian model
of the form \r{GaussianModel}, where the field is compactified
according to $\Phi\equiv\Phi+4\pi$. In addition there is a marginally
irrelevant interaction of spin currents of the form
\r{HJJ}. Bosonizing the staggered field term by means of \r{Sboso}
gives a sine-Gordon like Hamiltonian, but with a sine potential rather
than a cosine. Shifting the Bose field 
\be
\Phi\longrightarrow \Phi-\pi\, ,
\label{shiftPhi}
\ee
while keeping the dual field unchanged and dropping the marginally
irrelevant term then leads to \r{MFSGM}. The shift \r{shiftPhi}
changes the bosonized form of the lattice spin operator $S^z_j$ to
\be
S^z_j\sim\frac{a_0}{4\pi}\,\partial_x\Phi+ca_0^\frac{1}{2} (-1)^j
\cos\left(\frac{\Phi}{2}\right)+\ldots\ .
\ee

At the particular value of $\beta$, the spectrum is formed by
scattering states of four particles: soliton, antisoliton and the
first breather $B_1$ all with gap $\Delta$ and a second breather
with gap $\sqrt{3}\Delta$.
As described in Ref.~\cite{oa2}, the expectation value of the 
staggered magnetization can be calculated from the results of
Ref.~\cite{LZ}:
\bea 
m_0&\simeq&
\frac{2^\frac{2}{3}}{3\sqrt{3}\pi}\left[\frac{\Gamma(\frac{3}{4})} 
{\Gamma(\frac{1}{4})}\right]^\frac{4}{3}
\left[\frac{\Gamma(\frac{1}{6})} {\Gamma(\frac{2}{3})}\right]^2
\left(\frac{h}{J}\right)^\frac{1}{3}\
\left[\ln\left(\frac{J}{h}\right)\right]^\frac{1}{3} .
\label{m0}
\eea
Equation (\ref{m0}) is the self-consistency equation
of the mean field approximation (recall that $h=m_0J^\prime$) and has
the solution \cite{andrey}
\bea
m_0&\simeq&
\frac{\sqrt{2}}{3^\frac{7}{4}\pi^\frac{3}{2}}
  \left[\frac{\Gamma(\frac{3}{4})}{\Gamma(\frac{1}{4})}\right]^2
\left[\frac{\Gamma(\frac{1}{6})}{\Gamma(\frac{2}{3})}\right]^3
\left[\frac{J^\prime}{J}\ln\left(\frac{2.58495J}{J^\prime}\right)
\right]^\frac{1}{2}\  . 
\label{m0b}
\eea
We note that the constant $2.58495$ should not be taken too seriously
as we have ignored subleading logarithmic corrections. The result
(\ref{m0b}) is found to be in good agreement (for small $J^\prime/J$)
with a phenomenological expression obtained from quantum Monte-Carlo
simulations in Ref.~\cite{Sandvik99}. The soliton gap as a function of
the staggered field $h$ has been calculated in
Refs.~\cite{Essler99,oa2}. Expressing $h$ in terms of $m_0$ by
(\ref{HMF}) and then using (\ref{m0b}) we obtain 
\bea
  \frac{\Delta}{J}&\simeq& 
\frac{1}{3\pi}\left[\frac{\Gamma(\frac{3}{4})}
{\Gamma(\frac{1}{4})}\right]^2 
\left[\frac{\Gamma(\frac{1}{6})}{\Gamma(\frac{2}{3})}\right]^3
\  \frac{J^\prime}{J}\left[\ln\left(\frac{2.58495J}{J^\prime}
  \right)\right]^\frac{1}{2}.
\label{MFgap}
\eea
Let us now turn to the dynamical susceptibilities. A very useful
observation is that because $\beta=\frac{1}{2}$ in the sine-Gordon
description of the mean-field Hamiltonian \r{HMF}, the correlation
functions of the staggered magnetizations are related by the SU(2)
symmetry \r{SU2}. Hence\,\footnote{\,Recall that due to the shift
\r{shiftPhi} the staggered component of $S^z_j$ is now the cosine
rather than the sine.\\[-3mm]~}$^,$%
\footnote{\,The same constant enters both
transverse and longitudinal correlations because the spin rotational
symmetry is restored at high energies.\\[-3mm]~} 
\bea
&&\langle S^x_{j+1}(t)S^x_1(0)\rangle\simeq c^2a_0
(-1)^j\bigl\langle\sin\bigl(\frac{1}{2}\Phi(t,x)\bigr)
\sin\bigl(\frac{1}{2}\Phi(0,0)\bigr)\bigr\rangle\ ,\nn
&&\langle S^y_{j+1}(t)S^y_1(0)\rangle\simeq c^2a_0
(-1)^j\bigl\langle\sin\bigl(\frac{1}{2}\Phi(t,x)\bigr)
\sin\bigl(\frac{1}{2}\Phi(0,0)\bigr)\bigr\rangle\ ,\nn
&&\langle S^z_{j+1}(t)S^z_1(0)\rangle\simeq c^2a_0
(-1)^j\bigl\langle\cos\bigl(\frac{1}{2}\Phi(t,x)\bigr)
\cos\bigl(\frac{1}{2}\Phi(0,0)\bigr)\bigr\rangle\ .
\eea
The dynamical susceptibilities are
calculated as in section \ref{DSF}. However, now we are interested in
both the real and imaginary parts. Taking all contributions from
intermediate states with at most two particles into account one
finds \cite{ETD} 
\bea 
\tilde{\chi}^{\rm xx}_{\rm 1d}(\omega,\frac{\pi}{a_0}+q)&\approx&
\tilde{C}\Biggl\{
\frac{2\pi Z_1}{s^2-\Delta^2+i\epsilon}
+\int_0^\infty{d\theta}\frac{2|F^{\rm
sin}_{+-}(\theta)|^2} {s^2-[2\Delta\cosh(\theta/2)]^2+i\epsilon}\nn
&&+\int_0^\infty{d\theta}\frac{2|F^{\rm sin}_{12}(\theta)|^2}
{s^2-4\Delta^2(1+\frac{\sqrt{3}}{2}\cosh\theta)+i\epsilon}\Biggr\},
\label{chixx}
\eea
where $s^2=\omega^2-v^2q^2$.
The functions $Z_1$, $F^{\rm sin}_{+-}(\theta)$ and $F^{\rm sin}_{12}(\theta)$
are given by \r{Z1}, \r{fsinpm} and \r{fsin12} respectively, where
$\beta=\frac{1}{2}$ and hence $\xi=\frac{1}{3}$.
Similarly one finds
\bea
\tilde{\chi}^{\rm zz}_{\rm
1d}(\omega,\frac{\pi}{a_0}+q)&\approx&\tilde{C}\Biggl\{ 
\frac{2\pi Z_2}{s^2-3\Delta^2+i\epsilon}\
+\int_0^\infty{d\theta}\frac{2|F^{\rm cos}_{+-}(\theta)|^2
+ |F^{\rm cos}_{11}(\theta)|^2}
{s^2-[2\Delta\cosh(\theta/2)]^2+i\epsilon}\nn
&&+\int_0^\infty{d\theta}\frac{|F^{\rm
cos}_{22}(\theta)|^2}
{s^2-[\sqrt{12}\Delta\cosh(\theta/2)]^2+i\epsilon}\Biggr\} .
\eea
Here $Z_2$, $F^{\rm cos}_{+-}(\theta)$ and $F^{\rm cos}_{11}(\theta)$
are given by \r{Z2}, \r{fcospm} and \r{fcos11} respectively, where
$\beta=\frac{1}{2}$ and hence $\xi=\frac{1}{3}$. The contribution from
$B_2B_2$ two-particle states is given in \cite{ETD}.\footnote{\,Our
normalizations are related to those of \cite{ETD} by $Z=4\pi^3\tilde{C}$}

\subsubsection{Random Phase Approximation for the Interchain Coupling}
The mean-field approach clearly does not give a good description of
the magnetic response in the ordered phase: spin excitations have a
gap and there are no Goldstone modes. In order to reproduce the latter
it is necessary to go beyond the mean-field approximation. It is
possible to develop a perturbative expansion in the interchain
coupling along the lines of \cite{boies,Schulz96,Bocquet02}.
This gives
\bea 
{\chi}_{\rm  3d}^{\alpha\alpha}(\omega,\mathbf{p},k)&=&
\frac{\tilde{\chi}_{\rm  1d}^{\alpha\alpha}(\omega,k)
+\Sigma^{\alpha\alpha}(\omega,\mathbf{p},k)}
{1-2J^\prime(\mathbf{p})[\tilde{\chi}_{\rm 1d}^{\alpha\alpha}
(\omega,q)+\Sigma^{\alpha\alpha}(\omega,\mathbf{p},k)]}\ ,
 \label{RPA}
 \eea
where $\alpha=x,y,z$. In Eq.~(\ref{RPA}) $\Sigma^{\alpha\alpha}$
are the self-energies that are expressed in terms of
integrals involving three-point, four-point {\it etc} correlation
functions of spin operators. The analogous expressions in the
disordered phase were derived in Refs.
\cite{Bocquet02,Irkhin00}. To date, the relevant multipoint
correlation function have not been calculated for the SGM.
Furthermore, there is no small parameter in the expansion \r{RPA}. The
reason is that the mean-field gap is generated by the interchain
coupling itself and as a result the gap \r{MFgap} is of the same order
as the interchain coupling. Setting this issue aside, a simple
Random-Phase Approximation in the interchain coupling \r{RPA} was
suggested by Schulz in \cite{Schulz96}. Its essence to simply neglect the
self-energies in \r{RPA}. In other words, one sets
\bea
\Sigma^{\alpha\alpha}=0. 
\eea
One problem is that in this approximation the
transverse susceptibility will not have a zero-frequency spin wave
pole at the 3D magnetic zone-center, as it should, spin waves
being the Goldstone modes of the magnetically ordered state. In
order for the pole to be exactly at $\omega=0$ the full
self-energy $\Sigma^{xx}$ must be included. A work-around was
suggested in \cite{Schulz96}. If the RPA is a good approximation,
there will almost be a zero frequency pole 
\bea 
1\approx 2J'(0,\pi)\,
\tilde{\chi}_{\rm 1d}^{\rm xx}(0,\textstyle{\frac{\pi}{a_0}})\ . \label{pole1}
\eea
The idea is to replace \r{pole1} by an equality
\bea 
1= 2J'(0,\pi)\, \tilde{\chi}_{\rm  1d}^{\rm xx}(0,\textstyle{\frac{\pi}{a_0}})\, , 
\label{pole2}
\eea
and then use \r{pole2} to fix the overall normalization of
$\tilde{\chi}_{\rm 1d}^{\rm}$. Following this logic, we may carry out
the integral in \r{chixx} numerically and obtain  
\bea 
\tilde{C}\approx 0.0645\,\frac{\Delta^2}{|J^\prime|}\,. 
\eea 
Now it is a simple matter to determine ${\chi}_{\rm
  3d}^{\alpha}(\omega,{\bf p},\frac{\pi}{a_0}+q)$ by evaluating the 1D
susceptibilities numerically and then inserting them into \r{RPA}.

As discussed in Refs.~\cite{Schulz96,ETD}, the resulting dynamic
susceptibility for transverse spin fluctuations $\chi_{\rm 3d}^{\rm
xx}$, $\chi_{\rm 3d}^{\rm yy}$ in the coupled chains model contains a
pair of spin wave excitations that disperse perpendicular to the spin
chains, and are, by design, gapless. At higher energies there is an
incoherent scattering continuum with threshold $2\Delta$, independent
of the transverse momentum. This momentum independence is likely to be
an artefact of the approximation made, rather than being a genuine
feature of the model. 
The RPA result for the longitudinal susceptibility $\chi_{\rm 3d}^{\rm
zz}$ exhibits a gapped, coherent mode that disperses perpendicular to
the chains as well. Above a threshold of $2\Delta$ an incoherent
scattering continuum occurs. Like for the transverse susceptibility
the threshold for the continuum does not depend on the transverse
momentum. Within the RPA the longitudinal mode is sharp, i.e. it has
an infinite lifetime. This is certainly a feature of the approximation
made: a decay of the longitudinal mode into a pair of spinwaves is
permitted both by the quantum numbers of the excitations involved and
phase space. Such decay processes are simply not taken into account in
the RPA.
The physical picture that emerges in the RPA has been found to be in
good agreement with inelastic neutron scattering experiments on ${\rm
KCuF_3}$ \cite{Bella}. There a damped longitudinal mode has been
observed. It occurs approximately at the energy predicted by the RPA
and its spectral weight is found to be close to the RPA result. On the
other hand, the RPA does rather poorly when applied to
\ba\cite{Zheludev02,andrey}.

\section{O(3) Non-linear Sigma Model and Gapped, Integer Spin Chains}
\setcounter{equation}{0}

\def\obd{{\CO (e^{-\bd})}}
\def\otbd{{\CO (e^{-2\bd})}}
\def\hb{{\hat\beta}}

The existence of a gap in one-dimensional, integer-spin, Heisenberg
antiferromagnets was first predicted by Haldane \cite{haldane}.  He found
that such spin chains can be mapped onto a gapped integrable field theory,
the O(3) non-linear sigma model (NLSM), in 
the large-spin, continuum 
limit.  It is the goal of this section to explore the properties of the O(3) NLSM
and its relevant predictions for the physics of gapped spin chains.

While the O(3) NLSM is derived in a large $S$ limit, 
a variety of checks imply
that this behavior persists down to spin $S=1$.  A spin-$1$ chain
with a specific $(\vec{S}\cdot\vec{S})^2$ coupling has been rigorously
shown to exhibit a spin gap \cite{aklt}.  While at a differing value of the
$(\vec{S}\cdot\vec{S})^2$ coupling, the spin chain 
is gapless \cite{babu,kulisha,takhtajan}, this critical point is believed to be
unstable in the two-parameter space of couplings.  Gapless behavior
thus only arises as a product of fine-tuning.
Numerous numerical studies carried out
on spin-$1$ chains observe a gap 
\cite{campos,deisz,golinelli,haas,meshkov,moreo,parkinson,sakai,sorensen,Takahashi,Takahashib,white,whitea,yamamoto}.  
Experimentally,
inelastic neutron scattering studies on a number of 
quasi-one-dimensional, spin-$1$ chain 
materials are consistent with a finite spin
gap \cite{buyers,ma,ma1,morra,mutka,renard,steiner,tun,Igor94}.

Here we will suppose that the spin chain is described by a minimal
Heisenberg Hamiltonian and
so ignore (for the most part) the affects of
anisotropies upon the physics.  These can take (at least) two forms.
Easy-axis anisotropies,
\begin{equation}\nonumber
\Delta H = D_x 
\sum_i (S^x_{i})^2 + D_y \sum_i (S^y_{i})^2 + D_z \sum_i (S^z_{i})^2,
\end{equation}
of varying strengths are often
found in spin-$1$ chain materials.
Furthermore, actual spin chain materials never take the form of isolated
chains.  Rather the chains exist in three dimensional arrays with weak
but non-zero interchain couplings, $J'$.  Thus the chains are at best
quasi-one-dimensional.  With a finite $J'$, there will be some
correspondingly finite N\'{e}el temperature, $T_N$.  Below $T_N$
the (ultra low-energy) physics, consisting of higher dimensional
magnon modes, will be dramatically different than that described
here.

${\rm CsNiCl_3}$ was the first material for which evidence of a Haldane
gap was found \cite{buyers,morra,tun,Igor94}.  ${\rm CsNiCl_3}$ has only a 
weak easy axis anisotropy while possessing
an interchain coupling, $J'$, of strength  $J'/J \sim .017$.
This coupling, when combined with the large number (6) of nearest neighbour chains,
is sufficient to induce N\'{e}el order at $T\sim 5K$.
Provided however one is interested in physics 
at energies scales around the gap ($\Delta \sim .4J$), the absence
or presence of long range order plays an unimportant role.
Another material in which a Haldane gap is present
is $AgVP_2S_6$\cite{mutka}.  It has an extremely small
interchain coupling, $J'/J \sim 10^{-5}$ and a similarly
small easy axis anisotropy, $D_z/\Delta \sim 10^{-2}$.  However
it possesses a comparatively large gap, $\Delta \sim 320K$.

Other Haldane gap integer spin chain materials are studied experimentally, but these compounds
are characterized by large easy plane anisotropies.  One 
such material is $\rm Ni(C_2H_8N_2)_2NO_2ClO_4$
(NENP)\cite{regnault,renard,ma,ma1,Igor98}. 
It has an easy-plane anisotropy given by
$D_z/J \sim .16; ~D_z/\Delta \sim 2/5$ ($D_x \sim D_y \sim 0$)\cite{Igor98}.
For NENP, the ratio $J'/J \sim 8\times10^{-4}$ is sufficiently
small that 3D N\'{e}el order has not been observed down to
temperatures $\sim 1.2K$.
Related materials $\rm Ni(C_5H_{14}N_2)_3(PF_6)$
(NDMAP)\cite{Hagiwara,Zheludev01,andrey1,andrey2} and  
$\rm Ni(C_5H_{14}N_2)_2N_3(ClO_4)$ (NDMAZ)\cite{honda1,honda2,Zheludev01a} 
share similar easy-plane anisotropies. These latter compounds share
the additional feature of field-induced antiferromagnetism
\cite{chen,honda1,honda2,honda3,honda4,Zheludev02a}. The Luttinger
liquid that results from magnetic fields large
enough to extinguish the Haldane gap leads to quasi-long range
antiferromagnetic correlations.  With a small finite $J'$,
these quasi-long range correlations are promoted to full fledged
long range order.  The corresponding N\'{e}el temperature increases
with applied magnetic fields.  
In this review we will not explore this phenomena nor related behavior occurring
in the presence of a magnetic field
(but see 
\cite{affleck90,affleck91,fath,furusaki,gia,Haldane81a,loss,konik1,mila,normand,normanda,shelton,Takahashia,totsuka}).

The physics underlying the Haldane gap is particularly robust:
related systems such as two-leg spin-1/2 or Hubbard ladders also
exhibit a gap to spin excitations \cite{Dagotto,furusaki,gia,konik1,mila}.  (We will see this gapped
behavior again when we discuss ladder materials in the context of SO(8) Gross--Neveu).  
Roughly speaking, integer-spin composites form across the rungs of the
ladder making it into an effective integer-spin chain.
Both the ability to fabricate these materials 
and their relationship to high $T_c$ cuprate superconductors
have made them the focus of intense theoretical and experimental
study \cite{bala,duffy,fab,doped,so8,kuroki,lina,noack1,noack2,noack3,schulz96a}.

In this section we first turn to how to derive the O(3) NLSM.  We will then
provide a description of its spectrum, scattering matrices, and form factors.
We then turn to using these form factors to compute correlators both at zero
and finite temperature.  Our goal at finite temperature will be to analyze
whether transport in the O(3) NLSM is ballistic or diffusive.

\subsection{From Integer Spin Heisenberg Chains to the O(3) Non-linear
Sigma Model} 
~\\[-15mm]
\begin{figure}[ht]
\vskip .2in
\begin{center}
\epsfxsize=0.9\textwidth
\epsfbox{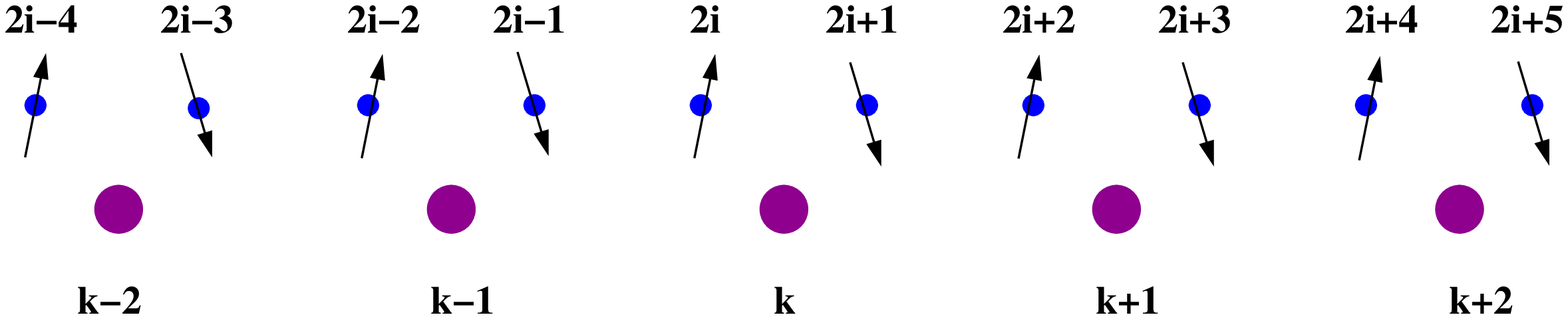}
\end{center}
\caption{The Heisenberg spin chain}
\label{figIVi}
\end{figure}

We begin by deriving the O(3) NLSM from the Hamiltonian of the Heisenberg 
spin chain\,\footnote{\,This development of the O(3) NLSM follows I. Affleck's in 
Ref. \cite{Affleck89b}.}
\begin{equation}\label{eIVi}
H = J\sum_i {\bf S}_i\cdot {\bf S}_{i+1}\,,
\end{equation}
where $S$ is a spin of arbitrary size.
We begin the derivation by decomposing each spin, $S^a_i$ (a=1,2,3), in terms of two 
operators, $n^a$ and $L^a$,
via
\begin{eqnarray}\label{eIVii}
S^a_{2i} = sn^a_{2i+1/2} + L^a_{2i+1/2}\,,
\end{eqnarray}
where\\[-7mm]~
\begin{eqnarray}\label{eIViii}
n^a_{2i+1/2} &=& (S^a_{2i+1}-S^a_{2i})/2s\,;\nonumber\\[1mm]
L^a_{2i+1/2} &=& (S^a_{2i+1}+S^a_{2i})/2\,.
\end{eqnarray}
Roughly speaking, the $n^a$ and $L^a$ operators govern antiferromagnetic and ferromagnetic 
fluctuations respectively.
For convenience we introduce a new lattice indexed by k (see Fig.\,\ref{figIVi}) 
with sites at $2i+1/2$ and so with twice the lattice spacing
of the original lattice.
Indexed in terms of this lattice, the operators, $n^a$ and $L^a$, have
the following commutations relations
\begin{eqnarray}\label{eIViv}
\big[ L^a_k,L^b_{k'}\big] &=& i\epsilon^{abc}L^c_k \delta_{kk'}\,;\nonumber\\[2mm]
\big[ L^a_k,n^b_{k'}\big] &=& i\epsilon^{abc}n^c_k \delta_{kk'}\,;\nonumber\\[1mm]
\big[ n^a_k,n^b_{k'}\big] &=& i\epsilon^{abc}\frac{1}{s^2}L^c_k \delta_{kk'}\,,
\end{eqnarray}
where $\delta_{kk'}=\delta_{ii'}/2$, $\delta_{ii'}$ being the Kronecker delta function on the original lattice,
the factor of two taking into account the change in lattice spacing.
The middle commutation relation indicates that in the continuum limit ${\bf L}$ becomes the generator of rotations
for the field ${\bf n}$.
We can also easily show that these operators obey the relations
\begin{eqnarray}\label{eIVv}
{\bf n}\cdot {\bf L} &=& 0\,;\nonumber\\
{\bf n}\cdot {\bf n} &=& 1 + \frac{1}{s} - \frac{{\bf L}\cdot {\bf L}}{s^2}\ .
\end{eqnarray}
In the large $s$ limit we see ${\bf n}\cdot {\bf n} = 1$.  Our assumption that ${\bf n}$ satisfies this
relation in general will be the main approximation of this derivation.

We are now in position to recast the Hamiltonian.  Noting that
\begin{eqnarray}\label{eIVvi}
\hspace{-0.5cm}{\bf S}_{2i}\cdot{\bf S}_{2i+1} &=& 2{\bf L}_{2i+1/2}\cdot{\bf L}_{2i+1/2} + {\rm const.}\,;\nonumber\\[2mm] 
\hspace{-0.5cm}{\bf S}_{2i}\cdot{\bf S}_{2i-1} &=& 2{\bf L}_{2i+1/2}\cdot{\bf L}_{2i+1/2} -2s^2{\bf n}\del^2_x{\bf n} + 
2s({\bf L}\cdot\del_x{\bf n} + \del_x{\bf n}\cdot {\bf L}) \nonumber\\[1mm]
&& \hskip 2.5in + \, {\rm const.}\,,
\end{eqnarray}
we can write the original Hamiltonian as
\begin{eqnarray}\label{eIVvii}
H &=& J \int \frac{dx}{2}\bigg\{ 4{\bf L}\cdot{\bf L} + 2s^2(\del_x{\bf n})^2 + s({\bf L}\cdot\partial_x{\bf n} + 
\partial_x{\bf n}\cdot{\bf L})\bigg\}\cr\cr
&=& \frac{v}{2}\int dx \, \bigg\{g\,\big[{\bf L}+ \frac{\theta}{4\pi}\partial_x{\bf n}\big]^2 + \frac{1}{g}\,(\partial_x{\bf n})^2\bigg\}\,,
\end{eqnarray}
where the spin velocity, $v$, equals $2Js$, the coupling constant $g$ is given by $2/s$ and the parameter,
$\theta$, the theta angle, equals $2\pi s$.  In the above two equations, we have set the (original)
lattice spacing to 1.  We henceforth also set $v=1$.

This Hamiltonian corresponds to the Lagrangian
\begin{equation}\label{eIVviii}
{\cal L} = \frac{1}{2g}\,\del_\mu{\bf n}\cdot \del^\mu {\bf n} - \frac{\theta}{8\pi}\epsilon^{\mu\nu}{\bf n}\cdot 
(\del_\mu{\bf n}\times\del_{\nu}{\bf n})\,,
\end{equation}
subject to the constraint that ${\bf n}\cdot{\bf n}=1$.  
This model is known as the O(3) NLSM with theta angle, $\theta$.
The Lagrangian is given solely in terms of the field ${\bf n}$.
$L$, the generator of rotations for ${\bf n}$, is expressible in terms 
of ${\bf n}$ and it conjugate momentum, ${\bf p}$.
${\bf p}$ is given by
\begin{equation}\label{eIVix}
{\bf p} = \frac{1}{g}\,\del_t {\bf n} + \frac{\theta}{4\pi}\,{\bf n}\times\del_x{\bf n}\,.
\end{equation}
${\bf L}$ is then readily found to be
$$
{\bf L} =  {\bf n}\times{\bf p}.
$$
We have that ${\bf L}\cdot{\bf n} = 0$ in accordance with Eq. (\ref{eIVv}).
Using this form of ${\bf L}$ we verify that
$H = {\bf p}\cdot{\bf \dot{n}} - {\cal L}$ is consistent with Eq. (\ref{eIVvii}).

The theta angle plays a crucial role in determining the properties of the above Lagrangian.  The term to which
it serves as a coupling constant, $\epsilon^{\mu\nu}{\bf n}\cdot(\del_\mu{\bf n}\times\del_\nu{\bf n})/8\pi$, 
counts (when integrated)
the number
of times the field ${\bf n}$ winds itself about the sphere (i.e. the second homotopy group of the sphere).
As it is always an integer, this term in the Lagrangian only affects the physics if $s$ is a half integer
(and so $\theta$ is an odd multiple of $\pi$).  In this case this Lagrangian describes a massless theory
which at low energies is equivalent to the $SU(2)_1$ Wess-Zumino-Witten theory.  If on the other hand
$s$ is integer, the theta term has no affect on the physics and the theory is one of gapped bosons.
It will be this latter case in which we will be interested.

\subsection{Basic Description of the O(3) Non-linear Sigma Model} 

With $\theta = 0$, the low energy theory of integer spin chains is then simply 
\begin{equation}\label{eIVx}
S = \frac{1}{2g} \int dtdx\, (\del^\nu{\bf n}\cdot\del_\nu{\bf n}).
\end{equation}
In terms of $L^a$ and $n^a$, the original spins of the theory appear as
\begin{eqnarray*}
S^a_{2i+1} &=& sn^a_{2i+1/2} + L^a_{2i+1/2}\cr\cr
&=& s n^a(x) + \frac{s}{2} \partial_x n^a (x) + L^a(x) + \frac{1}{2}\partial_x L^a(x);\cr\cr
S^a_{2i} &=& -sn^a_{2i+1/2} + L^a_{2i+1/2}\cr\cr
&=& -s n^a(x) + \frac{s}{2} \partial_x n^a (x) + L^a(x) - \frac{1}{2}\partial_x L^a (x).
\end{eqnarray*}
The staggered component of the spin (governing fluctuations near $k\sim \pi$)
is then $s{\bf n}(x) + \partial_x{\bf L}(x)/2 \sim s{\bf n}(x)$ (keeping
into mind we are working at large $s$) while the corresponding smooth component (governing fluctuations near $k \sim 0$)
on the spin (termed {\bf M}),
is given by
$$
{\bf M}(x) \equiv {\bf L}(x) + \frac{s}{2}\partial_x {\bf n} = \frac{1}{g}{\bf n}(x)\times\partial_t{\bf n}(x),
$$
We thus see ${\bf M}(x)$ is quadratic in the field $n(x)$.
Unlike its ancestral theory, the O(3) NLSM is integrable \cite{zamo,zam92,wiegmann}.
Local conserved charges were constructed in Ref. \cite{polyakov,goldschmidt} while
non-local conserved charges were first found by L\"uscher \cite{lus}.

The low energy excitations in the \ot take the form of a triplet of bosons.
The bosons have a relativistic dispersion relation given by
$$
E (p) = (p^2+\Delta^2)^{1/2} .
$$
Here $\Delta$ is the energy gap or mass of the bosons related to the
bare coupling, $g$, via $\Delta \sim Je^{-2\pi/g}$.  The dispersion
relations of all three bosons are identical as the model has a global
SU(2) symmetry.  The exact eigenfunctions of the \ot Hamiltonian are then
multi-particle states made up of mixtures of the three bosons.  Scattering
between the bosons is described by the S-matrix \cite{zamo}:
\begin{eqnarray}\label{eIVxii}
S^{a_3a_4}_{a_1a_2} (\th ) = 
\delta_{a_1a_2}\delta_{a_3a_4}\sigma_1 (\th ) &+&
\delta_{a_1a_3}\delta_{a_2a_4}\sigma_2 (\th ) +
\delta_{a_1a_4}\delta_{a_2a_3}\sigma_3 (\th )\,;\cr\cr
\sigma_1 (\th ) &=& \frac{2\pi i \th}{(\th + i\pi)(\th - i2\pi)}\,;\nonumber\\[1mm]
\sigma_2 (\th ) &=& \frac{\th (\th - i\pi)}{(\th + i\pi)(\th - i2\pi)}\,;\nonumber\\[1mm]
\sigma_3 (\th ) &=& \frac{2\pi i (i\pi - \th)}{(\th + i\pi)(\th - i2\pi)}\,.
\end{eqnarray}
Here again $\th$ parameterizes a particle's energy/momentum via 
$E=\Delta\cosh(\th )$, $P =\Delta \sinh (\th )$.  
These S-matrices are relativistically invariant.
While we stress
that this relativistic invariance is a natural feature of the low
energy structure of the spin chain, we do point out for
spin-1 chains, $\Delta \sim .4 J$.  As $J$ serves as the cutoff for
the theory, the low energy sector of the theory in this case is not 
unambiguously defined.\\[-7mm]~

\subsubsection{Form Factors for the Staggered Magnetization Operator, ${\bf n}$}

We will be interested in computing correlators involving the field, $\bf{n}$,
describing excitations near $k\sim \pi$.
Form factors for the field $n(x,t)$ have
been computed by both Smirnov \cite{smirnov} 
and Balog and Niedermaier \cite{balog}.
However Ref. \cite{balog} presents them in a more amenable form, 
possible in this particular case
because of the simple structure of the S-matrix
of the $O(3)$ sigma model.

The staggered magnetization operator is the fundamental field in the
theory. It thus has non-zero overlap with single particle excitations.
By the $Z_2$ symmetry, ${\bf n} \rightarrow -{\bf n}$, it also only
couples to odd numbers of particles. 
Using the axioms of Section 2, Ref. \cite{balog} finds for the one and
three particle form factors:
\begin{eqnarray}\label{eIVxiii}
f^{n_a}_{b}(\th_1) 
&\equiv& \langle 0 | n_a (0,0) |A_b(\theta )\rangle = \delta_{ab}\,;\\[1mm]
f^{n_a}_{bcd}(\th_1,\th_2,\th_3) &\equiv& \langle 0 |n_a(0,0) | 
A_d(\th_3)A_c(\th_2) A_b (\th_1)\rangle \nonumber\\[1mm]
&&\hspace{-2.3cm} =-\frac{\pi^3}{2}\psi(\th_{12})\psi(\th_{13})\psi(\th_{23})
\bigg(\delta_{ab}\delta_{cd}\theta_{23} 
+ \delta_{ac}\delta_{bd}(\theta_{31}-2\pi i) + \delta_{ad}\delta_{bc}\theta_{12}\bigg),\nonumber
\end{eqnarray}
where $\theta_{ij}\equiv \theta_i-\theta_j$ and $\psi(\th )$ is
defined in Eq.(\ref{eIVxv}).

\subsubsection{Form Factors for the Magnetization Operator, ${\bf M}$}

\def\ep{\epsilon}

We will also be interested in computing correlators involving the magnetization, ${\bf M}$.
This field, $\bf{M}\equiv {\bf M}_0$, is part of a Lorentz two-current 
$({\bf M}_0,{\bf M}_1)$.
As this current is topological, both components of the current can be written in terms
of a single Lorentz scalar field, $m(x,t)$:
\begin{equation}\label{eIVxiv}
{\bf M}_\mu(x,t) = \epsilon_{\mu\nu}\del^\nu {\bf m}(x,t)\,.
\end{equation}
We note that only even numbers of particles
couple to  the magnetization current and density operators.  This is a consequence 
of ${\bf M}$ being
expressible as a bilinear in the (fundamental) field $\bf{n}$, the field which 
creates the excitations.

Given Eq.(\ref{eIVxiv}), it is sufficient to compute the form
factors of the scalar field ${\bf m}$:  the form factors of ${\bf M}$ 
are then given by
\begin{equation}\nonumber
f^{M_\mu}_{a_1\cdots a_n} (\th_1,\cdots ,\th_n) =
\epsilon_{\mu\nu} P^\nu (\th_i ) 
f^{m}_{a_1\cdots a_n} (\th_1,\cdots ,\th_n)\,,
\end{equation}
where
$P^0 = \sum_i \Delta \cosh (\th_i )$ and 
$P^1 = \sum_i \Delta \sinh (\th_i )$.
Ref. \cite{balog}
then finds the following for the two and four particle form factors of ${\bf m}$:
\begin{eqnarray}\label{eIVxv}
&&f^{m_a}_{a_1a_2}(\th_1,\th_2) = i \frac{\pi^2}{4} \ep^{aa_1a_2}
\psi(\th_{12})\,, ~~~~~\psi (\th ) = \frac{\tanh^2(\th/2)}{\th}
\frac{i\pi + \th}{2\pi i + \th}\,;\cr\cr
&&f^{m_a}_{a_1a_2a_3a_4}(\th_1,\th_2,\th_3,\th_4 )
=\ -\frac{\pi^5}{8}\prod_{i<j} \psi (\th_{ij}) G^{m_a}_{a_1a_2a_3a_4}
\nonumber\\
&&
\qquad= -\frac{\pi^5}{8}\prod_{i<j} \psi (\th_{ij})\times
\bigg( \delta^{a_4a_3}\ep^{aa_2a_1} g_1(\th_i) 
+\delta^{a_4a_2}\ep^{aa_3a_1} g_2(\th_i)\nonumber\\[-1mm]
&& \hspace{4.5cm}+ \delta^{a_4a_1}\ep^{aa_3a_2} g_3(\th_i)
 +\delta^{a_3a_2}\ep^{aa_4a_1} g_4(\th_i)\nonumber\\
&&  \hspace{4.5cm}+\delta^{a_3a_1}\ep^{aa_4a_2} g_5(\th_i)
+\delta^{a_2a_1}\ep^{aa_4a_3} g_6(\th_i)\bigg);\cr\cr
&&
\left(
\begin{matrix}
g_1(\th_i ) \cr
g_2(\th_i ) \cr
g_3(\th_i ) \cr
g_4(\th_i ) \cr
g_5(\th_i ) \cr
g_6(\th_i )
\end{matrix}
\right)
= i\left(
\begin{matrix}
-i\pi (\ut^2+\vt^2 -i\pi\ut-i\pi\vt+2\pi^2) \cr
(\ut-i\pi)\vt(\vt-i\pi) \cr
(\ut-i\pi)(\ut+i2\pi)(i\pi-\vt) \cr
\ut\vt(3\pi i-\vt) \cr
\ut(\ut-i\pi)\vt \cr
2\pi i(i\pi -\ut)\vt \cr
\end{matrix}\right) \\[2mm]
&& 
+ i(\wt\!-\!i\pi)\!
\left(\begin{matrix}
-4\pi^2\!\! -\!i\pi(\ut\!+\!\vt)\! -\! (\ut\!-\!\vt)^2 \cr
-2\pi^2 - 3\pi i\vt + \vt^2 \cr
-4\pi^2 + i\pi(\ut-2\vt)-\ut^2 \cr
2\pi^2 + i\pi (\ut+2\vt ) -2\ut\vt \cr
-i\pi(2\ut+\vt)+2\ut\vt \cr
-2\pi^2 +i\pi(\ut-3\vt )
\end{matrix}\right)
\!\!+\! i(\wt\!-\!i\pi)^2 \left(
\begin{matrix}
0 \cr
0 \cr
0 \cr
-\ut \cr
\vt\! -\! 2\pi i \cr
\ut\! -\! \vt 
\end{matrix}\right).\nonumber
\end{eqnarray}
\noindent 
The reader should note however that we use a different
particle normalization than Ref. \cite{balog} 
and so the above presented results differ by multiplicative constants.

\subsection{Zero Temperature Correlators}

In this section we consider the zero temperature dynamic spectral functions of 
the spin operator near $k\sim 0$ and $k\sim \pi$.  These spectral functions govern
the response in inelastic neutron scattering experiments.

\subsubsection{$k\approx 0$ Magnetization Correlations}

This problem was first studied theoretically in Ref. \cite{affwes}.
From this work, we expect at $k\sim 0$ to see a two-magnon continuum.  
A high resolution study mapping out this continuum has been carried out in
Ref.\cite{igor01}.  Observing the two-magnon continuum is a difficult task as
the scattering function vanishes at small momentum (as we will demonstrate).

The spectral function for an integer spin chain near $k\sim 0$ is
given by  
\bea\label{eIVxvi}
S(\omega ,k )
&=& -\frac{1}{\pi}\, \text{Im}\bigg[ \int^\infty_{-\infty} dx 
\int^\infty_{-\infty}
d\tau e^{i\omega \tau-i k x}\nn
&&\qquad\quad\times
\big\{-\langle T (M^3(x,\tau)M^3(0,0))\rangle\big\}
\big|_{\omega\rightarrow \epsilon-i\omega}
\bigg].
\eea
Here $\langle T(M^3M^3)\rangle$ is an imaginary time-ordered correlator.
We evaluate this correlator using two particle form-factors.  As the next
form factor that contributes is a four-particle one, our result
will be exact for energies, $\omega < 4\Delta$.  Moreover we can expect
the contribution to the overall spectral
weight of the four-particle form factors and beyond to be small.  

\begin{figure}[ht]
\vskip -.1in 
\begin{center}
\psfrag{w}{$\omega/\Delta$}
\psfrag{k}{$k/\Delta$}
\epsfxsize=0.6\textwidth
\epsfysize=0.6\textwidth
\rotatebox{0}{\epsfbox{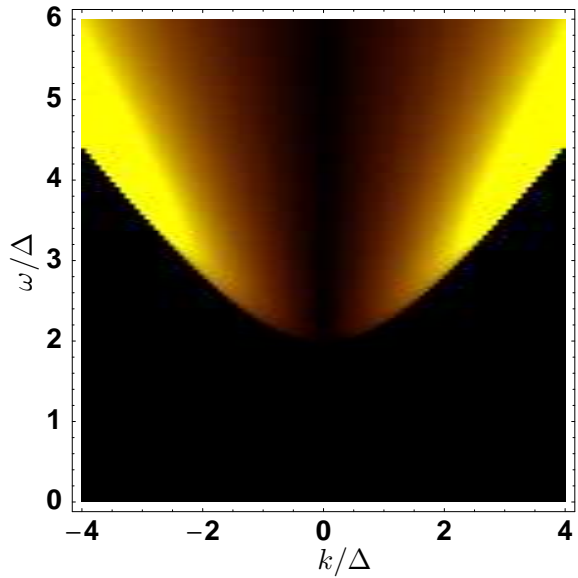}}
\end{center}
\caption{A plot of two magnon contribution to 
the spectral weight of the spin-spin correlator around $k=0$ in the 
$O(3)$ NLSM.  This plot is exact (within the confines of this model) for 
energies less than $4\Delta$.}
\label{figIVii}
\end{figure}

The time-ordered correlator evaluated to the two-particle level equals
\begin{eqnarray}\label{eIVxvii}
&&G(x,\tau) = -\Theta (\tau) \langle M^3(x,\tau)M^3(0,0)\rangle
-\Theta(-\tau)\langle M^3 (0,0) M^3(x, \tau)\rangle \nonumber\\[1mm]
&& 
~= -\Theta(\tau )\frac{1}{2}
\int d\hat\th_1d\hat\th_2 \sum_{a_1,a_2=1}^3 |f^{M^3}_{a_1a_2}(\th_1\th_2 )|^2 
e^{-m\tau(c(\th_1)+c(\th_2))+imx(s(\th_1)+s(\th_2))}\nonumber\\[1mm]
&& 
~~~-\Theta(-\tau )\frac{1}{2}\int d\hat\th_1d\hat\th_2 \sum_{a_1,a_2=1}^3 
|f^{M^3}_{a_1a_2}(\th_1\th_2 )|^2 e^{m\tau(c(\th_1)+c(\th_2))-imx(s(\th_1)+s(\th_2))}\,,\cr &&
\end{eqnarray}
where $\hat\th \equiv \th/(2\pi)$, $c(\th ) \equiv \cosh(\th)$, and
$s(\th ) \equiv \sinh(\th )$.
If we now take Fourier transforms in space and time, make the necessary
analytic continuation in $\omega$, and use the results for the form factors
in the previous section, we find
\begin{eqnarray}\label{eIVxviii}
S (\omega>0, k) &=& k^2 \,\frac{\pi^3}{16}\, \frac{1}{\sqrt{\omega^2-k^2}}\,
\frac{1}{\sqrt{\omega^2-k^2-4\Delta^2}}\,
\frac{\tanh^4(\theta_{12}/2)}{\theta_{12}^2}\,
\frac{\theta^2_{12}+\pi^2}{4\pi^2+\th_{12}^2}\,;\nonumber\\[2mm]
\theta_{12} &=& \cosh^{-1}\Big(\frac{\omega^2-k^2-2\Delta^2}{2\Delta^2}\Big).
\end{eqnarray}
We have only given the spectral function for $w>0$.  
This response function of an incoherent two magnon continuum 
is plotted in Fig.\,\ref{figIVii}.  We can observe there the lack of
spectral weight at small momentum\cite {affwes}.  Ref. \cite{igor01} compares
these theoretical predictions with a neutron scattering study of ${\rm CsNiCl_3}$.

\subsubsection{$k\approx\pi$ Magnetization Correlations}
We now consider the dynamic response function for wavevectors
near $\pi$.  While at $k=0$ the response is governed by the 
magnetization operator, ${\bf M}$, here at $k=\pi$ the relevant
operator is ${\bf n}$.  We expect then that the response will be dominated
by a coherent single mode (the magnon).  Beyond single magnon excitations
exists a three magnon continuum.  The continuum's contribution to
the response function has been studied in the context of the O(3) NLSM in Ref. \cite{horton,essler3mag}
where it was found to be small.  This accords with the intuition built up in Section 2.5 of this
review that higher particle form-factors make negligible contributions to their corresponding
spectral function.

The computation of the dynamic response function
in terms of the O(3) NLSM amounts to computing
\begin{equation}\label{eIVxix}
S(\omega ,k )
= -\frac{1}{\pi} \int^\infty_{-\infty} dx \int^\infty_{-\infty}d\tau e^{i\omega \tau-i k x}
\big(-\langle T (n^3(x,\tau)n^3(0,0))\rangle\big)|_{\omega \rightarrow -i\omega + \epsilon}.
\end{equation}
We compute contributions to this correlator up to and including three magnon form-factors.
The next contribution comes at five magnons and so this result will be exact
for frequencies, $\omega$, up to $5\Delta$.
The time-ordered correlator evaluated to the three-particle level is given by
\begin{eqnarray}\label{eIVxx}
G(x,\tau) &=& -\Theta (\tau) \langle n^3(x,\tau)n^3(0,0)\rangle-\Theta(-\tau)\langle n^3 (0,0) 
n^3(x,\tau)\rangle \nonumber\\
&=& -\Theta(\tau )\bigg\{
\int d\hat\th \sum_{a_1=1}^3 |f^{n_3}_{a_1}(\th )|^2 
e^{-m\tau c(\th)+imxs(\th )}\nonumber\\[-1mm]
&& 
\hskip -1.5cm + \frac{1}{3!}\int d\hat\th_1d\hat\th_2d\hat\th_3 
\sum_{a_1,a_2,a_3=1}^3|f^{n_3}_{a_1a_2a_3}(\th_1,\th_2,\th_3)|^2
e^{\sum_{i=1}^3(-m\tau c(\th_i)+i m xs(\th_i))}\bigg\}\nonumber\\ 
&& -\, (\tau,x \rightarrow -\tau,-x).
\end{eqnarray}
If we again Fourier transform, using the form-factor results of the previous
section we
find for positive frequencies \cite{essler3mag}
\begin{eqnarray}\label{eIVxxi}
\hspace{-6mm}
S(\omega > 0, \pi+k) &=& S_1(\omega,k) + S_3(\omega,k) \nonumber\\[1mm]
&=& \frac{1}{\sqrt{k^2+\Delta^2}}
\delta (\omega - \sqrt{k^2+\Delta^2})\nonumber\\[1mm]
&+& \frac{2\pi^4}{3}\int^{z_0}_0\big(3\pi^2+3z^2+Y^2)\big|\psi(2z)\psi(z+Y)\psi(z-Y)\big|^2
\nonumber\\
&\times& \frac{1}{\sqrt{(\omega^2\!-\!k^2\!-\!\Delta^2\!-\!4\Delta^2\cosh^2(z))^2\!-\!16\Delta^4\cosh^2(z)}}\ ,~~
\end{eqnarray}
where 
\begin{eqnarray}\label{eIVxxii}
z_0 &=& \cosh^{-1}[(x-1)/2]\,;\nonumber\\[1mm]
Y &=& \cosh^{-1}(\frac{x^2-1-4\cosh^2(z)}{4\cosh(z)})\,;\nonumber\\[1mm]
x^2 &=& \frac{\omega^2-q^2}{\Delta^2}\,.
\end{eqnarray}
The $\delta$-function term, $S_1(\omega,k)$, in the above expression 
for $S(\omega,k)$ corresponds to the coherent
one magnon contribution.  The next term, $S_3(\omega,k)$, 
arises from the incoherent three magnon
continuum.  From Lorentz invariance, both are a function of $s = \sqrt{w^2+k^2}$.

\begin{figure}[ht]
\vskip -.2in 
\begin{center}
\epsfxsize=0.5\textwidth
\rotatebox{270}{\epsfbox{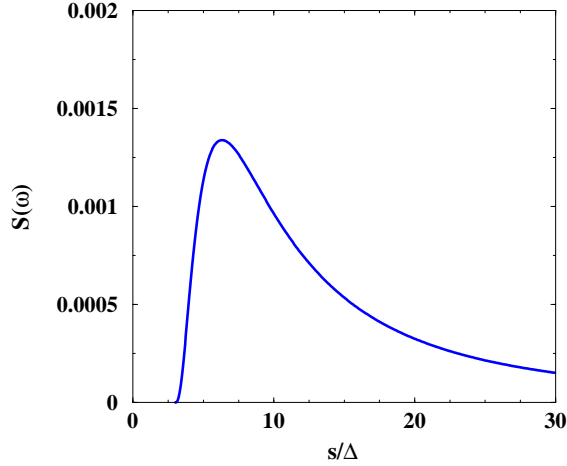}}
\end{center}
\caption{A plot of the three 
magnon contribution to the spectral weight of the spin-spin correlator near 
wavevector $\pi$ in the $O(3)$ NLSM as a function of $s$.}
\label{figIViii}
\end{figure}

We can evaluate the three magnon contribution numerically.  The result is plotted
in Fig.\,\ref{figIViii} as a function of $s$.  It is zero
for $s < 3\Delta$.  At threshold, i.e. 
$s-3\Delta$ small and positive, $S(\omega,k+\pi)$ behaves as 
$(s-3\Delta)^3$.  We see that it
peaks at around $s\sim 6\Delta$.  Furthermore it is clear that its contribution to the 
response function is small.  We can make this observation more qualitative.
If we define 
\begin{equation}\label{eIVxxiii}
I_i (k+\pi) = \int^{30\Delta}_0 d\omega S_i(\omega,k+\pi)\,,
\end{equation}
we find that
\begin{equation}\label{eIVxxiv}
\frac{I_3 (\pi)}{I_1(\pi)} = 0.018\,.
\end{equation}
Thus 98\% of the total weight at wavevector $\pi$ is found in the coherent 
one magnon contribution\cite{essler3mag,horton}.  

While the three magnon continuum's contribution to the spectral function is 
small, we do note that there are reports of its observation 
in the Haldane gapped material, ${\rm CsNiCl_3}$
\cite{Kenzelmann01,Kenzelmann02a}.  These reports
suggest that the three magnon continuum is larger than computed in the
O(3) NLSM.  Ref. \cite{Kenzelmann01} finds instead that the ratio in
Eq.\,(\ref{eIVxii}) is given by $I_3 (\pi)/I_1(\pi) \sim 0.1$.  That
the O(3) NLSM might underestimate the three magnon weight is not an
impossibility.  It is a low energy description of an integer spin
chain.  Given the spin gap, $\Delta$, equals $0.4 J$ where $J$, the 
spin-spin interaction strength, is an effective UV-cutoff, we would
not necessarily expect accurate predictions for energies beyond
$\omega > J$. On the other hand, the NLSM calculations are in fair
agreement with numerical computations \cite{Taka90,kw}, which suggests
that the large continuum reported experimentally has its origin in
physical mechanisms not included in the description by means of a
spin-1 Heisenberg model. The effects of one such candidate, a
biquadratic interaction of spins \cite{tsv3maj}, were investigated in
\cite{essler3mag}. A calculation of the corresponding response function gives
$\frac{I_3(\pi )}{I_1 (\pi )} \sim .2$. 
\vskip 3mm

~\\[-18mm]~
\subsection{Low Temperature Properties of Correlation Functions}

In this section we show that finite temperature correlation functions admit low temperature
expansions as discussed in Section 2.  We begin by computing the finite DC susceptibility
of the system.  As we can access this quantity exactly through other means, we are able
to test the methodology.  We will find that the form factors reproduce precisely the known
exact results.  We then turn to computing the NMR relaxation rate as well as the finite field spin conductance.
These quantities both probe the nature of transport in integer spin chains as described by the 
O(3) NLSM.  We will thus be able to argue that transport here is ballistic not diffusive.

\subsubsection{Magnetization}

\def\cor{{\langle M^3_0(x,\tau ) M^3_0 (0,0) \rangle}}
\def\mm{{M^3_0(x,\tau ) M^3_0 (0,0)}}
\def\mo{{M^3_0(x,\tau )}}
\def\mt{{M^3_0(0,0)}}
\def\eb{{e^{-\beta E_{s_n}}}}
\def\ebt{{e^{-\beta \Delta \cosh (\th )}}}
\def\ebtot{{e^{-\beta \Delta (\cosh (\th_1 )+\cosh(\th_2))}}}
\def\dh{{2\pi\delta (0)}}

The susceptibility, $\chi$, at $H=0$ can be computed from
the magnetization-magnetization operator using a Kubo formula:
\begin{eqnarray}\label{eIVxxv}
\hspace{-6mm}\chi_(H=0) &=& C(\om = 0,k = 0)\,,\nonumber\\[1mm]
\hspace{-6mm}C(\om ,k) &=& \bigg[ \int^\infty_{-\infty} dx \int^\beta_0
d\tau e^{iw_n\tau}e^{ikx} 
\langle T(M^3_0(x,\tau )M^3_0(0,0))
\rangle\bigg]_{\om_n \rightarrow -i\om+\delta}\!\!\! \!\!\!\!\!\!\!\!\!\!\!\!.
\end{eqnarray}
To evaluate this correlator we first expand out the thermal trace:
\begin{eqnarray}\label{eIVxxvi}
\langle M^3_0(x,\tau )M^3_0(0,0)\rangle &=& \frac{1}{\cal Z}\, 
{\rm Tr}(e^{-\beta H} {\cal O}(x,t) {\cal O}(0,0))\nonumber\\[1mm]
&=& \frac{\sum_{n s_n} e^{-\beta E_{s_n}}
\langle n,S_n|{\cal O}(x,t){\cal O}(0,0)|n,S_n\rangle}{\sum_{n S_n} e^{-\beta E_{s_n}} \langle n,S_n|n,S_n\rangle}\,.~~~
\end{eqnarray}
Here $|n,S_n\rb$ is a state of n excitations with spins described
by $S_n=\{s_1,\cdots,s_n\}$.  In writing the above we 
have suppressed sums over the energy
and momenta of the excitations.
A term in the thermal trace with $n$ excitations is weighted by
a factor of $e^{-n\beta\Delta}$.
Thus, as discussed in Section 2, at low temperatures it is
a good approximation to truncate this trace.
For this computation we keep only terms with one and two excitations,
i.e. $n=1,2$.
To evaluate the matrix elements appearing in Eq.\,(\ref{eIVxxvi}), we insert
a resolution of the identity in between the two fields.
As we only consider matrix elements involving one and two excitations
from the thermal trace, we thus have
\begin{eqnarray}\label{eIVxxvii}
\lb s_1 | M^3_0(x,\tau )M^3_0(0,0) | s_1 \rb 
&=& \sum_{mS_m}
\lb s_1 | M^3_0(x,\tau )|mS_m\rb
\lb mS_m | M^3_0(0,0) | s_1\rb\cr\cr
&=& \sum_{s'_1}
\lb s_1 | M^3_0(x,\tau )|s'_1\rb
\lb s'_1|M^3_0(0,0)| s_1\rb + \cdots ~;~\cr\cr
\lb s_1 s_2|M^3_0(x,\tau )M^3_0(0,0) | 
s_2 s_1  \rb &=& \sum_{mS_m}  
\lb s_1 s_2 |M^3_0(x,\tau )| m S_m\rb 
\cr
&& \hskip 1in \times
\lb mS_m | M^3_0(0,0) | s_2 s_1\rb\nonumber\\[1mm]
&& \hskip -4cm 
=\bigg(\sum_{s'_1s'_2}
\lb s_1 s_2 |M^3_0(x,\tau )|s'_1s'_2\rb
\lb s'_2s'_1 | M^3_0(0,0) | s_2s_1\rb\bigg) + \cdots .
\end{eqnarray}
In the above we have truncated the sum arising from the resolution
of the identity.  With the first matrix element 
of the thermal trace, we only keep
terms from the resolution of identity with one excitation.  We 
are interested in the behavior of the susceptibility
at $\omega = 0$ and this term provides the only contribution\cite{konik}.
Similarly, the only term arising from the second matrix element of the thermal
trace contributing to the DC susceptibility
comes from keeping the term from the resolution of the identity involving
two excitations.
We can then evaluate $C(\om,k)$ with the result
\begin{equation}\label{eIVxxviii}
C(\om , k) = C_1(\om ,k) + C_2(\om ,k)\,.
\end{equation}
For $C_1(\om ,k)$, we then have the corresponding expression
\begin{eqnarray}\label{eIVxxix}
&& \hskip -.25in C_1(x,\tau) 
= \sum_{s_1s_2} \int \frac{d\th}{2\pi}\,\dto \,e^{-\Delta\beta\cosh (\th )} \lb A_{s_1}(\th)|\mo| A_{s_2} (\th_1 )\rb \cr
&& \hskip 1.75in \times
\lb A_{s_2} (\th_1 )|\mt|A_{s_1} (\th )\rb \nonumber\\[1mm]
&& \hskip -.25in =
\sum_{s_1s_2} \int \frac{d\th}{2\pi}\,\dto \,e^{-\Delta\beta\cosh (\th )} e^{-\tau \Delta (\cosh (\th_1 ) -\cosh (\th ))
+ ix\Delta (\sinh (\th_1 ) -\sinh (\th ))} \cr
&& \times \lb \mt| A_{s_2}(\th_1) A_{s_1} (\th - i\pi )\rb
\lb \mt| A_{s_1}(\th ) A_{s_2} (\th_1 - i\pi )\rb \nonumber\\[1mm]
&&\hskip -.25in = \sum_{s_1s_2} \int \frac{d\th}{2\pi}\,\dto \,e^{-\Delta\beta\cosh (\th )} e^{-\tau \Delta (\cosh (\th_1 ) -\cosh (\th ))
+ ix\Delta (\sinh (\th_1 ) -\sinh (\th ))} \nonumber\\[-2mm]
&& \hskip 1in \times f^{M^3_0}_{s_1s_2}(\th -i\pi,\th_1 ) f^{M^3_0}_{s_2s_1}(\th_1-i\pi,\th ).
\end{eqnarray}
We have used crossing symmetry in the second line.  From Section 4.2.2,
the form factor $f^{M^3_0}_{aa_1}(\th ,\th_1 )$ is
given by
\begin{equation}\label{eIVxxx}
f^{M^3_0}_{aa_1}(\th ,\th_1 ) = 
i  \frac{\pi^2\Delta}{4} \ep^{3aa_1}(\sinh (\th )+\sinh(\th_1))\psi(\th-\th_1).
\end{equation}
Then upon Fourier transforming the above in $x$ and $\tau$ and continuing
$\om_n \rightarrow -i\om + \delta$, we obtain
\begin{equation}\label{eIVxxxi}
C_1(\omega = 0, k = 0) =  \frac{\beta \Delta}{\pi}
\int d\th \cosh (\th ) e^{-\beta \Delta \cosh (\th )} = 
\frac{2\beta\Delta}{\pi}K_1(\beta\Delta),
\end{equation}
where $K_1$ is a modified Bessel function.
This has the expected small temperature behavior,
$C_1(\omega = 0, k= 0) \sim T^{-1/2} e^{-\beta \Delta} $.

On the other hand $C_2(\om = 0,k=0)$  is given by 
\begin{eqnarray}\label{eIVxxxii}
\hskip -6mm C_2(x,\tau) &=&
\frac{1}{4}\sum_{s_1s_2s_3s_4} \int \dto\dtt \dttr \dtf \nn[1mm]
&& \hskip .5in \times 
\lb A_{s_1}(\th_1)A_{s_2}(\th_2) | \mo | A_{s_3}(\th_3 ) A_{s_4}(\th_4 )\rb 
\nonumber\\[1mm]
&& \hskip .5in \times
\lb A_{s_4}(\th_4)A_{s_3}(\th_3) | \mt | A_{s_2}(\th_2 ) A_{s_1}(\th_1 )\rb \nn[1mm]
&& \hskip .5in - C_1(x,\tau)
\sum_a \int \dt \,\ebt \lb A_a (\th ) | A_a (\th ) \rb .
\end{eqnarray}
The last term is disconnected (proportional to $C_1(x,\tau)$) 
and will ultimately cancel (good, as it is proportional to $\delta (0)$).  It arises from the expansion of the 
partition function in Eqn. \ref{eIVxxvi}.
The four particle matrix elements
appearing in the above takes the form (as per Section 2.7.1):
\begin{eqnarray}\label{eIVxxxiii}
\lb A_{s_1}(\th_1)A_{s_2}(\th_2) | \mo | A_{s_3}(\th_3 ) A_{s_4}(\th_4 )\rb 
&& \nonumber\\[1mm]
&& \hskip -2in =\delta_{s_1s_4} 2\pi\delta(\th_1-\th_4)
f^{M^3_0}_{\bar{s}_2,s_3}(\th_2-i\pi,\th_3) \nonumber\\[1mm]
&& \hskip -2in +\,\delta_{s'_3s'_2} 2\pi \delta(\th_3-\th_2)
S^{s'_1s'_2}_{s_1s_2}(\th_{12})S^{s'_3s'_4}_{s_3s_4}(\th_{34})
f^{M^3_0}_{\bar{s}'_1,s'_4}(\th_1-i\pi,\th_4)\nonumber\\[1mm]
&& \hskip -2in + \,\delta_{s'_2s_4} 2\pi \delta(\th_2-\th_4)
S^{s'_1s'_2}_{s_1s_2}(\th_{12})
f^{M^3_0}_{\bar{s}'_1,s_3}(\th_1-i\pi,\th_3)\nonumber\\[1mm]
&& \hskip -2in +\, \delta_{s_1s'_3} 2\pi \delta(\th_1-\th_3)
S^{s'_3s'_4}_{s_3s_4}(\th_{34})
f^{M^3_0}_{\bar{s}_2,s'_4}(\th_2-i\pi,\th_4)\nonumber\\[1mm]
&& \hskip -2in + f^{M^3_0}_{\bar{s}_2,\bar{s}_1,s_4,s_3}
(\th_2-i\pi,\th_1-i\pi,\th_4,\th_3)_{\rm c}\,,
\end{eqnarray}
where $f_c$ refers to a connected form-factor.
Substituting this expression into the above and using the form factors of Section 4.2.2,
we find (see Ref. \cite{konik} for details)
\begin{eqnarray}\label{eIVxxxiv}
C_2(\om = 0, k = 0) &=& 
-\frac{6\beta\Delta}{\pi}\, K_1(2\beta\Delta ) \cr
&& \hskip -1in +\, \frac{2\beta \Delta}{\pi}\int d\th_1d\th_2 
e^{-\beta\Delta (\cosh(\th_1)+\cosh(\th_2))} \cosh(\th_1)
\,\frac{11\pi^2+2\th_{12}^2}{\th_{12}^4+5\pi^2\th_{12}^2+4\pi^4}\cr
&& \hskip -1in = -\frac{6\beta\Delta}{\pi}\, K_1(2\beta\Delta )
+ \frac{22 \beta \Delta}{\pi^3}K_0(\beta\Delta)K_1(\beta\Delta) 
+ {\cal O}\Big(\frac{T}{\Delta}e^{-2\beta\Delta}\Big)\,,
\end{eqnarray}
where $\th_{12} = \th_1-\th_2$ and $K_n$ are standard 
modified Bessel functions.  
The first term in $C_2$ is a `disconnected' contribution
related to $C_1$.  The second term is a connected contribution and
as such is genuinely distinct from $C_1$.

\begin{figure}
\label{figIViv}
\begin{center}
\epsfxsize=0.55\textwidth
\rotatebox{270}{\epsfbox{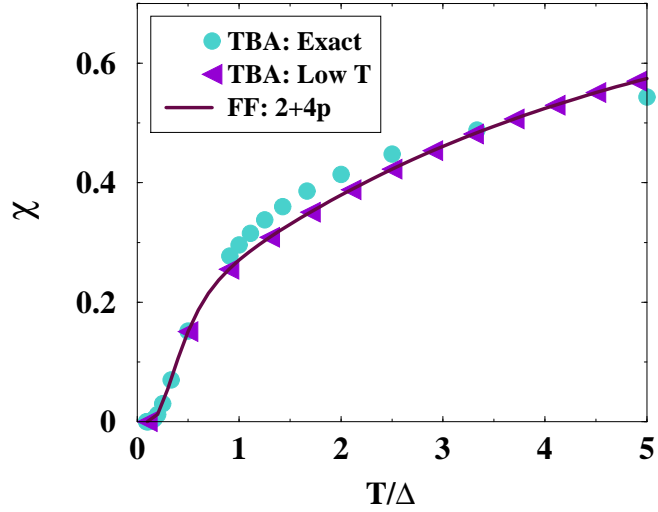}}
\caption{Plots of the zero-field
susceptibility computed both from the TBA equations
and from the form factor expansion.
The first of these is an
exact numerical solution of the TBA equations for the O(3) non-linear sigma model.  
The second is arrived at from a
small temperature expansion in powers of $e^{-\beta\Delta}$ of these same
equations.  
The final plot gives the form factor computation of the
susceptibility.  We have truncated the form factor expansion
at the four particle level.}
\end{center}
\end{figure}

It is possible in the case of the $O(3)$ sigma model to arrive
at exact expressions (in the form of coupled integral equations)
for the zero-field susceptibility \cite{tsvelik,wiegmann}.  These
equations, in their most compact form, appear as\\[-8mm]~
\begin{eqnarray}\label{eIVxxxv}
\chi (H=0) &=& - \frac{\Delta}{2\pi} \int d\th \cosh (\th ) 
\frac{\del^2_H \ep (\th )|_{H=0}}{1+e^{\beta\ep(\th )}};\cr
\ep (\th ) &=& \Delta\cosh(\th ) - T \int d\th'
\log (1+e^{\beta\ep_2(\th')})s(\th-\th');\cr
\ep_n (\th ) &=& T \int d\th '
s(\th-\th ')\bigg\{
\log (1+e^{\beta\ep_{n-1}(\th')}) \cr
&& \hskip 1.25in +\log (1+e^{\beta\ep_{n+1}(\th ')})
+\delta_{2n}\log (1+e^{\beta\ep(\th')})\bigg\}
\cr
H &=& \lim_{n\rightarrow \infty} \frac{\ep_n(\th )}{n} \,.
\end{eqnarray}
These equations admit
a closed form low temperature expansion.  The details of this
expansion may be found in Ref.\cite{tsvelik}.  Here we just quote the results
\begin{eqnarray}\label{eIVxxxvi}
\chi &=& \frac{2\beta \Delta}{\pi} K_1(\beta\Delta)
-\frac{6\beta\Delta}{\pi} K_1(2\beta\Delta ) \cr\cr
&& \hskip -.7in + \frac{2\beta \Delta}{\pi}\int d\th_1d\th_2 
e^{-\beta\Delta (\cosh(\th_1)+\cosh(\th_2))} \cosh(\th_1)\,
\frac{11\pi^2+2\th_{12}^2}{\th_{12}^4+5\pi^2\th_{12}^2+4\pi^4}\,.
\end{eqnarray}
Remarkably, we see this expansion agrees exactly with the corresponding
expression derived with the aid of form factors.  Thus
the form factor expansion at finite temperature meets an
important test.

In Fig.\,17
are plotted the susceptibilities computed via an exact numerical
analysis of the TBA equations, a low temperature expansion
of the same equations, and a computation based upon the two and
four particle form factors.
We see that as indicated previously that
the form factor computation and the low temperature expansion
match exactly.  Moreover these two computations track
the exact susceptibility over a considerable range of temperatures
despite the fact
these computations are truncated low temperature expansions.

\subsubsection{NMR Relaxation Rate}

In this section we compute the NMR relaxation rate, $\tn$.
We are interested in computing this rate in order to compare it
to the experimental data found in Ref. \cite{takigawa} on the relaxation
rate of the quasi one-dimensional spin chain, $AgVP_2S_6$.
For temperatures in excess of $100K$ (the gap, $\Delta$,
in this compound is on the order of $320K$),
the experimental data \cite{takigawa}
shows the relaxation rate to have an inverse dependence
upon $\sqrt{H}$:
$$
\tn \propto \frac{1}{\sqrt{H}}\ .
$$
This dependence is nicely reproduced by the semi-classical methodology
in Refs. \cite{damle1,damle2}.  Moreover the semi-classical computation
reproduces the activated behavior of $\tn$ in this same
temperature regime:
$$ 
\tn \propto e^{-3\bd/2}.
$$
We are interested in determining whether a calculation in the fully
quantum O(3) NLSM can reproduce these results.  To this end we
compute $\tn$ using a form factor expansion.  Sagi and Affleck \cite{sagi}
have already done such a computation to lowest order in $e^{-\bd}$.
But they do not find the above behavior.  Rather they see
$$
\tn \propto \log (H) ; ~~~~~ \tn \propto e^{-\bd} .
$$
We continue this computation one further step, computing to $\CO (e^{-2\bd})$.
Given the behavior, $\tn \sim H^{-1/2}$, appears only as T is increased beyond
100K (i.e. $T/\Delta \sim 1/3$), 
it is not unreasonable to suppose higher order terms in a low
temperature expansion of $\tn$ are needed to see this singularity.

\begin{figure}
\begin{center}
\epsfxsize=0.5\textwidth
\rotatebox{270}{\epsfbox{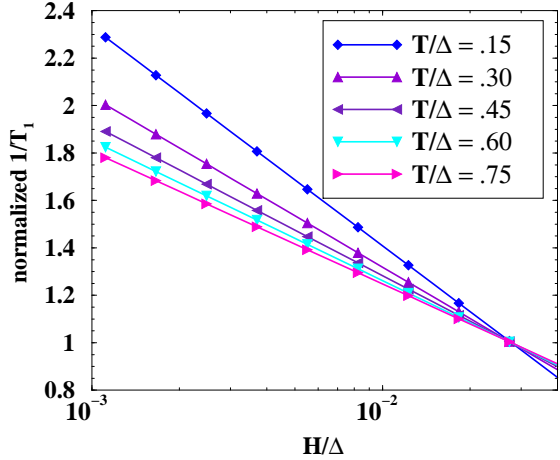}}
\caption{In this log-linear
plot we present the form factor
computation of the NMR relaxation rate, $1/T_1$, as a function of H
for a variety of temperatures.  We plot a normalized
rate, the ratio of $1/T_1(H)$ with $1/T_1 (H=\Delta/36)$.}
\end{center}
\label{figIVv}
\end{figure}

To proceed with the computation of $\tn$, we review its constituent
elements.  $\tn$ can be expressed in terms of the spin-spin
correlation function \cite{sagi}:
\begin{equation}\label{eIVxxxvii}
\tn = \sum_{\genfrac{}{}{0pt}{}{\alpha = 1,2}{\beta = 1,2,3}}
\int \frac{dk}{2\pi} A_{\alpha\beta}(k)A_{\alpha\gamma}(-k)
\lb M^\beta_0M^\gamma_0 \rb (k,\om_N)\,,
\end{equation}
where $\om_N = \gamma_N H$ is the nuclear Lamour frequency with 
$\gamma_N$ the nuclear gyromagnetic ratio and the $A_{\alpha\beta}$
are the hyperfine coupling constants.  In the above we assume H is
aligned in the 3-direction.
The above integral is dominated
by values of $k$ near 0 \cite{sagi}.  Moreover in the relevant experiment,
the hyperfine couplings are such that only $\lb M^1_0M^1_0 \rb$
contributes.  Hence
\begin{equation}\label{eIVxxxviii}
\tn \propto \lb M^1_0M^1_0 \rb (x=0,\om_N \sim 0)\,.
\end{equation}
We now proceed to compute $\lb M^1_0M^1_0 \rb$.

To compute $\lb M^1_0M^1_0 \rb$, we again employ a form factor
expansion.  Akin to the computation of the susceptibility and 
the spin conductance, this computation amounts to a low temperature
expansion of $\lb M^1_0M^1_0 \rb$,
$$
\lb M^1_0M^1_0 \rb = a_1 e^{-\bd} + a_2 e^{-2\bd} + \cdots ,
$$
where we are able to compute $a_1$ and $a_2$.
For the details of this computation we refer the reader to Ref. \cite{konik}.
We find there
\begin{eqnarray}\label{eIVxxxix}
\lb M^1_0M^1_0 \rb (x=0,\om = 0) &=& \bigg(\frac{2\Delta}{\pi}\,e^{-\bd}
\big(\log (\frac{4T}{H}) - \gamma\big) 
-\frac{6\Delta}{\pi}\,e^{-2\bd}\big(\log (\frac{2T}{H}) - \gamma\big)\nonumber\\[1mm]
&& \hskip -1.7in + {\Delta}e^{-2\bd}\big(\log (\frac{4T}{H})-\gamma\big)
\sqrt{\frac{2\pi}{\bd}}\Big(24\pi + \frac{17}{\pi^3}\Big)\bigg)
\big(1 + \CO (H/T) + \CO (T/\Delta)\big)\,,\cr &&
\end{eqnarray}
where $\gamma=.577\ldots$ is Euler's constant.
We are interested in the regime $H \ll T \ll \Delta$ (the
regime where it is expected spin diffusion produces singular behavior
in $\tn$).  The terms that we have dropped do not affect this behavior.
In principle there is no difficulty in writing down the exact expression
(to $\CO (e^{-2\bd})$); it is merely unwieldy.
This expression for $\tn$ is plotted in Fig. 18 for a variety 
of values of the ratio $T/\Delta$.

We see that we do not obtain the same behavior 
as found in Refs. \cite{damle1,damle2}.
Going to the next order in $\CO (e^{-2\bd})$ produces a behavior in $\tn$
as $H\rightarrow 0$ identical to the lower order computation of $\obd$:
we again find a logarithmic behavior consistent with ballistic transport.
An alternative comparison we might make to the results of Refs. \cite{damle1,damle2}
is to perform a low temperature expansion (in $\obd$) of the semi-classical
computation of $\lb M^1_0M^1_0 \rb (x=0,\om = 0)$.
Doing so by
treating $Te^{-\bd}/H$ as a small parameter, we find
\begin{eqnarray}\label{eIVxl}
\lb M^1_0M^1_0 \rb (x=0,\om = 0) &\propto & \Delta e^{-\bd}
\Big(\log (\frac{4T}{H}) - \gamma + \Big(\frac{\pi}{4} - \frac{1}{2}\Big)\,\frac{T^2}{\pi H^2}\,
e^{-2\bd}\cr\cr
&&\hskip 1.5in  + \CO (e^{-3\bd})\Big)\,.
\end{eqnarray}
We see that the low temperature expansion of the
semi-classical result agrees to leading order with
our computation but afterward differs.  
It possesses no term of $\CO(e^{-2\bd})$.  The next term rather
appears at $\CO (e^{-3\bd})$ and possesses a $1/H^2$ divergence.
That the small $H$ behavior
should be $1/\sqrt{H}$ does suggest the importance
of summing up terms.  But the lack
of a term of $\CO (e^{-2\bd})$ in the semi-classical result nonetheless
hints that the two results are genuinely different.

\subsubsection{Drude Weight of Spin Conductance at Finite Field}

In this section we compute the spin conductivity, $\sigma_s$.
The spin conductivity gives the response of the spin chain to
a spatially varying magnetic field.  It is defined via
\begin{equation}\label{eIVxli}
j_1 (x,t) = \sigma_s \nabla H\, ,
\end{equation}
and so can be expressed in terms of a Kubo formula,
\begin{equation}\label{eIVxlii}
{\rm Re}\thinspace\sigma_s (k,\om )= -\frac{1}{k}
\int dx 
dte^{ikx+i\om t}\thinspace{\rm Im}\lb j_0(x,t)j_1(0,0)\rb_{\rm retarded}\,.
\end{equation}
In the notation used here the spin current $j_1$ is synonymous
with $M^3_1$ of Section 4.2.2. (for a field in the 3-direction), the Lorentz current counterpart of the uniform magnetization,
$M^3_0\equiv j_0$.  We will focus primarily on computing the Drude weight, D,
of ${\rm Re}\thinspace\sigma_s$, i.e. computing the term in $\sigma_s (k,\om )$
of the form
\begin{equation}\label{eIVxliii}
\sigma_s (k=0,\om ) = D \delta (\om )\,.
\end{equation}
However we are able to compute $\sigma_s$ for general $k,\om$.  We
find that for $\omega \ll 2\Delta$, $k=0$, the spin conductivity is
described solely by the Drude weight.  In particular, we find no indication
of a regular contribution to $\sigma_s (k=0,\om )$.

\begin{figure}
\begin{center}
\epsfxsize=0.5\textwidth
\rotatebox{270}{\epsfbox{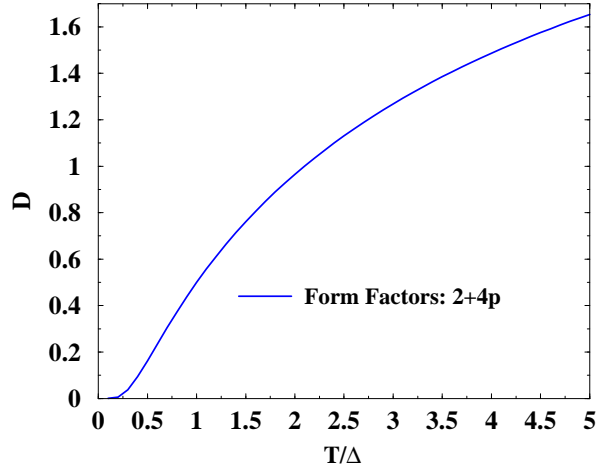}}
\caption{In this plot we present the form factor
computation of the zero field ($H=0$) Drude weight, D, of the spin conductance.}
\end{center}
\label{figIVvi}
\end{figure}

To evaluate $\sigma_s$, we employ the identical form factor
expansion to that used in computing the susceptibility.  And like the
susceptibility, our result is an exact low temperature expansion
of $D$,
$$
D = \sum_n D_n e^{-n\bd}.
$$
Here we will compute $D_1$ and $D_2$ exactly.  As the details of the
computation are nearly identical to that of the susceptibility, we merely
write down the results:
\begin{eqnarray}\label{eIVxliv}
D(H=0) &=& {\beta \Delta} \int d\th e^{-\bd\cosh(\th)}
\frac{\sinh^2(\th)}{\cosh(\th)}(1-3e^{-\bd\cosh(\th)}) \cr\cr
&&\hskip -.6in + 2\bd \int d\th_1d\th_2
e^{-\beta\Delta (\cosh(\th_1)+\cosh(\th_2))} 
\frac{\sinh^2(\th_2)}{\cosh(\th_2)}
\frac{11\pi^2+2\th_{12}^2}{\th_{12}^4+5\pi^2\th_{12}^2+4\pi^4}\cr\cr
&& \hskip 2in + \CO (e^{-3\bd}) \cr
&& \hskip -.2in 
= e^{-\beta\Delta} \sqrt{\frac{2\pi}{\beta\Delta} }\,
\Big(1 +{\cal O}\Big(\frac{T}{\Delta}\Big)\Big)\cr\cr
&& \hskip -.2in - e^{-2\beta\Delta} \sqrt{\frac{1}{\beta\Delta}}
\Big(\frac{3}{2}\sqrt{\pi} - \frac{11}{\pi}\sqrt{\frac{T}{\Delta}}
+{\cal O}\Big(\frac{T}{\Delta}\Big)\Big)
+ {\cal O}(e^{-3\beta\Delta})\,.
\end{eqnarray}
This expression, like the susceptibility, involves only the two and four particle form
factors.
We plot this result in Fig.\,\ref{figIVvi} as a function of $T/\Delta$.

We first observe that $D(H=0) \neq 0$.  This
is in accordance with Ref. \cite{fuji} where $D$ is computed using an argument
involving the finite size scaling of 
the thermodynamic Bethe ansatz equations.  (We do note that the 
computation of $D$ at $H=0$ in Ref. \cite{fuji} appears only as a 
note added in proof and so is decidedly sketchy.  
However the equations governing $D$ developed
in Ref. \cite{fuji} are manifestly positive with the consequence $D$ cannot
vanish.)
But our results do differ from
the semi-classical computation of Ref. \cite{reply} where it was found that
$D$ vanishes at $H=0$.
We find as well no additional regular contributions
to $\sigma_s (\om , k=0)$ near $\om = 0$ -- only the Drude term is present
in contrast to Refs. \cite{damle1,damle2}.
(There will, however, be regular contributions at higher
frequencies, in particular for $\om > 2\Delta$, which persist
even in the zero temperature limit).

We have only given the spin conductivity at $H=0$.  However it is extremely
straightforward to generalize the form factor computation to finite H.
As H couples to the total spin, a conserved quantity, the form
factors, $f^\CO (x,t)$, 
of an operator, $\CO (x,t)$, carrying spin s, are altered
via the rule
$$
f^\CO (t) \rightarrow e^{iHst} f^\CO (t)\,.
$$
(In the case of the spin conductance, the spin currents, $j_\mu=M^3_\mu$, carry
no spin and so are not altered at all.)  The only remaining change
induced by a finite field is to the Boltzmann factor appearing in the
thermal trace.  If an excitation with rapidity, $\th$, carries spin s,
its Boltzmann factor becomes
$$
e^{-\beta(\Delta\cosh (\th )-sH)}.
$$
For example we find $D$ as a function of $H$ (to $\CO(e^{-\bd})$)
to be
\begin{equation}\label{eIVxlv}
D(H) = {\beta \Delta} \cosh (\beta H) \int d\th\, e^{-\bd\cosh(\th)}
\,\frac{\sinh^2(\th)}{\cosh(\th)}\ .
\end{equation}
Again this in agreement with Ref. \cite{fuji}.  Indeed Ref. \cite{fuji}
computes $D(H)$ at large $H/T$ (but $H\ll \Delta$) to be
\begin{equation}\label{eIVxlvi}
D = \frac{\beta\Delta}{4\pi}\, e^{\beta H} \int d\th\, \frac{\sinh^2 (\th )}{\cosh (\th )}\, e^{-\beta\Delta\cosh (\th)} + {\cal O}(e^{-2\beta\Delta})\,.
\end{equation}
Up to a factor of $2\pi$, this expression is in exact agreement with
\ref{eIVxlv}.
In this particular case our derivation of $D(H)$ agrees
with the semi-classical computation
\cite{reply} (provided $T \ll H \ll \Delta$).  The symmetries
in the semi-classical model that lead $D(H=0)$ to vanish are broken
for finite $H$.


\subsection{Transport: Ballistic or Diffusive}
We have compared our transport calculations to the semi-classical
computations in Refs. \cite{damle1,damle2}.  The essence of this
method lies in treating the spin-chain as a Maxwell-Boltzmann gas of
spins which interact with one another through the low energy limit of
the scattering of the \ots,
\begin{equation}\label{eIVxlvii}
S^{cd}_{ab} (\th =0 ) = - \delta_{ad}\delta_{cb}\, .
\end{equation}
In contradistinction to the semi-classical
computation, we have found that the Drude weight of the spin conductance
is finite in the limit of zero external field.  Our results for
the NMR relaxation rate, $\tn$, indicate a similar
discrepancy.  We, like Ref. \cite{sagi}, find that $\tn$ is characterized 
by ballistic logarithms.  These logarithms are relatively
robust: they continue to appear at higher orders in the low
temperature expansion.
We do not, however, see diffusive behavior
in the relaxation rate, i.e. $\tn \sim 1/\sqrt{H}$, nor does our low
temperature expansion match the low temperature expansion of the
semi-classical computation of the correlator.

To come to some sort of judgment between the form-factor and
the semi-classical approaches, an
understanding is needed of the differences between
our computations of the spin conductance and the NMR relaxation
rate.  In the case
of the first quantity, it is likely this difference is real and
not an artefact of our methodology.  The data
that goes into the spin conductance is identical to that needed to
compute the susceptibility and we know that we can match the low
temperature expansion of the susceptibility with a similar expansion
coming from the exact free energy.  Moreover we know that the Drude weight of
$\sigma_s (H=0)$ has been found to be finite from an approach
\cite{fuji} independent of ours.

In generic systems, the Drude weight, $D$, at finite
temperatures will be zero.  It is then the integrability  of
the O(3) NLSM and the attendant existence of an infinite number of conserved
quantities that leads to a finite weight.  The existence of these
quantities can be directly related to a finite $D$.  As discussed in 
Refs. \cite{castella,zotos1,zotos2}, $D$ is bounded from below via an inequality developed
by Mazur: 
\begin{equation}\label{eIVxlviii}
D \geq c \sum_n \frac{\langle J Q_n \rangle^2}{\langle Q_n^2\rangle}\,,
\end{equation}
where $J$ is the relevant current operator, $Q_n$ are a set of
orthogonal conserved quantities, i.e. 
$\langle Q_n Q_m \rangle = \delta_{nm}\langle Q_n^2 \rangle$, and
$c$ is some constant.
For a finite Drude weight, we then require that at least one matrix
element, $\langle J Q_n \rangle$, does not vanish.  While we do no
direct computations, we can obtain an indication of whether the matrix
elements vanish by examining the symmetries of the model.
Under the discrete ($Z_2$) symmetries of the O(3) NLSM, the 
spin current, $J$, transforms via
$$
Z_2(J) \rightarrow \pm J\,.
$$ 
In order that the matrix element, $\langle JQ_n \rangle$, not vanish
we require that
$$
Z_2(Q_n) \rightarrow \pm Q_n\,.
$$ 
The $Z_2$ symmetries in the O(3) NLSM include 
$n_a \rightarrow -n_a$, $a=1,2,3$,
parity, and time reversal.  The spin current we are interested in
transforms under rotations as a pseudo-vector.  Thus any charge, $Q_n$,
coupling to the current must also transform as such.  From the
work by L\"uscher \cite{lus}, there is at least one conserved
pseudo-vectorial quantity such that $\langle JQ_n \rangle$ does not vanish due
to the action of one of the above $Z_2$ symmetries.
For the sake of completeness we exhibit it.  Rewriting 
the magnetization and spin current, $M_{0,1}$, explicitly as antisymmetric
tensors,
$$
M^{ab}_\mu = n^a\partial_\mu n^b - n^b\partial_\mu n^a\, ,\quad\mu = 0,1\,;
$$
the conserved quantity takes the form,
$$
Q^{ab}(t) = \sum_c\int dx_1 dx_2\, {\rm sgn} (x_1-x_2)M^{ac}_0(t,x_1)M^{cb}_0(t,x_2)
-\int dx M^{ab}_1(t,x)\,.
$$
While the first term of $Q^{ab}$ does not contribute to the matrix element,
$\langle JQ^{ab} \rangle$, as it is bilinear in the currents, $M^{ab}$,
the second term does.  We point out that $Q^{ab}$ is an exotic
object inasmuch as it is a {\it non-local} conserved quantity.
As pointed out in Ref. \cite{lus}, it is the first in a series of
non-local charges.

While the structure of the conserved quantities in the O(3) NLSM
seem to be consistent with the existence of a finite Drude
weight, this is not the case in the semi-classical approach.
The dynamics of the semi-classical approximation used in 
Refs. \cite{damle1,damle2} are also governed by an infinite number of conserved quantities.
Importantly however, these are different than those appearing in the fully
quantum model.  In particular, the semi-classical approximation
does not admit non-local conserved quantities.
As shown in Refs. \cite{damle1,damle2}, the structure of the $Z_2$ symmetries
in the semi-classical approach is such that all matrix elements,
$\langle J Q_n \rangle$, vanish.  It would thus seem the absence
of a Drude weight in the semi-classical case is a consequence
of differences in the symmetries between the semi-classical and fully
quantum models.

To understand the discrepancies in 
the case of the NMR relaxation rate, $1/T_1$, is not as simple.
However if we believe that the spin conductance demonstrates finite 
temperature ballistic behavior, it is hardly surprising to find
the NMR relaxation rate characterized by ballistic logarithms.  Again
the difference between the fully quantum treatment and the semi-classical
approach will lie 
in the differences between the models' conserved quantities.  Nonetheless
one possibility that we must consider is that merely going to 
$\CO (e^{-2\bd})$ in the computation of $1/T_1$
is insufficient.  It is possible that we need 
to perform some resummation of contributions from all orders to see
the desired singular behavior, $\tn \sim 1/\sqrt{H}$.  While this would
belie our experience with computing the susceptibility and 
the spin conductance via the correlators,
the data that goes into
the two computations is not exactly identical.  Thus the possibility
that the
low temperature expansion of $\tn$ is not well controlled cannot be
entirely ruled out.

The differences in the nature of the conserved quantities between
the O(3) NLSM and the semi-classical model of Refs. \cite{damle1,damle2}
suggest the latter is not equivalent to the O(3) NLSM,
even at low energies.  An indication of this lack of equivalency
may lie in the universal nature of the ultra low energy S-matrix.
This quantity is the primary input of the semi-classical model.
The semi-classical model imagines a set of classical
spins interacting via
$$
S^{cd}_{ab} (\th = 0)  = -\delta_{ad}\delta_{cb}\, ,
$$
i.e. in the scattering of two spins, the spins exchange their 
quantum numbers.  However this specification may be insufficient
to adequately describe the O(3) NLSM.  Even beyond the quantum interference
effects which are neglected by the semi-classical 
treatment, it is not clear that the zero-momentum S-matrix
is enough to determine the model.  

In this light it is instructive
to consider the sine-Gordon model in its repulsive regime.  
The sine-Gordon model is given by the action,
\begin{equation}\label{eIVxlix}
S = \frac{1}{16\pi} \int dx dt\, \big( \partial_\mu \Phi \partial^\mu \Phi
+ \lambda \cos ({\beta}\Phi )\big)\,,
\end{equation}
where $\beta$ falls in the range $0 < \beta < 1$.
The model is generically gapped.  Its repulsive regime occurs in
the range, $1/2 < \beta^2 < 1$.  The model's
spectrum in this same range then consists solely of a doublet of solitons carrying
U(1) charge.  It is repulsive in the sense that the solitons have no bound
states.
The sine-Gordon
model has a similar low energy S-matrix to the O(3) NLSM,
$$
S^{cd}_{ab} (\th = 0)  = -\delta_{ad}\delta_{cb}\, ,
$$
where here the particle indices range over $\pm$, the two solitons
in the theory.  Thus we might expect that sine-Gordon model
to possess identical low energy behavior over its entire repulsive
regime. 

This is likely to be in general untrue. {
For example we might consider the behavior
of the single particle spectral function.  We might thus want
to compute a correlator of the form
$$
\langle \psi_+ (x,t) \psi_- (0,0) \rangle \,,
$$
where $\psi_\pm$ are Mandelstam fermions given by
\begin{eqnarray}\label{eIVl}
\psi_\pm (x,t) &=&
\exp \bigg( \pm \frac{i}{2}\Big(\frac{1}{\sqrt{2}\beta}+\sqrt{2}\beta\Big)\phi_L (x,t)
\mp \frac{i}{2}(\frac{1}{\sqrt{2}\beta}-\sqrt{2}\beta)\phi_R (x,t)\bigg);\cr\cr
\phi_{L/R} &=& \frac{1}{2}\bigg(\Phi (x,t) 
\pm i \int^x_{-\infty} dy \partial_t \Phi (y,t)\bigg)\, .
\end{eqnarray}
Now consider the finite temperature spectral function corresponding to this correlator with energy, $\omega$, 
fixed at just above the one-soliton gap.  This spectral function has a contribution from a one particle form factor.
This contribution's $\beta$-dependence, while present, is trivial in that it only appears in the overall normalization.
But because we are working at finite temperature, higher particle form factors will also contribute.
Their contribution will depend upon $\beta$ in a more complex fashion than
an overall normalization.  We thus expect the low energy properties of this spectral function
to have an overall non-trivial dependence upon
$\beta$.  This accords with intuition.
$\beta$ determines the compactification radius
of the boson in the model and so is related in a fundamental
way to the model's properties.

It is useful to point out that Mandelstam fermions are the unique
fields that create/destroy solitons that carry Lorentz spin 1/2, i.e.
a spin that is independent of $\beta$.  They would then be the only
fields with a chance of matching any semi-classical computation.
However there are other soliton creation fields, for example,
$$
e^{\pm i\phi_{L,R}/(\sqrt{2}\beta)},
$$
for which one could determine the corresponding spectral density.
As these fields carry spin that varies as a function of $\beta$,
their spectral functions will depend upon more than the ultra low
energy soliton S-matrix.  In general, the semi-classical treatment
of the sine-Gordon model cannot capture its full quantum field content.

As with the O(3) NLSM, the conductance of the fully quantum
model differs from that of the semi-classical treatment.
If one were to compute the conductance at finite temperature in
the sine-Gordon model one would again find a finite Drude weight, $D$,
while the semi-classical approach yields $D=0$ \cite{rosch}.
The notion of under-specificity appears here again.  The semi-classical
approach for the sine-Gordon model equally 
well describes the Hubbard model at half-filling (the solitons are replaced
by particle/hole excitations in the half-filled band).  But it fails to 
give the correct Drude weight.  An analysis of finite size corrections
to the free energy in the presence of an Aharonov--Bohm flux 
\cite{fuji1} again finds a finite Drude weight in the half-filled
Hubbard model at finite temperature.

Interestingly however, there are certain properties at low energies
that seem to be independent of $\beta$.  For example, if one were to
compute the low temperature static charge susceptibility, the term
of $\CO (e^{-\bd})$ would be independent of $\beta$.  However at the
next order, $\CO (e^{-2\bd})$, this would almost certainly cease to be
true.  And the energy/temperature ranges we are interested in exploring
do not permit dropping terms of $\CO (e^{-2\bd})$.

It is important to stress we do not question the agreement between the
semi-classical model and experiment.  What we do question is whether
the fully quantum O(3) NLSM exhibits spin diffusivity.  If we are then
to understand spin diffusion in terms of the O(3) NLSM, it is possible
we need to include additional physics explicitly such as an easy axis spin
anisotropy (weakly present in the experimental system, $AgVP_2S_6$),
inter-chain couplings, or a spin-phonon coupling (as done in \cite{fuji}).

Beyond these, another mechanism that might lead to diffusive behavior are 
small integrable breaking perturbations of the O(3) NLSM.  Generically any
physical realization of a spin chain will possess such perturbations,
even if arbitrarily small.  Such perturbations may introduce the necessary
ergodicity into the system, ergodicity that is absent in the integrable
model because of the presence of non-trivial conserved charges, and so
lead to diffusive behavior.  As discussed in the semi-classical context
by Garst and Rosch \cite{rosch}, such perturbations introduce an additional
time scale, $\tau$, governing the decay of conserved quantities
in the problem.  For times, $t < \tau$, the behavior of
the system is ballistic and the original conserved quantities do not decay.
For times, $t > \tau$, the behavior is then diffusive.  Consequently the
Drude weight in the purely integrable model is transformed into a peak
in $\sigma (\omega )$ at $\omega \sim 1/\tau$.

Now the difference in the physics between the O(3) NLSM and its semi-classical
variant is not that of integrable breaking perturbations.  
As demonstrated in Refs. \cite{damle1,damle2},
their semi-classical model is classically integrable.  
However as discussed above the models do possess different conserved
charges.  It might then seem for certain transport quantities, 
the semi-classical model cures the lack of ergodicity present
in its quantum counterpart.

\section{U(1) Thirring Model and Quasi One Dimensional Mott
  Insulators} 
\setcounter{equation}{0}

The Mott metal-insulator transition \cite{Mott,Mottbook,Gebhardbook}
is a zero temperature (``quantum'') phase transition between a gapless
metallic phase and a gapped insulating one. It occurs at some critical
ratio of the strength of the electron-electron interaction ``$U$'' to the
kinetic energy, which is usually measured in terms of the tunneling
amplitude $t$ of an appropriate tight-binding model. In practice $U/t$
can be varied by applying pressure to a crystal, which results in a
better overlap between the orbitals of the conduction band and leads
to an increase in $t$. In high-energy physics the Mott transition
corresponds to the phenomenon of dynamical mass generation.

The intrinsic difficulty in describing the Mott transition
quantitatively in the general case is that it occurs when the kinetic
and potential energies are of the same order of magnitude. This regime
is difficult to access from either the ``band'' limit, where one
diagonalizes the kinetic energy first and then takes electron-electron
interactions into account perturbatively, or the ``atomic'' limit, in
which the electron-electron interaction is diagonalized first and the
electron hopping is taken into account perturbatively. Interestingly,
the Mott transition on lattices where tunneling along one direction is
much larger than along all others can be understood in some detail by
employing methods of integrable quantum field theory. 
The systems under consideration can
be thought of in terms of weakly coupled chains of electrons and will
be referred to as {\sl quasi one dimensional Mott insulators}. When
the band is half-filled and the interchain tunneling is switched off, 
Umklapp processes dynamically generate a spectral gap $M$ and we are
dealing with an ensemble of uncoupled 1D Mott insulating chains. 
The same Umklapp scattering mechanism can generate gaps at {\sl any}
commensurate filling e.g. quarter filling, but only if the
interactions are sufficiently strong. There are two questions we want
to address: 

\vskip .1in 

{\bf(i)} What is the dynamical response of the uncoupled
Mott-insulating chains system? 

\vskip .05in 

{\bf (ii)}  What are the effects of a weak
interchain tunneling?

\vskip .1in 

\noindent As far as point (i) is concerned we will
concentrate on the optical conductivity and the single particle
spectral function. Results are also available for the density-density
response function \cite{CE02} and the dynamical structure factor
\cite{BEG03}. 
Examples of materials that are believed to fall
into the general category of quasi-1D Mott insulators are the
Bechgaard salts \cite{review} and chain cuprates like ${\rm
  SrCuO_2}$, ${\rm Sr_2CuO_3}\,$\,\footnote{\,More precisely, these 
  compounds are considered to be   charge-transfer insulators.} or
${\rm PrBa_2Cu_3O_7}$. They have been found to exhibit very rich and
unusual physical properties such as spin-charge separation
\cite{kim96,kobayashi99,fujisawa99,Igor99,Igor04,p123arpes}.

\subsection{Lattice Models of Correlated Electrons}

The simplest models used in the description of one dimensional Mott
insulators are ``extended'' Hubbard models of the form 
\bea
\hspace{-8mm}
{\hat H}&=&-t\sum_{n,\sigma}
\left[c^\dagger_{n,\sigma}c_{n+1,\sigma}+{\rm h.c.}\right]
+U\sum_k n_{k,\uparrow}n_{k,\downarrow}
+\sum_{j\geq 1}V_j\sum_k n_{k}n_{k+j}\ ,
\label{Hamiltonian}
\eea
where $n_{k,\sigma}=c^\dagger_{k,\sigma}c_{k,\sigma}$
and $n_k=n_{k,\uparrow}+n_{k,\downarrow}$ are electron number
operators. The electron-electron interaction terms mimic the effects
of a screened Coulomb interaction. Two cases are of particular
interest from the point of view of application to, for example, the Bechgaard
salts \cite{review}: 
\begin{enumerate}
\item{} half filling (one electron per site);
\item{} quarter filling (one electron per two sites). 
\end{enumerate}
We will discuss both
these cases and emphasize similarities and differences in their
dynamical response.

\subsection{Field Theory Description of the Low Energy Limit}
The field theory limit is constructed by keeping only the low-energy
modes in the vicinity of the Fermi points $\pm k_F$. We may express
the lattice electron annihilation operator in terms of the slowly
varying (on the scale of the lattice spacing $a_0$) right and left
moving electron fields $R(x)$ and $L(x)$,  
\begin{equation}
c_{\,l,\sigma}\longrightarrow \sqrt{a_0} \left[\exp(ik_F x)\
  R_\sigma(x)+\exp(-ik_F x)\ L_\sigma(x)\right] .
\label{cpsi}
\end{equation}
Here $k_F=\pi/2a_0$ for the half-filled band and $x=l a_0$. The
resulting fermion Hamiltonian can then be bosonized by standard
methods \cite{GNT}. In order to make our presentation reasonably
self-contained we review the relevant steps in the half-filled case
next.

\subsubsection{Half Filled Band}
Using \r{cpsi} in \r{Hamiltonian} we arrive at the following
low-energy effective theory
\begin{eqnarray}
{\mathcal H}&=& \sum_{\sigma} v_F\! \int\! dx\, :\!\left[
L^\dagger_\sigma\ i\partial_x L_\sigma - R_\sigma^\dagger\ 
i \partial_x R_\sigma\right]\!:
-\frac{g_0}{6}\!\int\! dx\,:\!\left[{\bf J}\cdot {\bf J} +
{\bf \bar{J}}\cdot {\bf\bar{J}} \right]\!:\nn
&&-g_s\!\int\! dx\,:\! {\bf J}\cdot {\bf\bar{J}}\!:
+\frac{g_c^\perp+g_c^\parallel-g_0}{6}
\!\int dx\, :\!\left[{\bf I}\cdot{\bf I} + {\bf
\bar{I}}\cdot {\bf\bar{I}}\right]\!:\nn
&&+\int dx\left[\frac{g_c^\perp}{2}:\!({I}^+{\bar{I}}^-+
{I}^-{\bar{I}}^+)\!:+g_c^\parallel :\!{I}^z{\bar{I}}^z\!:\right].
\label{hamil1}
\end{eqnarray}
Here $v_F=2ta_0$ is the Fermi velocity,
\be
g_0=2U a_0,\quad g_s=g_c^\perp=2a_0(U-2V_1+2V_2),\quad
g_c^\parallel=2a_0(U+6V_1+2V_2),
\ee
and $\bf J$, ${\bf I}$ are the chiral components of SU(2) spin and
pseudospin currents,
\begin{eqnarray}
\bar{I}^z&=&\frac{1}{2}:\!\left(L^\dagger_\up L_\up+L^\dagger_\downarrow 
L_\downarrow\right)\!:\ ,\quad
\bar{I}^+=(\bar{I}^-)^\dagger=L^\dagger_\uparrow L^\dagger_\downarrow\
,\nonumber\\ 
I^z&=&\frac{1}{2}:\!\left(R^\dagger_\uparrow R_\uparrow 
+R^\dagger_\downarrow R_\downarrow\right)\!:\ ,\quad
{I}^+=(I^-)^\dagger=R^\dagger_\uparrow R^\dagger_\downarrow\,\nonumber\\ 
\bar{J}^z&=&\frac{1}{2}:\!\left(L^\dagger_\uparrow L_\uparrow
-L^\dagger_\downarrow L_\downarrow\right)\!:\ ,\quad
\bar{J}^+=(\bar{J}^-)^\dagger=
L^\dagger_\uparrow L_\downarrow\ ,\nonumber\\
{J}^z&=&\frac{1}{2}:\!\left(R^\dagger_\uparrow R_\uparrow
-R^\dagger_\downarrow R_\downarrow\right)\!:\ ,\quad
{J}^+=(J^-)^\dagger=R^\dagger_\uparrow R_\downarrow\ .
\label{currents}
\end{eqnarray}
Here ``$:$'' denotes normal ordering of point-split expressions
\cite{Affleck86b}. The ``kinetic'' terms in the Hamiltonian \r{hamil1}
can be expressed as normal ordered bilinears of currents as well
\cite{Dashen75a,Affleck86b,GNT}
\bea
\frac{2\pi}{3}\int dx\ :\left[
{\bf I}\cdot {\bf I}+{\bf J}\cdot {\bf J}\right]:&=&
-\int dx\left[\sum_\sigma:R^\dagger_\sigma\ i\partial_x\
  R_\sigma:\right],\nn
\frac{2\pi}{3}\int dx\ :\!\left[
\bar{\bf I}\cdot \bar{\bf I}+\bar{\bf J}\cdot \bar{\bf J}\right]\!:&=&
\int dx\left[\sum_\sigma:L^\dagger_\sigma\ i\partial_x\
  L_\sigma:\right].
\label{currentskinetic}
\eea
Using \r{currentskinetic} the Hamiltonian (\ref{hamil1}) can now be
split into two parts, corresponding to the spin and charge sectors
respectively
\begin{eqnarray}
{\mathcal H}&=& {\mathcal H}_c+{\mathcal H}_s\ ,\nonumber\\
{\mathcal H}_c&=& \frac{2\pi v_c'}{3}\int dx\ :\left[
{\bf I}\cdot {\bf I}+\bar{\bf I}\cdot \bar{\bf I}\right]:\nn
&&+\int dx\left[ \frac{g_c^\perp}{2}:\![{I}^+{\bar{I}}^-+
{I}^-{\bar{I}}^+]\!:+g_c^\parallel :\!{I}^z{\bar{I}}^z\!:\right],\nn
{\mathcal H}_s&=& \frac{2\pi v_s}{3}\int dx\ :\left[
{\bf J}\cdot {\bf J}+\bar{\bf J}\cdot \bar{\bf J}\right]:
-g_s\int dx\ :{\bf J}\cdot \bar{\bf J}:\ .
\label{su2thi}
\end{eqnarray}
Here $v_s=v_F-Ua_0/2\pi$ and $v_c'=v_F+(U+4V_1+4V_2)a_0/2\pi$. 
The Hamiltonian \r{su2thi} with $v_s=v_c'$ and $g_s=0$ is known as the
U(1) Thirring model in the literature. As we will see the difference
in velocities can be accommodated and as will show now the
current-current interaction in the spin sector is marginally
irrelevant and will only lead to logarithmic corrections at low
energies. The 1-loop renormalization group equation for $g_s$ is
\be
r\,\frac{\partial g_s}{\partial r}=-\frac{g_s^2}{2\pi v_s}\ .
\ee
Hence $g_s$ diminishes under renormalization and the current-current
interaction in the spin sector is marginally irrelevant. In order to
keep things simple we will drop it from now on. Keeping it would lead
to logarithmic corrections in many of the formulas below.
The Hamiltonian \r{su2thi} can now be bosonized using
\bea
L^\dagger_\sigma(x)&=&\frac{\eta_\sigma}{\sqrt{2\pi}}e^{if_\sigma\pi/4}
\exp\left(-\frac{i}{2}\bar{\phi}_c\right)
\exp\left(-\frac{if_\sigma}{2}\bar{\phi}_s\right),\nonumber\\[2mm]
R^\dagger_\sigma(x)&=&\frac{\eta_\sigma}{\sqrt{2\pi}}e^{if_\sigma\pi/4}
\exp\left(\frac{i}{2}\phi_c\right)
\exp\left(\frac{if_\sigma}{2}\phi_s\right),
\label{LR0}
\eea
where $\eta_a$ are Klein factors with $\{\eta_a,\eta_b\}=2\delta_{ab}$
and where $f_\uparrow=1$, $f_\downarrow=-1\,$.\,\footnote{\,The phase factors in
\r{LR} have been introduced in order to ensure the standard
bosonization formulas for the staggered magnetizations.}
The canonical Bose fields $\Phi_{s,c}$ and their respective dual
fields $\Theta_{s,c}$ are given by 
\bea
\Phi_a=\phi_a+\bar{\phi}_a\ ,\qquad
\Theta_a=\phi_a-\bar{\phi}_a\ ,\ a=s,c\ ,
\eea
where the chiral boson fields $\phi_a$ and $\bar{\phi}_a$ fulfill the
following commutation relations
\be
[\phi_a(x),\bar{\phi}_a(y)]=2\pi i\ ,\quad a=c,s.
\ee
We choose a normalization such that for $|x-y|\longrightarrow 0$ the
following operator product expansion holds ($a=s,c$)
\bea
\exp\left(i\alpha\Phi_a(x)\right)\exp\left(i\beta\Phi_a(y)\right)
\longrightarrow
{|x-y|^{4\alpha\beta}}\exp\left(i\alpha\Phi_a(x)\!+\!i\beta\Phi_a(y)\right).~~~~~
\eea
Applying the bosonization identities we obtain the following bosonic
form of the low energy effective Hamiltonian (we recall that we have set
$g_s=0$) 
\begin{eqnarray}
{\cal H}_c&=& \frac{v_c'}{16\pi}\int dx\left[
(\partial_x\Phi_c)^2+(\partial_x\Theta_c)^2\right]\nn
&&-\frac{g_c^\perp}{(2\pi)^2}\int dx\ \cos(\Phi_c)
+\frac{g_c^\parallel}{(8\pi)^2}\int dx\left[
(\partial_x\Phi_c)^2-(\partial_x\Theta_c)^2\right]\ ,\nn
{\cal H}_s&=&\frac{v_s}{16\pi}\int dx\left[
(\partial_x\Phi_s)^2+(\partial_x\Theta_s)^2\right].
\label{su2thiboso}
\end{eqnarray}
Finally we carry out a canonical transformation on the charge boson
\be
\Phi_c\longrightarrow\beta\Phi_c\ ,\qquad
\Theta_c\longrightarrow\frac{1}{\beta}\Theta_c\ ,
\ee
where
\be
\beta=\left[\frac{1-g_c^\parallel/4\pi v_c'}
{1+g_c^\parallel/4\pi v_c'}\right]^\frac{1}{4}.
\ee
The transformation property of the dual field follows from the fact
that $-\partial_x\Theta_c$ is the momentum conjugate to $\Phi_c$.
In terms of the rescaled fields the charge sector takes the form of a
sine-Gordon model
\begin{eqnarray}
{\mathcal H}_{c}&=&\frac{v_{c}}{16\pi}\int dx
\left[(\partial_x\Theta_{c})^2+(\partial_x\Phi_{\rm c})^2
\right]-\frac{g_c^\perp}{4\pi^2}\int dx\, \cos\beta\Phi_{c}\ ,
\label{HamiltFT}
\end{eqnarray}
where $v_c=(v_c'+g_c^\parallel/4\pi)\beta^2$. The entire procedure we have been
following can be summarized as follows. We first project the lattice
Hamiltonian to the low-energy degrees of freedom using \r{cpsi} and
then bosonize the resulting fermion Hamiltonian by means of the identities
\bea
L^\dagger_\sigma(x)&=&\frac{\eta_\sigma}{\sqrt{2\pi}}e^{if_\sigma\pi/4}
\exp\left(-\frac{i}{4}\left[\beta\Phi_c-\frac{1}{\beta}\Theta_c\right]
\right)
\exp\left(-\frac{if_\sigma}{4}\left[\Phi_s-\Theta_s\right]\right),\nonumber\\[1mm]
R^\dagger_\sigma(x)&=&\frac{\eta_\sigma}{\sqrt{2\pi}}e^{if_\sigma\pi/4}
\exp\left(\frac{i}{4}\left[\beta\Phi_c+\frac{1}{\beta}\Theta_c\right]
\right)
\exp\left(\frac{if_\sigma}{4}\left[\Phi_s+\Theta_s\right]\right).
\label{LR}
\eea

The Hamiltonian (\ref{HamiltFT}) exhibits spin-charge separation:
${\mathcal   H}_{\rm c,s}$ describe collective charge and spin degrees
of freedom respectively, which are independent of one another.
The pure Hubbard model corresponds to the limit $\beta\to 1$ and the 
effect of $V_j$ is to decrease the value of $\beta$. 
{}From the form \r{HamiltFT} we can deduce a number of important
properties. Firstly, the spin sector is gapless and is described by a
free bosonic theory. Hence correlation functions involving (vertex
operators of) the spin boson $\Phi_s$ and its dual field $\Theta_s$
can be calculated by standard methods \cite{GNT}. Excitations in the
spin sector are scattering states of gapless, chargeless spin
$\frac{1}{2}$ objects called {\sl spinons}. The charge sector of
\r{HamiltFT} is a SGM. Here excitations in the regime
$\beta>\frac{1}{\sqrt{2}}$ are scattering states of gapped 
charge $\pm e$ soliton and antisoliton excitations respectively. In
the context of the half-filled Mott insulator these are also known as
the {\sl holon} and {\sl antiholon}. The soliton and antisoliton have massive
relativistic dispersions with velocity $v_{\rm c}$ and single-particle
gap $\Delta$,  
\begin{equation}
E(P)=\sqrt{\Delta^2+v_{\rm c}^2 P^2}\ .
\label{holondispersion}
\end{equation}
For $V_j=0,\ j\geq 3$ the gap 
can be determined by renormalization group methods\\[-8mm]~
\begin{equation}
\Delta\approx\frac{8t}{\sqrt{2\pi}}
\sqrt{g(1+x)}\left(\frac{1-x}{1+x}\right)^{(gx+2)/{4gx}}\ ,
\label{FTgap1}
\end{equation}
where we have fixed the constant factor by comparing to the exact 
result for the Hubbard model \cite{LW}, and where
\begin{eqnarray}
x&=&\left[1-\left(\frac{U-2V_1+2V_2}{U+6V_1+2V_2}\right)^2\right]^{1/2}
\ ,\nonumber\\[6pt]
g&=&(U+6V_1+2V_2)/2\pi t\ .
\label{FTgap2}
\end{eqnarray}
We note that the gap vanishes on the critical surface
$U-2V_1+2V_2=0$ separating the Mott-insulating phase
with gapless spin excitations from another phase
with a spin gap.

In the regime $0<\beta <1/\sqrt{2}$, soliton and antisoliton
attract and can form bound states. In the SGM these are known as
``breathers'' and correspond to excitons in the underlying
extended Hubbard lattice model. There are 
\begin{equation}
N= \left[ \frac{1-\beta^2}{\beta^2}\right]
\label{Nex}
\end{equation}
different types of excitons, where $[ x ]$ in~(\ref{Nex}) denotes
the integer part of~$x$.
The exciton gaps are given by
\begin{equation}
\Delta_n=2\Delta \sin(n\pi\xi/2)\ ,\quad n=1,\ldots ,N\; ,
\label{exgap}
\end{equation}
where we have defined
\begin{equation}
\xi=\frac{\beta^2}{1-\beta^2} \; .
\label{xi}
\end{equation}
We note that in the weak coupling analysis summarized above $\beta$
cannot be appreciably smaller than 1. However, numerical
Dynamical Density Matrix Renormalization Group (DDMRG) computations
\cite{EGJ,eric} indicate that the U(1) Thirring model description
remains valid in an extended region of parameter space where
$U,V_1,V_2$ are not small compared to $t$, but the model is still in a
Mott insulating phase. A rough criterion for the applicability of the
U(1) Thirring model to the description of the low-energy physics is
that the charge gap should be small compared to the electronic band
width $4t$. When applying the U(1) Thirring model in the extended
region of parameter space, the gap $\Delta$ and the parameter $\beta$ have
to be determined numerically.

\subsubsection{Quarter Filled Band}
In the quarter filled case there are no simple Umklapp processes that
can open a gap in the charge sector. As a result the quarter-filled
extended Hubbard model is metallic in the weak coupling regime,
i.e. both charge and spin sectors are gapless. However, integrating
out the high energy degrees or freedom in a path-integral formulation
generates ``double Umklapp'' 
processes involving four electron creation and annihilation operators
each \cite{4umklapp,yoshioka}. For small $U,V_j$ these processes are
irrelevant, but increasing $U,V_j$ decreases their scaling dimension.
This suggests that for sufficiently large $U,V_j$ the double Umklapp
terms eventually become relevant and open up a Mott gap in the charge
sector. Such a scenario is indeed supported by numerical computations
\cite{PencMila}.
Assuming that a bosonized description remains valid beyond the
weak coupling region, the low-energy effective Hamiltonian is
identical to \r{HamiltFT}. However, the relations between the Fermi 
operators and the Bose fields are different
\bea
L^\dagger_\sigma(x)&=&\frac{\eta_\sigma}{\sqrt{2\pi}}\,e^{if_\sigma\pi/4}
\exp\left(-\frac{i}{4}\left[\frac{\beta}{2}\Phi_c-\frac{2}{\beta}
\Theta_c\right]\right)
\exp\left(-\frac{if_\sigma}{4}\left[\Phi_s-\Theta_s\right]\right),~~~\nonumber\\[1mm]
R^\dagger_\sigma(x)&=&\frac{\eta_\sigma}{\sqrt{2\pi}}\,e^{if_\sigma\pi/4}
\exp\left(\frac{i}{4}\left[\frac{\beta}{2}\Phi_c+\frac{2}{\beta}
\Theta_c\right]\right)
\exp\left(\frac{if_\sigma}{4}\left[\Phi_s+\Theta_s\right]\right).
\label{LRquart}
\eea
Although the low-energy effective Hamiltonian is the same as for the
half-filled Mott insulator, the physical properties in the
quarter-filled case are rather different. Firstly, the insulating state
emerging for sufficiently large $U,V_j$ at quarter filling
is generated by a different physical mechanism (double Umklapp
scattering) as compared to half filling (Umklapp scattering)
and concomitantly is referred to as a $4k_F$ charge-density wave
insulator in the literature \cite{nakamura}. We adopt this
terminology here. Secondly, the quantum numbers of elementary excitations
in the charge sector are different. Like for the half-filled case the
elementary excitations in the charge sector are a soliton/antisoliton
doublet, but now they carry {\sl fractional charge} $\pm\frac{e}{2}$.
A simple way to see this is to recall that the conserved topological
charge in the SGM is defined as
\be
Q=\frac{\beta}{2\pi}\int_{-\infty}^\infty dx\ \partial_x\Phi_c\ .
\ee
The soliton has topological charge $-1$ and the antisoliton $+1$.
A simple calculation shows that the right moving fermion creates two 
solitons
\bea
QR^\dagger_\sigma(x)|0\rangle=-2R^\dagger_\sigma(x)|0\rangle\ .
\eea
This implies that fermion number 1 corresponds to topological charge
2 and hence solitons have fractional charge.
The elementary excitations in the spin sector are again a pair of
gapless, chargeless spin $\pm\frac{1}{2}$ spinons.

\subsection{Correlation Functions}
Due to spin charge separation a general local operator ${\mathcal O}(t,x)$
can be represented as a product of a charge and a spin piece
${\mathcal O}={\mathcal O}_c{\mathcal O}_s$. As a result correlation functions
factorize as well
\bea
\hskip -4mm
\langle 0|{\mathcal O}^\dagger(\tau,x)\, {\mathcal O}(0)|0\rangle=
{_c\langle} 0|{\mathcal O}^\dagger_c(\tau,x)\, {\mathcal O}_c(0)|0\rangle_c\ 
{_s\langle} 0|{\mathcal O}^\dagger_s(\tau,x)\, {\mathcal O}_s(0)|0\rangle_s\,,\
\eea
where $|0\rangle_{s,c}$ are the vacua in the spin and charge sectors
respectively. Correlation functions in the spin sector are easily
evaluated as we are dealing with a free theory. In what follows we
only need correlators of vertex operators, which in our normalizations
are given by
\bea
{}_s{\langle}\exp\left(i\alpha\phi_s(\tau,x)\right)
\exp\left(-i\alpha\phi_s(0)\right)\rangle_s
&=&\frac{1}{(v_s\tau-ix)^{2\alpha^2}}\, ,\nn 
{}_s{\langle}\exp\left(i\alpha\bar{\phi}_s(\tau,x)\right)
\exp\left(-i\alpha\bar{\phi}_s(0)\right)\rangle_s
&=&\frac{1}{(v_s\tau+ix)^{2\alpha^2}}\, .
\eea

\subsubsection{Correlation Functions in the Charge Sector}
As the charge sector of the U(1) Thirring model is equal to a SGM, we
may use the form factor approach to determine the charge pieces of
correlation functions. In the parameter regime of interest the
spectrum of the SGM consists of soliton, antisoliton and several
breather bound states, which we denote by labels $s,\bar{s},B_1,B_2,\ldots
B_N$. Energy and momentum are then parametrized in terms of the
rapidity variable $\theta$ as
\begin{eqnarray}
E_{\alpha}(\theta)&=&\Delta_\alpha\cosh\theta\; , \quad
P_{\alpha}(\theta)=\frac{\Delta_\alpha}{v_{\rm c}}\sinh\theta\; ,
\end{eqnarray}
where 
\be
\Delta_s=\Delta_{\bar{s}}\equiv\Delta\ ,\quad
\Delta_{B_n}\equiv\Delta_n\ ,
\ee
and the breather gaps, $\Delta_n$, are given by~(\ref{exgap}).

\subsection{Optical Conductivity}
The optical conductivity was calculated in Refs.
\cite{JGE,cet,cet2,EGJ}. In the field-theory limit, the electrical
current operator is related to the fermion current $J$ by  
\begin{equation}
J_{\rm el}(\tau,x) = -ea_0 J(\tau,x) = 
-i\,\frac{ea_0{\cal A}}{2\pi}\,\partial_\tau \Phi_{\rm c} \; ,
\label{currentFT}
\end{equation}
where ${\cal A}$ is a non-universal constant. For $U,V_j\ll t$ we have
${\cal A}\approx 1$. The expression for
the current operator is the same for the half-filled and the quarter
filled bands. As seen from Eq.~(\ref{currentFT}), the current operator
does not couple to the spin sector. This shows that spinons do not
contribute to the optical conductivity in the field theory
limit. Hence, the calculation of the optical conductivity has been
reduced to the evaluation of the retarded current-current correlation
function in the charge sector. The real part of the optical
conductivity ($\omega>0$) has the following spectral representation
\begin{eqnarray}
{\rm Re}\ \sigma(\omega)&=& \frac{2 \pi^2e^2}{a_0^2\omega}
\sum_{n=1}^\infty\sum_{\epsilon_i}\int 
\frac{d\theta_1\ldots d\theta_n}{(2\pi)^n n!}
\left |
f^{J}(\theta_1\ldots\theta_n)_{\epsilon_1\ldots\epsilon_n}
\right | ^2 \label{expansion2}\nn
&&\qquad\quad\quad\times\ \delta\Big(\sum_k\frac{\Delta_{\epsilon_k}}{v_{\rm
    c}}\sinh\theta_k\Big) \ \delta (\omega - \sum_k 
\Delta_{\epsilon_k}\cosh\theta_k)\; .~~~~~~~~
\end{eqnarray}
In
Refs.
\cite{smirnov,karowski78,fring93,lukyanov95,lukyanov97,babujian99}
integral representations for the form 
factors of the current operator $J(\tau,x)$ in the SGM were
derived. Using these results we can determine the first few terms of
the expansion~(\ref{expansion2}). {}From (\ref{expansion2}) it is easy
to see for any given frequency $\omega$ only a finite number of
intermediate states will contribute (as per Section 2): 
the delta function forces the sum of single-particle gaps
$\sum_j\Delta_{\epsilon_j}$ to be less than $\omega$.
Expansions of the form (\ref{expansion2}) are usually found
to exhibit a rapid convergence (Section 2.5), which can be understood in terms of
phase space arguments~\cite{oldff1,mussardo.school}. Therefore we
expect that summing the first few terms in~(\ref{expansion2}) will
give us good results over a large frequency range.

Using the transformation property of the current operator under charge
conjugation one finds that many of the form factors in
(\ref{expansion2}) actually vanish. In particular, only the ``odd'' 
breathers $B_1, B_3,\ldots$ (assuming they exist, i.e., $\beta$ is
sufficiently small) couple to the current operator. The first few
non-vanishing terms of the spectral representation~(\ref{expansion2})
are 
\begin{equation}
{\rm Re}\ \sigma(\omega) =\left(\frac{{\cal A}e}{2a_0\beta}\right)^2 
v_c\left[\sum_{n=1}^{[(1+\xi)/2\xi]}\sigma_{B_{2n-1}}(\omega)
+\sigma_{s\bar{s}}(\omega)+\sigma_{B_1B_2}(\omega)+\ldots\right].
\label{sigmaSR}
\end{equation}
Here $\sigma_{B_n}(\omega)$, $\sigma_{s\bar{s}}(\omega)$ and
$\sigma_{B_1B_2}(\omega)$ are the contributions of the odd breathers,
the soliton-antisoliton continuum and the $B_1B_2$ breather-breather
continuum respectively. The latter of course exists only if $\xi\leq
\frac{1}{2}$.  
We find
\begin{eqnarray}
\sigma_{B_{2n-1}}(\omega)&=&
\pi\ f_{2n-1}\delta(\omega -\Delta_{2n-1})\nonumber\\[1mm]
f_{m}&=&\frac{2\xi^2}{\tan\bigl(\frac{m\pi\xi}{2}\bigr)}\prod_{n=1}^{m-1}
\tan^2\bigl(\frac{\pi n\xi}{2}\bigr)\nonumber\\[1mm]
&&\times\exp\left[-2\int_0^\infty
\frac{dt}{t}\frac{\sinh\bigl(t(1-\xi)\bigr)}{\sinh(t\xi)\cosh(t)}
\frac{\sinh^2(mt\xi)}{\sinh 2t}\right]\; .
\label{excitonsigma}
\end{eqnarray}
The soliton-antisoliton contribution is \cite{cet} 
\begin{eqnarray}
\sigma_{s\bar{s}}(\omega)&=&
\frac{4\sqrt{\omega^2-4\Delta^2}\ \Theta(\omega-2\Delta)}
{\omega^2[\cos\bigl(\frac{\pi}{\xi}\bigr)+\cosh\bigl(\frac{\theta_0}{\xi}\bigr)]}
\label{2part}\label{continuumsigma}\nn
&\times&\exp\left(\int_0^\infty
\frac{dt}{t}\frac{\sinh(t(1-\xi))\left[1-\cos\bigl(\frac{2t\theta_0}{\pi}\bigr)
\cosh 2t\right]}{\sinh(t\xi)\cosh(t)\sinh 2t}
\right)\, ,
\end{eqnarray}
where $\theta_0=2 {\rm arccosh} (\omega/2\Delta)$.
The result for the $B_1B_2$ breather-breather continuum is given in
Ref. \cite{EGJ}. As a function of the parameter $\beta$, the
optical conductivity behaves as follows.
\begin{itemize}
\item{} $1 \geq \beta^2 > 1/2$:

In this regime the optical spectrum consists of a single ``band''
corresponding to (multi) soliton-antisoliton states above a threshold
$2\Delta$. The absorption band increases smoothly above the
threshold $2\Delta$ in a universal square root fashion
\begin{equation}
\sigma(\omega) \propto \sqrt{\omega - 2\Delta} \quad \hbox{for}
\quad \omega \to 2\Delta^{+} \ .
\label{thressigma}
\end{equation}
In Fig.\,\ref{fig:sigma} we plot the leading contributions for the case
$\beta^2=0.9$. Clearly, the four-particle contribution is negligible
at low frequencies.
\begin{figure}[ht]
\begin{center}
\epsfxsize=0.6\textwidth
\epsfbox{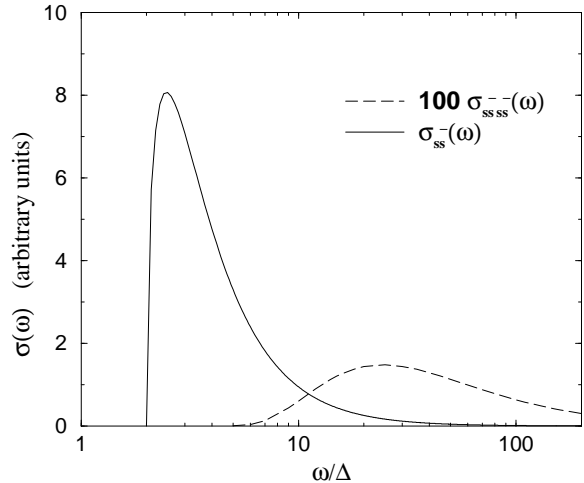}\qquad
\end{center}
\caption{Optical conductivity for $\beta^2=0.9$. Shown are the
  dominant contributions at low frequencies: the soliton-antisoliton
  part $\sigma_{s\bar{s}}(\omega)$ and the two soliton - two antisoliton
  contribution $\sigma_{ss\bar{s}\bar{s}}(\omega)$.}
\label{fig:sigma}
\end{figure}

\item{} $1/2 \geq \beta^2 > 1/3$:

Here the optical spectrum contains one band and one excitonic breather
peak below the optical gap $2\Delta$ at the energy $\omega_{B_1}= \Delta_1$. 
The optical weight is progressively transferred from the band to the
breather as $\beta^2$ decreases down to $1/3$. The absorption band
again increases in a square-root fashion \r{thressigma} above the
threshold for all values of $\beta^2$ except $\beta^2 = 1/2$,  where
the breather peak merges with the band. In this case $\sigma(\omega)$
shows a square-root divergence at the absorption threshold 
\begin{equation}
\sigma(\omega) \propto \frac{1}{\sqrt{\omega - 2\Delta}}
\quad \hbox{for} \quad
\omega \to 2\Delta^{+} \; (\beta^2=1/2)\; .
\end{equation}
\end{itemize}
The field theory results discussed here have been compared to
DDMRG computations (see e.g. Ref. \cite{eric2} and references therein)
computations of  
$\sigma(\omega)$ for extended Hubbard models in Refs.
\cite{JGE,EGJ,eric} and good agreement has been found in the
appropriate regime of parameters. We note that the DDMRG method can
also deal with parameter regimes in the underlying
lattice model, to which field theory does not apply.
Let us discuss the above results from the point of view of an
application to optical conductivity measurements in the Bechgaard
salts \cite{schwartz98,vescoli98,henderson99}. There it is found that
up to $99\%$ of the total spectral weight is concentrated in a
finite-frequency feature, which has been attributed to Mott physics of
the type discussed here \cite{schwartz98}. A comparison of \r{sigmaSR}
to the experimental data gives satisfactory agreement at high
frequencies, but the detailed peak structure at low frequencies is not
reproduced \cite{cet}. A likely source for these differences is the
interchain tunneling. 

\subsection{Spectral Function}

\subsubsection{Half-Filled Mott Insulator}
The zero-temperature spectral function of the half-filled Mott
insulator has been studied in many previous works. There have been
extensive numerical studies on finite size $t$-$J$ and Hubbard models
e.g. Refs. \cite{zacher98,bannister00,senechal00,kim96}. The limit
where the single-particle gap is much larger than the bandwidth was
treated in Refs. \cite{Penc96,Parola96,Parola98}. This regime is
complementary to the case we address here. The weak-coupling limit we
are interested in was studied in Refs. \cite{vigman,voit,starykh},
where a conjecture for the spectral function was put forward. Here we
derive these results by means of an exact, systematic method. In what
follows we will for simplicity fix $\beta=1$, i.e. deal with the
Hubbard model only. 

\subsubsection{Zero Temperature}
The single particle Green's function is calculated by following the
steps outlined above \cite{ET02b}. The creation and annihilation operators for
right and left moving fermions factorize into spin and charge pieces
upon bosonization \r{LR}. The spin part is easily calculated: in
imaginary time we have
\bea
\left\langle e^{-\frac{if_\sigma}{4}\left[\Phi_s+\Theta_s\right]}\
e^{\frac{if_\sigma}{4}\left[\Phi_s+\Theta_s\right]}\right\rangle&=&
\frac{1}{\sqrt{v_s\tau-ix}}\ .
\eea
The correlation function in the charge sector is calculated by means
of the form factor bootstrap approach. Taking into account only
processes involving one soliton we obtain 
\bea
\left\langle e^{-\frac{i}{4}\left[\Phi_c+\Theta_c\right]}\
e^{\frac{i}{4}\left[\Phi_c+\Theta_c\right]}\right\rangle&\simeq&
\frac{Z_0}{\sqrt{v_c\tau-ix}}\exp\left(-\frac{\Delta}{v_c}\sqrt{x^2+v_c^2\tau^2}
\right) ,
\label{GFhalf1}
\eea
where $Z_0\approx 0.9218$ \cite{LukZam01}. The form of \r{GFhalf1} is
fixed by the transformation properties of the operator
$\exp(-\frac{i}{4}\left[\Phi_c+\Theta_c\right])$ under Lorentz
transformations. The corrections to \r{GFhalf1} involve intermediate
states with three particles and are negligible at long distances/low
energies \cite{Doyon}. The imaginary time Green's functions of left
and right moving electrons are then given by
\bea
-\langle T_\tau\ R_\sigma(\tau,x) R^\dagger_\sigma(0,0)\rangle &\simeq&
-\frac{Z_0}{2\pi}\,\frac{\exp\Big[- \Delta\sqrt{\tau^2 +
x^2v_c^{-2}}\Big]}{\sqrt{v_s\tau-ix}\sqrt{v_c\tau-ix}}\ ,\\[2mm]
-\langle T_\tau\ L_\sigma(\tau,x) L^\dagger_\sigma(0,0)\rangle &\simeq& 
-\frac{Z_0}{2\pi}\,\frac{\exp\Big[-
\Delta\sqrt{\tau^2 + x^2v_c^{-2}}\Big]}{\sqrt{v_s\tau +ix}\sqrt{v_c\tau+ix}}\ .
\eea
Fourier transforming and analytically continuing to real
frequencies (we suppress the spin index $\sigma$ in the formulas
below) we arrive at the following result for the retarded
single-particle Green's function
\bea
&&G^{(R)}(\omega, k_F+q) \simeq -Z_0\sqrt{\frac{2v_c}{v_c+v_s}}\,
\frac{\omega+v_c q}{\sqrt{\Delta^2+v_c^2q^2-\omega^2}}\nn
&&\times\left[\left(\Delta+\sqrt{\Delta^2+v_c^2q^2-\omega^2}\right)^2
-\frac{v_c-v_s}{v_c+v_s}(\omega+v_c q)^2\right]^{-\frac{1}{2}} .
\label{GFhalf}
\eea
We note that the charge velocity $v_c$ is larger than the spin velocity
$v_s$. The spectral function is obtained from the imaginary part of
the single particle Green's function \r{GFhalf}. In the case
$v_s=v_c=v$ it takes the simple form
\be
A(\omega,k_F+q)
=
-\frac{1}{\pi}{\rm Im}G^{(R)}(\omega,k_F+q)
=
\frac{Z_0\ \Delta}{\pi|\omega - vq|}\, \frac{\Theta\big(|\omega|-\sqrt{\Delta^2+v^2q^2}\big)}{
\sqrt{\omega^2 - \Delta^2-v^2q^2}}\ . 
\label{arr}
\ee

In Fig.\,\ref{fig:SFhalf} we plot the spectral function in the case
$v_s=0.8511v_c$ and $\beta=1$ in a series of
constant $q$ scans.
\begin{figure}[ht]
\begin{center}
\epsfxsize=0.65\textwidth
\epsfbox{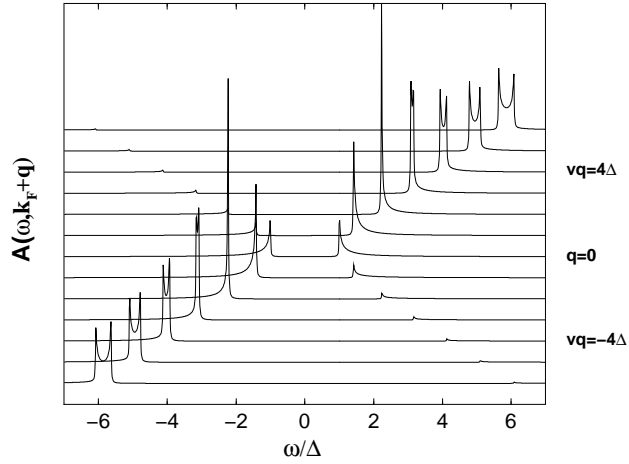}
\end{center}
\caption{Spectral function for a half-filled Mott insulator with
  $v_s=0.8511 v_c$ and $\beta=1$, corresponding to the half-filled
Hubbard model with $U=t$.}
\label{fig:SFhalf}
\end{figure}
There is a continuum of states above the Mott gap, which is smallest at
$k_F$. The most striking feature is the presence of two distinct
``peaks'' dispersing with 
velocities $v_s$ and $v_c$ respectively. Most of the spectral weight
is concentrated in these features, which are a direct manifestation of
spin-charge separation. The higher (lower) energy feature corresponds
to the situation where the momentum carried by the spin (charge)
sector is held constant. Concomitantly the high/low energy feature is
referred to as antiholon/spinon peak. Neither peak is sharp but has
intrinsic width (more precisely, they correspond to square root
divergences). The threshold at $\omega>0$ in the vicinity of $k_F$ is 
\cite{ET02b}
\bea
E_{\rm thres}(k_F+q)
&=&\begin{cases}
\sqrt{\Delta^2+v_c^2 q^2} & \text{if}\ q\leq Q\,,\\[1mm]
v_sq+\Delta\sqrt{1-\alpha^2} & {\rm if}\ q\geq Q\,,\\
\end{cases}
\eea
where
\be
\alpha=\frac{v_s}{v_c}\ ,\quad
Q = \frac{v_s\Delta}{v_c\sqrt{v_c^2 - v_s^2}}\ .
\ee
We see that for $q<Q$ the threshold follows the antiholon dispersion,
whereas for $q>Q$ it follows the (linear) spinon dispersion (shifted by
a constant).

\subsubsection{Finite Temperature $T\ll \Delta$}
It is possible to extend the results for the spectral function to
small temperatures $T\ll \Delta$ \cite{ET03} by the method described
in Ref. \cite{konik} and summarized in section 2.7. As $T\ll \Delta$
the effects of temperature on the charge piece of the correlation
function are small, but the spin piece can be affected quite strongly
because the spinons are gapless. The spectral function in the vicinity
of $k_F$ can be represented as 
\bea
\hspace{-1cm}
A(\omega,k_F+q)&\approx&
{\mathcal A}\int_{-\infty}^\infty dz\ e^{z/2}
\biggl[\tilde{g}_s(\omega-\Delta c(z),q-\frac{\Delta}{v_c} s(z))\nn
&+& e^{-\Delta c(z)/T}\tilde{g}_s(\omega+\Delta c(z),q+\frac{\Delta}{v_c}s(z)) +
\left\lbrace
\begin{matrix}
&\omega\to -\omega\\ 
& q\to -q\\
\end{matrix}
\right\rbrace\biggr],
\eea
where ${\mathcal A}=\sqrt{\frac{\pi\Delta}{v_c}}\frac{Z_2}{(2\pi)^3}$,
$c(z)=\cosh z$, $s(z)=\sinh z$ and
\bea
\hskip -8mm
\tilde{g}_s(\omega,q)&=&\sqrt{\frac{8\pi v_s}{T}}
{\rm
Re}\left[\sqrt{-2i}B\left(\frac{1}{4}-i\frac{\omega+v_sq}{4\pi T},\frac{1}{2}\right)
\right]\delta(\omega-v_sq)\,.
\label{gc}
\eea
\begin{figure}[ht]
\vskip 3mm
\begin{center}
\epsfxsize=0.65\textwidth
\epsfbox{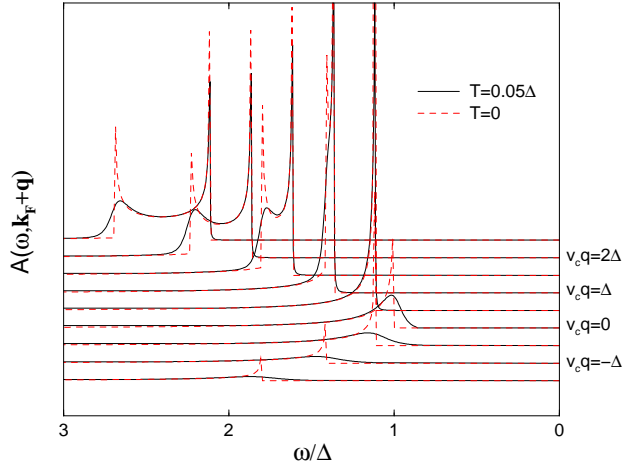}
\end{center}
\caption{
Spectral function for a half-filled Mott insulator with $v_s=0.5 v_c$,
$T=0$ (dotted) and $T=0.05\Delta$ (solid). The different curves are
constant-q scans with $v_cq/\Delta=-1.5,-1,\ldots 2.5$ (from bottom to
top), which have been offset.}
\label{fig:finiteT}
\end{figure}
In Fig.\,\ref{fig:finiteT} we plot the spectral function for
$v_s=0.5v_c$ and $\beta=1$ at a temperature of $0.05$ times
the single particle gap. A significant temperature dependence of the
``charge'' peak is apparent. It gets damped rather strongly at
temperatures that are still small compared to the gap. This may make
an unambiguous detection of SC-separation by ARPES more difficult as
the experiments are done at elevated temperatures (room temperature
for ${\rm Sr_2CuO_3}$) in order to avoid charging effects. 

\subsubsection{Quarter-Filled CDW Insulator}
The single-particle Green's function in the quarter-filled case can be
determined by the same method \cite{ET02a} employed in the half-filled
case. The spin sector is again
gapless and the spin part of the Green's function is easily
determined. The charge piece is analyzed by means of the form factor
bootstrap approach. The difference to the half-filled case is that the
charge piece of the single-electron operator \r{LRquart} now couples
to at least two (anti)solitons. Neglecting contributions of four or
more particles in the charge sector we obtain the following result for
the single-particle Green's function in the vicinity of $k_F$ \cite{ET02a}
\bea
\hspace{-4mm}
G^{(R)}(\omega, k_F+q) &=&
-{\cal C}\sqrt{\frac{2}{1+\alpha}}
\int_{-\infty}^\infty d\theta\
\frac{E(2\theta)\sinh^2(\theta)}{\sqrt{\cosh(\theta)}}
\frac{\omega+v_cq}{\sqrt{c^2(\theta)-s^2}}\nn
&&\times
\Biggl[\left(c(\theta)+\sqrt{c^2(\theta)-s^2}\right)^2-\frac{1-\alpha}{1+\alpha}
{(\omega+v_cq)^2}\Biggr]^{-\frac{1}{2}},~~~~
\label{G}
\eea
where $E(\theta)$ is given in \r{Eoftheta}, $s^2=\omega^2-v_c^2q^2$,
$c(\theta)=2\Delta\cosh\theta$, and $\alpha = v_s/v_c$. ${\cal C}$
is a numerical constant. In Fig.\,\ref{fig:constq} we plot the
spectral function $A(\omega,k_F+q)$ as a function of $\omega$ for
$v_s=0.8  v_c$ and different values of $q$. We observe that the spectral
function is rather featureless and there are no
singularities. Furthermore, in contrast to the half-filled Mott
insulator discussed above, there are no dispersing features associated
separately with $v_c$ and $v_s$. The absence of any distinct features
is clearly related to the fact that the electron has ``fallen apart''
into at least three pieces. 

Just above the threshold at $s^2=\omega^2-v^2q^2=4\Delta^2$ one finds
for $v_c=v_s$ 
\bea
A(\omega,q) \propto\frac{1}{|\omega -vq|}\left(\frac{s}{2\Delta}-1\right). 
\eea
\begin{figure}[ht]
\begin{center}
\epsfxsize=0.7\textwidth
\epsfbox{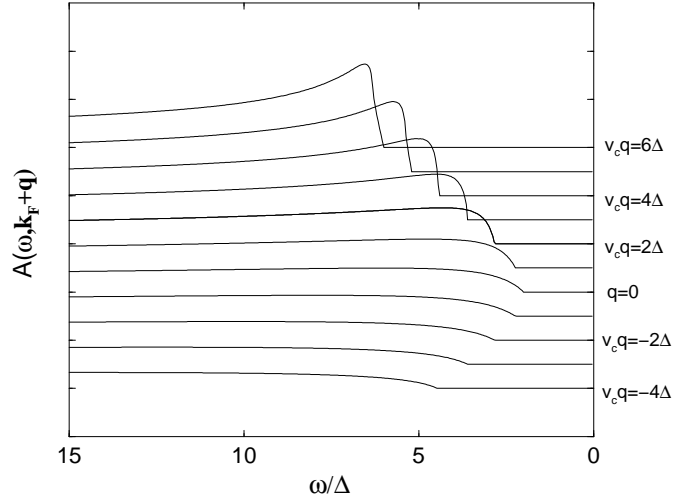}
\end{center}
\caption{\label{fig:constq}
$A(\omega,k_F+q)$ as a function of $\omega/\Delta$ for $v_s=0.8v_c$. The
curves for different $q$ are offset.
}
\end{figure}
Thus the spectral weight increases linearly with $s-2\Delta$ above the
threshold. The tunneling density of states is
\bea
\rho(\omega)&=& -\frac{1}{\pi}\int_{-\pi}^\pi\frac{dk}{2\pi}
{\rm Im}\ G^{(R)}(\omega,k)\nn
&=&\frac{2{\cal C}}{\pi\sqrt{v_cv_s}}\int_0^{{\rm arccosh}(\omega/2\Delta)}
d\theta\ \frac{E(2\theta)\sinh^2(\theta)}{\sqrt{\cosh\theta}}\ ,
\eea
and displays a roughly linear increase after an initial
$(\omega-2\Delta)^{3/2}$ behavior just above the threshold at
$\omega=2\Delta$. 

Let us now turn to a comparison with experiments. ${\rm
PrBa_2Cu_3O_7}$ (P123) is a quarter-filled quasi-1D cuprate, to which
the theory presented here may have some relevance. The ARPES data for
P123 (Fig.\,3 of \cite{p123arpes}) show a single, very broad, dispersing
feature that is asymmetric around $k_F$. If we interpret the
underlying increase in intensity in the data as background, the signal
has a form similar to Fig.\,\ref{fig:constq}. In order to assess whether
the theory presented here is indeed relevant to P123, it would be
interesting to extract a value $\Delta_{\rm PE}^{(\frac{1}{4})}$ for
the gap from the ARPES data and compare it to gaps seen in optical
measurements $\Delta^{(\frac{1}{4})}_{\rm opt}$ and the thermal
activation gap $\Delta^{(\frac{1}{4})}_{\rm T}$ extracted e.g. from
the dc conductivity. The theory presented here predicts 
\be
\Delta^{(\frac{1}{4})}_{\rm
  opt}=2\Delta^{(\frac{1}{4})}_{\rm T}=\Delta^{(\frac{1}{4})}_{\rm PE}\ .
\ee
This is in contrast to the case of the half-filled Mott insulator,
where one has 
\be
\Delta^{(\frac{1}{2})}_{\rm opt}=2\Delta^{(\frac{1}{2})}_{\rm T}
=2\Delta^{(\frac{1}{2})}_{\rm PE}\ .
\ee

\subsection{A Remark on Luttinger's Theorem}
In all cases we have discussed, the Green's functions have branch cuts
but no poles. In particular there are no poles at zero frequency and
hence no Fermi surface. Nevertheless Luttinger's theorem holds as we
will now show following \cite{ET03}. Luttinger's theorem reads
\cite{AGD,Dz} 
\bea
\frac{N}{V} = 2\int_{G(0,k) >0} \frac{d^Dk}{(2\pi)^D}\,,
\label{lthm}
\eea
where $N$ is the total number of electrons, $V$ is the volume and the
integration is over the interior of the region defined by either
singularities or zeroes of the single-particle Green's 
function. The former is the case for a Fermi liquid whereas the latter 
is the case at hand. The interest in the equality \r{lthm} derives
from the fact that it implies that the integral on the right hand side
is independent of electron-electron interactions. In a Fermi liquid
this means that the volume of the Fermi surface is unaffected by
interactions. The Green's functions we have discussed above all fulfill
\r{lthm} by virtue of them having zeroes at the position of the
non-interacting Fermi surface. 

The mechanism underlying this fact can be understood as follows. For
simplicity let us concentrate on the Lorentz invariant case, where
$v_s=v_c$. Then
\begin{itemize}
\item{}
As the spin sector is gapless, we have $\langle
R_\sigma(\tau,x) L^\dagger_\sigma (0)\rangle=0$;
\item{}
Lorentz invariance of the low-energy effective theory implies that
\bea
\langle \Psi_\sigma(\tau,x) \Psi_\sigma^\dagger
(0)\rangle&\sim&\exp(\pm i\phi){\mathcal R}(r);\quad \Psi=R,L\,.
\label{rr}
\eea
Here $r$ and $\phi$ are polar coordinates and ${\mathcal R}$ denotes the 
radial part of the correlation function. 
\end{itemize}
As we
are dealing with an insulating state we have ${\mathcal R}(r)\propto
\exp(-\Delta r)$ at large distances and hence $\int dr\
{\mathcal R}(r)r$ is finite. Thus
\bea
G(0,\pm k_F)=\int_{-\pi}^\pi d\phi\,\exp(\pm i\phi)\int dr\ {\mathcal
R}(r)r=0\,. 
\eea
For a metallic state the $r$ integral would diverge and the Green's
function would have a singularity rather than a zero.

\subsection{Interchain Tunneling}
Let us now consider a quasi one dimensional situation of Hubbard
chains weakly coupled by a long-ranged interchain tunneling
\bea
H&=&\sum_l H^{(l)} + \sum_{l\neq m;n;\sigma}t_\perp^{lm}\
{c^{(l)\dagger}_{n,\sigma}} c^{(m)}_{n,\sigma} +{\rm h.c.}\nn
H^{(l)}&=&-t\sum_{n,\sigma}
{c^{(l)\dagger}_{n,\sigma}}
c^{(l)}_{n+1,\sigma}+{\rm h.c.}+U\sum_n
n^{(l)}_{j,\uparrow}n^{(l)}_{j,\downarrow}\ .
\eea
We allow the interchain tunneling to be long-ranged in order to
be able to carry out a controlled expansion in the case where the
Fourier transform $\tilde{t}_\perp(\vec{k})$ of the interchain tunneling
becomes of the same order as the 1D Mott gap $\Delta$ (see the discussion
below). For simplicity we take $t_\perp$ long-ranged only in the
direction perpendicular to the chains, although it is straightforward
to generalize all formulas to an interchain tunneling of the form
\be
\sum_{l\neq m;n;p;\sigma}[t_\perp]^{lm}_{np}\
{c^{(l)\dagger}_{n,\sigma}} c^{(m)}_{p,\sigma} +{\rm h.c.}\ .
\ee
\subsubsection{Expansion Around Uncoupled Chains}
Following the analogous treatment for the case of coupled Luttinger
liquids \cite{wen90,boies,arrigoni00} we take the interchain tunneling into
account in a perturbative expansion. The building blocks of this
expansion are the $n$-point functions of fermion operators for
uncoupled chains, which are represented pictorially in
Fig.\,\ref{fig:build}. 
\begin{figure}
\begin{center}
\epsfxsize=0.4\textwidth
\epsfbox{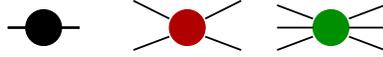}
\end{center}
\caption{Elements of the diagrammatic perturbation theory in the
  interchain tunneling.}
\label{fig:build}
\end{figure}
The full single-particle Green's function is given in terms of a~diagrammatic 
expansion, the first few terms of which are shown in
Fig.\,\ref{fig:expansion} 

\begin{figure}
\begin{center}
\epsfxsize=0.4\textwidth
\epsfbox{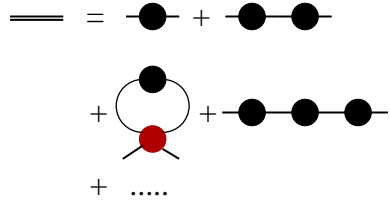}
\end{center}
\caption{Diagrammatic expansion for the single-particle Green's
  function of coupled chains.}
\label{fig:expansion}
\end{figure}

Unlike for Luttinger liquids, it is extremely difficult to determine
four-point functions for 1D Mott insulators. On the other hand, it is
straightforward to sum all diagrams involving only two-point functions of
uncoupled chains. This approximation is known as the random phase approximation (RPA) and has a long
history \cite{RPA}. The small parameter making RPA a controlled
approximation for any form of the interchain tunneling is the ratio
of the interchain tunneling to the Mott gap (at zero
temperature)\cite{boies}. This is because the only energy scale entering the
n-point functions is the dynamically generated gap
$\Delta$. Dimensional analysis then shows that the diagrams neglected
in RPA are suppressed by powers of $t_\perp/\Delta$. Within RPA the
single-particle Green's function is given by 
\bea
G_{\rm 3D}(\omega,q,\vec{k})&=&\frac{G_{1D}(\omega,q)}
{1-t_\perp(\vec{k})\ G_{1D}(\omega,q)}\,,\nonumber\\[1mm]
t_\perp(\vec{k})&=&\sum_{m}t^{lm}_\perp\exp(i\vec{k}\cdot[\vec{R}_{l}
-\vec{R}_{m}])\ .
\label{G3D}
\eea
The RPA Green's function $G_{\rm 3D}(\omega,q,\vec{k})$  has the
interesting property that it has a pole, which corresponds to a bound
state of an antiholon and a spinon with the quantum numbers of an
electron. For small interchain tunneling this bound state still has a
gap. Fig.\,\ref{fig:G3D} is a density plot of $G_{\rm
  3D}(\omega,q,\vec{k})$ as a function of $\omega$ and $q$ for a fixed
value of ${\vec{k}}$ and hence a fixed value of $t_\perp(\vec{k})$.
\begin{figure}
\begin{center}
\epsfxsize=0.5\textwidth
\epsfbox{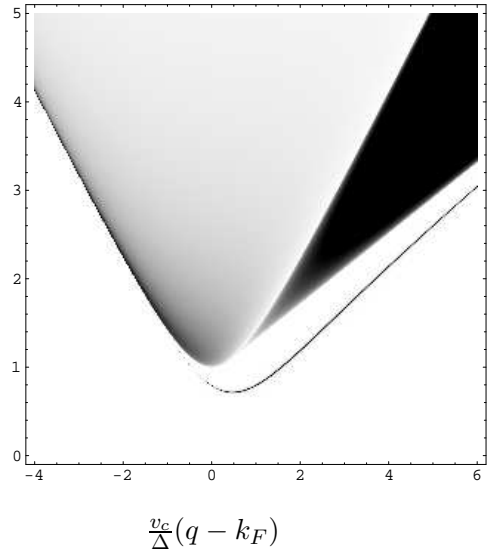}
\end{center}
\hskip 5cm$\frac{v_c}{\Delta}(q-k_F)$
\caption{The spectral function for fixed momentum perpendicular to the
chains $\vec{k}$. The electronic bound state is visible as a
delta-function peak below the antiholon-spinon continuum}
\label{fig:G3D}
\end{figure}
At energies above the Mott gap there is a continuum of states, which
is similar in nature to the result for uncoupled chains. Below the Mott
gap the coherent electronic mode is visible.

\subsubsection{Formation of a Fermi surface}

As long as the ``binding energy'' of the electronic bound state is
small, RPA is a controlled approximation for any form of the
interchain tunneling \cite{boies}. However, in the most interesting
situation when the gap of the bound states becomes very small, RPA
becomes uncontrolled; there is no longer any small expansion
parameter for a generic $t_\perp(\vec{k})$. An exception is the case
of a long-ranged interchain tunneling: here the support of
$t_\perp(\vec{k})$ becomes very small, so that any integration over
the transverse momentum generates a small volume factor
proportional to the inverse range of the hopping. Recalling that RPA
takes into account all terms not involving any integrations over the
transverse hopping (i.e. ``loops''), we conclude that RPA is the
leading term in a controlled loop expansion. Increasing
$t_\perp(\vec{k})$ in RPA reduces the gap of the electronic bound
state, until it eventually vanishes and electron and hole pockets are
formed: we have crossed over from weakly coupled 1D Mott insulators to
an anisotropic Fermi liquid. As an example, let us consider a 2D
square lattice with interchain tunneling between nearest neighbor
chain only. Here RPA is an uncontrolled approximation, but we still
find it instructive to discuss its predictions. 
\begin{figure}
\begin{center}
\epsfxsize=0.5\textwidth
\epsfbox{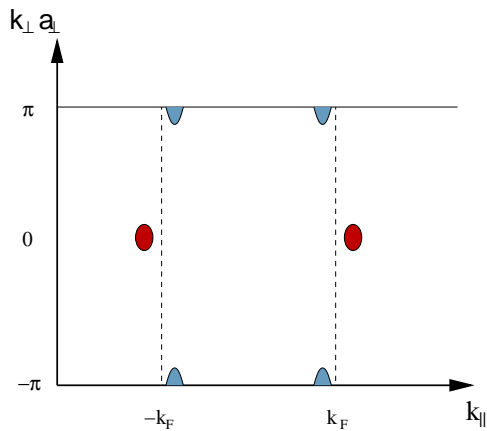}
\end{center}
\caption{Fermi surface predicted by the RPA for a 2D square lattice
  with nearest-neighbor interchain tunneling.}
\end{figure}
In this case electron pockets are formed in the vicinity of the points
$(\pm k_F,{\bf   0})$ and hole pockets form around $(\pm k_F,\pm{\bf
  \pi})$. In the particle-hole symmetric case the volume of the electron
pockets is precisely the same as the volume of the hole pockets.

\section{Hubbard Ladders and Carbon Nanotubes: SO(8) Gross--Neveu model}
\setcounter{equation}{0}

\def\ccr{c^\dagger_{Rj\sigma}}
\def\car{c_{Rj\sigma}}
\def\ccl{c^\dagger_{Lj\sigma}}
\def\calf{c_{Lj\sigma}}
\def\del{\partial}
\def\js{J_s}
\def\rs{R_s}
\def\bjs{\bar{J}_s}
\def\brs{\bar{R}_s}
\def\jnm{J^{n,m}_s}
\def\rnm{R^{n,m}_s}
\def\prnm{R^{n',m'}_s}
\def\bjnm{\bar{J}^{n,m}_s}
\def\brnm{\bar{R}^{n,m}_s}
\def\pbrnm{\bar{R}^{n',m'}_s}
\def\bt{\beta}
\def\bs{{\hat{\beta}^2}}
\def\tcor{\langle T(G^{12}_1(x,\tau)G^{12}_1(0,0))\rangle}
\def\cor{\langle G^{12}_1(x,t)G^{12}_1(0,0)\rangle}
\def\CL{{\cal L}}

In this section we address another set of examples 
of quasi one-dimensional interacting electronic systems,
two-leg Hubbard ladders and single-walled carbon nanotubes.

Two-leg ladders have been the focus of much
theoretical (i.e.\cite{azaria,bal,bala,Balents01,Barnes94,Dagotto,Dagottoa,fab,gia,Johnston96,Johnston00,lin,lina,doped,Nersesyan97,so8,Rice,Schulz86,Schulz98,Wang00,White94})
and experimental activity (i.e. \cite{Hiroi91,Hiroi95,Hiroi96,Johnston00,LaPlaca93,McCarron,Siegrist}).  
At half-filling
they are Mott insulators, exhibiting gaps to all excitations,
and in particular, a spin gap.
They are typical
examples of `spin-liquids'.  Upon doping, the gaps
to all excitations except for those with charge-two survive \cite{bala,fab,doped,lin,lina,Schulz98}.
The gapless charge modes induce
quasi long-range
superconducting pairing correlations, with approximate
$d$-wave symmetry, reminiscent of underdoped cuprate superconductors.

Carbon nanotubes are novel materials whose mechanical and
electronic properties promise potential for new technological
applications \cite{Ebbesen}.
They are formed by wrapping graphite sheets into cylinders of nanoscale
dimensions.  They support electronic excitations, which,
for a prominent member of the nanotube family, the armchair $(n,n)$ type,
can be described by the same theoretical model as that
used for the two-leg Hubbard ladders \cite{bal,kro,lin,lina,Nersesyan03}.  Even though
these systems would be one-dimensional band metals
in the absence of interactions,
they become Mott insulators at half-filling due to the presence of
short-ranged electronic interactions, which play an important role due
to their one-dimensional nature.\,\footnote{\,If forward scattering interactions
are assumed predominant, the metallic nanotube is a Luttinger liquid, not a 
Mott insulator\cite{Egger97,Egger98,Kane97,Levitov03,Schulz98,lin}.}  It is these interaction
effects that we analyze exactly in this section using form factors.

After experimental techniques
had been developed to fabricate long single-walled
nanotubes with high yields in the laboratory, this field
of material science has seen a explosive
development \cite{dresselhaus,bockrath,rao,wildoer}.
Electronic
properties can be measured relatively easily by attaching
metallic leads \cite{twoterminal}
or by  tunneling into these materials with 
scanning tunneling microscope (STM)
tips.  The practical feasibility of such tunneling
experiments from an STM tip into an individual single-walled
nanotube placed upon a gold substrate (screening long-range
Coulomb forces) has been demonstrated in a number of experimental reports
\cite{Janssen01,Janssen02,Venema00,wildoer}.

As said, both systems, two-leg ladders and single-walled armchair nanotubes,
would be one-dimensional metals in the absence of electron
interactions. These
are described theoretically (on scales
much smaller than the non-interacting band width) by two species
of spinful massless
Dirac fermions in $(1+1)$ dimensions.  (These two species
have equal Fermi velocities due to particle-hole symmetry
present at half-filling.)  For the ladder compounds,
the two species arise in an obvious way from the two legs of
the ladder, whereas for the nanotubes they arise from the
particular band structure of the underlying hexagonal
graphite lattice, characterized by two Fermi points in the Brillouin
zone \cite{bal,kro}.  These massless  Fermi surface excitations
interact with short-range interactions whose
detailed nature is determined by non-universal
microscopic considerations.

A remarkable low energy feature of these system was observed by Lin, Balents, 
and Fisher \cite{lin}.
These authors argued
that within an 1-loop RG any such model with generic, non-chiral,
short range interactions
flows at half-filling into a theory
with an immense symmetry, namely the $SO(8)$ symmetric Gross--Neveu
model.  This model not only has a large 
$SO(8)$ global symmetry, which encompasses an one-dimensional
version of $SO(5)$ advocated by 
S. C. Zhang~\cite{ZhangSO5},
but in addition has an infinite number of hidden
conservation laws, which are a consequence of the
model's integrability.

Motivated by this work, we study in this section the low energy properties of nanotubes and
two-leg ladders through computing correlation functions via form factors in SO(8) Gross--Neveu.
In order to be pedagogical, we initiate our discussion by explaining why the low energy excitation
spectrum of armchair carbon nanotubes and Hubbard ladders (four Dirac fermions) are identical.
We then move on by reviewing the RG argument 
(and its limitations) by which these systems flow onto the SO(8) Gross--Neveu
model.  Having done this, we consider the excitation spectrum of SO(8) Gross--Neveu together with the connection
between the fields in the theory and the original lattice operators.  Finally we compute a number of
physically relevant correlation functions for these systems.

\subsection{Armchair Carbon Nanotubes and Hubbard Ladders: Identical Low Energy Behavior}

\begin{figure}[ht]
\begin{center}
\epsfxsize=.4\textwidth
\epsfbox{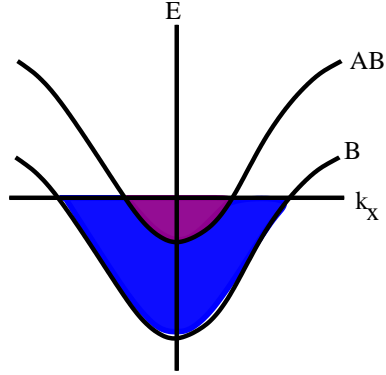}\quad
\end{center}
\caption{The bonding and anti-bonding bands of a Hubbard ladder.}
\label{figVIi}
\end{figure}

We begin by showing that the set of low energy excitations in a non-interacting Hubbard
ladder is equivalent to the low energy sector in an armchair carbon nanotube\cite{lin,linb}.
The Hubbard ladder Hamiltonian is
\begin{eqnarray}\label{eVIi}
H_0 = -\sum_{x,\alpha }
\bigl(
&& ta^\dagger_{1\alpha}(x+1)a_{1\alpha}(x)
+ ta^\dagger_{2\alpha}(x+1)a_{2\alpha}(x) \cr
&& ~~~~~ + t_{\perp}a^\dagger_{1\alpha}(x)a_{2\alpha}(x) + {\rm h.c.}
\bigr).
\end{eqnarray}
Here the $a_l$/$a_l^\dagger$ are the electron annihilation/creation
operators for the electrons on rung $l$ of the ladder, $x$ is a discrete
coordinate along the ladder, and $\alpha = \uparrow ,\downarrow$
describes electron spin.  $t$ and $t_\perp$ describe respectively
hopping between and
along the ladder's rung.
The first step in the map is to reexpress the $a$'s of $H_0$ in terms
of bonding/anti-bonding pairs:
\begin{equation}\label{eVIii}
c_{j\alpha} = \frac{1}{\sqrt{2}}(a_{1\alpha} + (-1)^j a_{2\alpha})\,.
\end{equation}
With this transformation, the Hamiltonian can be diagonalized in
momentum space in terms of two bands, the bonding (B) and anti-bonding bands (AB).
Their dispersions are given by
\begin{equation}\label{eVIiii}
E_{B/AB}(k_x) =-2t\cos (k_x) \mp t_\perp\,.
\end{equation}
These bands are pictured as half-filled in Fig.\,\ref{figVIi}.
Working at this filling, 
particle-hole
symmetry dictates that the Fermi velocities, $v_{Fj}$, of the two bands,
$j=1,2$, are equal.  As we are interested in the low energy behavior of
the theory, the $c_{j\alpha}$'s are
linearized about the Fermi surface, $k_{Fj}$:
\begin{equation}\label{eVIiv}
c_{j\alpha} \sim \car e^{ik_{Fj}x} + \calf e^{-ik_{Fj}x},
\end{equation}
where $L,R$ correspond to right and left moving modes.
With this $H_0$ becomes,
\begin{equation}\label{eVIv}
H_0 = v_F \int dx \sum_{j\alpha}
\bigl [
 \ccr i\del_x \car - \ccl i\del_x \calf 
\bigr ]\,.
\end{equation}
Thus the lower energy description of a non-interacting Hubbard ladder is that of
four Dirac fermions (once spin degeneracy is taken into account).  We now show
that the low energy spectrum of a non-interacting armchair carbon nanotube is
exactly the same.

\begin{figure}[ht]
\begin{center}
\epsfxsize=0.9\textwidth
\epsfbox{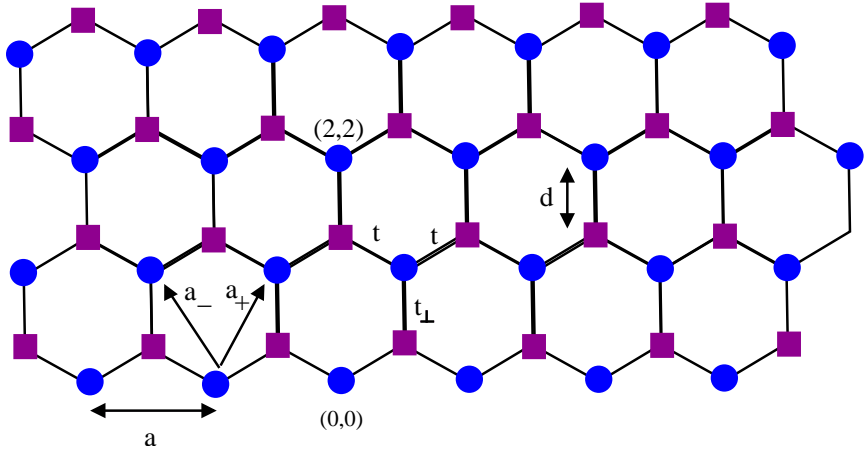}\quad
\end{center}
\caption{The graphite lattice is composed of two interpenetrating triangular lattices.}
\label{figVIii}
\end{figure}

A single-walled armchair carbon nanotube is made by rolling up a sheet of graphite\cite{lin,hamada,mintmire,saito}.
As such we first focus upon the graphite lattice, picture in Fig.\,\ref{figVIii}.  This lattice
consist of two interpenetrating triangular lattices marked as blue circles and violet
squares in Fig.\,\ref{figVIii}.  A tight-binding hopping model of electrons on this lattice
is described by the Hamiltonian
\begin{eqnarray}\label{eVIvi}
H &=& \sum_{{\bf r \in R},\sigma}\bigg\{-t a^\dagger_{1\sigma}({\bf r})a_{2\sigma}({\bf r+a_-+d})
-t a^\dagger_{1\sigma}({\bf r})a_{2\sigma}({\bf r+a_++d})\cr\cr
&& \hskip 1in -t_\perp a^\dagger_{1\sigma}({\bf r})a_{2\sigma}({\bf r+d}) + {\rm h.c.}\bigg\}\,.
\end{eqnarray}
Here ${\bf R}$ is the set of lattice vectors generated by the bases $a_-$ and $a_+$ where
$a_\pm = a(\pm 1/2,\sqrt{3}/2)$.  The two triangular lattices are displaced by a vector
$d = a(0,1/\sqrt{3})$. Here again $a_{\bf r}/a^\dagger_{\bf r}$ create and destroy
an electron at site $\bf r$.

\begin{figure}[ht]
\begin{center}
\epsfxsize=.6\textwidth
\epsfbox{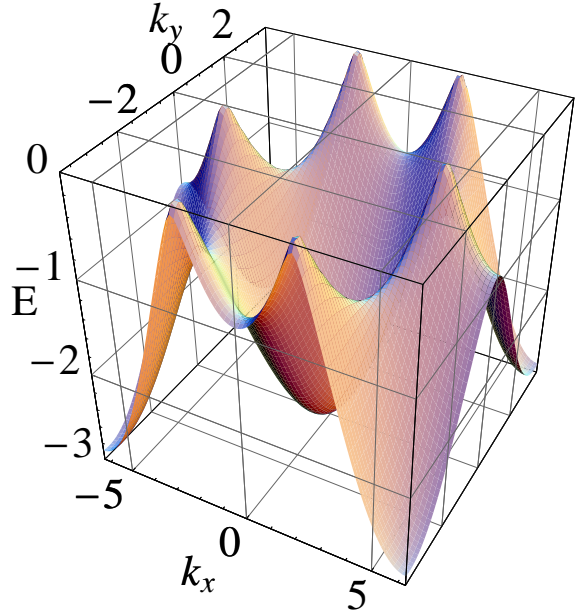}\quad
\end{center}
\caption{The dispersion of the bonding band of graphite (with $a=1$).}
\label{figVIiii}
\end{figure}

In the absence of interactions, it is possible to diagonalize this Hamiltonian.
One looks for eigenfunctions of the form
\begin{equation}\label{eVIvii}
|\psi_{\bf k\sigma}\rangle = \sum_{\bf r \in R}(a^\dagger_{1\sigma}+\beta({\bf k})a^\dagger_{2\sigma})
e^{i{\bf k\cdot r}}\,.
\end{equation}
Solving $H|\psi_{k\sigma}\rangle = E({\bf k})|\psi_{k\sigma}\rangle$, one then finds the following solution
\begin{eqnarray}\label{eVIviii}
\Gamma({\bf k}) &=& 2t\cos (k_x a/2)e^{ik_y a/\sqrt{12}} + t_\perp e^{-ik_y a/\sqrt{3}}\,;\nonumber\\[1mm]
E({\bf k}) &=& \pm |\Gamma({\bf k})|\,;\nonumber\\[1mm]
\beta^2({\bf k}) &=& \Gamma({\bf k})\Gamma^*({\bf k})\,.
\end{eqnarray}
The two signs in the above equation for the energy, $E(k)$, correspond to the bonding and antibonding bands.
We plot the dispersion of the bonding band in Fig.\,\ref{figVIiii}.

We see that the dispersion is characterized by a series of cones whose tips extend to zero energy.
The positions of these tips are known as Dirac points.  The Dirac points are located at
${\bf k} = (\pm 4\pi/3a,0)$ (and equivalent reciprocal lattice vectors).  In Fig.\,\ref{figVIiv}
we mark with violet squares the position of the Dirac points in reciprocal space provided
$t = t_\perp$.
At half-filling the bonding band is completely filled and there will be a set of low energy
excitations.  These excitations, however, have a vanishing density of states at the Fermi
surface.  Consequently graphite is known as a semi-metal.

At this point we want to consider the effects of rolling up the sheet of graphite on the low
energy spectrum.  These effects will depend on how precisely this is done.  We will
only be interested when the sheet is rolled up such that an $(N,N)$ armchair carbon
nanotube is obtained.  This tube has a circumference of $\sqrt{3}a N$. 
Referring to Fig.\,\ref{figVIii}, a $(2,2)$ tube can be obtained by rolling
the sheet so that the two lattice points labelled $(0,0)$ and $(2,2)$ are identified.
When the sheet is rolled up, momentum transverse to the direction of the resulting tube
will be quantized.  For an armchair tube, $k_y$ momentum is quantized.  For an $(N,N)$ tube,
there will be $N$ distinct allowed values of $k_y$ given by 
\begin{eqnarray}\label{eVIix}
k_y &=& \frac{2\pi}{\sqrt{3}a} \frac{n}{N}, ~~~~ n = 0, \pm 1, \pm 2, \cdots , \pm \frac{(N-1)}{2}\,, ~~~~~~~~~~{\rm N~odd};\cr\cr
k_y &=& \frac{2\pi}{\sqrt{3}a} \frac{n}{N}, ~~~~ n = 0, \pm 1, \pm 2, \cdots , \pm \frac{(N-2)}{2},N/2\,, ~~~{\rm N~even}.
\end{eqnarray}
The allowed values of discrete $k_y$-momentum are shown in Fig.\,\ref{figVIiv} for a $(3,3)$ tube.

When the tube is rolled up, one generically expects the hopping parameters to change.  If once $t=t_\perp$,
we then expect $t\neq t_\perp$.  With such a change, the Dirac points are shifted.  How these Dirac points
are shifted is pictured in Fig.\,\ref{figVIiv} for a $(3,3)$ tube supposing that $t_\perp$ is now less than $t$.
We see the shift involves only $k_x$ and not $k_y$.  Thus rolling up an armchair nanotube leaves the Dirac points (and
so the presence of low energy excitations) in the allowed spectrum.
The linearization of the spectrum about each of the two Dirac points gives a single Dirac fermion.  Thus
with spin degeneracy, the low energy spectrum of an armchair carbon nanotube is four Dirac fermions, identical
to that of a Hubbard ladder.

\begin{figure}[ht]
\begin{center}
\epsfxsize=.4\textwidth
\epsfbox{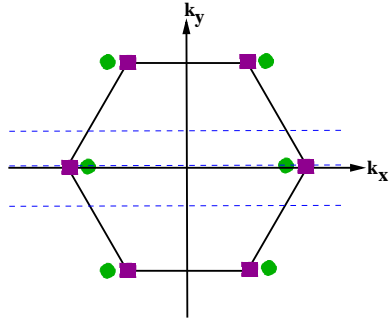}\quad
\end{center}
\caption{The Brillouin zone for the graphite lattice.  Pictured are Dirac points for both the case
$t=t_\perp$ (violet squares) and $t>t_\perp$ (green circles).  The blue dashed lines mark the allowed
values of quantized $k_y$-momenta for a $(3,3)$ armchair carbon nanotube.  
For the armchair tubes, the Dirac
points always lie at an allowed ${\bf k}$ value.}
\label{figVIiv}
\end{figure}

\subsection{Weak Coupling Flow Onto SO(8) Gross--Neveu}

There exist a number of treatments of interactions in carbon nanotubes.
As carbon nanotubes are quasi-one dimensional, a number of authors
have employed bosonization, in combination with the renormalization group,
to understand the effects of correlations \cite{fab,Egger97,Egger98,Kane97,lin,Schulz98}.
We will be interested in treatments of half-filled nanotubes with short range interactions
where all excitation are gapped as opposed to nanotubes with long range, unscreened
Coulomb interactions.  In the latter case, the nanotubes are Luttinger liquids
\cite{Egger97,Egger98,Kane97,Levitov03,Schulz98,lin}.  In particular, we will focus
upon  Ref. \cite{lin} where it was shown that 
weak interactions lead the interacting electrons to flow in an RG sense onto the SO(8)
Gross-Neveu model.

To set up this analysis, we consider the low-energy, non-interacting Hamiltonian for the ladders/tubes:
\begin{equation}\label{eVIx}
H_0 = v_F \int dx \sum_{\sigma~j=1,2}
\bigl [
 \ccr i\del_x \car - \ccl i\del_x \calf 
\bigr ]\,.
\end{equation}
We now consider adding all possible allowed interactions to this Hamiltonian.  To organize this addition,
we introduce various $SU(2)$ scalar and vector currents\\[-8mm]~
\begin{eqnarray}\label{eVIxi}
J_{ij} &=& c^\dagger_{i\sigma}c_{j\sigma}\,; 
~~~ {\bf J}_{ij}=\frac{1}{2}c^\dagger_{i\sigma}{\boldsymbol{\sigma}}_{\sigma\sigma'}c_{j\sigma'}\,;\cr
I_{ij} &=& c^\dagger_{i\sigma}\epsilon_{\sigma\sigma'}c_{j\sigma'}\,; 
~~~ {\bf I}_{ij}=\frac{1}{2}c^\dagger_{i\sigma}(\epsilon{\boldsymbol{\sigma}})_{\sigma\sigma'}c_{j\sigma'}\,.
\end{eqnarray}
Here ${\bf\sigma}$ are the Pauli matrices and $\epsilon_{\sigma\sigma'}$ is the $\epsilon$-tensor.
The crystal-momentum conserving interactions divide themselves into forward and backward scattering terms:
\begin{eqnarray}\label{eVIxii}
H_{B} &=& \sum_{i,j=1,2} b^{\rho}_{ij}J^R_{ij}J^L_{ij}-b^{\sigma}_{ij}{\bf J}^R_{ij}\cdot {\bf J}^L_{ij}\,;\cr\cr
H_{F} &=& \sum_{i\neq j=1,2} f^{\rho}_{ij}J^R_{ii}J^L_{jj}-f^{\sigma}_{ij}{\bf J}^R_{ii}\cdot {\bf J}^L_{ij}\,.
\end{eqnarray}
Here $f$ and $b$ are the forward and backward scattering amplitudes.
From hermiticity and parity, we have $b_{12}=b_{21}$ and $f_{12}=f_{21}$.  Furthermore
at half-filling, $b_{11}=b_{22}$.  Thus we obtain
six independent amplitudes.  The Umklapp interactions take the form
\begin{eqnarray}\label{eVIxiii}
H_{U} &=& \sum_{i,j=1,2} u^{\rho}_{ij}I^R_{ij}I^{L\dagger}_{\hat i\hat j}-
\sum_{i\neq j=1,2} u^{\sigma}_{ij}{\bf I}^{R\dagger}_{ij}\cdot {\bf I}^L_{\hat i\hat j} + {\rm h.c.}\,,
\end{eqnarray}
where $\hat 1 = 2,\hat 2 = 1$.  As ${\bf I_{ij}},I_{ij}$ are anti-symmetric, one may choose $u_{12}=u_{21}$
leading to three additional independent couplings.
For this set of generic interactions, the one-loop RG equations were derived in Ref. \cite{lin}. They
show that $H_0 + H_{\rm int}$ flow onto the SO(8) Gross--Neveu model. 

To express this remarkable result compactly we must invoke a change to variables.  We begin by bosonizing
the $c$'s:
\begin{equation}\label{eVIxiv}
c_{Pj\alpha} = \kappa_{j\alpha} e^{i\phi_{Pj\alpha}},
~~~~ {\rm P = +,- = R,L}\,.
\end{equation}
Here $\kappa_{j\alpha}$ are Klein factors satisfying
\begin{equation}\label{eVIxv}
\{\kappa_{j\alpha},\kappa_{i\beta}\} = 2 \delta_{ij}\delta_{\alpha\beta}\,.
\end{equation}
In terms of these four Bose fields, four new Bose fields are defined
(effectively separating charge and spin):
\begin{eqnarray}\label{eVIxvi}
\phi_{P1} &=& \frac{1}{2}(\phi_{P1\uparrow} + \phi_{P1\downarrow} +
\phi_{P2\uparrow} + \phi_{P2\downarrow})\, ;\nonumber\\[1mm]
\phi_{P2} &=& \frac{1}{2}(\phi_{P1\uparrow} - \phi_{P1\downarrow} +
\phi_{P2\uparrow} - \phi_{P2\downarrow})\, ;\nonumber\\[1mm]
\phi_{P3} &=& \frac{1}{2}(\phi_{P1\uparrow} - \phi_{P1\downarrow} -
\phi_{P2\uparrow} + \phi_{P2\downarrow})\, ;\nonumber\\[1mm]
\phi_{P4} &=& \frac{P}{2}(\phi_{P1\uparrow} + \phi_{P1\downarrow} -
\phi_{P2\uparrow} - \phi_{P2\downarrow}) .
\end{eqnarray}
Note that $\phi_{P4}$ has a relative sign between the right and left movers.
This sign effectively masks the $SO(8)$ symmetry in the original 
Hamiltonian.
From these chiral bosons, one can define pairs of conjugate bosons in
the standard fashion\\[-8mm]~
\begin{eqnarray}\label{eVIxvii}
\varphi_{i} &=& \phi_{Ri}+\phi_{Li}\,;\nonumber\\[1mm]
\theta_{i} &=& \phi_{Ri}-\phi_{Li}\,,
\end{eqnarray}
which obey the commutation relations,
\begin{equation}\label{eVIxviii}
[\varphi (x),\theta (x')] = -i4\pi\Theta (x-x')\,.
\end{equation}
where $\Theta (x-x')$ is the Heaviside step function.

In terms of these variables the free part of the Hamiltonian can be written
\begin{equation}\label{eVIxix}
H_0 = \frac{v_F}{8\pi} \sum_a \big\{ (\del_x\theta_a)^2+(\del_x\varphi_a)^2\big\}\,.
\end{equation}
The interacting momentum conserving interacting term can be written as
\begin{eqnarray}\label{eVIxx}
H_B + H_F &=& \frac{1}{16\pi^2}\sum^4_{a=1} A_a\big\{ (\del_x\theta_a)^2 - (\del_x\varphi_a)^2\big\}\nonumber\\[1mm]
&& - 2b^\sigma_{12}\cos(\theta_4)\cos(\theta_2) + \cos(\theta_2)(2b^\sigma_{11}\cos(\theta_3)+2f^\sigma_{12}\cos(\varphi_3))\nonumber\\[1mm]
&& - \cos(\theta_4)((b^\sigma_{12}+4b^\rho_{12})\cos(\theta_3)+(b^\sigma_{12}-4b^\rho_{12})\cos(\varphi_3))\,,
\end{eqnarray}
where the coefficients $A_a$ are equal to $A_{1/4}=2(c^\rho_{11} \pm f^\rho_{12})$ and 
$A_{2/3} = -(c^\sigma_{11}\pm f^\sigma_{12})/2$.
The Umklapp interactions in these variables takes the form
\begin{eqnarray}\label{eVIxxi}
H_U &=& -16u^\rho_{11}\cos (\theta_1)\cos(\theta_4) - 4u^\sigma_{12}\cos(\theta_1)\cos(\theta_2) \nonumber\\[1mm]
&& -\cos(\theta_1)(2(u^\sigma_{12}+4u^\rho_{12})\cos(\theta_3)+2(u^\sigma_{12}-4u^\rho_{12})\cos(\varphi_3))\,.
\end{eqnarray}
As demonstrated in Ref. \cite{lin}, the Klein factors combine in such a fashion so as to be equal to 1.

For generically repulsive scattering amplitudes, the various couplings $f,b,$ and $u$
flow to fixed ratios under the RG \cite{lin}:
\begin{eqnarray}\label{eVIxxii}
b^\rho_{12} = \frac{1}{4}b^\sigma_{12} = f^\rho_{12} = -\frac{1}{4}b^\sigma_{11} = 2u^\rho_{11} = 2u^{\rho}_{12} 
= \frac{1}{2}u^\sigma_{12} = g > 0\,.
\end{eqnarray}
With these values, the interaction Hamiltonian dramatically simplifies to
\begin{eqnarray}\label{eVIxxiii}
H_{\rm int} &=& H_B + H_F + H_U\cr
&=&  -\frac{g}{2\pi^2}\sum^4_{a=1} \partial_x \phi_{Ra}\partial_x\phi_{La}
-4g\sum^4_{a\neq b=1}\cos (\theta_a)\cos (\theta_b)\,.
\end{eqnarray}
We now write this Hamiltonian in fermionic form.
We thus refermionize the bosons $\phi_{Pa}$, $a=1,\cdots,4$, i.e.,
\begin{eqnarray}\label{eVIxxiv}
\Psi_{Pa} &=& \kappa_a e^{i\phi_a}\,, ~~~~ {\rm a = 1,\cdots,4}\,;\nonumber\\[1mm]
\Psi_{P4} &=& P\kappa_4 e^{i\phi_4}\,,\end{eqnarray}
where the Klein factors are given by
\begin{equation}\label{eVIxxv}
\kappa_1 = \kappa_{2\uparrow}, ~~~ \kappa_2 = \kappa_{1\uparrow}, ~~~
\kappa_3 = \kappa_{1\downarrow}, ~~~ \kappa_4 = \kappa_{2\downarrow}\,.
\end{equation}
We then find the free Hamiltonian can be written as
\begin{equation}\label{eVIxxvi}
H_0 = \int dx \sum_a 
\bigl [
\Psi^\dagger_{aL}i\del_x \Psi_{aL} -
\Psi^\dagger_{aR} i\del_x \Psi_{aR} 
\bigr ]\,,
\end{equation}
where the Fermi velocity, $v_F$, has been set to 1, while
the interaction Hamiltonian can be written as
\begin{equation}\label{eVIxxvii}
H_{int} =
g \big[\sum^4_{a=1} 
(i\Psi^\dagger_{aL}\Psi_{aR} - i\Psi^\dagger_{aR}\Psi_{aL})\big]^2 .
\end{equation}
This is, of course, $H_{int}$ for the $SO(8)$ Gross--Neveu model.

It will sometimes prove convenient to recast the theory in terms of 
Majorana fermions, $\psi_{aP}$.  In terms of the Dirac fermions, 
$\Psi_{aP}$,
they are given
by
\begin{equation}\label{eVIxxviii}
\Psi_{aP} = \frac{1}{\sqrt{2}}(\psi_{2a,P} + i\psi_{2a-1,P})\,,
\qquad (a=1,...,4)\,.
\end{equation}
In this basis, $H_{int}$ can be recast as
\begin{equation}\label{eVIxxix}
H_{int} = g G^{ab}_R G^{ab}_L\,, \qquad (a>b = 1,...,8)\,,
\end{equation}
where $G^{ab}_P = i \psi_{aP}\psi_{bP}$ is one of the 28
$SO(8)$ Gross--Neveu
currents.

\subsection{Limitations of the RG Analysis}

In the previous section we have sketched the argument by which Ref. \cite{lin}
demonstrated that generically interacting Hubbard ladders/armchair carbon nanotubes
behave at low energies according to the SO(8) Gross--Neveu.  In this section
we want to consider how precise a statement this is. 

The analysis in \cite{lin} is based on an 1-loop RG.
As such the initial microscopic (bare)
interactions must be small enough so that
the integrable $SO(8)$ invariant RG trajectory is approached sufficiently
closely after a number of RG steps, before leaving the
range of validity of the 1-loop RG equations.
Whenever this is the case, it can be argued that 
the integrable model is approached
independently of the (sufficiently weak) values of the bare interactions.
The situation for the 2-leg ladder
is thus similar in spirit
to that of the point contact device encountered
in \cite{FLS,FLS1,FLS2},  where only a single operator was relevant, and
this relevant operator
was integrable. In the latter case all other interactions were irrelevant
in the RG sense, and could in principle be treated perturbatively.

The requirement of the 
RG that the interactions be short-ranged is natural
in the case of the Hubbard ladders.  However it may not seem so
in the case of the carbon nanotubes.  Both theory 
\cite{Egger97,Egger98,Kane97} and experiment \cite{bockratha,cobden}
have discussed the case where long-ranged Coulomb forces drive Luttinger
liquid behavior in single-walled carbon nanotubes.  
However we do not have such
situations in mind for the paper at hand.  Rather we want to consider
situations such as those found in the experiments
\cite{wildoer} where the long range forces
are screened.

Although the restriction to such experiments in the case of the carbon
nanotubes
places us upon safe ground, it is not 
inconceivable that experiments where
the long-ranged forces are present would nevertheless see 
behavior indicative of the SO(8) symmetry.  
An unscreened force translates into an unusually
large bare coupling (in comparison with other bare couplings)
in the forward scattering direction.
However this does not mean the RG is inapplicable.  The RG still 
indicates a potential
enhancement in the symmetry.  Because of the large
bare coupling, the RG must be  
run a longer time before any enhancement
would be seen 
but nevertheless an enhancement may well occur at some
low energy scale.  In terms of the experiments in \cite{bockratha,cobden}, this would
mean that at medium energy scales, 
Luttinger liquid behavior would predominate,
while at much lower energy scales, SO(8) behavior would be expected.
However at current standing, the material science is not advanced
to the point where it is possible to 
accurately probe the very low energy behavior.
But the potential for advancement in this area is ever present.

The RG analysis further requires the bare couplings to be weak.  With
Hubbard ladder compounds, this condition will not be generically met,
although it certainly will not be universally violated.  However with
(N,N)-armchair carbon nanotubes, the bare couplings are naturally 
weak.  It is one of the hallmarks of the physics of the (N,N) armchair
carbon nanotubes that the electrons are delocalized around the circumference
of the tube.  This in turn leads to a scaling of the effective short-ranged
interaction by 1/N, making it naturally small \cite{bal}.

It can, however, be questioned on a more fundamental level
whether an 1-loop RG adequately describes
the system's behavior.  Difficulties with the analysis in \cite{lin} 
take two forms.
As a first objection,
the authors of Ref. \cite{tsv} point out that an RG flow can imply
a symmetry restoration which in fact does not occur.  As an example
they consider
a U(1) symmetric Thirring model,
\begin{equation}\label{eVIxxx}
{\cal{L}} = 
\bar\Psi_\alpha \gamma^\mu\partial_\mu\Psi_\alpha + \frac{1}{4}g_\parallel
(j_z)^2 + \frac{1}{4}g_\perp [(j_x)^2+(j_y)^2]\,,
\end{equation}
where $j^\mu_a = \bar\Psi\gamma^\mu\sigma_a\Psi$.
Although the 1-loop RG equations for this model 
seems to indicate a generic
symmetry 
restoration to a more symmetric SU(2) case (i.e. $g_\perp = g_\parallel$), 
this in fact only occurs in a certain region of coupling space.
For $\pi - |g_\perp| > -g_\parallel > |g_\perp| > 0$, 
the U(1) model maps onto the sine-Gordon model with interaction 
$\cos (\beta\phi)$
\cite{Wieg}, where $\beta$ is given by
\begin{equation}\label{eVIxxxi}
8\pi\beta^2 = 8\pi - 8\mu ; ~~~~~ \mu = 
{\cos ^{-1} [\cos (g_\parallel ) / \cos (g_\perp)]}\,.
\end{equation}
The value of $\beta$ completely characterizes the model.  While
$g_\perp$ and $g_\parallel$ flow under the RG, the particular combination
of these parameters forming $\beta$ does not.
Thus for this particular region of parameter space
the model moves no closer to the SU(2) symmetric point
under an RG flow.  In other regions however (for example 
$|g_\perp| > |g_\parallel|$), 
the situation is better; the effect of the anisotropy in the 
couplings is exponentially suppressed.

However 
it is reasonably clear that such pessimism is not warranted in the analysis
of the RG of Ref. \cite{lin}.  A salient criticism of Ref. \cite{tsv} is that in
considering the action of the renormalization group, they fail
to consider the consequences of working in the scaling limit.  The
scaling limit is exactly the limit in which a field theory becomes
available.  In turn, the scaling limit places constraints upon the
possible range of bare couplings consistent with a field theory.
In the case of sine-Gordon, the underlying integrability/solvability
of the theory allows explicit investigation of this question.
It is found that the allowed range is such that even moderate
anisotropic deviations are forbidden \cite{anis}.  
The scaling limit, in other words,
enforces isotropy.  In cases where there are RG flows indicating an
enhancement in symmetry (including the case at hand), this enforcement turns out to be a general phenomena and
it leads to an expanded notion of symmetry 
restoration \cite{anis}.

We can see however that this conclusion is a double edged sword.
Strongly anisotropic models will not have sensible continuum limits and so will
not be able to be described by a field theory.  If the Hubbard ladder/carbon
nanotube is strongly anisotropic, the continuum theory is not a good starting point rendering the RG analysis
irrelevant.  However if interactions are
only weakly anisotropic, the continuum theory can be used and the weakly anisotropic Hamiltonian is better
able to flow under the RG to the SO(8) symmetric theory.

On a more concrete level, the breaking of the SU(2) symmetry considered in
Ref. \cite{tsv} is a rather special case.  
Ultimately, the parameter, $\beta$, in the
sine-Gordon model is protected under an RG flow by the presence of a 
quantum group symmetry arrived at by deforming a Yangian symmetry present 
at the SU(2) point.  There is, however, no such known way to
deform the Yangian in $SO(8)$ Gross--Neveu.  Indeed the natural 
generalization 
of the sine-Gordon
model to $SO(8)$ is not to $SO(8)$ Gross--Neveu but to an affine toda 
$SO(8)$ theory where such a deformation of the Yangian symmetry is 
possible \cite{leclair}.

Another question that one must ask in looking at the analysis in Ref. \cite{tsv}
is how the choice of the symmetry breaking terms affects the symmetry
restoration.  The sine-Gordon model still possesses a U(1) symmetry.  
However
it is certainly possible to consider perturbations that break this U(1).
Such perturbations would destroy the quantum group symmetry of sine-Gordon
and thus might lead to symmetry restoration.  
This would be perhaps closer to the RG analysis
of Ref. \cite{lin} where a large number (nine) of marginal perturbations 
were included.  We in fact consider exactly such a situation in Ref. \cite{anis}
and find that indeed there is symmetry restoration.
We also note in passing that the authors of Ref. \cite{tsv} consider
an anisotropic Gross--Neveu model, a model of direct relevance
to the situation at hand.  They conclude
through a mean-field/large N limit 
computation that the model is intrinsically
anisotropic thus throwing doubt upon the analysis in Lin et al. \cite{lin}.
However we would point to how the
bare couplings are scaled
in the large N limit.  The anisotropic model they consider
has three bare couplings.  One is chosen to not scale at all, one scales
as $1/N$, and the last scales as $1/N^{d/2}$, $d<1$.  With 
this scaling \cite{tsv}
the model is intrinsically anisotropic.  However 
this model possesses a diverging bare anisotropy in the large N limit
while the RG is not allowed to act.
As such, we believe this example is not so directly telling.

The second objection to the analysis of Ref. \cite{lin} is 
its omission of chiral interactions
that alter the Fermi velocities \cite{kiv}.  
Such interactions, although they
are absent from the 1-loop RG, likely play a role 
at higher order.  However their effect is less drastic than envisioned
in Ref. \cite{kiv}.  There a scenario was considered where the
invariant RG trajectory of higher symmetry was inherently unstable to
perturbations.  However the $SO(8)$ RG ray in Ref. \cite{lin} has a basin of 
attraction of finite measure.  The effect of chiral interactions is to then
slightly alter the direction of the ray.
In turn the ratio of
masses of the various fundamental excitations will be slightly perturbed away from one.

In taking into account of these objections, prudence suggests
a modification in the understanding of the RG analysis of Ref. \cite{lin}.  
This analysis
tells us that while the RG flow does
not restore an exact symmetry, it leaves us close to the symmetrical 
situation.  In particular, it indicates that while the masses in the actual
system may differ from their $SO(8)$ values, they do not wildly diverge.
One then understands the $SO(8)$ Gross--Neveu
theory, not as precisely
representative of the actual system, but in near perturbative vicinity
of it, that is,
as an excellent starting
point about which to perform perturbation theory in the
non-integrable interactions breaking $SO(8)$ in much
the same spirit as done for
a non-critical Ising model in the presence of a magnetic
field \cite{simo}.

We can state with some confidence that there are scenarios where the effects of these perturbations will
be small.  A related Hubbard lattice model possessing an
$SU(4)$ symmetry has been studied in Ref. \cite{azaria}.  Provided that the interaction
strength was not overly large ($U/t < 3$), it was found using
quantum Monte Carlo that the low energy sector of the $SU(4)$ theory was enhanced to $SO(8)$ Gross-Neveu.
In particular it was found that the charge gap, the spin gap, and the single particle gap satisfy
the corresponding $SO(8)$ ratios.

\subsection{Excitations and Physical Fields in SO(8) Gross--Neveu}

We now consider the basic excitations of SO(8) Gross--Neveu together with identifying
the fields in the theory that correspond to operators of physical interest.

\subsubsection{Excitation Spectrum}

The Gross--Neveu $SO(8)$ model has 
an exceedingly rich spectrum.  There are
24 fermionic particles of mass $m$ organized into one eight dimensional
vector representation and two eight dimensional spinor representations.
We denote the particles of the vector representation by $A_a$,
$a = 1,\ldots ,8$.  The $A_a$'s are the Majorana fermions of Eq.\,(\ref{eVIxxviii}).
We will often refer to these particles as Gross--Neveu fermions.  
The kink particles, in turn, will be denoted by $A_\alpha$.  Here $\alpha$
is of the form $\alpha = (\pm 1/2, \pm 1/2,\pm 1/2,\pm 1/2)$ and so takes
on 16 values.  These 16 particles decompose into the two eight-dimensional
spinor representations.  The division is affected by the chirality (either even or odd) of the kinks.
An even chirality kink has an even number of $+1/2$'s in its corresponding $\alpha$ while an odd
chirality kink has a correspondingly odd number.

Beyond the eight dimensional representations, there are
$29$ bosonic particle states of mass $\sqrt{3} m$,
transforming as a  rank-two tensor of dimension $28$
and a singlet.  Together they form a representation of the $SO(8)$ Yangian
symmetry.  
These particles can be thought of as
bound states of either two kinks or two fundamental fermions.

As $SO(8)$ is a rank 4 algebra, the $SO(8)$ Gross--Neveu 
model has four Cartan
bosons (i.e. the $\phi_{Pa}$ , $a = 1, \ldots 4$) and so its excitations
are characterized by four
quantum numbers, $N_i$, $i = 1,\ldots ,4$.
With the Majorana fermions, the combination
\begin{equation}\label{eVIxxxii}
A_{2a} \pm iA_{2a-1}\,,
\end{equation}
carries quantum number $N_a=\pm  1$, $N_b=0, b \not =  a$. 
The quantum numbers carried by the
kinks $A_\alpha$ are directly encoded in $\alpha$.
If $\alpha = (a_1,a_2,a_3,a_4)$, $a_i=\pm 1/2$, the $A_\alpha$ carries
the quantum numbers, $N_i = a_i$.  The quantum numbers carried by the rank
two tensor states can be directly deduced from the particles forming the
bound state.  As we will always think of the bound states in this way,
we will not list their quantum numbers directly.

The last thing needed in the section is to identify the relationship 
between the quantum numbers, $N_i$, and the physical quantum numbers of the
system: the
z-component of spin, $S_z$, the charge, $Q$, the difference in
z-component of spin
between the two bands, $S_{12}$, and the ``relative
band chirality'', $P_{12}$, defined as
$P_{12} = N_{R1} - N_{L1} - N_{R2} + N_{L2}$, where $N_{Pj}$ is the
number electrons in band $j$ with chirality $P$.
We have
\begin{eqnarray}\label{eVIxxxiii}
(N_1=1,0,0,0) &\leftrightarrow& (Q=2, S_z = 0, S_{12} = 0, P_{12} = 0)\,;
\nonumber\\[1mm]
(0,N_2=1,0,0) &\leftrightarrow& (Q=0, S_z = 1, S_{12} = 0, P_{12} = 0)\,;
\nonumber\\[1mm]
(0,0,N_3=1,0) &\leftrightarrow& (Q=0, S_z = 0, S_{12} = 1, P_{12} = 0)\,;
\nonumber\\[1mm]
(0,0,0,N_4=1) &\leftrightarrow& (Q=0, S_z = 0, S_{12} = 0, P_{12} = 2)\,.
\end{eqnarray}
With this assignment, we can see that the vector representation of 
fundamental
fermions corresponds to states of two electrons in the original 
formulation.
For example, the fermion $A_2 \pm iA_1$ carries charge $\pm 2$
and no spin (the cooperons),
and the fermion $A_4 \pm iA_3$ carries spin, $S_z = 1$,
and no charge (the magnons). 
The spinor representations, the kinks, in turn correspond to single 
particle
excitations as their quantum numbers are combinations of $N_i/2$.\\[-6mm]~

\subsubsection{Relationship between Lattice Operators and Gross--Neveu Fields}

In this section we make contact between the fields of the 
$SO(8)$ Gross--Neveu model
and the original fields of the Hubbard ladders.  This identification is 
crucial if we are to compute physically relevant correlation functions.

As we have 
already discussed,
the fundamental (Dirac)
fermions of the vector representation are given by
\begin{eqnarray}\label{eVIxxxiv}
\Psi_{aP} &=& \kappa_a e^{i\phi_{aP}}\,, \nonumber\\[1mm]
\Psi_{aP} &=& P\kappa_a e^{i\phi_{aP}} \,,\end{eqnarray}
and carry quantum numbers corresponding to two electronic excitations.
However the $\Psi_{aP}$ are fermionic, whereas such excitations are
bosonic.  As such, $\Psi_{aP}$ are not simply related to a fermionic
bilinear of the original electrons but must be a fermion bilinear
multiplying some non-local field (a Jordan--Wigner string).  As we
will not compute correlators involving such fields in this review, we
will not elaborate upon this.

As discussed previously, the kinks correspond to single particle 
excitations.
Thus we expect to find that the kink fields are related to the original
electron operators.  This is true in part.  There are 32 kinks in total
(counting both left and right movers), but only sixteen electron
operators, the $c$'s and $c^{\dagger}$'s (four for each of the 
four Fermi points).  So
we expect only $1/2$ of the kinks to correspond to actual electron 
operators.

We represented the fundamental fermions in terms of the four Cartan bosons.
There is a corresponding representation for the kink fields:
\begin{equation}\label{eVIxxxv}
\psi_{\alpha P} \sim e^{i\alpha \cdot \bar{\phi}_P} ,
\end{equation}
where $\bar{\phi} = (\phi_1 , \phi_2 , \phi_3 ,\phi_4 )$.
The kink fields that then correspond to the electron operators
$c$'s are as follows:
\begin{eqnarray}\label{eVIxxxvi}\cr
c_{R1\uparrow} &\sim& e^{i(\phi_{1R} + \phi_{2R} + \phi_{3R} + \phi_{4R})/2};
\nonumber\\[1mm]
c_{R2\uparrow} &\sim& e^{i(\phi_{1R} + \phi_{2R} - \phi_{3R} - 
\phi_{4R})/2} ;\nonumber\\[1mm]
c_{R2\downarrow} &\sim& e^{i(\phi_{1R} - \phi_{2R} + \phi_{3R} - 
\phi_{4R})/2} ;\nonumber\\[1mm]
c_{R1\downarrow} &\sim& e^{i(\phi_{1R} - \phi_{2R} - \phi_{3R} + 
\phi_{4R})/2} ;\nonumber\\[2mm]
&& {\rm (even~chirality)}\cr
\cr\cr
c_{L1\uparrow} &\sim& e^{i(\phi_{1L} + \phi_{2L} + \phi_{3L} - 
\phi_{4L})/2} ;\nonumber\\[1mm]
c_{L2\uparrow} &\sim& e^{i(\phi_{1L} + \phi_{2L} - \phi_{3L} + 
\phi_{4L})/2} ;\nonumber\\[1mm]
c_{L2\downarrow}
&\sim& e^{i(\phi_{1L} - \phi_{2L} + \phi_{3L} + \phi_{4L})/2} ;\nonumber\\[1mm]
c_{L1\downarrow}
&\sim& e^{i(\phi_{1L} - \phi_{2L} - \phi_{3L} - \phi_{4L})/2} ; \nonumber\\[2mm]
&& {\rm (odd~chirality)} .
\end{eqnarray}
With hermitian conjugates, this totals to sixteen fields.
The $\sim$ sign is meant to indicate that these equivalences hold up to
Klein factors.  The $c_{Pj\alpha}$'s, of course, are fermionic.  However
the kink fields as defined are not.

The last set of fields that are of concern to us are the currents.  The
electric current of the ladder has the lattice representation
\begin{equation}\label{eVIxxxvii}
J \sim -i \sum_{l\alpha}
\biggl [
 a^\dagger_{l\alpha}(x)a_{l\alpha}(x+1)
- a^\dagger_{l\alpha}(x+1)a_{l\alpha}(x) 
\biggr ],
\end{equation}
where we have summed over the contribution coming from each spin ($\alpha$)
and each leg ($l$) of the ladder. 
Taking the continuum limit, $J$ equals, in Gross--Neveu language, 
\begin{equation}\label{eVIxxxviii}
J \sim i \sin k_{F1} ~\del_t \phi_1 \sim G_{12} ,
\end{equation}
where $G_{12}$ is one of the $SO(8)$ currents discussed in Eq.\,(\ref{eVIxxix}).

\subsection{Form Factors in SO(8) Gross--Neveu}

\def\fcf{_\mu f^{ab}_{cd}}
\def\fcft{_\mu f^{ab}_{cd}(\th_1,\th_2)}
\def\fck{_\mu f^{ab}_{\alpha\beta}}
\def\fckt{_\mu f^{ab}_{\alpha\beta}(\th_1,\th_2)}
\def\cop{G^{ab}_\mu}
\def\cs{(C \sigma^{ab})_{\alpha\beta}}
\def\sco{(\sigma^{ab}C)_{\alpha\beta}}
\def\fk{\psi^\alpha_\pm}
\def\kfk{{_\pm f}^\alpha_{a\beta}}
\def\kfkt{{_\pm f}^\alpha_{a\beta}(\th_1,\th_2)}
\def\kkf{{_\pm f}^\alpha_{\beta a}}
\def\kkft{{_\pm f}^\alpha_{\beta a}(\th_1,\th_2)}
\def\ff{\psi^a_\pm}
\def\fkk{_\pm f^a_{\alpha\beta}}
\def\fkkt{_\pm f^a_{\alpha\beta}(\th_1,\th_2)}
\def\cg{(C\gamma^a)_{\alpha\beta}}
\def\gc{(\gamma^aC)_{\alpha\beta}}
\def\fpm{{f_\pm (\th_1,\th_2)}}
\def\fo{f_0}
\def\kk{{_\pm}f^\alpha_\beta}

Having identified the fields that we need to compute physical correlators, we now turn to
the corresponding form factors of these fields.  Here we only state the needed form factors.
Their derivation may be found in \cite{so8} and \cite{karowski78}.

\subsubsection{One and Two Particle Current Form-Factors}

At the two particle level, two kinks of the same chirality or two GN fermions can couple to one
of the Gross--Neveu current operators (see Eq.\,(\ref{eVIxxix})).  The corresponding form-factors are
\begin{eqnarray}\label{eVIxxxix}
\fcft &\equiv& \langle G_\mu^{ab} (0) A_b (\th_2) A_a (\th_1) \rangle
=
i A_G (\delta_{ac}\delta_{bd} - \delta_{ad}\delta_{bc})
f_\mu (\th_1,\th_2)\, ,\cr\cr
\fckt &\equiv& \langle G_\mu^{ab} (0) A_\beta (\th_2) 
A_\alpha (\th_1) \rangle
=
i\, \frac{A_G}{2}\, \cs f_\mu (\th_1,\th_2)\, ,
\end{eqnarray}
where
\begin{eqnarray}\label{eVIxl}
f_{\mu}(\th_1,\th_2) &=&
(e^{(\th_1+\th_2)/2} - (-1)^\mu e^{-(\th_1+\th_2)/2})
\,\frac{s(\th_{12}/2)}{c(\th_{12}) - 1/2}
\nonumber\\[1mm]
&& \times \exp \bigg[\int^\infty_0 \frac{dx}{x} \frac{G_c(x)}{s(x)}
\sin^2(\frac{x}{2\pi}(i\pi + \th_{12}))\bigg]\, ,\nonumber\\[2mm]
G_c(x) &=& 2\, \frac{c(x/6) - s(x/6)e^{-2x/3}}{c(x/2)}\,.
\end{eqnarray}
Here the current $G^{ab}_\mu$ is a particular combination, $G^{ab}_R - (-1)^\mu G^{ab}_L$, of the chiral currents of
Eq.\,(\ref{eVIxxix}).  $A_G$ is some arbitrary (real) normalization constant.

The current operators also possess one particle form factors.  The currents are able to couple to
one of the 29 bound state excitations.  The corresponding form factor can be computed following the
discussion in Section 2.6 of this review.  We thus have
\begin{eqnarray}\label{eVIxli}
_\mu f^{ab}_{\{cd\}} (\th ) &\equiv&
\langle G^{ab} (0) A_{\{cd\}} (\th ) \rangle\nonumber\\[1mm]
&=& i A_G
(\delta_{ac}\delta_{bd} - \delta_{ad}\delta_{bc})(e^\th - 
(-1)^\mu e^{-\th})
\,\frac{1}{\sqrt{3}}\,
\bigg(\! 2\sqrt{3\pi}\, \frac{\Gamma (2/3)}{\Gamma (1/6)}\bigg)^{-1/2}
\nonumber\\[1mm]
&&\hskip -.2in 
\times \exp \bigg[-\int^\infty_0 \frac{dx}{x}\, \frac{G_c(x)}{s(x)} \,
s^2(x/3)\bigg]\, .
\end{eqnarray}
Notice that the normalization constant $A_G$ is the same as that which appears in the two-particle current form factor.

\subsubsection{One and Two Particle Kink Form-Factors}

We recall that the single electron excitations of the theory are represented by kink fields.
A kink field, $\fk$, of a given chirality couples to a combination of excitations consisting of
a kink particle of opposite chirality and a Gross--Neveu fermion.  The corresponding form factor
is given by
\begin{eqnarray}\label{eVIxlii}
\kfkt &=& \kkft \equiv \langle \psi^\alpha_\pm (0) A_\beta (\th_2) 
A_a (\th_1)
\rangle \nonumber\\[1mm]
&=& -A_F e^{\pm i\pi/4} \cg \fpm\, ,
\end{eqnarray}
where
\begin{eqnarray}\label{eVIxliii}
\fpm &=& \frac{e^{\pm (\th_1 + \th_2)/4}}{c(\th_{12} ) + 1/2}
\exp \bigg[\int^\infty_0 \frac{dx}{x}\,
\frac{G_f(x)}{s(x)} \sin^2\Big(\frac{x}{2\pi}(i\pi + \th_{12})\Big)\bigg]\, ,
\nonumber\\[1mm]
G_f(x) &=& \frac{2 c(x/6) + e^{-7x/6}}{c(x/2)}\ .
\end{eqnarray}
Here $A_F$ is an arbitrary (real) normalization parameter.

The kink field of course also couples to a single kink.  The associated form factor can be computed
again using the bound state relations of Section 2.5 (a kink of a given chirality is a bound state of
a kink of opposite chirality together with a Gross--Neveu fermion).  The result of the calculation is
\begin{eqnarray}\label{eVIxliv}
\kk &\equiv& \langle \psi^\alpha_\pm (0) A_\alpha ( \th ) \rangle =
c_\pm e^{\pm \th/2} C_{\alpha\beta}\, ,
\end{eqnarray}
where the constant $c_\pm$ is
\begin{eqnarray}\label{eVIxlv}
c_\pm &=& \frac{4}{\sqrt{3}}\, e^{\pm i\pi/4}
A_F \bigg(\!\sqrt{3\pi}\, \frac{\Gamma (5/3)}{\Gamma (7/6)}\bigg)^{\!-1/2}
\hskip -5mm \exp \bigg[ -\!\! \int^\infty_0\! \frac{dx}{x}\,\frac{G_f(x)}{s(x)}\, s^2(x/6)\bigg]\,.~~~~~~
\end{eqnarray}
Again $A_F$ is the same normalization constant appearing in the above two-particle form factor.

\subsection{Exact Low Energy Correlation Functions in SO(8) Gross--Neveu}

In this section we study various aspects of carbon nanotubes 
and Hubbard ladders using the previously stated form factors.
In particular we study the optical conductivity of a ladder system, the single particle spectral
function of a ladder/nanotube, and finally the associated differential conductance arising from
tunneling into a ladder/nanotube via a scanning tunneling microscope (STM).
We again emphasize that although we calculate these quantities with a small, finite
number of form factors, the results are exact up to some energy scale.

\subsubsection{Behavior of Optical Conductivity in a Hubbard Ladder}

We first consider the response of the ladder system to an
electric field polarized along the legs.  Apart from the
meanfield treatment in Ref. \cite{lin}, this problem
has been examined previously,
both theoretically \cite{scalp} and experimentally
\cite{Hiroi95}.  However
these two latter papers did not consider undoped
ladders at zero temperature.

In linear response, the optical conductivity is given by
\begin{equation}\label{eVIxlvi}
{\rm Re} \big[\sigma (\om ,k)\big]
= {\rm Im} \Big[ \frac{\Delta (\om ,k)}{\om}\Big]\,,
\end{equation}
where $\Delta$ is the current-current correlator
\begin{equation}\label{eVIxlvii}
\Delta (\om ,k) = \int dx d\tau e^{i\om \tau}e^{i x k} 
\langle T(J(x,\tau)J(0,0))
\rangle |_{\om \rightarrow -i\om + \delta}\,.
\end{equation}
$J$ is given by Eq.\,(\ref{eVIxxxviii})
\begin{equation}\label{eVIxlviii}
J \sim G^{12}_1.
\end{equation}
As explained in Section 2, to compute the correlator, $\tcor$, we insert a resolution of the identity
between the two $J's$, turning the correlator into a form factor sum.
We then have
\begin{eqnarray}\label{eVIxlix}
\tcor 
&&
\nonumber\\[1mm]
&& \hskip -1.5in 
=\sum^\infty_{n=0}\sum_{a_1,\cdots ,a_n} \int \frac{d\th_1}{2\pi}
\cdots \frac{d\th_n}{2\pi} \,
\langle G^{12}_1 (0)|A^\dagger_{a_1}(\th_1) \cdots
A^\dagger_{a_n}(\th_n) \rangle \nonumber\\[1mm]
&& \hskip -1.5in \times \langle A_{a_n}(\th_n) \cdots
A_{a_1}(\th_1)| G^{12}_1 (0)\rangle\nonumber\\[1mm]
&& \hskip -1.5in \times\exp \big(-|\tau| \sum^n_{i=1} m_{a_i}\cosh (\th_i)
+ i{\rm sign} (\tau) x\sum^n_{i=1} m_{a_i}\sinh (\th_i)\big)\,,
\end{eqnarray}
\noindent where 
the first sum $\sum_n$ runs over the number of particles in the form
factor expansion and the second sum $\sum_{a_i}$ runs over 
the different particle
types.  We have also extracted the spacetime dependence of each term.

We do not compute this sum in its entirety but truncate
at the two particle level:
\begin{eqnarray}\label{eVIl}
\tcor &=& \int \frac{d\th_1}{2\pi}\, \langle G^{12}_1(0) | A^\dagger_{12}
(\th_1)\rangle
\langle A_{12}(\th_1) | G^{12}_1(0)\rangle \nonumber\\[1mm]
&& \hskip -.7in \times \exp\big(-|\tau|\sqrt{3}m\cosh (\th_1)\big)
+ i{\rm sign}(\tau)x\sqrt{3}m\sinh (\th_1)) \nonumber\\[1mm]
&& \hskip -1.75in + \frac{1}{2}\! \int\! \frac{d\th_1}{2\pi} \frac{d\th_2}{2\pi}
\exp\big(\!\sum_{i=1,2}\big(-|\tau|m\cosh(\th_i)
\!+\! i{\rm sign}(\tau)xm\sinh(\th_i)\big)\big)
\cr\cr
&&
\hskip -1in \times \Big( \sum_{ab} \langle G^{12}_1(0)|A^\dagger_a(\th_1)A^\dagger_b
(\th_2)\rangle
\langle A_b(\th_2)A_a(\th_1)|G^{12}_1(0)\rangle \cr
&& \hskip -1in +
\sum_{\alpha\beta} \langle G^{12}_1(0)|A^\dagger_\alpha (\th_1)
A^\dagger_\beta (\th_2)\rangle
\langle A_\beta(\th_2)A_\alpha(\th_1)|G^{12}_1(0)\rangle \Big)\, .
\end{eqnarray}
The first term gives 
the single particle contribution to the correlation
function.  The only particle that contributes here is $A_{12}$,
denoting one of the particles
belonging to the rank 2 tensor multiplet.  At the two particle level a
variety of contributions are non-zero.  The second term in \ref{eVIl} gives
the contribution of two Majorana fermions while the third term gives
the contribution of kinks with the same chirality.

As discussed in Section 2, this truncation of the form factor
sum is better than it may at first seem.  Because the correlator 
is evaluated
at zero temperature in a massive system, the higher order terms make
contributions only at higher energies, $\om$.  That is, the massiveness
of the system leads to particle thresholds.  
The next contribution comes from a three particle combination of
even kink/fermion/odd kink that carries mass $3m$.  Thus for $\om < 3m$,
this term gives no contribution to ${\rm Re} [\sigma (\omega,k)]$
for arbitrary $k$.
Hence our result for ${\rm Re} [\sigma (\omega ,k)]$ is exact 
for $\omega <3m$. 

In the case when $\om$ does exceed $3m$, we expect the higher particle
form factors to make only a small contribution to $\sigma (\om )$.
As we saw both in Section 2.5 and Section 4.3.2,
terms in the form factor sum involving higher numbers of particles
make only negligible contributions to the spectral function at any
given energy.

Using the results for the form factors of Section 6.5, we can put everything
together and write down an expression for ${\rm Re} [\sigma (\om )]$:
\begin{eqnarray}\label{eVIli}
{\rm Re} [\sigma (\om )] &=& \delta(w-\sqrt{3m^2+k^2})
~\Big(\frac{A_G}{m}\Big)^2 \frac{2}{9}\sqrt{\frac{\pi}{3}}\,
\frac{\Gamma (1/6)}{\Gamma (2/3)}\nonumber\\[1mm]
&& \hskip 1in \times \exp\Big[-2\int^\infty_0 \frac{dx}{x}\, \frac{G_c(x)}{s(x)}\, s^2 (x/3)\Big]\cr\cr
&& \hskip -.5in + \theta (w^2 - k^2 - 4m^2)\, \frac{24 m^2 A_G^2}{(\om^2 - k^2 - 3m^2)^2}\,
\frac{\om {\sqrt{\om^2-k^2-4m^2}}}{(\om^2 - k^2)^{3/2}} \nonumber\\[1mm]
&&  \hskip .25in\times
\exp \Big[ \int^\infty_0 \frac{dx}{x}\frac{G_c(x)}{s(x)}
(1-c(x)\cos (\frac{x\th_{12}}{\pi}))\Big]\,,
\end{eqnarray}
\noindent where $s(x) = \sinh (x)$, $c(x) = \cosh (x)$, and
\begin{equation}\label{eVIlii}
\th_{12} = \cosh^{-1} \Big[\frac{\om^2-k^2-2m^2}{2m^2}\Big]\,.
\end{equation}
As indicated in Section 6.5, $A_G$ is an arbitrary constant
normalizing all current form-factors while $G_c(x)$ can be found
in Eq.\,(\ref{eVIxl}).

\begin{figure}[ht]
\begin{center}
\epsfxsize=0.5\textwidth
\rotatebox{270}{\epsfbox{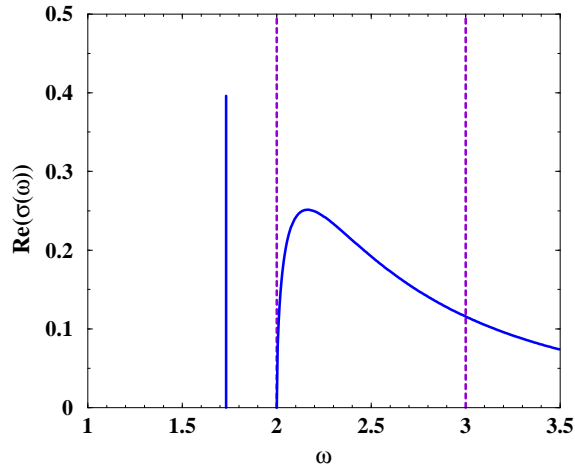}}
\end{center}
\vskip -5mm
\caption{Plot of the optical conductivity at wavevector $k=0$.}
\label{figVIv}
\end{figure}

In Fig.\,\ref{figVIv} we plot the regular real part of the optical 
conductivity for wavevector $k=0$.\footnote{\,We will not comment here whether
a finite Drude weight exists beyond saying that it is a possibility as the model
is integrable.}
We see there is an exciton type peak at $\om = \sqrt{3}m$ corresponding
to the single particle form factor contribution. 
The first vertical
dashed line marks out the beginning of the two particle
form factor contribution to the conductivity.
The onset of the two particle
contribution behaves as $\sqrt{\omega -2m}$ and not 
as $1/\sqrt{\omega -2m}$ as would
be expected in a free theory due to the divergence in the
density of states, the van-Hove singularity,
that occurs in one dimensional systems.  This singularity
is removed by the corresponding current matrix element which behaves
as $(\om-2m)$ with $\om \rightarrow 2m^+$ 
as the low energy behavior becomes strongly renormalized in
the presence of even weak interactions.

The optical conductivity was computed 
in Ref. \cite{lin} using the large $N$ limit
of $SO(2N)$, or in an alternate language, an RPA approximation.
In such an approximation, the model becomes equivalent to a theory
of four massive, non-interacting Dirac fermions.  Hence Ref. \cite{lin} finds
that the van-Hove singularity is present.

The second vertical dashed line in Fig.\,\ref{figVIv}
at $3m$ marks the point where the three particle form factors would begin
to make a contribution.  Up to this point, the result is exact.  
We note that the three particle contribution is strictly a consequence of
interactions.  In $SO(8)$ language, two kinks of opposite chirality
together with a fermion will couple to the current operator.
In a free theory these different particles 
would not all exist and there would be
no three particle contribution.

If we were to compute the three particle contribution, three possibilities
present themselves.  
The three particle
density of states approaches a constant as $\om \rightarrow 3m^+$.  
If the corresponding matrix element vanishes
as $\om \rightarrow 3m^+$, the contribution will open up gradually,
leaving $\sigma (\om )$ continuous at $\om = 3m$.
If the three
particle matrix element also approaches a constant value as
$\om \rightarrow 3m^+$, the conductivity
will be marked by a jump at $\om = 3m$.  But if the matrix element 
diverges in this limit, we expect to find a corresponding
divergence in the conductivity at $\om = 3m$.
Of these scenarios, it is our belief that the first is most likely.  Moreover we expect that the
total spectral weight in such a contribution in comparison to the two-particle
contribution will be extremely small.

\subsubsection{Single Particle Spectral Function}

In this section we
compute the single-particle 
spectral function of the electrons of the ladder/nanotube. 
To do so we first consider the correlator,
\begin{eqnarray}\label{eVIliii}
G (k_x,k_y,\tau) &=& 
\sum_{l=1,2} \int^\infty_{-\infty} dx ~e^{-ik_y l - ik_x x}
\lb T(a_{l\alpha}(x,\tau)a^\dagger_{l\alpha}(0,0))\rb\, .
\end{eqnarray}
Here $k_y$ takes on the values $0,\pi$.
We then define the particle/hole spectral functions, $A_{p/h}$, as follows:
\begin{eqnarray}\label{eVIliv}
A_p({\bf k},\om ) + A_h(-{\bf k},-\om ) &=& 
{\rm Im} \int^\infty_{-\infty} d\tau~ e^{-i\om \tau} G(k_x,k_y,\tau)
\big|_{\om\rightarrow -i\om + \delta}\, .~~~~
\end{eqnarray}
We note that we have not explicitly summed over spin, $\alpha$.

As described in Section 6.4.2,
electronic excitations around the Fermi point
correspond in the Gross--Neveu
language  to low energy excitations of kinks.  We thus expect to recast
the Green's function, $G$, above in terms of kink correlators.
This in fact can be done with the result,
\begin{eqnarray}\label{eVIlv}
G (Pk_{Fi} + k,k_{yi},\tau) &=& 
\int^\infty_{-\infty} dx e^{ikx}
\lb T(c_{Pi\alpha}(x,\tau)c^\dagger_{Pi\alpha}(0,0))\rb\, ,
\end{eqnarray}
where $i=1,2$ and $k_{yi} = (2-i)\pi$.  The $c$'s, the bonding-anti-bonding
electrons are in turn related to the various kinks
via Eq.\,(\ref{eVIxxxvi}).
The Greens function on the r.h.s. of Eq.\,(\ref{eVIlv}) is thus equal to 
\begin{equation}\label{eVIlvi}
\lb T(c_{Pi\alpha}(x,\tau)c^\dagger_{Pi\alpha}(0,0))\rb = 
\langle T(\kappa_\alpha\psi_\pm^\alpha (x,\tau)
\kappa_{\bar{\alpha}}\psi_\pm^{\bar{\alpha}} (0,0)) \rangle\,,
\end{equation}
where $\alpha$ ($\bar{\alpha}$ being its charge conjugate) is 
the particular kink corresponding
to the Fermi point $(k_{Fi},k_{yi})$.
The $\kappa_\alpha$ are Klein factors included to ensure the $\psi^\alpha$
are anti-commuting.
Because of the $SO(8)$ symmetry together with its associated triality
symmetry, 
$\langle T(\kappa_\alpha\psi_\pm^\alpha (x,\tau)
\kappa_{\bar{\alpha}}\psi_\pm^{\bar{\alpha}} (0,0)) \rangle$ 
turns out to be independent of the type $\alpha$  of kink. 
It is only
sensitive to whether the kink field is right ($+$) or left ($-$) moving.

To compute this correlator, we again expand to the two lowest contributions:
\begin{eqnarray}\label{eVIlvii}
&&\langle T(\kappa_\alpha\psi_\pm^\alpha (x,\tau > 0)
\kappa_{\bar{\alpha}}\psi_\pm^{\bar{\alpha}} (0,0)) \rangle
=\int^\infty_{-\infty} \frac{d\th_1}{2\pi}
\langle \psi_+^\alpha(x,\tau) A^\dagger_{\bar{\alpha}}(\th_1) \rangle
\lb A_{\bar{\alpha}}(\th_1)\psi_+^{\bar{\alpha}}(0)\rb \nonumber\\[1mm]
&&~+ \frac{1}{2}\sum_{a\beta} \int^\infty_{-\infty} \frac{d\th_1}{2\pi}
\frac{d\th_2}{2\pi} 
\langle \psi_+^\alpha(x,\tau) A^\dagger_\beta (\th_2)
A^\dagger_a (\th_1) \rb 
\lb A_a(\th_1) A_\beta (\th_2) \psi_+^{\bar{\alpha}}(0)\rb\, .
\end{eqnarray}
The first contribution, the one particle contribution, 
comes from the kink excitation,
$A_{\bar{\alpha}}$, destroyed
by the field, $\psi_\alpha$.  The second contribution, 
a two particle contribution,
arises from kinks, $A_\beta$, of opposite chirality to $A_\alpha$, and
Majorana fermions, $A_a$. 
(This reflects the group theoretical fact that
the tensor product of an $SO(8)$ spinor representation 
with an $SO(8)$ vector representation
gives the other $SO(8)$ spinor representation \cite{sla}.)  
The first contribution not included
is a bound state-kink pair.  
It begins to contribute at $\om = (1+\sqrt{3})m$.

From the form factor expressions from Section 6.5, we can then write
down the expression for the spectral functions, $A_{p/h}(\om ,k)$,
\begin{eqnarray}\label{eVIlviii}
A_p(\om ,Pk_{Fi}+k,k_{yi}) &=& A_h(\om ,-Pk_{Fi}+k,k_{yi})\nonumber\\[1mm]
&=& \frac{\pi |c_P|^2}{m}\, \frac{\om + Pk}{\sqrt{k^2+m^2}}\,
\delta (\om - \sqrt{k^2+m^2}) \nonumber\\[1mm]
&& \hskip -1in +\, \theta (\om - \sqrt{k^2+4m^2})\,
\frac{32 m^4 A_F^2}{\om - P k }\,\frac{1}{(\om^2-k^2-m^2)^2}\,
\frac{1}{\sqrt{\om^2-k^2-4m^2}}\nonumber\\[1mm]
&& \hskip -.4in \times\exp\Big[\int^\infty_0\frac{dx}{x}\,\frac{G_f(x)}{s(x)}\,
\Big(1-c(x)\cos \Big(\frac{x\th_{12}}{\pi}\Big)\Big)\Big]\,,
\end{eqnarray} 
\noindent where $A_F$ 
is the (unspecified) normalization of the two particle
kink form-factor, $G_f(x)$ 
is given in Eq.(\ref{eVIxliii}), and $c_\pm$ is found in Eq.\.(\ref{eVIxlv}).
For $P=R=+$ (i.e. right-moving electrons/kinks), this function
is plotted in Fig.\,\ref{figVIvi}.

\begin{figure}[ht]
\begin{center}
\epsfxsize=0.7\textwidth
\epsfysize=0.8\textwidth
\epsfbox{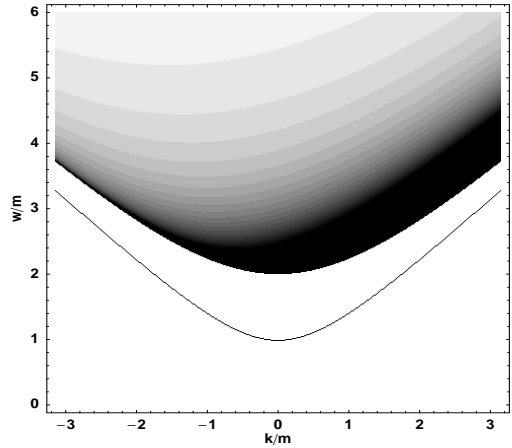}
\end{center}
\vskip -1.4in 
\caption{Plot of the single particle 
spectral function for right moving kinks.  The more darkly shaded region
corresponds to greater spectral weight.}
\label{figVIvi}
\end{figure}

The parabolic line in Fig.\,\ref{figVIvi} arises from the single particle form
factor contribution, and represents the standard dispersion relation
of a particle of mass, $m$.  Above this curve comes the two particle
form factor contribution to the spectral function.  This contribution
is bounded by the curve, $\om = \sqrt{k^2 + 4m^2}$, and so the single
particle states do not cross into the two particle region.  As
can be seen from Eq.\,(\ref{eVIlviii}), the two particle contribution opens up
at threshold with a square-root singularity, indicative of the 
van-Hove singularity in the density of states.
 
The plot is manifestly chiral with weighting greater for $k>0$ than
for $k<0$.  This is to be expected as we are plotting the excitations
linearized about the Fermi momentum, $+k_{Fi}$.  The heavier weighting
for $k>0$ indicates that is easier to create excitations above the Fermi
sea than below it.
It is interesting indeed that excitations below
the Fermi surface can be created at all and is a mark that interactions
are at play.

\subsubsection{STM Tunneling Current}

In this section we study the tunneling between a metallic 
lead and the carbon nanotube/Hubbard ladder through a point 
contact.  Our starting point is a Lagrangian
describing the nanotube/ladder, the metallic lead, and the
tunneling interaction:
\begin{equation}\label{eVIlix}
\CL = \CL_{SO(8)} + \CL_{lead} + \CL_{tun}\, .
\end{equation}
$\CL_{SO(8)}$ is the Lagrangian of the $SO(8)$ Gross--Neveu model.

The electron gas in the lead is, in general, three dimensional.
However, in the context of tunneling through a
point contact, the electron gas can be mapped onto an one dimensional
chiral fermion (see for example
\cite{chfr,lud,ludd}).
The general idea is well illustrated by its application to the
Kondo problem.  There an electron scatters off a spin impurity at $x=0$.
The scattering is determined by the electron operator, $\psi (x=0)$.
As $\psi (0)$ only depends on its spherically
symmetric, $L=0$, mode, one can consider the scattering electron in terms of
an ingoing and outgoing radial model defined on the half-line,
$r\in [0,\infty ]$.  Unfolding the system onto the full line leaves one with
a chiral fermion.  We emphasize however that the map requires no
special symmetry; the result is exact regardless of particular anisotropies
\cite{ludd}.
As a consequence, we write $\CL_{lead}$ as
\begin{equation}\label{eVIlx}
\CL_{lead} = \frac{1}{8\pi}\, \Psi^\dagger \partial_{\bar{z}} \Psi\, ,
\end{equation}
where $\Psi$ is a massless, left moving fermion, and $z=(\tau+ix)/2$.

\def\pla{\psi_-^\alpha (\tau )}
\def\plab{\psi_-^{\bar{\alpha}}(\tau )}
\def\pras{\psi_+^\alpha (\tau )}
\def\prab{\psi_+^{\bar{\alpha}}(\tau )}
\def\ps{\Psi (\tau )}
\def\psd{\Psi^\dagger (\tau )}
\def\plant{\psi_-^\alpha}
\def\plabnt{\psi_-^{\bar{\alpha}}}
\def\prant{\psi_+^\alpha}
\def\prabnt{\psi_+^{\bar{\alpha}}}
\def\psnt{\Psi}
\def\psdnt{\Psi^\dagger}

It remains to specify $\CL_{tun}$.  In order to preserve charge, the
electrons must couple to the kinks of the $SO(8)$ 
Gross--Neveu model, the excitations
with the quantum numbers of the electron.  Thus
\begin{eqnarray}\label{eVIlxi}
\CL_{tun} &=& ~ g_L
\bigl[
 \psd \pla + \plab\ps 
\bigr ]
 \delta (x) \nonumber\\[1mm]
&& + g_R 
\bigl [
\psd\pras + \prab\ps 
\bigr ]
 \delta (x)\, .
\end{eqnarray}
Here we have coupled the lead electrons to both the right and left moving
fields creating the kink, $\alpha$, and have allowed the two couplings,
$g_L$ and $g_R$, to be unequal.  However as we will work  
to lowest non-vanishing order in the tunneling matrix elements $g_{L/R}$,
the tunneling current will depend upon the sum,
$g_L^2 + g_R^2$, that is, the contribution of the left and right channels
to tunneling will add linearly.  Similarly, permitting other kinks to
couple to the lead electrons will give lowest order
contributions which simply add.

\begin{figure}[ht]
\begin{center}
\epsfxsize=0.5\textwidth
\rotatebox{270}{\epsfbox{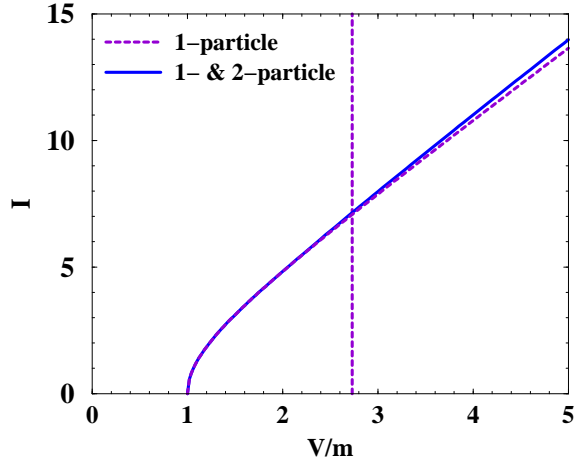}}
\end{center}
\caption{Plot of the tunneling current
as a function of applied voltage.  The dashed curve describes
tunneling into a non-interacting fermionic system of mass, m.  The vertical
dashed line marks where a second set of two particle states begins 
to make a contribution.}
\label{figVIvii}
\end{figure}

The calculation of the current to lowest non-vanishing order in $g_{L/R}$
follows the standard route.
The tunneling current operator is given by
\begin{eqnarray}\label{eVIlxii}
I(\tau ) &=& i g_R(\prab\ps - \psd\pras) \nonumber\\[1mm]
&& + i g_L(\plab\ps - \psd\pla)\, .
\end{eqnarray}
In order to induce current flow, $\lb I(\tau) \rb$, through the point
contact, one biases the lead with a voltage, V.
This bias can be taken into account via
a gauge transformation,
\begin{equation}\label{eVIlxiii}
\ps \rightarrow e^{iV\tau}\ps \,.
\end{equation}
In effect we have shifted the energy levels of the electrons.
Treating the couplings, $g_L/g_R$, with linear response theory, we find
\begin{eqnarray}\label{eVIlxiv}
\lb I(\om )\rb &=& g_L^2 {\rm Re} \bigg\{ \int d\tau\, i e^{i\om \tau}
\big[
e^{-iV\tau}\lb \pla \plabnt (0)\rb\lb\psd\psnt (0)\rb \nonumber\\[1mm]
&& -\,e^{iV\tau}\lb \plab \plant (0)\rb\lb\ps\psdnt (0)\rb
\big]|_{\genfrac{}{}{0pt}{}{\om \rightarrow -i\om + \delta}
{V \rightarrow - iV}} \bigg\}\nonumber\\[1mm]
&& +\, (L\rightarrow R, \plant /\plabnt 
\rightarrow \prant /\prabnt)\,.
\end{eqnarray}
The lead electron correlator $\lb \psd \psnt (0)\rb$ is well known:
\begin{equation}\label{eVIlxv}
\lb \psdnt (\tau ,x) \psnt (0)\rb = \lb \psnt (\tau ,x) \psdnt (0)\rb = 
\frac{1}{\tau +ix}\ .
\end{equation}
With this it is straightforward to express the dc current, 
$\lb I(\om = 0)\rb$,
in terms of the single particle kink spectral function,
\begin{eqnarray}\label{eVIlxvi}
\lb I(\om = 0)\rb &=& \frac{1}{2\pi} \int^V_{-V} d \omega
\int^\infty_{-\infty} dk
\bigl [ g_L^2 A_-(\om ,k)
+ g_R^2 A_+(\om ,k)
\bigr ],
\end{eqnarray}
where 
$A_\pm(\om ,k) = 
A_p(\om,\pm k_{Fi}+k,k_{yi}) + A_h(-\om,\mp k_{Fi}-k,k_{yi})$, 
and $A_{p/h}$ are
the spectral functions given in Eq.\,(\ref{eVIxlviii}).
We note that
as a technical point, in deriving the above equation we have displaced,
$\Psi$, the lead electron operator, slightly from $x=0$.  In this way
we cure the UV divergence attendant as $\tau\rightarrow 0$.  At the end
of the calculation we then take $x$ to 0.

In the previous section we have computed $A_\pm(\om ,k)$ exactly for 
energies $\om < (\sqrt{3} + 1)m$.
Inserting Eq.\,(\ref{eVIlviii}) into Eq.\,(\ref{eVIlxvi}), we find $\lb I (0)\rb$ takes
the form
\begin{eqnarray}\label{eVIlxvii}
\lb I(0) \rb &=& \frac{ |c_\pm |^2}{m} (g_R^2 + g_L^2) (V^2 - m^2)^{1/2}
\theta (|V| -m) {\rm sgn} (V) \nonumber\\[1mm]
&& 
+ \theta (|V| - 2m){\rm sgn} (V) \times 
{\rm two~particle~contribution}\,.
\end{eqnarray}
We see that for $|V| < 2m$, the system behaves as a gapped free fermion.
The first sign that there is any interaction comes for $|V| > 2m$ where
the voltage begins to probe the two particle states, a signature of
interacting fermions.

We explicitly plot $\lb I(0)\rb$ in
Fig.\,\ref{figVIvii}.  The square root behavior near $V/m = 1$ and subsequent linear
form is typical of a gapped fermion.  At $V/m = 2$, the two particle
states begin to contribute leading to a small
change in the slope of the $I-V$
curve.  At $V/m = \sqrt{3} + 1$ (marked by the the vertical dashed line),
a second set of two 
particle states (a mass $\sqrt{3}m$ bound state together with a kink)
begin to contribute and at this point the result
ceases to be exact.  However as with the current correlators, we expect
this higher energy contribution to be small.

\def\djv{\partial_V \lb I(0) \rb}

The change in slope in the $I-V$ curve at $V/m = 2$ can be 
explicitly computed.
To do so we consider $\djv$.  This quantity is given by
\begin{equation}\label{eVIlxviii}
\djv = \frac{1}{\pi}\, (g_R^2 + g_L^2)\int^\infty_{-\infty} dk ~A_\pm(V,k)\, .
\end{equation}
We can thus see $\djv$ directly measures the local density of states at
$x=0$ of the nanotube/ladder system.

We plot $\djv$ in Fig.\,\ref{figVIviii}.  The square root singularity at $V=m$ signals
the singularity of the density of states in an one dimensional system.
At $V=2m$ we see a sudden jump, indicative of the onset of the two
particle contribution.  The height of the jump can be determined
exactly:
\begin{eqnarray}\label{eVIlxix}
&& \djv (V/m=2^+) - \djv (V/m=2^-)\nonumber\\[1mm]
&& ~~~ =\frac{16 A_F^2}{9 m}\, (g_R^2 + g_L^2)
\exp \Big[ \int^\infty_0 \frac{dx}{x}\,\frac{G_f(x)}{s(x)}\, (1-c(x))\Big]\,.
\end{eqnarray}
The region $m<V<2m$ of $\djv$ completely determines
$m$ (by the location of the jump),
as well as  an overall scale (the product of
$(g_L^2+g_R^2)$ and the constant $A_F$, normalizing the spectral
function). 
Dividing out these non-universal quantities leaves a universal
number, characterizing the magnitude of the jump:
\begin{equation}\label{eVIlxx}
\frac{16}{9}\,\exp\Big[\int^\infty_0 \frac{dx}{x}\, \frac{G_f(x)}{s(x)}\, (1-c(x))\Big]\,.
\end{equation}
This number represents a definite prediction based upon
the integrability of the model.

\begin{figure}[ht]
\vskip -.2in 
\begin{center}
\epsfxsize=0.5\textwidth
\rotatebox{270}{\epsfbox{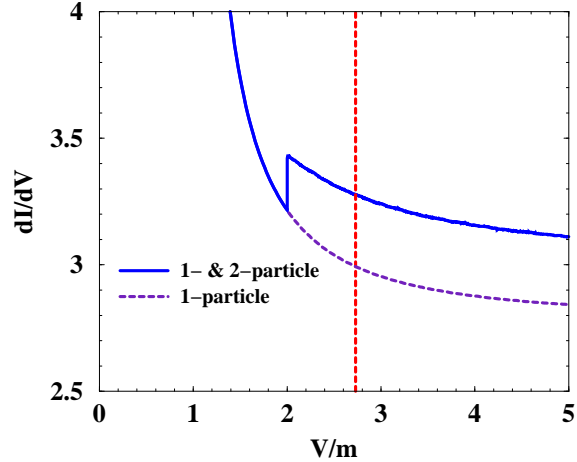}}
\end{center}
\caption{Plot of the differential conductance
as a function of applied voltage.  The dashed curve marks
the single particle contribution to this quantity while the solid curves give both
the single and two particle contribution.  The latter plot is
exact up to $V/m = \sqrt{3}+1$ (indicated by the dashed vertical line) where
a bound state-kink pair begins to make a contribution.}
\label{figVIviii}
\end{figure}

\subsection{Effect of Integrable Breaking Perturbations}

It can now be asked how perturbations to $SO(8)$ Gross--Neveu
will affect the various computations
discussed in the previous section.  We consider this in the broadest terms by
focusing upon how the spectrum of $SO(8)$ Gross--Neveu is changed under a
perturbing term.  We 
do so through straightforward stationary perturbation theory, in
the same spirit that Ref. \cite{simo} treated the off-critical Ising model in
a magnetic field.  The most
general possible perturbation takes the form
\begin{equation}\label{eVIlxxi}
H_{\rm pert} = \lambda G_{ab}G_{cd}\, ,
\end{equation}
where $G_{ab}$, $G_{cd}$ are $SO(8)$ currents (of unspecified chirality).
For such a perturbation it is necessary to consider degenerate perturbation
theory.  Thus in a given particle multiplet (for example, the fundamental
fermions in the vector representation), the perturbed energies arise
through diagonalizing the matrix
\begin{equation}\label{eVIlxxii}
M_{ij} = \langle A_i (\theta ) H_{\rm pert} A_j^\dagger (\theta )\rangle ,
\end{equation}
where here the index $i,j$ indicates the particles $A_i$,$A_j$ belong to
the multiplet of concern.
In the case that $G_{ab} = G_{cd}$, $M_{ij}$ is 
necessarily diagonal, i.e. nondegenerate perturbation theory is sufficient.

It is important to emphasize that this procedure can be handled in the
context of integrability.  The expression in Eq.\,(\ref{eVIlxxii}) is no more than
a form-factor which can readily be computed.  
Moreover as the theory is massive,
perturbation theory is well controlled.
We expect the unperturbed theory to describe all qualitative features
of the model while the perturbations to only introduce small quantitative
changes.

We first consider the consequences of such a perturbation on 
the optical conductivity.  They are 
two-fold.  We expect the exciton peak (found, for 
example, in Fig.\,\ref{figVIv} at $k=0$ and $\omega = \sqrt{3}$) 
to split.  In the unperturbed model
the peak results from a single rank-two bosonic bound state coupling to the
current operator.  When the matrix
$M_{ab}$ above is diagonalized, this particular state should be 
mixed into many
others resulting in several states that couple to the current operator.
However we do not expect the functional forms of the exciton peaks 
to change: they should
remain delta functions.  They must do so provided the perturbation is
not so large as to push the exciton peak 
past the threshold of two particle
states where it then conceivably could decay.  As there is a gap between
the excitonic peak and the two particle threshold, this will not
happen for small perturbations.  Experimentally changes to the exciton
peak may
not be detectable.  Given that any experiment will be conducted at finite
temperature, the excitonic peak will be thermally broadened, perhaps
washing out any splitting of the original zero temperature peak.

We also expect the perturbation to affect the onset of
the two particle threshold, although in a less dramatic fashion.  
Like the unperturbed case, there will be several two particle contributions 
to the optical conductivity.  However unlike the unperturbed case,
the thresholds of the two particle contributions will not all occur
at $\omega = 2m$ but be distributed about this energy.  Thus the
two particle contribution is arrived at (approximately) by superimposing
several slightly shifted two-particle contributions similar to that
found in Fig.\,\ref{figVIv}.  But given
the optical conductivity vanishes at threshold, the qualitative picture
remains effectively unchanged (i.e. the superimposed contributions will
appear nearly identical to the original picture).  
That the optical conductivity vanishes
at the two particle threshold is a result of the vanishing of the
relevant matrix element at threshold.  This should be robust under
perturbation as it is ultimately a consequence of the mere presence of
interactions and not some particular type of interactions.

We can also consider in a similar spirit the approximate effect of perturbations
breaking integrability on the tunneling conductance.  As 
the tunneling conductance is determined directly from 
the single particle spectral function, we can deduce how the 
former is affected
from how the latter is changed.  At a given momentum, the single particle 
contribution to the single particle spectral function under perturbation 
comes at a discrete set of energies.  In terms of the tunneling conductance,
we expect a series of closely spaced square root divergences (a sawtooth
behavior) about $V=m$ indicative of a series of van-Hove
singularities.  As the perturbation is removed these singularities would 
collapse on top of one another leaving the original picture in Fig.\,\ref{figVIviii}.

In the unperturbed case the two-particle threshold is characterized
by a jump in the differential conductance.  Under a perturbation,
this jump would become a staircase or a series of smaller,
closely spaced jumps.  This is a reflection of the series of
van-Hove singularities found about $\omega = \sqrt{k^2 + 4m^2}$
in the two particle contribution to the single particle spectral function.

\section*{Acknowledgments}
We thank I. Affleck, L. Balents, C. Broholm, R. Coldea, D. Controzzi,
N. d'Ambrumenil, G. Delfino, P. Fendley, M.P.A. Fisher, F. Gebhard,
E. Jeckelmann, M. Kenzelmann, A. LeClair, H. H. Lin, A.W.W. Ludwig,
S. Lukyanov, G. Mussardo, A.A. Nersesyan, E. Orignac, H. Saleur,
F. Smirnov, A.M. Tsvelik, I. Zaliznyak, and A. Zheludev for
enlightening discussions over the course of the work appearing in this
review.


\newpage
\begin{thebibliography}{300}
\addcontentsline{toc}{section}{References}

\bibitem{Abada97}
A. Abada, A.H. Bougourzi and B. Si-Lakhal, Nucl. Phys. B {\bf 497}, 733
(1997). 

\bibitem{AGD}
A.A. Abrikosov, L.P. Gorkov and I.E. Dzyaloshinski,
{\em Methods of Quantum Field Theory in Statistical Physics},
(Dover, New York, 1975),  page 168.

\bibitem{Affleck86b}
I. Affleck, Nucl. Phys. B {\bf 265}, 409 (1986).

\bibitem{aklt}
I. Affleck, T. Kennedy, E. Lieb and H. Tasaki,
Comm. Math. Phys. {\bf 115}, 477 (1988).

\bibitem{Affleck89b}
I. Affleck, {\rm in} {\em Fields, Strings and Critical Phenomena},
            {\rm eds E. Br\'ezin and\\ J. Zinn-Justin},
      (Elsevier, Amsterdam, 1989).

\bibitem{Affleck89}
I. Affleck, D. Gepner, H. J. Schulz and T. Ziman,
J. Phys. A {\bf 22}, 511 (1989).

\bibitem{affleck} 
I. Affleck, Phys. Rev. B {\bf 41}, 6697 (1990).

\bibitem{lud} 
I. Affleck and A.W.W. Ludwig, Nucl. Phys. B {\bf 330}, 641 (1991).

\bibitem{affleck90} 
I. Affleck, Phys. Rev. B {\bf 41}, 6697 (1990).

\bibitem{affleck91}
I. Affleck, Phys. Rev. B {\bf 43}, 3215 (1991).

\bibitem{affwes} 
I. Affleck and R. Weston, Phys. Rev. B {\bf 45}, 4667 (1992).

\bibitem{ludd} 
I. Affleck, A.W.W. Ludwig and
B. Jones, Phys. Rev. B {\bf 52}, 9528 (1995).

\bibitem{AffleckHalperin}
I. Affleck and B.I. Halperin, J. Phys. A {\bf 29}, 2627 (1996).

\bibitem{Affleck98}
 I. Affleck, J. Phys A {\bf 31}, 4573 (1998).

\bibitem{oa2}
I. Affleck and M. Oshikawa, Phys. Rev. B {\bf 60}, 1038 (1999).

\bibitem{Ajiro00}
Y. Ajiro, T. Asano, Y. Inagaki, J.P. Boucher, H. Nojiri, S. Luther,
T. Sakon and M.~Motokawa, J. Phys. Soc. Jpn {\bf 69}, 297 (2000).

\bibitem{Alcaraz89}
F. Alcaraz and W.F. Wreszinski, J. Stat. Phys. {\bf 58}, 45 (1989).

\bibitem{arrigoni00}
E. Arrigoni, Phys. Rev. B {\bf 61}, 7909 (2000).

\bibitem{Asano00}
T. Asano, H. Nojiri, Y. Inagaki, J.P. Boucher, T. Sakon, Y. Ajiro and
M. Motokawa, Phys. Rev. Lett {\bf 84}, 5880 (2000).

\bibitem{Asano02}
T. Asano, H. Nojiri, W. Higemoto, A. Koda, R. Kadono and Y. Ajiro,
J. Phys. Soc. Jpn {\bf 69}, 594 (2002). 

\bibitem{tsv} 
P. Azaria, P. Lecheminant and A.M. Tsvelik, cond-mat/9806099 (unpublished).

\bibitem{azaria}
R. Assaraf, P. Azaria, E. Boulat, M. Caffarel, and P. Lecheminant,
Phys. Rev. Lett. {\bf 93}, 016407 (2004).

\bibitem{babu} H. M. Babujian, Phys. Lett. A {\bf 90}, 479 (1982);
ibid. Nucl. Phys. B {\bf 215}, 317 (1983). 

\bibitem{babujian99}
H. Babujian, A. Fring, M. Karowski and A. Zapletal, Nucl. Phys. B {\bf
538}, 535 (1999).

\bibitem{babujian02a}
H. M. Babujian and M. Karowski, Nucl. Phys. B {\bf 620}, 407 (2002).

\bibitem{babujian02b}
H. M. Babujian and M. Karowski, J. Phys. A {\bf 35}, 9081 (2002).

\bibitem{bala}
L. Balents and M.P.A. Fisher, Phys. Rev. B {\bf 53}, 12133 (1996); 

\bibitem{bal} 
L. Balents and M.P.A. Fisher, Phys. Rev. B {\bf 55}, R11973 (1997).

\bibitem{Balents01}
L. Balents and R. Egger, Phys. Rev. B {\bf 64}, 035310 (2001).

\bibitem{balog1} 
J. Balog, Nucl. Phys. B {\bf 419}, 480 (1994).

\bibitem{balog} 
J. Balog and M. Niedermaier, Nucl. Phys. B {\bf 500}, 421 (1997).

\bibitem{bannister00}
R.N. Bannister and N. d'Ambrumenil, Phys. Rev. B {\bf 61}, 4651 (2000).

\bibitem{Barnes94}
T. Barnes and J. Riera, Phys. Rev. B {\bf 50}, 6817 (1994).

\bibitem{BarzykinAffleck} 
V. Barzykin and I. Affleck, J. Phys. A {\bf 32}, 867 (1999).

\bibitem{Barzykin00a} 
V. Barzykin, J. Phys. Cond. Mat. {\bf 12}, 2053 (2000).

\bibitem{leclair} 
D. Bernard and A. LeClair, Commun. Math. Phys. {\bf 142}, 99 (1991).

\bibitem{double} 
D. Bernard and A. LeClair, Nucl. Phys. B {\bf 399}, 709 (1993).

\bibitem{Bethe31}
H. Bethe, Z. Phys. {\bf 71}, 205 (1931).

\bibitem{Bergknoff}
H. Bergknoff and H. Thacker, Phys. Rev. D {\bf 19}, {3666} (1979).

\bibitem{berg}
B. Berg, M. Karowski and P. Weisz, Phys. Rev. D {\bf 19}, 2477 (1979).

\bibitem{BEG03}
M.J. Bhaseen, F.H.L. Essler and A. Grage, 
{\em Itineracy Effects on Spin Correlations in 1D Mott Insulators}, 
cond-mat/0312055.

\bibitem{bockrath}
M. Bockrath, D. Cobden, P. McEuen, N. Chopra, A. Zettl, A. Thess and
R. Smalley, Science {\bf 275}, 1922 (1997);

\bibitem{bockratha}
M. Bockrath, D. H. Cobden, J. Lu, A. G.
Rinzler, R. Smalley, L. Balents and P.~McEuen, Nature {\bf 417}, 725 (2002).

\bibitem{Bocquet02}
M. Bocquet, Phys. Rev. B {\bf 65}, 184415 (2002).

\bibitem{boies}
D. Boies, C. Bourbonnais and A.-M. S. Tremblay, Phys. Rev. Lett. {\bf
74}, 968 (1995).

\bibitem{Bougourzi96}
A.H. Bougourzi, M. Couture and M. Kacir, Phys. Rev. B {\bf 54}, 12669
(1996). 

\bibitem{Bougourzi97}
A. H. Bougourzi, M. Karbach and G. M\"uller, Phys. Rev. B {\bf 57},
11429-11438 (1998). 

\bibitem{review} 
C. Bourbonnais and  D. Jerome, in {\em Advances in Synthetic Metals, 
Twenty years of Progress in Science and Technology}, 
ed. by P. Bernier, S. Lefrant and G. Bidan (Elsevier, New York, 1999), 
pp. 206-301 and references therein. See also cond-mat/903101.  

\bibitem{buyers} W.J.L. Buyers, R.M. Morra, R.L. Armstrong, M.J. Hogan,
P. Gerlach and K. Hirakawa, Phys. Rev. Lett. {\bf 56}, 371 (1986).

\bibitem{campos}
L. Campos Venuti, E. Ercolessi, G. Morandi, P. Pieri and M. Roncaglia,
{\em Spin Chains in an External Magnetic Field. Closure of the Haldane Gap  and Effective Field Theories}, cond-mat/9908044.

\bibitem{capraro}
 F. Capraro and C. Gros, Eur. Phys. J. {\bf B29}, 35 (2002).

\bibitem{castro}
O. A. Castro-Alvaredo and A. Fring, Nucl. Phys. B {\bf 636}, 611 (2002).

\bibitem{cardmuss}
J. Cardy and G. Mussardo, Nucl. Phys. B {\bf 340}, 387 (1990).

\bibitem{oldff1}
J. Cardy and G. Mussardo, Nucl. Phys. B {\bf 410}, 451 (1993).

\bibitem{CEL}
J.-S. Caux, F.H.L. Essler and U. L\"ow, Phys. Rev. B {\bf 68}, 134431 (2003).

\bibitem{castella} H. Castella, X. Zotos and P. Prelovsek, Phys. Rev. Lett.
{\bf 74}, 972 (1995).

\bibitem{chfr} C. Chamon and E. Fradkin, Phys. Rev. B {\bf 56}, 2012 (1997).

\bibitem{cheianov04a}
V.V. Cheianov and M. B. Zvonarev,
Phys. Rev. Lett. {\bf 92}, 176401 (2004).

\bibitem{cheianov04b}
V.V. Cheianov and M. B. Zvonarev,
J. Phys. {\bf A37}, 2261 (2004).

\bibitem{chen} Y. Chen, Z. Honda, A. Zheludev, C. Broholm, K. Katsumata and S. M. Shapiro,
Phys. Rev. Lett. {\bf 86}, 1618 (2001).

\bibitem{cobden} D. H. Cobden, J. Nygard, M. Bockrath and P. McEuen,
{\em One-dimensional transport in bundles of single-walled carbon nanotubes},
Proceedings of IWEPNM 99 (Kirchberg), cond-mat/9904179.

\bibitem{cet} 
D. Controzzi, F.H.L. Essler and A.M. Tsvelik, Phys. Rev. Lett. {\bf
86}, 680 (2001).

\bibitem{cet2} 
D. Controzzi, F.H.L. Essler and A.M. Tsvelik, in {\em New Theoretical
Approaches to Strongly Correlated Systems}, ed. A.M. Tsvelik, NATO
Science Series II Vol. 23\\{} [cond-mat/0011439].  

\bibitem{CE02}
D. Controzzi and F.H.L. Essler, Phys. Rev. B {\bf 66}, 165112 (2002).

\bibitem{Dashen75a}
R. Dashen and Y. Frishman, Phys. Rev. D {\bf 11}, 2781 (1975).

\bibitem{Dashen75b}
R.F. Dashen, B. Hasslacher and A. Neveu, Phys. Rev. D {\bf 11}, 3424
(1975).

\bibitem{Dagotto} E. Dagotto and T. Rice, Science {\bf 271}, 618 (1996).

\bibitem{Dagottoa} E. Dagotto, J. Riera and D. Scalapino, Phys. Rev. B {\bf 45}, 5744 (1992).

\bibitem{damle2}  K. Damle and S. Sachdev, Phys. Rev. B {\bf 57}, 8307 (1998).

\bibitem{deisz}
J. Deisz, M. Jarrell and D. Cox, Phys. Rev. B {\bf 48}, 10227 (1993).


\bibitem{simo} G. Delfino, G. Mussardo and P. Simonetti,
Nucl. Phys. B {\bf 473}, 469 (1996).

\bibitem{delone} G. Delfino and G. Mussardo, Nucl. Phys B {\bf 455} 724 (1995).

\bibitem{deltwo} G. Delfino and J. Cardy, Nucl.Phys. B {\bf 519}, 551 (1998).

\bibitem{magn}
D.C. Dender, D. Davidovi\'c, D.H. Reich, C. Broholm, K. Lefmann and
G. Aeppli, Phys. Rev. B {\bf 53}, 2583 (1996).

\bibitem{dender}
D.C. Dender, P.R. Hammar, D.H. Reich, C. Broholm and G. Aeppli,
Phys. Rev. Lett. {\bf 79}, 1750 (1997).


\bibitem{ddv92}
C. Destri and H.J. de Vega, Phys. Rev. Lett. {\bf 69}, {2313}
(1992).

\bibitem{ddv95}
C. Destri and H.J. de Vega, Nucl. Phys. B {\bf 438}, {413} (1995). 

\bibitem{Dmitriev02a}
D.V. Dmitriev, V.Y. Krivnov and A.A. Ovchinnikov,
Phys. Rev. B {\bf 65}, 172409 (2002).

\bibitem{Dmitriev02b}
D.V.Dmitriev, V.Ya.Krivnov, A.A.Ovchinnikov, A.Langari,
JETP {\bf 95}, 538 (2002).

\bibitem{Doyon}
B. Doyon and S. Lukyanov, Nucl. Phys. B {\bf 644}, 451 (2002).

\bibitem{dresselhaus}
M. S. Dresselhaus, Nature {\bf 391}, 19 (1998).

\bibitem{duffy}
D. Duffy, S. Haas and E. Kim, Phys. Rev. B {\bf 58}, R5932 (1998).

\bibitem{DN78}
G. I. Dzhaparidze and A. A. Nersesyan, JETP Lett. {\bf 27}, 224
(1978).

\bibitem{Dzyaloshinskii58}
I.E. Dzyaloshinski, J. Phys. Chem. Solids {\bf 4}, 241 (1958).

\bibitem{Dz}
I.E. Dzyaloshinski, Phys. Rev. B {\bf 68}, 085113 (2003).

\bibitem{Ebbesen} 
T. Ebbesen, Phys. Today {\bf 49}, 26 (1996).


\bibitem{kiv} 
V. Emery, S. Kivelson and O. Zachar, Phys. Rev. B {\bf 59}, 15641 (1999).

\bibitem{Egger97}
R. Egger and A.O. Gogolin, Phys. Rev. Lett. {\bf 79}, 5082 (1998).

\bibitem{Egger98}
R. Egger and A.O. Gogolin, Eur. Phys. J. B {\bf 3}, 281 (1998).

\bibitem{efik95a}
F.H.L. Essler, H. Frahm, A.R. Its and V.E. Korepin,
Comm. Math. Phys. {\bf 174}, 191 (1995). 

\bibitem{efik95b}
F.H.L. Essler, H. Frahm, A.R. Its and V.E. Korepin,
Nucl. Phys. {\bf B446}, 448 (1995).

\bibitem{efik96}
F.H.L. Essler, H. Frahm, A.R. Its and V.E. Korepin,
J. Phys. {\bf A29}, 5619 (1996). 

\bibitem{ETD}
F.H.L. Essler, A.M. Tsvelik and G. Delfino, Phys. Rev. B {\bf 56},
11001 (1997). 

\bibitem{ET97}
F.H.L. Essler and A.M. Tsvelik, Phys. Rev. B {\bf 57}, 10592 (1998).

\bibitem{Essler99}
F.H.L. Essler, Phys. Rev. B {\bf 59}, 14376 (1999).

\bibitem{essler3mag}
F.H.L. Essler, Phys. Rev. B {\bf 62}, 3264 (2000).

\bibitem{EGJ}
F.H.L. Essler, F. Gebhard and E. Jeckelmann, Phys. Rev. B {\bf 64},
5119 (2001). 

\bibitem{ET02b}
F.H.L. Essler and A.M. Tsvelik, Phys. Rev. B {\bf 65}, 115117 (2002).

\bibitem{ET02a}
F.H.L. Essler and A.M. Tsvelik, Phys. Rev. Lett. {\bf 88}, 096403 (2002).

\bibitem{ET03}
F.H.L. Essler and A.M. Tsvelik, Phys. Rev. Lett. {\bf 90}, 126401 (2003).

\bibitem{EFH03}
F.H.L. Essler, A. Furusaki and T. Hikihara, Phys. Rev. B {\bf 68},
064410 (2003).

\bibitem{fab}
M. Fabrizio, Phys. Rev. B {\bf 48}, 15838 (1993).

\bibitem{FaddeevKorepin78}
L.D. Faddeev and V.E. Korepin, Phys. Rept. C {\bf 42}, 1 (1978).

\bibitem{Faddeev81}
L.D. Faddeev and L. Takhtajan, Phys. Lett. A {\bf 85}, 375 (1981).

\bibitem{Faddeev84}
L.D. Faddeev and L. Takhtajan, Jour. Sov. Math. {\bf 24}, 241
(1984).

\bibitem{fath}
G. F\'ath and P. Littlewood, Phys. Rev. B {\bf 58}, R14709 (1998).

\bibitem{FLS} P. Fendley, A.W.W. Ludwig and
H. Saleur, Phys. Rev. Lett. {\bf 74}, 3005 (1995).

\bibitem{FLS1}
P. Fendley, A.W.W. Ludwig and H. Saleur, {\bf 75}, 2196 (1995).

\bibitem{FLS2}
P. Fendley, A.W.W. Ludwig and H. Saleur, Phys. Rev. B {\bf 52}, 8934 (1995).

\bibitem{feyer}
R. Feyerherm \textit{et al}., J. Phys.: Condens. Matter {\bf 12},
 8495 (2000).

\bibitem{Fowler81}
M. Fowler and X. Zotos, Phys. Rev. B {\bf 24}, 2634 (1981),

\bibitem{Fowler82}
M. Fowler and X. Zotos, Phys. Rev. B {\bf 25}, 5806 (1982).

\bibitem{fred} F. Lesage and H. Saleur, Nucl. Phys. B {\bf 490}, 543 (1997).

\bibitem{fred1} F. Lesage and H. Saleur, Nucl. Phys. B {\bf 493}, 613 (1997).

\bibitem{fring93}
A. Fring, G. Mussardo and P. Simonetti, Nucl.Phys. B {\bf 393}, 413
(1993).

\bibitem{fuji1} 
S. Fujimoto and N. Kawakami, J. Phys. A {\bf 31}, 465 (1998).

\bibitem{fuji} 
S. Fujimoto, J. Phys. Soc. Jpn.  {\bf 68}, 2810 (1999).

\bibitem{fujisawa99}
H. Fujisawa et al, Phys. Rev. B {\bf 59}, 7358 (1999).

\bibitem{furusaki}
A. Furusaki and S.C. Zhang, Phys. Rev. B 
{\bf 60}, 1175 (1999).

\bibitem{rosch} 
M. Garst and A. Rosch, Europhys. Lett. {\bf 55}, 66 (2001) [cond-mat/0102109].

\bibitem{Gebhardbook}
F. Gebhard, {\it The Mott Metal-Insulator Transition}, 
(Springer, Berlin, 1997). 

\bibitem{4umklapp}
T. Giamarchi, Physica B {\bf 230-232}, 975 (1997).

\bibitem{gia} T. Giamarchi and A.M. Tsvelik, Phys. Rev. B {\bf 59}, 11398 (1999).

\bibitem{goehmann98a}
F. G\"ohmann, A.G. Izergin, V.E. Korepin and A.G. Pronko,
Int. J. Mod. Phys. {\bf B12}, 2409 (1998).

\bibitem{goehmann98b}
F. G\"ohmann, A.R. Its and V.E. Korepin, 
Phys. Lett. {\bf A249}, 117 (1998).

\bibitem{goehmann99}
F. G\"ohmann and V.E. Korepin, 
Phys. Lett. {\bf A260}, 516 (1999).

\bibitem{goldschmidt}
Y.Y. Goldschmidt and E. Witten, Phys. Lett. B {\bf 91}, 392 (1980).

\bibitem{GNT}
 A.O. Gogolin, A.A. Nersesyan and A.M. Tsvelik, {\it Bosonization in Strongly Correlated
Systems} (Cambridge University Press, 1999).

\bibitem{golinelli}
O. Golinelli, Th. Jolicoeur and R. Lacaze, Phys. Rev. B {\bf 46}, 10854 (1992).

\bibitem{haas}
S. Haas, J. Riera and E. Dagotto, Phys. Rev. B {\bf 48}, 3281 (1993).

\bibitem{Hagiwara}
M. Hagiwara, Z. Honda, K. Katsumata, A. K. Kolezhuk and H.-J. Mikeska,\\
Phys. Rev. Lett. {\bf 91} 177601 (2003).

\bibitem{hamada}
N. Hamada, S. Sawada and A. Oshiyama, Phys. Rev. Lett. {\bf 68}, 1579 (1992).

\bibitem{Haldane81a}
F.D.M. Haldane, Phys. Rev. Lett. {\bf 47}, 1840 (1981).

\bibitem{Haldane81b}
F.D.M. Haldane, J. Phys. C {\bf 14}, 2585 (1981).

\bibitem{haldane82}
F.D.M. Haldane, Phys. Rev. B {\bf 25}, 4925 (1982).

\bibitem{Haldane82b}
F.D.M. Haldane, J. Phys. A {\bf 15}, 507 (1982).

\bibitem{haldane} 
F.D.M. Haldane, Phys. Lett. A {\bf 93}, 464  (1983).

\bibitem{scalp} 
C. Hayward, D. Poilblanc and D. Scalapino, Phys. Rev. B {\bf 53}, 11721 (1996). 

\bibitem{henderson99}
W. Henderson, V. Vescoli, P. Tran, L. Degiorgi and G. Gr\"uner,
Eur. Phys. J. B {\bf 11}, 365 (1999).
 
\bibitem{HF01}
T. Hikihara and A. Furusaki, Phys. Rev. B {\bf 63}, 134438 (2001).

\bibitem{HF04}
T. Hikihara and A. Furusaki, Phys. Rev. B {\bf 69}, 064427 (2004).

\bibitem{Hiroi91}
Z. Hiroi, M. Azuma, M. Takano, Y. Bando, J. Solid State Chem. {\bf 95}, 230 (1991).

\bibitem{Hiroi95} 
Z. Hiroi and M. Takano, Nature {\bf 377}, 41 (1995).

\bibitem{Hiroi96}
Z. Hiroi, J. Solid State Chem. {\bf 123}, 223 (1996).

\bibitem{honda1} Z. Honda, K. Katsumata, H. Aruga Katori, K. Yamada,
T. Ohishi, T. Manabe and M. Yamashita, J. Phys.: Condensed Matter
{\bf 9}, L83 (1997).

\bibitem{honda2} Z. Honda, K. Katsumata, H. Aruga Katori, K. Yamada,
T. Ohishi, T. Manabe and M. Yamashita, J. Phys.: Condensed Matter
{\bf 9}, 3487 (1997).

\bibitem{honda3}
Z. Honda, H. Asakawa and K. Katsumata,
Phys. Rev. Lett. {\bf 81}, 2566 (1998). 

\bibitem{honda4}
Z. Honda, K. Katsumata, M. Hagiwara and M. Tokunaga, Phys. Rev. B {\bf 60},
9272 (1999).

\bibitem{horton}
M. Horton and I. Affleck, Phys. Rev. B {\bf 60}, 11891 (1999).

\bibitem{hulthen}
L. Hulth\'en, Arkiv Mat. Astron. Fysik {\bf 26A}, No. 11 (1938).

\bibitem{Irkhin00}
V.~Y. Irkhin and A.A. Katanin, Phys. Rev. B {\bf 61}, 6757 (2000).

\bibitem{its90}
A.~R. Its, A.~G. Izergin, V.~E. Korepin, and N.~A. Slavnov,
Int. J. Mod. Phys. \textbf{B4}, 1003 (1990).

\bibitem{JGE} E. Jeckelmann, F. Gebhard and F. H. L. Essler,
Phys. Rev. Lett. {\bf 85}, 3910 (2000). 

\bibitem{eric2} E. Jeckelmann, Phys. Rev. B {\bf 66}, 045114 (2002).

\bibitem{eric} E. Jeckelmann, Phys. Rev. B {\bf 67}, 075106 (2003).

\bibitem{Johnston96} D.C. Johnston, Phys. Rev. B {\bf 54}, 13009 (1996).

\bibitem{Johnston00} D.C. Johnston, M. Troyer, S. Miyahara, D. Lidsky, K. Ueda,
M. Azuma, Z. Hiroi, M. Takano, M. Isobe, Y. Ueda, M.A. Korotin, V.I. Anisimov, A.V. Mahajan
and L.L. Miller, {\it Magnetic Susceptibilities of Spin-1/2 Antiferromagnetic Heisenberg Ladders
and Applications to Ladder Oxide Compounds}, cond-mat/0001147.  See extensive references
therein.

\bibitem{Kane97} 
C. Kane, L. Balents and M. P. A. Fisher, Phys. Rev. Lett. {\bf 79},
5086 (1997). 

\bibitem{Janssen01} J.W. Janssen, S.G. Lemay, M. van den Hout, M. Mooij, L.P. Kouwenhoven 
and C. Dekker,
{\it Scanning Tunneling Spectroscopy on a Carbon Nanotube Buckle},
Conference Proceedings AIP 519, p. 293-297 (Kirchberg, March 2001).

\bibitem{Janssen02} J.W. Janssen, S.G. Lemay, L.P. Kouwenhoven and C. Dekker,
Phys. Rev. B {\bf 65}, 115423 (2002).

\bibitem{Wieg} 
G. E. Japaridze, A. A. Nersesyan and P. Wiegmann, Nucl. Phys. B {\bf
  230}, 511 (1984). 

\bibitem{Jimbo80}
M. Jimbo, T. Miwa, Y. Mori, M. Sato,
Physica \textbf{1D}, 80 (1980).

\bibitem{MiwaJimbo}
M. Jimbo and T. Miwa, {\sl Algebraic Analysis of Solvable Lattice
Models}, American Mathematical Society (1994).

\bibitem{JMC}
J.D. Johnson and B.M. McCoy, Phys. Rev. A {\bf 6}, 1613 (1972).

\bibitem{karbach}
M. Karbach, G. M\"uller, A.H. Bougourzi, A. Fledderjohann and
K.H. M\"utter, Phys. Rev. B {\bf 55}, 12510 (1997).

\bibitem{Karbach02}
M. Karbach, D. Biegel and G. M\"uller, Phys. Rev. B {\bf 66}, 054405
(2002). 

\bibitem{karowski78}
M. Karowski and P. Weisz, Nucl. Phys. B {\bf 139}, 455 (1978).

\bibitem{Kenzelmann01}
M. Kenzelmann, R.A. Cowley, W.J.L. Buyers, R. Coldea, J.S. Gardner,
M. Enderle, D.F. McMorrow, S.M. Bennington, Phys. Rev. Lett. {\bf 87},
017201 (2001).
 
\bibitem{KenzelmannZheludev01}
M. Kenzelmann, A. Zheludev, S. Raymond, E. Ressouche, T. Masuda,
P. B\"oni, K.~Kakurai, I. Tsukada, K. Uchinokura, and R. Coldea 
Phys. Rev. B {\bf 64}, 054422 (2001).

\bibitem{Kenzelmann02a}
M. Kenzelmann, R. A. Cowley, W. J. L. Buyers, Z. Tun, R. Coldea, M. Enderle,
Phys. Rev. B {\bf 66}, 024407 (2002).

\bibitem{Kenzelmann02} M. Kenzelmann, R. Coldea, D.A. Tennant,
D. Visser, M. Hofmann, P. Smeibidl and Z. Tylczynski, Phys. Rev. B {\bf
65}, 144432 (2002).

\bibitem{Kenzelmann}
M. Kenzelmann, Y. Chen, C. Broholm, D.H. Reich and Y. Qiu,
Phys. Rev. Lett. {\bf 93}, 017204 (2004).

\bibitem{kim96}
C. Kim, A.Y. Matsuura, Z.X. Shen, N. Montoyama, H. Eisaki, S. Uchida, 
T. Tohyama and S. Maekawa, Phys. Rev. Lett. {\bf 77}, {4054}(1996).

\bibitem{kit1}
N. Kitanine, J. M. Maillet, N. A. Slavnov, V. Terras,
Nucl.Phys. {\bf B641}, 487 (2002).

\bibitem{kit2}
N. Kitanine, J. M. Maillet, N. A. Slavnov, V. Terras,
J.Phys. {\bf A35}, L385 (2002).

\bibitem{kit3}
N. Kitanine, J. M. Maillet, N. A. Slavnov, V. Terras,
Nucl.Phys. {\bf B642}, 433 (2002).
 
\bibitem{korepin87}
V.E. Korepin,
Comm. Math. Phys. \textbf{113}, 177 (1987).

\bibitem{kuroki}
K. Kuroki and H. Aoki, Phys. Rev. Lett. {\bf 72}, 2947 (1994);

\bibitem{Klumper91}
A. Kl\"umper, Z. Phys. {\bf 91}, 507 (1993).

\bibitem{kobayashi99}
K. Kobayashi {\sl et. al.}, Phys. Rev. Lett. {\bf 82}, 803 (1999).

\bibitem{kohgi}
M. Kohgi, K. Iwasa, J.M. Mignot, B. Fak, P. Gegenwart, M. Lang,
A. Ochiai, H.~Aoki and T. Suzuki, Phys. Rev. Lett. {\bf 86}, 2439 (2001).

\bibitem{Koma87}
T. Koma, Prog. Theor. Phys. {\bf 78}, 1213 (1987).

\bibitem{ising} 
R. M. Konik, A. LeClair and G. Mussardo, Int. J. of Mod. Phys. A {\bf
  11}, 2765 (1996). 

\bibitem{doped} 
R. M. Konik, F. Lesage, A.W.W. Ludwig and H. Saleur,
Phys. Rev. B {\bf 64}, 155112 (2001). 

\bibitem{so8} 
R. M. Konik and A. W. W. Ludwig, Phys. Rev. B {\bf 64}, 155112 (2001).

\bibitem{anis} 
R. M. Konik, H. Saleur and A. W. W. Ludwig, Phys. Rev. B {\bf 66},
075105 (2002). 

\bibitem{konik1} R. M. Konik and P. Fendley, Phys. Rev. B {\bf 66}, 144416 (2002).

\bibitem{konik} 
R. M. Konik, Phys. Rev. B {\bf 68}, 104435 (2003).

\bibitem{kro}
Y. Krotov, D. Lee and S. Louie, Phys. Rev. Lett. {\bf 78}, 4245
(1997)\\{} [cond-mat/9611073].

\bibitem{Korepin79}
V.E. Korepin, Theor. Math. Phys. {\bf 41}, 169 (1979).

\bibitem{vladb}
V.E. Korepin, A.G. Izergin and N.M. Bogoliubov, {\em {Quantum Inverse
  Scattering Method, Correlation Functions and Algebraic Bethe Ansatz}}
  (Cambridge University Press, 1993).

\bibitem{kw}
T.D. K\"uhner and S.R. White, Phys. Rev. {\bf B60}, 335 (1999).

\bibitem{Bella}
B. Lake, D.A. Tennant and S.E. Nagler, Phys. Rev. Lett. {\bf 85}, 
832 (2000).

\bibitem{kulish}
P.P. Kulish, Theor. Math. Phys. {\bf 26}, 132 (1976);

\bibitem{kulisha}
P. Kulish, N. Reshetikhin,
and E. Sklyanin, Lett. Math. Phys. {\bf 5}, 393 (1981).

\bibitem{LaPlaca93} S.J. La Placa, J.F. Bringley, B.A. Scott and D.E. Cox,
Acta Crystallogr. C {\bf 49}, 1415 (1993).

\bibitem{lec} 
A. LeClair, F. Lesage, S. Lukyanov and H. Saleur, Phys. Lett. A {\bf
 235}, 203 (1997). 

\bibitem{leclaira1} A. LeClair, F. Lesage, S. Sachdev and H. Saleur,
Nucl. Phys. B {\bf 482}, 579 (1996).

\bibitem{leclaira} 
A. LeClair and G. Mussardo, Nucl. Phys. B {\bf 552}, 624 (1999).

\bibitem{Levitov03}
L.S. Levitov and A.M. Tsvelik, Phys. Rev. Lett. {\bf 90}, 016401 (2003).

\bibitem{les} 
F. Lesage, H. Saleur and S. Skorik, Nucl. Phys. B {\bf 474}, 602 (1996).

\bibitem{LW}
E.H. Lieb and F.Y. Wu, Phys.Rev. Lett. {\bf 20}, 1445 (1968).

\bibitem{lina}
H.H. Lin, L. Balents and M. Fisher, Phys. Rev. B {\bf 56}, 6569 (1997).

\bibitem{lin} 
H. L. Lin, L. Balents and M. Fisher, Phys. Rev. B {\bf 58}, 1794 (1998).

\bibitem{linb} 
H. L. Lin, Phys. Rev. B {\bf 58}, 4963 (1998).

\bibitem{loss} 
D. Loss and B. Normand, {\it Quantum Antiferromagnets in a Magnetic Field},\\ cond-mat/9804151.

\bibitem{Lou02}
J.Z. Lou, S.J. Qin, C.F. Chen, Z.B. Su and L. Yu,
Phys. Rev. B {\bf 65}, 064420 (2002).

\bibitem{lus} M. L\"uscher, Nucl. Phys. B {\bf 135}, 1 (1978).
The conserved charge here is non-local.  For a discussion
of local conserved charges in the model see \cite{goldschmidt,polyakov}.

\bibitem{lukyanov95}
S. Lukyanov, Comm. Math. Phys. {\bf 167}, 183 (1995);

\bibitem{lukyanov97}
S. Lukyanov, Mod. Phys. Lett. A {\bf 12}, 2911 (1997).

\bibitem{LZ}
S. Lukyanov and A. Zamolodchikov, Nucl. Phys. B {\bf 493}, 571 (1997).

\bibitem{LukyanovXXZ}
S. Lukyanov, Nucl. Phys. B {\bf 522}, 533 (1998).

\bibitem{LukZam01} 
S. Lukyanov and A. B. Zamolodchikov, Nucl. Phys. B {\bf 607}, 437 (2001).

\bibitem{LukTer}
S. Lukyanov and V. Terras, Nucl. Phys. B {\bf 654}, 323 (2003).

\bibitem{Luther75}
A. Luther and I. Peschel, Phys. Rev. B {\bf 12}, 3908 (1975).

\bibitem{Barry68}
B.M. McCoy and T.T. Wu, Il Nuovo Cimento {\bf LVI}, 311 (1968).

\bibitem{ma}
S. Ma, C. Broholm, D. H. Reich, B. J. Sternlieb and R. W. Erwin,
Phys. Rev. Lett. {\bf 69}, 3571 (1992).

\bibitem{ma1}
S. Ma, D. H. Reich, C. Broholm, B. J. Sternlieb and R. W. Erwin,
Phys. Rev. B {\bf 51}, 3289 (1995).

\bibitem{McCarron} E. M. McCarron, M. A. Subramanian, J. C. Calabrese and R. L. Harlow,
Mater. Res. Bull. {\bf 23}, 1429 (1988).

\bibitem{mila}
F. Mila, Eur. Phys. J. B {\bf 6}, 201 (1998).

\bibitem{mintmire}
J. Mintmire, B. Dunlap and C. White, Phys. Rev. Lett. {\bf 68}, 631 (1992).

\bibitem{p123arpes}
T. Mizokawa et. al., Phys. Rev. Lett. {\bf 85}, 4779 (2000).

\bibitem{meshkov}
S. Meshkov, Phys. Rev. B {\bf 48}, 6167 (1993).

\bibitem{moreo}
A. Moreo, Phys. Rev. B {\bf 35}, (1987) 8562.

\bibitem{morra}
R. Morra, W. Buyers, R. Armstrong and K. Hirakawa, Phys. Rev. B {\bf 38}, 543 (1988).

\bibitem{Moriya60}
T. Moriya, Phys. Rev. {\bf 120}, 91 (1960).

\bibitem{Mott} 
N. F. Mott, Proc. Roy. Soc. A {\bf 62}, 416 (1949);
Canad. J. Phys. {\bf 34}, 1356 (1964);\\ Phil. Mag. {\bf 6}, 287
(1961). 

\bibitem{Mottbook}
N. F. Mott, {\it Metal-Insulator Transitions}, $2^{nd}$
ed. (Taylor and Francis, London, 1990).

\bibitem{Mueller81}
G. M\"uller, H. Thomas, H. Beck and J.C. Bonner, Phys. Rev. B {\bf 24},
1429 (1981).

\bibitem{mussardo.school}
G. Mussardo, 
{\em Spectral representation of correlation functions in
two-dimensional quantum field theories}, hep-th/9405128.

\bibitem{mutka} 
H. Mutka, C. Payen, P. Molini\'{e}, J. L. Soubeyroux,
P. Colombet and A.D. Taylor, Phys. Rev. Lett. {\bf 67}, 497 (1991).

\bibitem{nakamura}
M. Nakamura, Phys. Rev. B {\bf 61}, 16377 (2000).

\bibitem{Fukuyama}
T. Nakano and H. Fukuyama, J. Phys. Soc. Jpn {\bf 50}, 2489 (1981).

\bibitem{Nersesyan97}
A.A. Nersesyan and A.M. Tsvelik, Phys. Rev. Lett. {\bf 78}, 3939 (1997).

\bibitem{Nersesyan03} 
A.A. Nersesyan and A.M. Tsvelik, Phys. Rev. B {\bf 68}, 235419 (2003).

\bibitem{noack1}
R. Noack, D. J. Scalapino and S. White, Europhys. Lett. {\bf 30}, 163 (1995).

\bibitem{noack2}
R. Noack, D. J. Scalapino and S. White, Physica C {\bf 270}, 281 (1996).

\bibitem{noack3}
R. Noack, D. J. Scalapino and S. White, Phil. Mag. B {\bf 74}, 485 (1996).

\bibitem{normand}
B. Normand, J. Kyriakidis and D. Loss, Ann. Phys.(Leipzig) {\bf 9}, 133 (2000)\\{}
[cond-mat/9902104].

\bibitem{normanda} 
B. Normand, Acta Phys. Polonica B {\bf 31}, 3005 (2000).

\bibitem{Orignac}
E. Orignac,
{\em Quantitative expression of the spin gap via bosonization for a dimerized  spin-1/2 chain}, cond-mat/0403175.

\bibitem{oa}
M. Oshikawa and I. Affleck, Phys. Rev. Lett. {\bf 78}, 1984 (1997).

\bibitem{Oshikawa99}
M. Oshikawa, K. Ueda, H. Aoki, A. Ochiai and M. Kohgi,
J. Phys. Soc. Jpn. {\bf 68}, 3181 (1999). 

\bibitem{OshikawaAffleck02}
M. Oshikawa and I. Affleck, Phys. Rev. B {\bf 65}, 134410 (2002).

\bibitem{oshima76}
K. Oshima, K. Okuda and M. Date, J. Phys. Soc. Jpn. {\bf 41}, 475
(1976), J. Phys. Soc. Jpn. {\bf 44}, 757 (1978).

\bibitem{oshima78}
K. Oshima, K. Okuda and M. Date, J. Phys. Soc. Jpn. {\bf 44}, 757 (1978).

\bibitem{Parola96}
A. Parola and S. Sorella, Phys. Rev. Lett. {\bf 76}, 4604 (1996).

\bibitem{Parola98}
A. Parola and S. Sorella, Phys. Rev. B {\bf 57}, 6444 (1998).

\bibitem{parkinson}
J.B. Parkinson, J.C. Bonner, Phys. Rev. B {\bf 32}, (1985) 4703.

\bibitem{PencMila}
K. Penc and F. Mila, Phys. Rev. B {\bf 49}, 9670 (1994).

\bibitem{Penc96}
K. Penc, K. Hallberg, F. Mila and H. Shiba, Phys. Rev. Lett. {\bf 77},
1390 (1996). 

\bibitem{schoutens} S. Peysson and K. Schoutens, J. Phys. A {\bf 35}, 6471 (2002).

\bibitem{Pokrovski79}
V. L. Pokrovsky and A. L. Talapov, Phys. Rev. Lett. {\bf 42},
65 (1979).

\bibitem{polyakov}
A. M. Polyakov, Phys. Lett. B {\bf 72}, 224 (1977).

\bibitem{rao}
A. Rao, E. Richter, S. Bandow, B. Chase, P. Eklund, K. Williams, S. Fang,
K.~Subbaswamy, M. Menon, A. Thess, R. Smalley, G. Dresselhaus
and M. Dresselhaus, Science {\bf 275}, 187 (1997);

\bibitem{regnault} L.-P. Regnault, I. Zaliznyak, J. P. Renard and C. Vettier;
Phys. Rev. B {\bf 50}, 9174 (1994).

\bibitem{renard} J.P. Renard, M. Verdaguer, L.P. Regnault,
W.A.C. Erkelens, J. Rossat-Mignod and W.G. Stirling,
Europhys. Lett. {\bf 3}, 945 (1987).

\bibitem{Rice} T. M. Rice, S. Gopalan and M. Sigrist, Europhys. Lett. {\bf 23}, 445 (1993).

\bibitem{damle1}  S. Sachdev and K. Damle, Phys. Rev. Lett. {\bf 78}, 943 (1997).

\bibitem{reply} 
S. Sachdev and K. Damle, J. Phys. Soc. Jpn. {\bf 69}, 2712 (2000).

\bibitem{sagi} 
J. Sagi and I. Affleck, Phys. Rev. B. {\bf 53}, 9188 (1996).

\bibitem{sakai} 
T. Sakai and M. Takahashi, Phys. Rev. B {\bf 43}, 13383 (1991).

\bibitem{Sandvik99}
A.W. Sandvik, Phys. Rev. Lett. {\bf 83}, 3069 (1999).

\bibitem{saito} 
R. Saito, M. Fujita, G. Dresselhaus and M. Dresselhaus, Appl. Phys. Lett. {\bf 60}, 2204 (1992).

\bibitem{salrep} 
H. Saleur, Nucl. Phys. B {\bf 567}, 602 (2000).

\bibitem{satija}
S.K. Satija, J.D. Axe, G. Shirane, H. Yoshizawa and K. Hirakawa,
Phys. Rev. B {\bf 21}, 2001 (1980).

\bibitem{Siegrist}
T. Siegrist, L.F. Schneemeyer, S.A. Sunshine and J.V. Waszczak, Mater. Res. Bull. {\bf 23}, 1429 (1988).

\bibitem{RPA}
D.J. Scalapino, Y. Imry and P. Pincus, Phys. Rev. B {\bf 11}, 2042
(1975).
 
\bibitem{Schulz80}
H.J. Schulz, Phys. Rev. B {\bf 22}, 5274 (1980).

\bibitem{Schulz86}
H.J. Schulz, Phys. Rev. B {\bf 34}, 6372 (1986).

\bibitem{Schulz96}
H.~J. Schulz, Phys. Rev. Lett. {\bf 77}, 2790 (1996).

\bibitem{schulz96a}
H.J. Schulz, Phys. Rev. B {\bf 53}, R 2959 (1996);

\bibitem{Schulz98}
H.J. Schulz,
{\it SO(N) symmetries in the two-chain model of correlated fermions}, cond-mat/9808167.

\bibitem{schwartz98}
A. Schwartz, M. Dressel, G. Gr\"{u}ner, V. Vescoli,
L. Degiorgi, T. Giamarchi, Phys. Rev. B {\bf 58}, 1261 (1998).

\bibitem{senechal00}
D. Senechal, D. Perez and M. Pioro-Ladiere, Phys. Rev. Lett. {\bf 84},
522 (2000).

\bibitem{shelton} D. Shelton, A.A. Nersesyan and A.M. Tsvelik,
Phys. Rev. B {\bf 53}, 8521 (1996). 

\bibitem{sla} R. Slansky, Phys. Rept. {\bf 79}, 1 (1981).

\bibitem{smirnov86b} 
F.A.~Smirnov, J. Phys. A {\bf 19}, L575 (1986).

\bibitem{smir} F. A. Smirnov, Commum. Math. Phys. {\bf 132}, 415 (1990).

\bibitem{smirnov} 
F.A.~Smirnov, {\sl Form Factors in Completely
Integrable Models of Quantum Field Theory} (World Scientific,
Singapore, 1992).

\bibitem{sorensen}
E. Sorensen and I. Affleck, Phys. Rev. B {\bf 49}, 13235 (1994).

\bibitem{starykh} 
O.A. Starykh, D.L. Maslov, W. H\"ausler and L.I. Glazman,
in {\sl Low-Dimensional Systems}, ed. T. Brandes, Lecture Notes in
Physics (Springer, 2000)

\bibitem{steiner}
M. Steiner, K. Kakurai, J. K. Kjems, D.~Petitgrand and R. Pynn,
J. Appl. Phys. {\bf 61} 3953 (1987).

\bibitem{Suzuki85}
M. Suzuki, Phys. Rev. B {\bf 31}, 2957 (1985).

\bibitem{takigawa} M. Takigawa, T. Asano, Y. Ajiro, M. Mekata and Y. Uemura,
Phys. Rev. Lett. {\bf 76}, 2173 (1996).

\bibitem{Taka71}
M. Takahashi, Prog. Theor. Phys. {\bf 46}, 401 (1971).

\bibitem{Taka72}
M. Takahashi and M. Suzuki, Prog. Theor. Phys. {\bf 48}, 2187 (1972).

\bibitem{Taka74}
M. Takahashi, Prog. Theor. Phys. {\bf 50}, 1519 (1974).

\bibitem{Taka90}
M. Takahashi, Phys. Rev. B {\bf 43}, 5788 (1990).

\bibitem{Taka94}
M. Takahashi, Phys. Rev. {\bf B50}, 3045 (1994).

\bibitem{Takahashi} M. Takahashi, Phys. Rev. Lett. {\bf 62}, 2313 (1989).

\bibitem{Takahashia} M. Takahashi and T. Sakai, J. Phys. Soc. of
Japan, {\bf 60}, 760 (1991).

\bibitem{Takahashib} M. Takahashi, Phys. Rev. B {\bf 48}, 311 (1993).

\bibitem{TalstraHaldane96}
J.C. Talstra and F.D.M. Haldane, Phys. Rev. B {\bf 54}, 12594 (1996).

\bibitem{takhtajan} L. Takhtajan, Phys. Lett. A {\bf 87}, 479 (1982).

\bibitem{twoterminal} S. Tans, 
M. Devoret, H. Dai, A. Thess, R. Smalley, L. Geerligs
and C. Dekker,\\  Nature {\bf 386}, 474 (1997).

\bibitem{Tennant95a}
D.A. Tennant, R. Cowley, S.E. Nagler and A.M. Tsvelik,
Phys. Rev. B {\bf 52}, 13368 (1995).

\bibitem{Tennant95b}
D.A. Tennant, S.E. Nagler, S. Welz, G. Shirane and K. Yamada,
Phys. Rev. B {\bf 52}, 13381 (1995).

\bibitem{Thun77}
H.-J.~Thun, T.T.\ Truong and P.H.\ Weisz, Phys.~Lett. B {\bf 67}, 321 
(1977).

\bibitem{totsuka}
K. Totsuka, Phys. Rev. B {\bf 57}, 3454 (1998).

\bibitem{tsvelik} A.M. Tsvelik, Sov. Phys. JETP {\bf 66}, 221 (1987).

\bibitem{tsv3maj} A.M. Tsvelik, Phys. Rev. B {\bf 42}, 10499 (1990).

\bibitem{tun} 
Z. Tun, W. Buyers, R. Armstrong, K. Hirakawa and B. Briat,
Phys. Rev. B {\bf 42}, 4677 (1990).

\bibitem{Venema00}
L.C. Venema, J.W. Janssen, M.R. Buitelaar, J.W.G. Wild\"oer, S.G. Lemay, L.P.~Kouwenhoven and C. Dekker,
Phys. Rev. B {\bf 62}, 5238 (2000).

\bibitem{vescoli98}
V. Vescoli, L. Degiorgi, W. Henderson, G. Gr\"{u}ner, K. P. Starkey and
L. K. Montgomery, {\sl Science} {\bf 281}, 1181 (1998).

\bibitem{voit} 
J. Voit, Eur. Phys. J. B {\bf 5}, 505 (1998). 

\bibitem{Wang00}
 Y.-J. Wang and A.A. Nersesyan, Nucl. Phys. B {\bf 583}, 671 (2000).

\bibitem{wen90}
X.G. Wen, Phys. Rev. B {\bf 42}, 6623 (1990).

\bibitem{white}
S. White, Phys. Rev. Lett. {\bf 69}, 2863 (1992).

\bibitem{whitea}
S. White and D. Huse, Phys. Rev. B {\bf 48}, 3844 (1993).

\bibitem{White94}
S. White, R. Noack and D. Scalapino, Phys. Rev. Lett. {\bf 73}, 886 (1994).

\bibitem{vigman}  
P. B. Wiegmann, Sov. Sci. Rev. Ser. A {\bf 2}, 43 (1980).

\bibitem{wiegmann} 
P. Wiegmann, JETP Lett. {\bf 41}, 95 (1985).

\bibitem{wildoer}
J. Wild\"oer, L. Venema, A. Rinzler, R. Smalley and C. Dekker,
Nature {\bf 391}, 59 (1998).

\bibitem{Wolter03a}
A.U.B. Wolter, P. Wzietek, F.J. Litterst, S. Sullow, D. Jerome,
R. Feyerherm and H.H. Klauss, Polyhedron {\bf 22}, 2273 (2003).

\bibitem{Wolter03b}
A.U.B. Wolter, H. Rakoto, M. Costes, A. Honecker, W. Brenig,
A. Kl\"umper, H.H.~Klauss, F.J. Litterst, R. Feyerherm, D. Jerome and
S. Sullow, Phys. Rev. B {\bf 68}, 220406 (2003).

\bibitem{yamamoto}
S. Yamamoto, Phys. Rev. Lett. {\bf 75}, 3348 (1995).

\bibitem{Yang66a}
C.N. Yang and C.P. Yang, Phys. Rev. {\bf 150}, 321 (1966).

\bibitem{Yang66b}
C.N. Yang and C.P. Yang, Phys. Rev. {\bf 150}, 327 (1966).

\bibitem{Yang66c}
C.N. Yang and C.P. Yang, Phys. Rev. {\bf 151}, 258 (1966).

\bibitem{yurov}
V. P. Yurov and Al. B. Zamolodchikov, Int. J. Mod. Phys. A {\bf 6}, 3419 (1991).

\bibitem{yoshioka}
H. Yoshioka, M. Tsuchizu and Y. Suzumura, J. Phys. Soc. Jpn {\bf 70},
762 (2001).

\bibitem{zacher98}
M.G. Zacher, E. Arrigoni, W. Hanke and J.R. Schrieffer,
Phys. Rev. B {\bf 57}, 6370 (1998).

\bibitem{Igor94}
I. A. Zaliznyak, L.-P. Regnault, and D. Petitgrand, Phys. Rev. B {\bf 50}, 15824 (1994).

\bibitem{Igor98} 
I. A. Zaliznyak, D. C. Dender, C. Broholm and D. H. Reich, Phys. Rev. B {\bf 57}, 5200 (1998).

\bibitem{Igor99}
I. A. Zaliznyak, C. Broholm, M. Kibune, M. Nohara and H. Takagi,
Phys. Rev. Lett. {\bf 83}, 5370 (1999).

\bibitem{igor01} I. A. Zaliznyak, S.-H. Lee and S. V. Petrov,
Phys. Rev. Lett. {\bf 87}, 017202 (2001); ibid. Phys. Rev. Lett. {\bf 91}, 039902
(2003).

\bibitem{Igor04}
I. A. Zaliznyak, H. Woo, T. G. Perring, C. L. Broholm, C. D. Frost
 and H. Takagi, Phys. Rev. Lett {\bf 93}, 087202 (2004).

\bibitem{zam77a}
A. B. Zamolodchikov, Comm. Math. Phys. {\bf 55}, 183 (1977).

\bibitem{Zam77} 
A. B. Zamolodchikov, JETP Lett.~{\bf 25}, 468 (1977).

\bibitem{zam78}
A. B. Zamolodchikov and Al. B. Zamolodchikov, Nucl. Phys. B {\bf 133}, 525 (1978).

\bibitem{zamo} A. B. Zamolodchikov and Al. B. Zamolodchikov, Ann. of
Phys. {\bf 120}, 253 (1979).

\bibitem{zam92}
A. B. Zamolodchikov and Al. B. Zamolodchikov, Nucl. Phys. B {\bf 379}, 602 (1992).

\bibitem{zam}
Al.B. Zamolodchikov, Int. J. Mod. Phys. A {\bf 10}, 1125 (1995).

\bibitem{ZhangSO5} S.C. Zhang, Science {\bf 275}, 1089 (1997).

\bibitem{Zheludev00}
A. Zheludev, M. Kenzelmann, S. Raymond, E. Ressouche, T. Masuda,
K. Kakurai, S.~Maslov, I. Tsukada, K. Uchinokura and A. Wildes, 
Phys. Rev. Lett. {\bf 85}, 4799 (2000).

\bibitem{ZheludevKenzelmann01}
A. Zheludev, M. Kenzelmann, S. Raymond, T. Masuda, K. Uchinokura and \mbox{S.-H.~Lee},
Phys. Rev. B {\bf 65}, 014402 (2002).

\bibitem{Zheludev01}
A. Zheludev, Y. Chen, C. L. Broholm, Z. Honda and K. Katsumata, Phys. Rev. B {\bf 63}, 104410 (2001).

\bibitem{Zheludev01a}
A. Zheludev, Z. Honda, K. Katsumata, R. Feyerherm and K. Prokes,
Europhys. Lett. {\bf 55}, 868 (2001).

\bibitem{Zheludev02}
A. Zheludev, K. Kakurai, T. Masuda, K. Uchinokura and K. Nakajima,
Phys. Rev. Lett. {\bf 89}, 197205 (2002).

\bibitem{Zheludev02a} 
A. Zheludev, Z. Honda, Y. Chen, C. L. Broholm, K. Katsumata and S. M. Shapiro,
Phys. Rev. Lett. {\bf 88}, 077206 (2002).

\bibitem{andrey}
A. Zheludev, S. Raymond, L.-P. Regnault, F. H. L. Essler, K. Kakurai,
T. Masuda and K. Uchinokura, Phys. Rev. B {\bf 67}, 134406 (2003).

\bibitem{andrey1} 
A. Zheludev, S. M. Shapiro, Z. Honda, K. Katsumata, B. Grenier, E. Ressouche, L.-P. Regnault, 
Y. Chen, P. Vorderwisch, H.-J. Mikeska and A. K. Kolezhuk,\\ Phys. Rev. B {\bf 69}, 054414 (2004).

\bibitem{andrey2} 
A. Zheludev, Z. Honda, C. L. Broholm, K. Katsumata, S. M. Shapiro, A. Kolezhuk, S. Park and Y. Qiu,
Phys. Rev. B {\bf 68}, 134438 (2003).
 

\bibitem{zotos1}
X. Zotos and P. Prelovsek, Phys. Rev. B {\bf 53}, 983 (1996). 

\bibitem{zotos2}
X. Zotos, F. Naef and P. Prelovsek, Phys. Rev. B {\bf 55}, 11029 (1997).





\end{thebibliography}
\end{document}